\RequirePackage{fix-cm}
\documentclass[smallextended]{svjour3}       % onecolumn (second format)
\smartqed  % flush right qed marks, e.g. at end of proof
\usepackage{graphicx}
\usepackage{braket}
\usepackage{amsmath, amssymb}
\usepackage{overpic}
\usepackage[table]{xcolor}
\usepackage{hyperref}
\usepackage{bm}
\usepackage{booktabs,multirow}
\usepackage{enumitem}
\usepackage{adjustbox}
\usepackage[normalem]{ulem}
\usepackage{aas_macros}
\usepackage{tikz}
\usetikzlibrary{arrows,positioning,decorations.pathmorphing}
\usepackage{subfig}

\usepackage{acronym}
\newacro{PSD}[PSD]{power spectral density}
\newacro{SNR}[SNR]{signal-to-noise ratio}
\newacro{GW}[GW]{gravitational wave}
\newacro{gw}[GW]{gravitational wave}
\newacro{GR}[GR]{General Relativity}
\newacro{pdf}[PDF]{probability density function}
\newacro{PN}[PN]{post-Newtonian}
\newacro{CBC}[CBC]{compact binary coalescence}
\newacro{NS}[NS]{neutron star}
\newacro{BH}[BH]{black hole}
\newacro{BNS}[BNS]{binary neutron star}
\newacro{BBH}[BBH]{binary black hole}
\newacro{NSBH}[NSBH]{neutron-star--black-hole}
\newacro{IMR}[IMR]{inspiral-merger-ringdown}
\newacro{NR}[NR]{numerical relativity}
\newacro{FT}[FT]{Fourier transform}
\newacro{FAR}[FAR]{false alarm rate}
\newacro{IFAR}[IFAR]{inverse false alarm rate}
\newacro{EM}[EM]{electromagnetic}
\newacro{GRB}[GRB]{gamma-ray burst}
\newacro{CNN}[CNN]{convo\-lutional neural network}
\newacro{PDF}[PDF]{probability density function}
\newacro{MCMC}[MCMC]{Markov-chain Monte Carlo}
\newacro{PE}[PE]{parameter estimation}
\newacro{ROQ}[ROQ]{Reduced Order Quadrature}
\newacro{CDF}[CDF]{cumulative distribution function}

\newcommand{\conj}[1]{\ensuremath{{#1}^{*}}}
\newcommand{\mat}[1]{\ensuremath{\mathrm{\mathbf{#1}}}}
\newcommand{\Mc}{\ensuremath{\mathcal{M}}}
\newcommand{\chieff}{\ensuremath{\chi_\mathrm{eff}}}
\newcommand{\chip}{\ensuremath{\chi_\mathrm{p}}}
\newcommand{\Mtot}{\ensuremath{M}}
\newcommand{\pvec}{\ensuremath{\vec{\theta}}}
\newcommand{\OptimalSNR}{\ensuremath{\rho_\mathrm{opt}}}
\newcommand{\Msun}{\ensuremath{M_\odot}}
\newcommand{\Nlive}{\ensuremath{N_\mathrm{live}}}

\journalname{Living Reviews in Relativity}
\renewcommand{\Re}{\operatorname{Re}}

\renewcommand{\vec}[1]{\ensuremath{\bm{#1}}}
\renewcommand{\d}{\ensuremath{\operatorname{d}\!}}

\begin{document}

\title{Compact binary coalescences: gravitational-wave astronomy with ground-based detectors
}

\titlerunning{Gravitational waves from compact binary coalescences} % if too long for running head

\author{K.~Chatziioannou \and
        T.~Dent        \and
        M.~Fishbach    \and
        F.~Ohme        \and
        M.~P{\"u}rrer \and
        V.~Raymond     \and
        J.~Veitch
}


\institute{K. Chatziioannou \at
                Department of Physics, California Institute of Technology, Pasadena, California 91125, USA
	    \and
	    T. Dent \at
            IGFAE, University of Santiago de Compostela, E-15782, Spain
              %\email{@example.com}           %  \\
            \and
        M. Fishbach \at
        	   Canadian Institute for Theoretical Astrophysics, 
        	   60 St George St, University of Toronto, Toronto, ON M5S 3H8, Canada
            \and
        F. Ohme \at
        Max-Planck-Institut für Gravitationsphysik, Albert-Einstein-Institut, Callinstr.~38, D-30167 Hannover, Germany
        \at Leibniz Universität Hannover, D-30167 Hannover, Germany
            \and
        M.~P{\"u}rrer \at
            Department of Physics and Center for Computational Research,
            University of Rhode Island,
            Kingston, RI 02881, USA
            %and
            %Max Planck Institute for Gravitational Physics (Albert Einstein Institute),
            %Am M\"uhlenberg 1, Potsdam 14476, Germany
            \and
        V. Raymond \at
            Cardiff University, Cardiff CF24 3AA, United Kingdom
        \and
            J. Veitch \at
                Institute for Gravitational Research,
                School of Physics and Astronomy,
                University of Glasgow,
                Glasgow G12 8QQ,
                United Kingdom
}

\date{Received: date / Accepted: date}

\maketitle

\begin{abstract}
The era of gravitational wave astronomy began in 2015 with the observation of the signal from the merger of two black holes by the LIGO detectors; %Since then, 
by 2021, almost 100 more such transient signals from coalescences of compact binaries of black holes and neutron stars were catalogued.
With improvements to the ground-based interferometer network consisting of LIGO, Virgo, and KAGRA 
now promising to bring the total number of detections into the hundreds, we review the observational signatures and analysis methods for the most prolific gravitational-wave source: 
%and the only class of transient signals detected to date: 
the coalescence of compact binaries. 
\keywords{Gravitational Waves \and Compact Binaries \and Data Analysis}
\end{abstract}

\tableofcontents

\newpage

\section*{Definitions and conventions}

Except where other units are specified, we use ``geometric units'' where $c = G = 1$. 
The following symbols and definitions are used throughout: 
\begin{itemize}
    \item $d$, $d(t)$ : detector strain data stream
    \item $n$, $n(t)$ : noise component of data stream
    \item $h$, $h(t)$ : gravitational-wave induced strain on the detector
    \item $\tilde{d}$, $\tilde{d}(f)$ : Fourier transform of $d$, etc.
    \item $h_+, h_\times$ : gravitational-wave plus and cross polarizations
    \item $H \equiv h_+ - ih_\times$ : complex combination of gravitational-wave polarizations
    \item $\tilde{x}(f) \equiv \int x(t) \exp(-i\, 2\pi ft)\, dt$ : forward Fourier transform %(no normalisation)
    \item $x(t) \equiv \int \tilde{x}(f) \exp(i\, 2\pi ft)\, df$ : inverse Fourier transform %(all normalisation)
    \item $\tilde{x}_k \equiv \sum_{j=0}^{N-1} x_j \exp(-i\, 2\pi j k/N)$ : forward discrete Fourier transform %(no normalisation)
    \item $x_j \equiv \frac{1}{N} \sum_{k=0}^{N-1} \tilde{x}_k \exp(i\, 2\pi j k/N)$ : inverse discrete Fourier transform %(all normalisation)
    \item $S_n$ : power spectral density (units Hz$^{-1}$) of noise; we use the one-sided density unless otherwise noted
%    \item $S_n^{(2)}$ : Two-sided power spectral density
    \item $\braket{a|b} \equiv 4 \Re \int_0^\infty \conj{\tilde a}(f)\tilde{b}(f) S_n(f)^{-1} \,df$ : noise-weighted inner product
    \item $P(A|B)$ : probability of a logical proposition $A$ given information $B$
    \item $p(a|B)$ : probability distribution %(PDF)
    over a parameter $a$ given information $B$
    \item $H_N$, $H_S$ : noise hypothesis and signal-plus-noise hypothesis
    \item $\pvec$ : vector of source parameters
    \item $\mat{C}$ : a matrix named C
    \item $m_1$, $m_2$ : primary and secondary component masses, $m_1\!\geq\!m_2$ by convention
    \item $\Mtot \equiv m_1 + m_2$ : total mass
    \item $\eta \equiv m_1m_2/M^2$ : symmetric mass ratio 
    \item $q \equiv m_2/m_1 \leq 1$ : mass ratio
    \item $\Mc \equiv M \eta^{3/5}$ : chirp mass
    \item $M_\odot$ : solar mass
    \item $\vec{S}_1, \vec{S}_2$ : dimensionful component spin vectors
    \item $\vec{\chi}_1 \equiv \vec{S}_1/m_1^2$ : dimensionless spin vector
    \item $\chi \equiv |\vec{\chi}|$ : dimensionless spin magnitude
    \item $\vec{L}$ : Newtonian orbital angular momentum vector
    \item $\cos\theta_1\equiv (\vec{S}_1\cdot \vec{L}) / (|\vec{S}_1||\vec{L}|)$ : spin tilt
    \item $\phi_1 $ : spin azimuthal angle in the $\vec{L}$ frame
    \item $\chi_{\rm eff} \equiv (m_1 \chi_1 \cos\theta_1 + m_2 \chi_2 \cos\theta_2)/M$ : effective aligned spin
    \item $\chi_p \equiv \mathrm{max}\left[\chi_1 \sin\theta_1,
    q(4q+3)/(4+3q)\chi_2 \sin\theta_2\right]$ : effective precessing spin
\end{itemize}

\newpage

\section{Introduction}
\label{sec:intro}

\subsection{(Pre-)history of compact binary GW sources}
\label{ss:history}
Coalescences of compact binaries involving neutron stars and black holes -- their inspiraling evolution due to \ac{GW} emission, and ensuing merger to form a remnant object -- are some of the most extreme astronomical phenomena known.  Each event can convert a few percent of the total binary mass to gravitational radiation, typically a few times the solar mass
for coalescences involving $\sim 30\,M_{\odot}$ black holes~\cite{LIGOScientific:2016wyt}, and the peak \ac{GW} luminosity at merger can reach
$\gtrsim 10^{56}$\,erg/s~\cite{LIGOScientific:2018mvr}.  Coalescences involving neutron stars, in addition to their \ac{GW} signature, can eject debris from the merger site that emits electromagnetic radiation across the spectrum, enabling spectacular multi-messenger observations~\cite{LIGOScientific:2017zic,LIGOScientific:2017ync}. 
However, while the calculation of \ac{GW} emission from gravitationally bound systems was promptly performed within Einstein's General Relativity~\cite{Lorentz1937,Einstein:1918btx,Eddington:1922ds,Einstein:1938yz,Peters:1963ux}, for decades it was thought that the effects would be too weak to ever be of observational interest.  This pessimistic estimate was based partly on considering astrophysical sources known at the time, i.e.\ planets and stars, and partly on the extremely weak expected effects, of order $10^{-20}$ or less even for optimistic assumptions. 

The order of magnitude of the \ac{GW} strain $h$ from a gravitationally bound binary of masses $m_1$, $m_2$ with orbital radius $r$, at distance $D$ from Earth, is 
%\kc{dimensional analysis cannot determine the factors of $2$, it is not clear where this equation comes from}
% TD : this is a simple rearrangement of eq. 3.332 from Maggiore using the Kepler formula for orbital frequency, the 4 is already there but in any case either we can have factors of 2 in the first expression or the second
\begin{equation}
 |h| \sim \frac{1}{r}\frac{1}{D} \frac{2Gm_1}{c^2} \frac{2Gm_2}{c^2} \equiv \frac{R_{S,1}R_{S,2}}{rD}\,,
\end{equation}
where $R_{S,\ast}$ denotes the component Schwarzschild radius.  
% Same notation seems to occur in the signals QNM discussion
For planetary or binary star systems at realistic distances, the strain is thus vanishingly small.  Furthermore, a bound system with total mass $M$ and size $R$ has a maximum dynamical frequency $\omega_d \sim \sqrt{(GM/R^3)} \sim (G\rho)^{1/2}$, where $\rho$ is the system's average density, and we expect the frequencies of \ac{GW} emission to be comparable.  Considering ``non-compact'' objects, this dynamical frequency may attain $\sim\!10^{-3}\,\mathrm{Hz}$ for rocky planets, a range impractical for terrestrial detector operation due to overwhelming seismic disturbances.  The discovery of white dwarfs -- coincidentally close in time to the formulation of General Relativity~\cite{Adams_1914} -- provided a source whose dynamical frequency could reach the decihertz to Hz range; indeed, white dwarf binaries are a major source for the future space-based detector LISA~\cite{LISA:2017pwj}. 
Only with the discovery of Galactic \acp{NS}~\cite{Baade:1934wuu,Hewish:1968bj}, as both isolated rotating (non-axisymmetric) sources and as binaries with emission frequencies up to $\sim\!10^3\,$Hz, was a credible \ac{GW} source for Earth-based detectors known to exist~\cite{Dyson_62,PhysRevLett.18.1071,Schutz:1989yw}. 
%(besides unbound systems such as core collapse supernova explosions)

A major impetus was given to the investigation of compact binaries by the discovery of a double \ac{NS} binary~\cite{Hulse:1974eb} and subsequent confirmation that its orbit decays as expected due to \ac{GW} emission, including \ac{PN} corrections to the lowest order of \ac{GW} emission~\cite{Taylor:1982zz}.  While known Galactic \acp{BNS} are very far from the point of merger, their existence puts a lower limit on the rate of detectable coalescences in the local universe~\cite{Phinney:1991ei,Kalogera:2001dz}.  Furthermore, the mass- and spin-dependence of \ac{PN} emission contributions carry nontrivial information about the properties of the source binary~\cite{Cutler:1992tc}.  This combination of detectability and science potential confirmed \ac{BNS} as a prime source for the first generation of interferometric detectors.  Binaries of \acp{BH} or mixed \ac{NSBH} binaries were also considered as possible sources, but with much more uncertain merger rates~\cite{LIGOScientific:2010nhs}.

\subsection{The Era of Interferometric Detectors}
\label{ss:detectors}

While resonant bar \ac{GW} detectors are sensitive at specific frequencies (see~\cite{Aguiar:2010kn} for a review), a broad-band antenna is necessary both to detect and characterize chirping \ac{CBC} signals~\cite{Abramovici:1992ah}.  The initial detector network of LIGO, GEO 600 and Virgo, as essentially Michelson interferometers with many enhancements that are beyond the scope of this review, realized a first technical sensitivity goal in their observations up to 2007~\cite{LIGOScientific:2007fwp,Grote:2010zz,VIRGO:2012dcp}.  
\ac{CBC} searches covering Initial data up to 2010 obtained null results~\cite{LIGOScientific:2011jth,LIGOScientific:2012vij}, as expected under all but the most optimistic rate predictions.  At the same time, a revolution in numerical methods for solving the Einstein equations for black-hole spacetimes enabled the full merger-ringdown \ac{GW} emission from \ac{BBH} systems to be investigated~\cite{Pretorius:2005gq,Campanelli:2005dd,Baker:2005vv} 
%Sperrhake 2014 review?] 
and combined with previous detailed knowledge of the \ac{PN} inspiral~\cite{Buonanno:1998gg,Buonanno:2000ef,Ajith:2007kx}; thus, in addition to searches for \ac{CBC} inspirals, higher-mass \ac{BBH} mergers could also be targeted~\cite{LIGOScientific:2011hqo}. 

\begin{figure}[tbp]
 \hspace{0.09\textwidth}
 \includegraphics[width=0.72\textwidth]{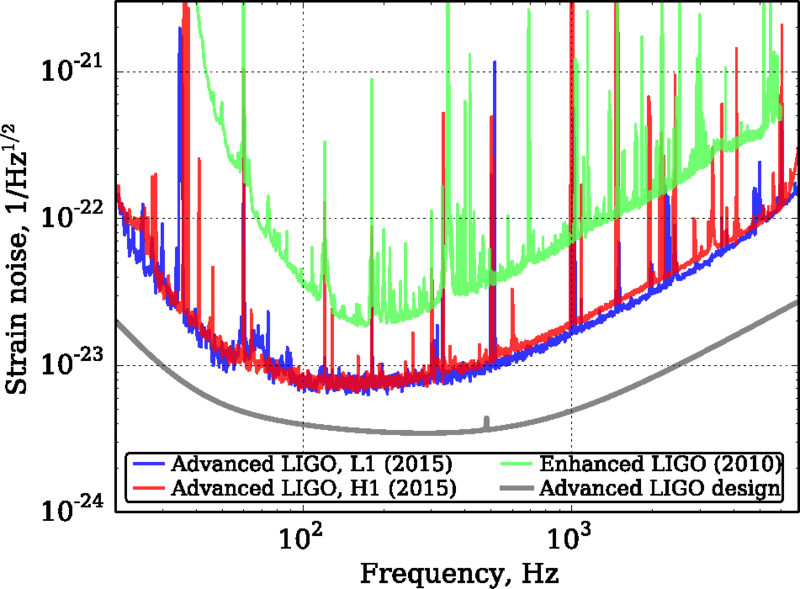}
 \caption{
 The strain sensitivity for the LIGO Livingston detector (L1) and the LIGO Hanford detector (H1) during the first observing run, O1. Also shown is the noise level for the Advanced LIGO design (gray curve) and the sensitivity during the final data collection run (S6) of the initial detectors.  Figure reproduced from~\cite{Abbott:2016xvh}. 
 %Typical power spectral densities for existing and upcoming detectors. Solid lines correspond to measured spectral from LIGO Hanford (LHO), LIGO Livingston (LLO), and Virgo during previous observing runs. Dashed lines correspond to proposed future detectors. \kc{find better PSDs that are not truncated}
}
 \label{fig:PSD}
\end{figure}
With the upgraded Advanced detectors coming online from 2015 onward~\cite{LIGOScientific:2014pky,VIRGO:2014yos}, improvements both in the response to \ac{GW}, and in suppression of noise over the full $\sim$\,$10$--$1000$\,Hz band~\cite{Abbott:2016xvh}, as shown in Fig.~\ref{fig:PSD}, were immediately rewarded with the discovery of a merging stellar-mass \ac{BBH} via its \ac{GW} signal, GW150914~\cite{LIGOScientific:2016aoc}.  Not only was this system the first of its kind observed, it also contained stellar-mass \ac{BH}s more massive than any confirmed via electromagnetic emission~\cite{LIGOScientific:2016vpg}.  Within two years, the first joint detection of a \ac{BNS} coalescence
through both \ac{GW}~\cite{LIGOScientific:2017vwq} and electromagnetic emission confirmed the exceptionally broad science potential of multimessenger astronomy with \ac{GW}s: yielding, among other results, confirmation that \ac{BNS} mergers source short \acp{GRB} and kilonovae, and that \ac{GW} propagate at the speed of light~\cite{LIGOScientific:2017zic,LIGOScientific:2017ync}. 
These overnight astrophysical revolutions were enabled by advances both in our understanding of \ac{GW} emission and in analysis methods, as we will review in detail. 

After the gamut of stellar-mass compact binary types was completed with the discovery of mixed \ac{NS}-\ac{BH} binaries~\cite{LIGOScientific:2021qlt} thanks again to increased detector sensitivities~\cite{aLIGO:2020wna}, with the Advanced era O4 run now underway, and as the community is designing future detectors to reach compact binary sources at the edge of the observable universe, this review revisits the basics of \ac{CBC}s and of current data analysis techniques.  We discuss the theory behind the signals we observe in 
Sec.~\ref{sec:cbc}; then in Sec.~\ref{sec:structure} we describe how we can use GW signals recorded in detector data to make statistical statements on the presence and properties of possible sources.  Based on this framework, we consider methods for \ac{CBC} signal detection in Sec.~\ref{sec:detection}, and parameter estimation in Sec.~\ref{sec:pe}. 
In Sec.~\ref{sec:combine} we generalize from one signal to the catalog of all detections, and describe how large sets of events can be analyzed while taking into account their uncertainties and the instrument selection effects. We conclude in Sec.~\ref{sec:outlook} with an outlook over upcoming \ac{CBC} observations in the near and medium term and discussion of the associated challenges for data analysis.

\newpage
\section{Compact binary signals}
\label{sec:cbc}

%\emph{Responsible:} Frank, Michael

Data analysis techniques for compact binaries consisting of \acp{BH} and \acp{NS} require 
detailed knowledge of the \ac{GW} signal. In principle this can be achieved by calculating
the orbital dynamics and waveform by solving Einstein's equations for a source binary.
This section starts out by providing an overview of the generation and emission of \acp{GW} 
under Newtonian and post-Newtonian assumptions in Secs.~\ref{sec:Newtonian_binary} and ~\ref{ss:PN},
including tidal effects and precession, as well as relevant cosmological effects on \ac{GW} propagation; we
then discuss salient characteristics of \ac{GW} signals in terms of the parameters describing compact
binary sources and the effect of these parameters on the \ac{GW} strain in Sec.~\ref{ss:signal_characteristics}.

The coalescence of compact binaries and the emitted \acp{GW} consist of several stages: (i) inspiral, where 
the orbital velocity is much smaller than the speed of light, and which is well described by post-Newtonian 
theory, (ii) the highly non-linear merger which requires full numerical solutions of Einstein's equations 
(see Sec.~\ref{ss:NR}), and, (iii), for \acp{BBH}, the ringdown stage which is governed by
quasi-normal modes emitted from a perturbed final \ac{BH} (see Sec.~\ref{ss:QNMs}). 
Modern waveform models may rely on semi-analytical techniques to model the complete \ac{GW} signal from these 
stages, as exemplified by the effective-one-body and phenomenological modeling frameworks discussed in
Sec.~\ref{sec:LEOBandPhenom}, or follow a full data-driven approach as used by surrogate models summarized in 
Sec.~\ref{sec:surrogate}. We close out with a comparison of the accuracy and efficiency of waveform models in Sec.~\ref{sub:comparison_of_waveform_accuracy_and_efficiency}.

\subsection{Emission of compact binaries: the quadrupole formula} \label{sec:Newtonian_binary}

%Compact binaries are the prime source of \ac{gw} radiation. 
We introduce the basic structure of the \ac{gw} signal from compact binaries
%, we %follow the approach that is taken by many textbooks introducing General Relativity. We 
by considering a weak perturbation of Minkowski space:
\begin{align}
 g_{\alpha \beta} = \eta_{\alpha \beta} + h_{\alpha \beta}\,,
\end{align}
where $\eta_{\alpha \beta} = \operatorname{diag}(-1, 1, 1, 1)$ is the Minkowski metric in Cartesian coordinates, and $\vert h_{\alpha \beta} \vert \ll 1$ is a small perturbation.\footnote{In the following we assume geometric units in which the gravitational constant and the speed of light are set to one, $G = c = 1$. In these units, lengths and times are measured in units of the total mass of the spacetime $M$. Conversion to SI units is discussed at the end of this section.} 
This perturbation tensor will represent \acp{GW} on a Minkowski background. The
trace-reversed perturbation tensor
\begin{align}
 \bar h_{\alpha \beta} \equiv h_{\alpha \beta} - \frac{1}{2} \eta_{\alpha \beta}{h_{\sigma}}^\sigma\,,
\end{align}
satisfies a simple wave equation if an appropriate coordinate system (gauge) is
chosen in which $\partial_\beta \bar{h}_{\alpha}^{\phantom{\alpha}\beta} = 0$.
This choice is commonly referred to as Einstein-de Donder gauge, or Lorenz gauge.\footnote{Named for Ludvig Lorenz in analogy to the gauge choice in electrodynamics.} With this choice of coordinates, the Einstein equation expanded to first
order in $\bar h_{\alpha \beta}$ becomes a flat-space wave equation,
\begin{align}
 \Box \bar h_{\alpha \beta} \equiv \left( - \frac{\partial^2}{\partial t^2} + \frac{\partial^2}{\partial x^2} + \frac{\partial^2}{\partial y^2} + \frac{\partial^2}{\partial z^2} \right) \bar h_{\alpha \beta} = -16 \pi T_{\alpha \beta}\,, \label{eq:GR_wave_eq}
\end{align}
with the stress-energy tensor $T_{\alpha \beta}$ as source term;
see, e.g.,~\cite{hartle2013gravity,schutz2009first} for a more detailed
derivation and discussion. The solution to Eq.~\eqref{eq:GR_wave_eq} for the spatial components of the metric perturbation can be
approximated far away from the source as
\begin{align}
 \bar h_{ij} (t, d) = \frac{2}{d} \frac{d^2}{dt^2} \int_{\mathcal S} x^i
x^j\rho(t - d, \vec x) \, d^3\vec x\,. \label{eq:quadrupole_wave}
\end{align}
Here %$t$ is the (flat-space) time coordinate, 
$\vec x$ are the spatial coordinates, $d$ is the distance between the observer
and the source, $\rho$ is the mass density, and the integral covers a
three-dimensional space $\mathcal S$ that fully contains the source. The time components of $\bar h_{\alpha \beta}$ can be set to zero in an appropriate gauge, as we discuss next.

The Einstein-de Donder gauge does not define a unique coordinate system.
The additional gauge freedom can be used to impose further constraints on $\bar
h_{\alpha \beta}$. A common choice is the transverse-traceless gauge, which can
only be imposed in vacuum away from the source. In this gauge,
all time components $\bar h_{0\beta}$ and the trace ${\bar{h}_{\alpha}}^{~\alpha}$
vanish. Therefore, if one has transformed to this gauge (a discussion on how to do that follows below), Eq.~(\ref{eq:quadrupole_wave}) specifies the complete metric perturbation and we can ignore the time components. In addition, we can drop the
distinction between $h_{\alpha \beta}$ and its trace-reversed counterpart $\bar
h_{\alpha \beta}$.

Equation~\eqref{eq:quadrupole_wave}, often referred to as the quadrupole formula of \ac{gw} emission, states that the \ac{gw} tensor $h_{ij}$ originates from the second time derivative of the mass quadrupole 
\begin{align}
 I^{ij}(t) &= \int_{\mathcal S} x^i x^j \rho(t, \vec x) \, d^3\vec x\,. 
\end{align}

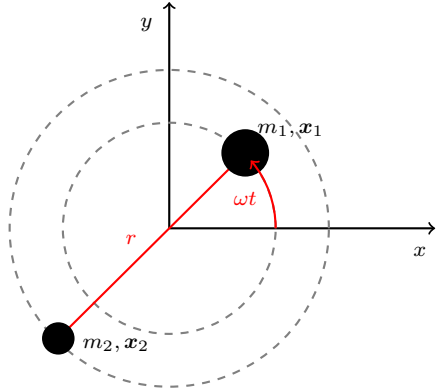
\begin{figure}[t]
 \centering
\begin{tikzpicture}[thick]
 \draw[->] (0,0) -- (3.5, 0);
 \draw[->] (0,0) -- (0, 3);
 \draw[dashed, gray] (0,0) circle (1.4cm);
 \draw[dashed, gray] (0,0) circle (2.1cm);
 \draw[red] (1, 1) -- (-1.46, -1.46);
 \draw[fill=black] (1, 1) circle (0.3cm);
 \draw[fill=black] (-1.46, -1.46) circle (0.2cm);
 \draw[red, ->] (1.4, 0) arc (0:40:1.4);
 \node at (3.3, -0.3) {$x$};
 \node at (-0.3, 2.7) {$y$};
 \node[red] at (-0.5, -0.15) {$r$};
  \node[red] at (1., 0.4) {$\omega t$};
   \node at (1.6, 1.3) {$m_1, \vec x_1$};
   \node at (-0.7, -1.55) {$m_2, \vec x_2$};
\end{tikzpicture}
 \caption{Illustration of a binary in an idealized circular orbit of radius $r$.
The two orbiting (point) masses are defined by their mass $m_i$ and coordinates
$\vec x_i$.\label{fig:fig_binary}}
\end{figure}

We now consider a binary system in a circular orbit in the $x$-$y$ plane, see
Fig.~\ref{fig:fig_binary}. Due to energy being carried away by the emitted
gravitational radiation, the binary orbit cannot remain perfectly closed and
circular; but considered as an idealized example, the quasicircular case serves
as a good starting point.
The system is characterized by the two masses $m_1\geq m_2$, which correspond to
the gravitational (rest) masses of the individual binary components in
isolation. As we consider widely separated orbits, the size of the objects is
negligible. For binary components orbiting with a separation $r$ and an orbital
frequency $\omega$, the location of each object in the center-of-mass coordinate
system is
\begin{align}
 \vec x_1 &= \frac{m_2 r}{M}\left( \cos ( \omega t), \sin(\omega t), 0 \right),\qquad
 \vec x_2 = - \frac{m_1}{m_2} \vec x_1\,, \label{eq:binary_coordinates}
\end{align}
where $M=m_1 + m_2$ is the total mass.
With the simplifying assumption of point masses, %simplify the problem to essentially two point masses, 
the mass density is %approximated as %via $\delta$-functions,
\begin{align}
 \rho(t, \vec x) &= m_1\delta\left[\vec x - \vec x_1(t)\right] +  m_2\delta\left[\vec x - \vec x_2(t)\right]\,.
\end{align}
The mass quadrupole moment and its second derivative are then
\begin{gather}
 I^{ij}(t) = m_1 x_1^i(t) x_1^j(t) + m_2 x_2^i(t) x_2^j(t)\,, \\
%  \begin{split}
 \frac{d^2 I^{ij}}{dt^2} \equiv \ddot I^{ij} = m_1 \left( \ddot x_1^i x_1^j + x_1^i \ddot x_1^j + 2 \dot x_1^i \dot x_1^j \right) + \left( 1 \leftrightarrow 2 \right)\,. \label{eq:quadmom_2nd_der}
%  \end{split}
\end{gather}
Using the explicit coordinates of the binary given in Eq.~\eqref{eq:binary_coordinates}, one can calculate the \ac{gw} tensor from Eq.~\eqref{eq:quadrupole_wave} as
\begin{align}
 h_{ij}(t, d) &= -\frac{4M \omega^2 r^2 \eta}{d}\left( \begin{array}{ccc}
                   \cos(2\omega t') & \sin(2\omega t') & 0 \\
                   \sin(2\omega t') & -\cos(2\omega t')  & 0 \\
                   0 & 0 & 0
                  \end{array} \right)\,, \label{eq:hij_binary}
\end{align}
where we use $t'$ to denote the retarded time,
\begin{align}
 t' &= t - d\,,
\end{align}
and the symmetric mass ratio, $\eta$, is defined by
\begin{align}
\eta &= \frac{m_1 \, m_2}{M^2}\,.
\end{align}

The \ac{GW} tensor in the transverse-traceless gauge only has two independent
components. In our example, we oriented the coordinate system such that the
binary orbits in the $x$-$y$ plane. This leads to the structure of $h_{ij}$
that can be observed in Eq.~(\ref{eq:hij_binary}). The tensor is traceless, because $h_{xx} = -h_{yy}$. For waves traveling along the $z$-direction, the ``transverse'' condition of the transverse-traceless gauge is also satisfied as only spatial components perpendicular to the direction of propagation contain
non-zero values.

Because $h_{ij}$ is a metric perturbation, it has to be
symmetric, i.e., $h_{xy} = h_{xy}$. The two independent components are
identified as the two polarizations of the \ac{GW}, $h_{xx} = h_+$, $h_{xy} =
h_\times$. In this idealised case of a circular binary, the polarizations are
harmonic oscillation whose frequency is twice that of the orbital motion; $h_+$
and $h_\times$ differ only by a phase offset of $\pi/2$.

An observer (e.g. a \ac{gw} detector such as a LIGO or Virgo)
will not necessarily be located along the $z$-axis (i.e., exactly above or below
the orbital plane). For more general scenarios, one has to
project $h_{ij}$ to the two-dimensional subspace perpendicular to the
direction of propagation to remain in the appropriate transverse-traceless
gauge. If $\vec{\hat{N}}$ is the unit vector from the source to the detector, then we
can find the projection operator $P_{kl}$ and the appropriately transformed
\ac{GW} perturbation in the transverse-traceless gauge as
\begin{align}
    P_{kl} &= \mathbb{I}_{kl} - \hat{N}_k \, \hat{N}_l, \\
 h'_{kl} &= {P_k}^i \, {P_l}^j \, h_{ij} - \frac{1}{2} P_{kl} \, P^{ij} \,
h_{ij}. \label{eq:projected_hlm}
\end{align}
Here, $\mathbb{I}$ denotes the three-dimensional identity matrix. The first
term in Eq.~(\ref{eq:projected_hlm}) projects the components of $h_{ij}$ to the
subspace perpendicular to $\vec\hat{N}$. The second term removes the trace.

The two polarizations can be read off the transformed \ac{GW} perturbation
tensor. An L-shaped interferometer will only be sensitive to a linear
combination of the two polarisations, depending on its orientation. We denote
this observable signal by $h$ and provide more details in
Sec.~\ref{ss:detector_response}. For the case of a binary signal in the form of
Eq.~(\ref{eq:hij_binary}), any linear combination of both polarizations can be
expressed by introducing an amplitude factor $\mathcal A$ and a phase offset $\phi_0$ that each depend on the
geometry of the source and the detector. Using the relative velocity $v = \omega r$ we may then
express the \ac{gw} strain due to stationary circular binary motion as
\begin{align}
 h(t, \vec \theta) = -\mathcal A \frac{ 4M v^2 \eta}{d} \cos\left( 2 \omega t +
\phi_0 \right)\,. \label{eq:h_circular}
\end{align}
Here, $\vec \theta = \{ M, v, \eta, d, \omega, \phi_0, \mathcal A \}$ is the
vector of parameters that characterizes the orbital motion and orientation of
the source.

Not all of those parameters are independent. According to Kepler's third law,
%which is based on the balance between (Newtonian) gravitational and centripetal forces, 
the total mass, velocity and orbital frequency of a system in a stable, circular orbit are related by
\begin{align}
 v^2 &= \frac{M}{r} \quad \Rightarrow \quad v^3 = M \omega\,. \label{eq:Keplers_law}
\end{align}
This relation will enable us to heuristically estimate the rate of change of the binary
velocity and separation.  Although we started by assuming
a stable, perfectly circular orbit, the \ac{gw} signal will carry energy
away from the system. The \ac{gw} flux is proportional
to the square of the time derivative of \ac{GW} tensor in the TT gauge, $\dot h^{ij} \, \dot h_{ij}$. The luminosity, $\mathcal L$, is the flux averaged over an orbit and integrated over a sphere of radius $d$. For our binary system, it is proportional to
\begin{align}
 \mathcal L \propto \vert d \, \dot h \vert^2 \propto M^2 v^4 \eta^2 \omega^2 = \eta^2 \, v^{10}\,. \label{eq:luminosity}
\end{align}
Deriving a general expression for the energy carried by a \ac{gw} is beyond the
scope of this introduction. It can be found in many excellent textbooks, e.g., 
\cite{carroll2019spacetime,hartle2013gravity}: the result reads
\begin{align}
 \mathcal L = \frac{32}{5} \eta^2 v^{10}\,. \label{eq:luminosity_exact}
\end{align}

The orbital energy of the binary system can
be written using Kepler's third law as
\begin{align}
 E &= - \frac{M \eta}{2} v^2\,. \label{eq:binary_energy}
\end{align}
Given the energy loss of the system due to \ac{GW} luminosity, Eq.~\eqref{eq:luminosity_exact}, and taking the mass
parameters $M$ and $\eta$ to be constant, one can use energy balance to calculate how the velocity %$v = v(t)$ 
is changing over time:
%The mass parameters $M$ and $\eta$ are assumed to be constant: the velocity evolution is then
\begin{align}
 \frac{dE(t)}{dt} &= - \mathcal L(t)\,, \\
 \Rightarrow ~~ \frac{dv(t)}{dt} &= - \frac{\mathcal L(t)}{dE(v) / dv} = \frac{32}{5M} \eta v^9(t) \\
 \Rightarrow ~~ v(t) &= \frac{1}{2} \sqrt[8]{\frac{5M}{\eta (t_c - t)}}\,.
\end{align}
%The result of this simplified calculation deserves a closer inspection. 
Thus, the
relative velocity of the binary increases over time as $(t_c - t)^{-1/8}$,
%power law, 
apparently diverging at the so-called coalescence time $t_c$,
although the simplifying assumptions made here are not valid
all the way up to the time of the binary's merger.
Following Eq.~\eqref{eq:Keplers_law}, the separation is
\begin{align}
 r(t) = 4M \sqrt[4]{\frac{\eta (t_c - t)}{5M}}\,.
\end{align}
Thus, the approximate time to coalescence for a binary with an initial separation of $r_0$ is
\begin{align}
 T = \frac{5 r_0^4}{256 M^3 \eta}\,. \label{eq:time_to_coalescence}
\end{align}
Finally, the orbital frequency, using Eq.~\eqref{eq:Keplers_law} again, behaves as
\begin{align}
 \omega(t) = \frac{v^3(t)}{M} = \frac{1}{8} \left(\frac{5}{M^{5/3} \eta 
  (t_c - t)}\right)^{3/8} 
  = \frac{1}{8 \Mc^{5/8}} \left(\frac{5}{t_c - t}\right)^{3/8}\,, \label{eq:orbital_freq_Newt}
\end{align}
where we have introduced the \emph{chirp mass} 
\begin{align}
 \Mc = M\eta^{3/5} = \frac{(m_1 \, m_2)^{3/5}}{(m_1 + m_2)^{1/5}}\,.
\label{eq:chirp_mass}
\end{align}
With an explicit expression for the orbital frequency, we can 
also calculate the orbital phase
\begin{align}
 \phi(t) &= \int \omega(t) \, dt = \phi_c - \left(\frac{t_c - t}{5 \Mc} \right)^{5/8} \,.
\end{align}
Here, we label the integration constant $\phi_c$ as it turns out to be the orbital phase at the divergence time $t_c$.

\begin{figure}[t]
 \includegraphics[width=\textwidth]{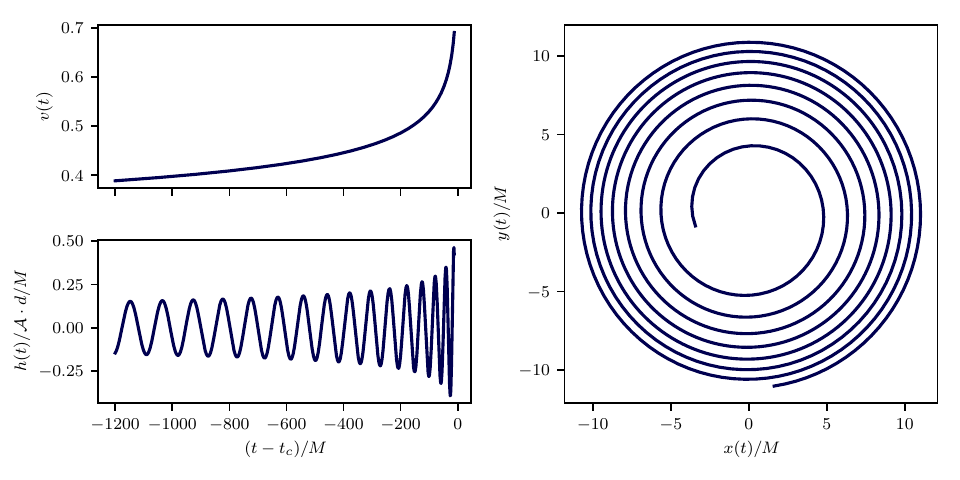}
 \caption{The evolution of the relative velocity $v(t)$, the (re-scaled) \ac{gw}
signal $h(t)$ and the inspiral track of a Newtonian binary in quasi-circular
orbits that evolve only due to the energy carried away by the \ac{gw}.}
 \label{fig:Newtonian_orbit}
\end{figure}

This is a remarkable result: the inspiral of a binary under the emission of
\acp{gw} is predominantly governed by one combination of the
component masses, the chirp mass $\Mc$. %defined in Eq.~\eqref{eq:chirp_mass}.
The functions $v(t)$, $h(t)$, as well as the binary separation
vector $(r \cos \phi, r \sin \phi)$ in the plane of the orbit
are shown in Fig.~\ref{fig:Newtonian_orbit}.

%We stress that 
The total mass $M$ acts as a trivial scale factor in the binary's dynamics and
GW signal. The signal amplitude scales with $M / d$, and the time, frequency,
and separation (at fixed velocity) are proportional to $M$. This will have
important consequences when considering binary sources at cosmological distances,
as we will discuss in Sec.~\ref{sss:redshift}.
The values in units of $M$ shown %on the axes 
in Fig.~\ref{fig:Newtonian_orbit} may be converted to physical (SI) units via % one has to employ 
%the simple transformations
\begin{align}
 t - t_c &= \left(\frac{t-t_c}{M}\right) \times M\frac{G}{c^3} \approx \left(\frac{t-t_c}{M} \right) \times 4.9\,\textrm{ms}\frac{M}{M_\odot}\,, \\
 x^i(t) &= \left( \frac{x^i(t)}{M} \right) \times M\frac{G}{c^2} \approx \left( \frac{x^i(t)}{M} \right) \times 1.5 \, \textrm{km} \frac{M}{M_\odot}\,,
\end{align}
explicitly reintroducing the gravitational constant $G$ and speed of light $c$.

\subsection{Post-Newtonian signal calculation}
\label{ss:PN}

The calculation presented above 
%in Sec.~\ref{sec:Newtonian_binary} 
%although (in some parts overly) simplistic, is nevertheless useful to 
illustrates the basic behaviour of a compact binary inspiral under
the emission of \acp{gw}.  Although General Relativity is a complex,
non-linear theory, for which even approximate solutions require much
careful work, the strategy used in Sec.~\ref{sec:Newtonian_binary}
to calculate the \ac{gw} emission of a binary can be
generalized to include non-linear effects by means of a series expansion.

%In particular, 
Considering energy and luminosity as functions of the 
binary's source parameters $\vec{\theta} = \{M, \eta, \ldots\}$ and of the
orbital velocity $v$, the \emph{\acf{PN}} framework may be applicable,\footnote{For
an introductory review of \ac{PN} theory see Ref.~\cite{Blanchet:2019zlt}
and Ref.~\cite{Blanchet:2013haa,Schafer:2018jfw} for a detailed discussion.} allowing us 
to express these quantities as a series in $v$ (assumed much smaller than
the speed of light), thus:
\begin{align}
 E(v, \vec \theta) \approx -\frac{M \eta}{2} v^2 \left( 1 + \sum_{k=2}^n
E_k(\vec \theta) v^{k} + \ln v \sum_{k=8}^n E_k^{(l)}(\vec \theta) v^{k} \right)\,, \label{eq:PNenergy} \\
 \mathcal L (v, \vec \theta) \approx \frac{32}{5} \eta^2 v^{10} \left( 1 +
\sum_{k=2}^n \mathcal L_k(\vec \theta) v^{k} + \ln v \sum_{k=6}^n {\mathcal L}_k^{(l)}(\vec \theta) v^{k} \right)\,. \label{eq:PNflux}
\end{align}
%\fo{TODO: Include log terms.}
% Blanchet, https://arxiv.org/pdf/1310.1528.pdf
% energy: Eq (233), p. 92
% flux: Eq (314), p. 119
The leading-order term in those expressions is often referred to as the
\emph{Newtonian order} term, while each additional contribution is known as the
$k/2$\,\ac{PN} order. In General Relativity, the 0.5\ac{PN} order term (i.e., $k=1$)
vanishes, and higher-order corrections include specific physical effects that
may not be present at lower orders.
The energy also contains logarithmic terms $\propto \ln v$ at 4\ac{PN}\ order
and beyond, and starting from 3\ac{PN}\ order in the flux.

\begin{table}[tb]
\centering
 \begin{tabular}{lccccccccccccc}
  \toprule
 \ac{PN}\ & 0 & 0.5 & 1 & 1.5 & 2 & 2.5 & 3 & 3.5 & 4 & 4.5 & 5 & 5.5 & 6\\
 $k$ & 0 & 1 & 2 & 3 & 4 & 5 & 6 & 7 & 8 & 9 & 10 & 11 & 12 \\
 \midrule
 energy &\checkmark & \checkmark & \checkmark & \checkmark & \checkmark &
\checkmark & \checkmark &
\checkmark & \checkmark & $\times$ & $\times$ & $\times$ & $\times$   \\
 flux & \checkmark & \checkmark & \checkmark & \checkmark & \checkmark &
\checkmark & \checkmark &
\checkmark & $\times$ & $\times$ & $\times$ & $\times$ & $\times$  \\[6pt]
 chirp mass & \cellcolor{gray!50!white} & & 
\multicolumn{11}{c}{\cellcolor{gray!50!white}}\\
 mass ratio &   &   & \multicolumn{11}{c}{\cellcolor{gray!50!white}} \\
 spin-orbit & & && \cellcolor{gray!50!white} & & 
\multicolumn{8}{c}{\cellcolor{gray!50!white}} \\
 spin-spin & & & && \cellcolor{gray!50!white} & & 
\multicolumn{7}{c}{\cellcolor{gray!50!white}}  \\
 tidal &&&&&&&&&& &{\cellcolor{gray!50!white}} & &{\cellcolor{gray!50!white}} \\[6pt]
 references & && \cite{Wagoner:1976am} & \cite{Kidder:1992fr,Wiseman:1992dv} &
\cite{Blanchet:1995fr,Blanchet:1995fg,Blanchet:1995ez} &
\cite{Damour:1981bh,Kidder:1995zr}
 \end{tabular}
 \caption{Overview of \ac{PN}\ orders and effects that enter the gravitational
waveform at this order. Here, \ensuremath{k} is the exponent of the orbital velocity above
leading order, as used in Eqs.~\eqref{eq:PNenergy} and~\eqref{eq:PNflux}.
\ac{PN}\ terms marked with an ``x'' are only partially unknown. Solid gray
boxes indicate which binary parameters
or physical effects contribute at the corresponding \ac{PN}\ order.
% \protect \linebreak
% \MP{Should we refer to table 6.1 of~\cite{Buonanno:2014aza}? Is there something more up-to-date?}
% \protect \linebreak
% \kc{Is the plan to add more references for higher order terms?}
}
\label{tab:PN_terms}
\end{table}
Examples of higher-order effects include the coupling of the individual
objects' spins with the orbital motion, which enters at 1.5\ac{PN}\ order
($k=3$); non-linear tail terms, also entering at 1.5\ac{PN};
 the self-spin and spin-spin coupling entering at 2.5\ac{PN}; horizon absorption
effects starting at 2.5\ac{PN}; and tidal effects for
compact objects that are not black holes, entering with a $v$-dependence corresponding to 5\ac{PN}\ order (see Sec.~\ref{sss:tidal}). A
list of effects and references is given in Table~\ref{tab:PN_terms}.
%The next section will provide some physical intuition about how the \ac{gw} signal 
%is affected by changes in specific binary parameters.
% TD : I don't know what this sentence about 'the next section' is pointing to
% or what the reader is supposed to make of it here.
% There is a lot of 'this section' still to come - including discussion of tidal
% and precession effects!

We conclude our brief introduction to the \ac{PN}\ framework by focussing 
on how the expressions of energy in Eq.~\eqref{eq:PNenergy} and flux in Eq.~\eqref{eq:PNflux} 
can be used directly to model the \ac{gw} signal of an inspiralling compact 
binary. Various
strategies have been developed to calculate the \ac{gw} strain $h$, written  
via a generalisation of Eq.~\eqref{eq:h_circular} as
\begin{align}  \label{eq:hoft_simple}
 h(t, \vec \theta) = \mathcal A \frac{ 4M v(t)^2 \eta}{d}
\cos\left[\phi_{\textrm{GW}}(t) +  \phi_0 \right]\,.
\end{align}
Here, we restrict the amplitude to the leading order term that is proportional to $v^2$. However, higher \ac{PN} corrections are included in the \ac{GW} phase and velocity as we discuss next.

The \ac{gw} phase, $\phi_{\rm GW}(t) = 2 \phi(t)$, is twice the orbital 
phase, which in turn can be calculated from the integrated orbital frequency 
$\omega$,
\begin{equation}
  \frac{d\phi}{dt}(t) = \omega (t) = \frac{v^3(t)}{M}\,, \label{eq:dphidt_PN}
\end{equation}
cf., Eq.~\eqref{eq:orbital_freq_Newt}. Finally, as in the Newtonian case, the 
relative velocity is given by an ordinary differential equation,
\begin{equation}
 \frac{dv}{dt} (t) = - \frac{\mathcal L(t)}{dE(v) / dv}\,. \label{eq:dvdt_PN}
\end{equation}
Unlike the Newtonian case, we now think of the energy $E$ and flux $\mathcal 
L$ as series expansions up to a given \ac{PN}\ order.

Different strategies to solve this system of equations are commonly referred to as
\ac{PN}\ \emph{approximants}: we name and summarize each such strategy 
below, see Ref.~\cite{Buonanno:2009zt} for a more detailed discussion and comparison.
% 

%\kc{Do we need to describe all these in detail? Beyond TaylorF2, most are no longer in frequent use, and their distinction is highly technical.}
\begin{description}[itemsep=6pt, leftmargin=1.85cm]
 \item[\textbf{TaylorT1}] Integrate Eqs.~\eqref{eq:dvdt_PN} and~\eqref{eq:dphidt_PN}
numerically.
 \item[\textbf{TaylorT2}] Solve the inverse of Eq.~\eqref{eq:dvdt_PN},
 \begin{align}
  \frac{dt}{dv} (v) = - \frac{dE(v) / dv}{\mathcal L(v)}\,,
 \end{align}
by re-expanding the expression into a series truncated at the same order as $E$ 
and $\mathcal L$ and integrate analytically to obtain $t(v)$. 
Equation~\eqref{eq:dphidt_PN} is solved in a similar way, by re-expanding
\begin{equation}
  \frac{d\phi}{dv}(v) =\frac{v^3(t)}{M} \frac{dt}{dv}(v)\,,
\end{equation}
into a series in powers of $v$ and integrating analytically. One then has 
analytical descriptions of time and phase, both parameterized by the relative 
velocity. Finding $\phi(\hat t)$ for a particular time $\hat t$ requires a 
root-finding algorithm to solve $t(v) = \hat t$.
\item[\textbf{TaylorT3}] This approach starts with the same steps as TaylorT2, 
but inverts $t(v)$ analytically (as a series expansion to consistent order) to 
obtain a series expression for $v(t)$. This is then used in $\phi(v)$ to obtain 
an analytical, closed-form series expansion for $\phi(t)$.
\item[\textbf{TaylorT4}] This approach is similar in spirit to TaylorT1 in that it integrates 
both differential equations numerically. However, Eq.~\eqref{eq:dvdt_PN} is first 
re-expanded and truncated, and this form is then integrated.
% MP: Added TF2 to 1.5\ac{PN}\ so I can refer to it later.
\item[\textbf{TaylorF2}] Under the \emph{stationary phase approximation}~\cite{Droz:1999qx} % missing!
the \ac{GW} phase in the Fourier 
domain can be derived from the time domain phase. A simplified expression for the dominant mode is 
given by~\cite{Poisson:1995ef,Arun:2004hn,Damour:2000zb,Damour:2002kr,Buonanno:2007yg,Buonanno:2009zt,Ajith:2011ec}
\begin{equation}
    \tilde h(f) = \mathcal{A} f^{-7/6} e^{-i \psi(f)}\,,
\end{equation}
where the amplitude 
\begin{equation}
    \label{eq:TF2_restricted_amplitude}
    \mathcal{A} \propto \mathcal{M}^{5/6} Q(\mathrm{angles}) / d\,,
\end{equation}
includes the response of the detector which depends on the location of the source with respect to the
reference frame of the \ac{GW} detector and the inclination angle under which the source is observed.
The TaylorF2 phasing is fully known up to 3.5\ac{PN}\ for the sake of illustration we only highlight the most important terms~\footnote{Formally the coalescence is reached at infinite frequency. In practice, a finite termination frequency and therefore a finite termination time and phase have to be used. Here, we stick to the usual notation $t_c, \phi_c$ which refers to the coalescence.}
\begin{equation}
    \label{eq:TF2_phase}
    \begin{split}
    \psi(f) &= 2\pi f t_c - 2\phi_c(\mathrm{angles}) - \frac{\pi}{4} +
    \frac{3}{128 \eta v^5} \biggl\{ 1 
        + v^2 \left[ \frac{3715}{756} + \frac{55}{9}\eta \right] \\
        &- v^3 \left[ 16\pi - \left( \frac{113}{3} - \frac{76}{3}\eta \right) \chi_s 
                                                  - \frac{113}{3} \delta_\mathrm{PN}\chi_a
              \right] \\
        &+ v^4 \left[ g(\eta) + \sigma \right]
        + \cdots 
        - v^{10} \frac{39}{2}\tilde\Lambda
        + \cdots
    \biggr\}\,,
    \end{split}
\end{equation}
where $v = (\pi M f)^{1/3}$ is the expansion parameter (assuming $c=1$, we expand in powers of $v/c$).
The Newtonian term (0\ac{PN}) only depends on binary parameters through the chirp mass $\Mc$ as
$1/(\eta v^5) = (\pi \Mc f)^{-5/3}$.
At 1.5\ac{PN}\ order ($v^3$ term \emph{within} the braces) spin-orbit contributions enter the phase which we write
as symmetric and anti-symmetric dimensionless spins aligned with the orbital 
angular momentum\footnote{
    In the notation of Sec.~\ref{ss:signal_characteristics} $\chi_i = c \vec S_i / (G m_i^2) \cdot \hat L_N$.
}, $\chi_s = (\chi_1 + \chi_2)/2$ and $\chi_a = (\chi_1 - \chi_2)/2$, as well as
$\delta_\mathrm{PN} = (m_1 - m_2) / M$ and $g(\eta)$ is an unspecified function.
Similarly, $\sigma$ represents the spin-spin contribution at 2\ac{PN}\ order.
The $v^{10}$ term describes tidal effects which we discuss below.
\end{description}

All these approximants correspond to different ways of obtaining expressions for the \ac{gw} given truncated series expressions 
for the binding energy and luminosity. In essence, they differ by the order in which they (numerically or analytically) 
solve the equations vs. re-expand
the series. Given the truncated nature of the \ac{PN}\ series, no method is \emph{a priori} preferred and their relative strengths
and weaknesses are determined by how well they approximate the full non-linear \ac{gw} calculation.

\subsubsection{Tidal effects}
\label{sss:tidal} 
An extended body in a spatially inhomogeneous external field will experience varying forces throughout its extent.
This is an example of a tidal interaction which is well understood in Newtonian gravity. In the following we define 
the tidal deformability of a \ac{NS} which quantifies how easily the star is deformed in an external field.
A static, spherically symmetric star of mass $m$ and radius $R$ in an external quadrupolar field $\mathcal{E}_{ij}$
will develop an induced quadrupole moment $Q_{ij}$. The (dimensionful) \emph{tidal deformability} of the star is 
given by the ratio~\cite{Flanagan:2007ix,Chatziioannou:2020pqz}
\begin{equation}
    \lambda = - \frac{Q_{ij}}{\mathcal{E}_{ij}}\,,
\end{equation}
which can be shown to be proportional to $R^5$. Hence, the \emph{dimensionless tidal deformability} is defined as
\begin{equation}
    \label{eq:tidal_deformability_def}
    \Lambda = \frac{\lambda}{m^5} = \frac{2}{3} k_2 \frac{R^5}{m^5}\,,
\end{equation}
where $k_2$ is the dimensionless \emph{Love number}; typically $k_2 \sim 0.2 - 0.3$.
For given $m$, both $k_2$ and $R$ 
%The dimensionless tidal deformability $\Lambda$ 
depend on the \emph{equation of state} which relates the interior pressure to the energy density and temperature of the \ac{NS}; the tidal deformability of a \ac{BH} is zero~\cite{Binnington:2009bb}.

In the TaylorF2 phase of Eq.~\eqref{eq:TF2_phase} we included the leading-order tidal contribution, which, though a Newtonian physical effect, has a velocity dependence equivalent to 5\ac{PN}\ expansion order, with a prefactor given by the effective tidal parameter~\cite{Flanagan:2007ix,Favata:2013rwa}
\begin{equation}
    \label{eq:Lambda_tilde}
    \tilde\Lambda = \frac{16}{3} \frac{(m_1 + 12 m_2) m_1^4 \Lambda_1 + (m_2 + 12 m_1) m_2^4 \Lambda_2}{(m_1 + m_2)^5}\,,
\end{equation}
where $\Lambda_i$ are %the dimensionless tidal deformabilities of the binary components, 
defined for each binary component via Eq.~\eqref{eq:tidal_deformability_def}. 
Although this tidal term enters at very high \ac{PN}\ order %in the expansion 
we include it here, since it enables the measurement of the neutron star equation of state from binary inspiral signals; this is possible for moderate \aclp{SNR} due to the large, $\mathcal{O}(100)$, value of $\tilde\Lambda$.
A detailed discussion of measuring tidal effects and the neutron star equation of state is given in Sec.~\ref{ss:matter_eos}.

\subsubsection{Precessing binaries}
\label{sss:precession}
% Maggiore II p. 254
% pretty much verbatim -- rewrite this a bit!
If the compact objects have spin vectors $\vec S_i$ that are not perfectly aligned with (equivalently, they have components orthogonal to)
the Newtonian orbital angular momentum of the binary,
then the spin vectors and the orbital angular momentum will change direction in space. 
The resulting motion resembles \emph{precession} around the total angular momentum, 
whose direction is approximately fixed when averaged over the timescale of precession.
This spin-precession effect enters through a combination of the 1.5\ac{PN}\ spin-orbit and 
2\ac{PN}\ spin-spin (and higher) interaction terms~\cite{Maggiore:2018sht}. 
The equation of motion for the spin of \ac{BH} $a$ is
\begin{equation}
    \frac{d \vec S_a}{dt} = \vec\Omega_a \times \vec S_a\,,
\end{equation} 
with $a \in{1, 2}$ denoting each binary component.
The angular velocity of the precessional motion for the two \acp{BH} labelled $a$ and $b$ is
\begin{equation}
    \vec\Omega_a = \frac{1}{r^3} \left[ \left( 2 + \frac{3 m_b}{2 m_a} \right) \vec L_N 
                                        - \vec S_b + 3(\hat x \cdot \vec S_b) \hat x  \right]\,,
\end{equation}
where $\vec L_N$ is the Newtonian orbital angular momentum, $\vec x$ the relative separation vector of
the two BHs, and $r = |\vec x|$, and $\hat x = \vec x / r$. 

The typical timescale of precession 
\begin{equation}
t_{\mathrm{pr}} \equiv \frac{|\vec S_i|}{|\dot{\vec S_i}|}\sim r^{5/2} \sim v^{-5}\,,
\end{equation}
is shorter than the radiation reaction timescale
\begin{equation}
t_{\mathrm{r}} \equiv \frac{E}{ \mathcal L} \sim v^{-8}\,,
\end{equation}
by 1.5\ac{PN}\ orders~\cite{Chatziioannou:2016ezg,Chatziioannou:2017tdw}.  %- missing in bib file
In the adiabatic limit we can approximately ignore radiation reaction over one precession
cycle. Then the total angular momentum $\vec J \simeq \vec L_N + \vec S_1 + \vec S_2$
is conserved, $d \vec J/dt = 0$, and therefore $d \vec L_N/dt = - d(\vec S_1 + \vec S_2)/dt$.
This implies that the orbital plane precesses with the same angular velocity as the total spin. In practice the total
angular momentum direction is conserved when averaged over spin-precession, while its magnitude shrinks due
to radiation reaction~\cite{Apostolatos:1994mx}.

% \MP{Comment on eccentric \ac{PN}\ and \ac{PN}\ precession equations?}

% \begin{table}[tb]
% \centering
%  \begin{tabular}{lllll}
%   \toprule
%  \ac{PN}\  & $k$ & physical effects & energy & flux \\
%  \midrule
%  0.0 & 0 & chirp mass && \\
%  0.5 & 1 & \multicolumn{3}{l}{\emph{vanishes in General Relativity}} \\
%  1.0 & 2 & mass ratio & & \\
%  1.5 & 3 & leading tail \& spin-orbit & & \\
%  2.0 & 4 & leading spin-spin & \cite{Blanchet:1995fr,Blanchet:1995fg} &
% \cite{Blanchet:1995fg,Blanchet:1995ez} \\
%  2.5 & 5 & 1\ac{PN}\ tail \& spin-orbit  \\
%  3.0 & 6 & tail-of-tail \& 1\ac{PN}\ spin-spin \\
%  3.5 & 7 & 5\ac{PN}\ tail \& spin-orbit   \\
%  4.0 & 8 & 5\ac{PN}\ spin-spin \\
%  4.5 & 9 & 5\ac{PN}\ tail \& spin-orbit, tail-of-tail-of-tail\\
%  && tidal deformability \\
%  5.0 & 10 & 5\ac{PN}\ spin-spin \\ \bottomrule
%  \end{tabular}
% \caption{Overview of \ac{PN}\ orders and effects that enter the gravitational
% waveform at this order. $k$ is the exponent of the orbital velocity above
% leading order, as used in \eqref{eq:\ac{PN}\energy} and \eqref{eq:\ac{PN}\flux}.}
% \end{table}

% TF2:
% * Damour:2000zb https://arxiv.org/pdf/gr-qc/0010009.pdf Eq 3.6
% SPA: Cutler:1994ys
% https://arxiv.org/pdf/0907.0700.pdf
% Blanchet https://doi.org/10.12942/lrr-2014-2 
% see e.g. Sec 2.2 of https://arxiv.org/pdf/0901.1628.pdf for convenient expressions
% Note that Sec 3.7 quotes the TF2 phase in order to compute the Fisher matrix

\subsubsection{Cosmological effects}
\label{sss:redshift}

So far, we have assumed that our signal is propagating on Minkowski spacetime.
However, \ac{CBC} sources can be observed at large enough distances (on the order of Gpc) that the 
effect of the cosmological expansion of the Universe becomes important.
Just as for electromagnetic radiation, the \ac{GW} signal is redshifted, reducing the instantaneous frequency
observed at the detector by a factor $1/(1+z)$ relative to frequencies in the source reference frame, where $z$ is the redshift. 
As we have seen in Sec.~\ref{sec:Newtonian_binary}, the phase evolution of a signal from a \ac{BBH} system (or any compact binary, in the point mass approximation) is determined by the combination $t/M$, thus a rescaling of the time coordinate is equivalent to rescaling the binary mass: for an observer with time coordinate defined by $dt_\mathrm{obs} = dt(1+z)$, the phase is a function of $t_\mathrm{obs}/(M(1+z))$. 
This scaling property also holds for numerical solutions of the full Einstein equations in vacuum (Sec.~\ref{ss:NR}), although it is broken by finite-size effects of neutron-star matter, which are strongly dependent on the \ac{NS} mass (Sec.~\ref{sss:tidal}, see also \cite{Messenger:2011gi}).  If such effects can be neglected, there is a complete degeneracy in phase between a source at nonzero redshift and one with a mass $(1+z)$ times greater~\cite{Krolak:1987ofj}. 

The amplitude or power of radiation propagating in cosmological models is described by the luminosity distance: a source's measured bolometric flux $F$ is to be related to an intrinsic bolometric luminosity $L$, which for an isotropically emitting source gives $d_L = \sqrt{L/(4\pi F)}$.  The luminosity distance may be derived as a function of redshift for any given model; for cosmologies with a spatially flat metric, it is related to the comoving distance $d_c$ by $d_L(z) = (1+z)d_c(z)$~\cite{Hogg:1999ad}. 
It can then be shown that the \ac{GW} signal received at Earth from a binary of masses $m_{1,2}$ at redshift $z$ and luminosity distance $d_L(z)$ is identical (again neglecting \ac{NS} matter effects) to a signal propagating in flat spacetime from a binary with masses $m_{1,2}(1+z)$ at distance $d_L$, e.g.~\cite{Flanagan:1997sx}.
A physical interpretation of luminosity distance is given in Sec.~\ref{sss:effect_of_params}.
%: \ac{GW} amplitude is determined by the luminosity distance. 

For convenience, it is common to define `detector frame', i.e.\ redshifted mass $M_z = M(1+z)$ as 
the parameter determining the observed phase of a signal. 
%Thanks to the mass-redshift degeneracy, 
This parameter and %the luminosity distance 
$d_L$ then suffice to determine the amplitude. 
% in exact analogy to a signal propagating in Minkowski space. 
%
%Since the amplitude of the signal also scales as $M/d=M(1+z)/d_L$, where $d_L$ is the luminosity distance to the source, one cannot disentangle $M$ from $(1+z)$ from the signal alone.
%
%Breaking the degeneracy requires additional knowledge about the mass of the source (for example from tidal effects~\cite{Messenger:2011gi}), or about the redshift (from e.g.\ identification of a host galaxy with measured redshift). 
%
Unless otherwise noted (e.g.\ in Sec~\ref{ss:populations_cosmo}), we will take masses, considered as waveform parameters in the context of data analysis, to be redshifted masses, but drop the $z$ subscript for simplicity. 
%neglect the effect of cosmology

\paragraph{Gravitational lensing}
The previous discussion assumed that \ac{GW} propagate through a homogeneous and isotropic Universe (for some choice of spatial slicing): in reality, this is not a perfect approximation.  Inhomogeneities in the distribution of matter along the line of sight to a source, and the associated perturbations in the cosmological metric, will affect the propagation of \ac{GW}.  The resulting effects go under the name of ``gravitational lensing'', by analogy with the distortion or magnification of optical images. 
Such effects may be broadly classified as ``weak'' lensing, relatively small (perturbative) changes to the \ac{GW} signal caused by the inhomogeneous large-scale structure of matter; and ``strong'' lensing, large but rare effects caused by the presence of a massive body -- most often considered as a galaxy or cluster of galaxies -- aligned with the source binary~\cite{Wang:1996as,Nakamura:1997sw}, which can significantly alter the signal morphology as well as the amplitude, and even give rise to multiple copies of a given transient, in analogy to multiple lensed optical images (e.g.~\cite{Oguri:2019fix}). 

The effects of weak lensing, as a perturbation to the \ac{GW} signal amplitude which averages to zero over many events, have generally been neglected for the ground-based network, though becoming significant for future observations~\cite{Hilbert:2010am,Congedo:2018wfn}.  The presence of strong lensing, while \emph{a priori} rare (e.g.~\cite{Ng:2017yiu}), will imply specialized strategies, models and methods, both for signal detection and parameter estimation as well for astrophysical and cosmological interpretation (see \cite{Hannuksela:2019kle,LIGOScientific:2021izm,LIGOScientific:2023bwz} and references therein).  Details of such analyses are beyond the scope of this review, thus unless otherwise stated we will consider only cases where strong lensing is absent.

\subsection{Signal characteristics}
\label{ss:signal_characteristics}

% extrinsic parameters: start analytically
% graphical illustrations for intrinsic parameters
% precession: mention Wigner rotation and co-precessing frame
% need beam patterns for polarization, sky position
% look at Sathya & Schutz LRR, Maggiore, Creighton

% \emph{Describe the effect of each source parameter on the \ac{gw} signal. In particular: distance, inclination, polarization, phase, time, total mass, mass ratio, aligned-spin components, perpendicular components. Also mention eccentricity.}

% \vr{Add also here the antenna beam pattern effects to complete the list of parameters? Or only bring those up
%  in~\ref{ss:detector_response}?}
% \MP{Currently, the parameters $(\alpha, \delta, \psi)$ are defined here, but not the beam pattern functions for IFOs
% $F_+, F_\times$.}
% \kc{The beam pattern functions depend on the detector configuration, so they probably belong where we discuss specifically the ground-based, 90-degree detectors.}
% \MP{Agreed. Added a comment in section 3.}

%%%%%%%%%%%%

\subsubsection{Description of binary parameters}
\label{sss:binary_params}

% Describe the parameters in the table
A \ac{BBH} undergoing quasi-circular inspiral can be described by the two component masses $m_i$ and the spin vectors $\vec S_i$ (or angular momenta) of its component Kerr black holes. We refer to these eight parameters as \emph{intrinsic} parameters. The mass space can alternatively be parametrized by the total mass $M$ or chirp mass 
$\Mc = M \eta^{3/5}$ and the (asymmetric) $q = m_2/m_1 \leq 1$ or symmetric mass-ratio $\eta = m_1m_2/M^2$.
% Spins
% Given a mass m [kg] multiply m^2 by G/c to get an angular momentum [kg m^2 / s]
Since the maximum spin a Kerr black hole of mass $m$ can reach is $|\vec S_\mathrm{max}| = (Gm^2)/c$ it is 
customary to work with \emph{dimensionless} spin vectors $\vec\chi_i = c \vec S_i / (G m_i^2)$. These dimensionless spin vectors are usually projected onto the (Newtonian) orbital angular momentum unit vector $\vec \hat L_N$ of the binary in order to separate the spin vectors into parallel and orthogonal components. As described in Sec.~\ref{sss:precession}, non-zero components in the
orbital plane
%with respect to the Newtonian orbital angular momentum vector
give rise to spin-precession of the orbital plane and the spin vectors around the total angular momentum $\vec J = \vec L + \vec S_1 + \vec S_2$~\cite{Apostolatos:1994mx,Kidder:1995zr}.
For precessing binaries the spin vectors and derived quantities evolve with time and therefore they 
are defined at a reference frequency (e.g.\ the time at which the dominant mode \ac{GW} frequency reaches
the reference frequency).

% effective spin parameters
Often, only the dominant dependence of the spins is of interest, motivating the use of 
the effective aligned spin
$\chi_\mathrm{eff} = (m_1 \vec\chi_1 + m_2 \vec\chi_2) \cdot \hat L_N / M$~\cite{Ajith:2009bn,Santamaria:2010yb} which is approximately constant 
during the inspiral to at least the 2\ac{PN}\ order~\cite{Racine:2008qv}. 
In a similar vein, the effective precession parameter~\cite{Schmidt:2014iyl}
\begin{equation}
  \label{eq:chip}
  \chi_\mathrm{p} = \max \left\{ \chi_1^\perp, \frac{q(4q + 3)}{4 + 3q} \chi_2^\perp \right\}\,,
\end{equation}
averages the in-plane spins (with $\chi_i^\perp$ being their magnitudes) over a number of precession cycles and assigns the effective
precession spin to the primary \ac{BH}; this is done because spins on the primary have a larger impact
on the waveform than spins on the secondary when the binary has unequal masses.
Generalizations of the $\chi_\mathrm{p}$ definition have also been proposed~\cite{Gerosa:2020aiw,Thomas:2020uqj}. %- missing in .bib
% PE parametrization
In parameter estimation, the spin vectors are often expressed in spherical polar coordinates with respect to to $\vec \hat L_N$
by spin magnitudes $\chi_i$ and polar (``tilt'') angles $\theta_i$ along with an appropriate 
choice of the inclination angle for precessing binaries (see below)~\cite{Farr:2014qka}. 
Instead of the azimuthal angles $\phi_i$, the angle $\phi_{12}$ between $\phi_2$ and $\phi_1$ and 
the angle $\phi_{JL}$ (the difference between the total and orbital angular momentum azimuthal angles)
are used.

% extrinsic parameters
In addition to the above set of eight intrinsic parameters seven more \emph{extrinsic} parameters are 
needed to describe a quasi-circular \ac{BBH} in relation to a \ac{GW} detector. These are the
sky location (right ascension $\alpha$ and declination $\delta$ defined in the equatorial celestial
coordinate system), polarization angle $\psi$, luminosity distance $d_L$, 
an orbital inclination angle, and the time $t_c$ and phase $\phi_c$ at coalescence.
% https://dcc.ligo.org/public/0152/T1800226/004/LAL_GW_Frames.pdf has a plot showing alpha, delta, but the entire slide is much too complicated overall.
%
The inclination of the binary's orbit as seen from the observer is traditionally parametrized by an 
angle $\iota$, the angle between $\vec L$ and the direction toward the observer.  %under which a binary is observed. 
Because $\vec L$ is not a stable direction for precessing binaries and the direction of the total
angular momentum $\vec \hat J$ is approximately fixed throughout the inspiral (except in the unusual
case of binaries undergoing transitional rather than simple precession~\cite{Apostolatos:1994mx}) we
prefer to instead use $\theta_\mathrm{JN}$, the angle between the total angular momentum and the unit vector $\hat N$ directed toward the observer as a measure of inclination.
Transitional precession~\cite{Apostolatos:1994mx,Zhao:2017tro} happens when the total angular momentum $\vec J$
goes through zero during inspiral, because $\vec S_1 + \vec S_2 \simeq - \vec L$, and thus flips its direction.
This can happen for sufficiently asymmetric precessing binaries with mass-ratios of $q \lesssim 0.3$.

% eccentricity
% e.g. https://arxiv.org/pdf/2112.06952.pdf Eq 14
% https://en.wikipedia.org/wiki/Eccentric_anomaly
% https://en.wikipedia.org/wiki/Orbital_elements#Keplerian
\ac{BBH}s do not necessarily undergo quasi-circular inspiral where the \ac{BH}s trace out
a sequence of shrinking circular orbits. Instead, the orbit is in general tangent to an ellipse at any
one point. Of course, the orbit is not a closed elliptical
orbit because of gravitational radiation. There is no unique definition of eccentricity in General Relativity. 
Locally we can choose a Keplerian parametrization of the binary's orbit $r = p / (1 + e \cos \nu)$, 
where $r$ is the radius, $p$ the semi-latus rectum, and we have introduced an eccentricity parameter $e$ 
and a radial phase parameter $\nu$ which describes the position of a point on an elliptical orbit. 
Eccentricity is radiated away during the inspiral and merger and most binaries are expected to 
circularize by the time they enter the frequency band of a \ac{GW} detector~\cite{Peters:1964zz}. Therefore, 
the eccentricity parameters must be defined at a reference frequency in a similar way as for spin parameters.

\begin{figure}
\begin{center}
\includegraphics{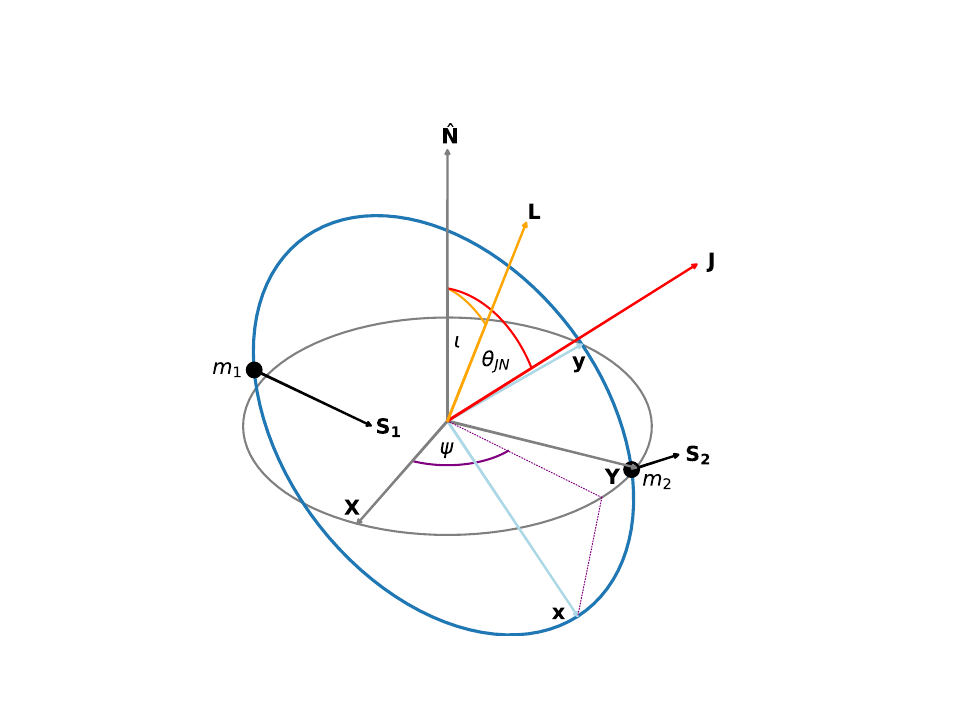}
\end{center}
\caption{\label{fig:cbc-coords-source} The orientation of the binary system, defined by the parameters $\vec{L}$, the orbital angular momentum; $\vec{J}$ the total angular momentum; $\vec{S}_{1,2}$ the spin angular momentum vectors for the two bodies with masses $m_{1,2}$; $\psi$ the polarisation angle; $\theta_{JN}$ the angle between the direction to the observer $\hat{N}$ and the total angular momentum. Also shown are the binary's \emph{source frame} $(x, y, z)$ with $\vec{L}$ pointing in the $z$-direction and \emph{wave frame} $(X, Y, Z)$ where $\hat{N}$ coincides with the $Z$-axis. Figure generated using~\cite{gwviz}. }
\end{figure}

For a generic, possibly eccentric and precessing, compact binary the relationship between the binary's
\emph{source frame} $(x, y, z)$ (spanned by the instantaneous orbital angular momentum vector and the orbital plane
at a reference frequency) and a \emph{wave frame} $(X, Y, Z)$ in which the emitted \ac{GW} propagates toward a \ac{GW} detector
is illustrated in Fig.~\ref{fig:cbc-coords-source} and the documentation of the LALSimulation package~\cite{lalsimulation-inspiral-orbital-elements}.
The $Z$-axis of the wave frame points toward Earth. Therefore, the $X-Y$ plane is the plane of the sky.
The angle between the $Z$-axis of the wave frame and the $z$-axis of the source frame is the inclination $\iota$.
To describe generic binaries we need to introduce three more (positional) \emph{orbital elements}~\cite{PoissonWillGravity} 
in addition to the inclination and reference phase. The \emph{longitude of the ascending node} is the angle on the plane 
of the sky from the X-axis of the reference direction in the wave frame to the \emph{ascending node}, which is 
the point at which the binary's orbit cuts the plane of the sky from below. The \emph{true anomaly} is the 
angle $\nu$ (introduced above) along the orbital plane from the periapsis or pericenter (the point of closest 
approach in the orbit) to the present position of the primary orbiting body.
In addition, the \emph{argument of the periapsis} needs to be specified to define the orientation of an elliptical orbit in the orbital plane. For non-eccentric orbits it is degenerate with the reference phase.
It is also worth mentioning that the longitude of the ascending node parameter is degenerate with the polarization angle (as both effect rotations on the plane of the sky).
% MP Note: In total 6 orbital elemnts are needed to describe the orbit of an eccentric Newtonian binary in a generic frame of reference
% as here the wave frame. We mention 4 positional orbital elemnts above. The two additional ones are eccentricity
% and semi-latus rectum. I think we don't need the latter for GWs because it leaves an imprint in the \ac{GW} frequency evolution.

% tidal effects
For binaries containing \ac{NS}s we need to take into account additional degrees of freedom 
related to their response to a tidal field as discussed in Sec.~\ref{ss:PN}.
The dominant quadrupolar ($\ell = 2$) tidal deformation is described by the dimensionless tidal 
deformability of each neutron star $\Lambda_i = (2/3)k_2 [ (c^2/G) (R_i/m_i) ]^5$
and the effective parameter $\tilde\Lambda$ is a mass-weighted linear combination of tidal deformabilities.

For reference we have collected all of the above intrinsic and extrinsic binary parameters discussed in Table~\ref{tab:parameters}.
For an extended table of parameters used by the \texttt{bilby} inference code see App. E of Ref.~\cite{Romero-Shaw:2020owr}.

\begin{table}[tbh]
\centering
 \begin{adjustbox}{width=1\textwidth}
 \begin{tabular}{{p{0.05\textwidth}p{0.25\textwidth}p{0.3\textwidth}p{0.3\textwidth}}}
  \toprule
 \multicolumn{4}{l}{Mass parameters}\\
 $m_1$ & primary mass & the larger of two binary masses & unit $M_\odot$\\
 $m_2$ & secondary mass & the smaller of two binary masses & unit $M_\odot$\\
 $\Mtot$ & total mass & $m_1 + m_2$ & unit $M_\odot$ \\
 $\eta$ & Symmetric mass ratio & $m_1m_2/M^2$ & dimensionless \\
 $q$ & Mass ratio & $m_2/m_1 \leq 1$ & Defined to be $\leq 1$ to avoid unbounded parameters \\
 $\Mc$ & Chirp mass & $M \eta^{3/5}$ & unit $M_\odot$\\
 %  = (m_1 m_2)^{3/5} / \Mtot^{1/5}
 \midrule
 \multicolumn{3}{l}{Spin parameters} & at reference frequency\\
 $\vec{S_i}$ & spin vector & $i = 1,2$ & dimensionful \\
 $\vec{\chi_i}$ & spin vector & $\vec S_i/m_i^2$ & dimensionless \\
 $\chi_{i,z}$ & aligned spin & $\vec{\chi_i} \cdot \hat L_N$ & dimensionless \\
 $\chi_i$ & spin magnitude & $|\vec{\chi_i}|$ & dimensionless \\
 $\theta_i$ & polar angle of $\vec{\chi_i}$ & $\arccos(\chi_{i,z}/a_i)$ & spin ``tilt'' angles\\
 $\phi_i$ & azimuthal angle of $\vec{\chi_i}$ &  & \\
 % $\arccos[\sgn(\theta_i) \frac{\chi_{ix}}{\sqrt{\chi_{ix}^2 + \chi_{iy}^2}}]$
 $\phi_{JL}$ & azimuthal angle of $\hat L_N$ on its cone about $\vec{J}$ & & \\
 % see https://lscsoft.docs.ligo.org/lalsuite/lalsimulation/group__lalsimulation__inference.html
 % SimInspiralTransformPrecessingWvf2PE() and SimInspiralTransformPrecessingNewInitialConditions()
 % https://git.ligo.org/lscsoft/lalsuite/-/issues/241
 $\phi_{12}$ & difference between azimuthal angles & $\phi_2 - \phi_1$ & \\
 $\chi_\mathrm{eff}$ & effective aligned spin & $(m_1 \chi_{1,z} + m_2 \chi_{2,z}) / M$ & dominant spin effect\\
 $\chi_p$ &  precession spin & See Eq.~\eqref{eq:chip} & avg. over precession cycles\\
 %\multicolumn{2}{l}{averaged over precession cycles and assigned to the primary} \\
 \midrule
% other intrinsic &  &  & \\
 \multicolumn{4}{l}{Tidal parameters}\\
 $\Lambda_i$ & dimensionless tidal deformability of object $i$ & $\frac{2}{3} k_2 (R_1 / m_1)^5$ & non-zero for neutron stars\\
 $\tilde\Lambda$ & effective dimensionless tidal deformability & See Eq.~\eqref{eq:Lambda_tilde} & dominant tidal effect\\
% $\delta\tilde\Lambda$ & relative difference in combined tidal deformability &  & subdominant tidal effect\\
 \midrule
 % \multicolumn{3}{l}{Eccentricity parameters} & at reference frequency\\
 % $e$ & eccentricity &  & no unique definition in GR \\
 % $\Omega$ & Longitude of ascending node & See \cite{lalsimulation-inspiral-orbital-elements} & degenerate with $\psi$\\
 % $\nu$    & True anomaly & See \cite{lalsimulation-inspiral-orbital-elements} & radial phase parameter\\
 % \midrule
 \multicolumn{4}{l}{Extrinsic parameters}\\
 $d_L$ & luminosity distance & \multicolumn{2}{l}{unit Mpc} \\
 $z$ & redshift & \multicolumn{2}{l}{can be computed from $d_L$ assuming a cosmological model}\\
 $\theta_\mathrm{JN}$ & inclination angle & \multicolumn{2}{l}{angle between total angular momentum $\vec{J}$ and direction to observer $\vec{N}$}\\
 $\iota$ & inclination angle & \multicolumn{2}{l}{angle between orbital angular momentum $\vec{L}$ and direction to observer $\vec{N}$}\\
 $t_c$ & time of coalescence &  \multicolumn{2}{l}{GPS reference time at the geocenter} \\
 $\phi_c$ & phase of coalescence & at reference frequency & \\
 $\alpha$ & right ascension &  \multicolumn{2}{l}{celestial equivalent of terrestrial longitude} \\
 $\delta$ & declination &  \multicolumn{2}{l}{celestial equivalent of terrestrial latitude} \\
 $\psi$   & polarization angle & \multicolumn{2}{l}{rotates polarizations in plane orthogonal to propagation of GW} \\
 \bottomrule
 \end{tabular}
 \end{adjustbox}
 \caption{Commonly used parameters to describe compact binary coalescences, as found in LIGO-Virgo-KAGRA parameter estimation releases.
 Mass parameters can be defined either as \emph{redshifted} (`detector frame') or \emph{source frame} masses, related by 
 $m_{zi} = m_i^\mathrm{src} (1 + z)$.
 %$m_i^\mathrm{src} = m_i^\mathrm{det} / (1 + z)$.
 Spins are labeled with an index $i = 1,2$ indicating the primary or secondary object.% they belong to.
 Most spin quantities evolve during the coalescence and are therefore defined at a reference frequency $f_\mathrm{ref}$.
 }
\label{tab:parameters}
\end{table}

\subsubsection{The effect of binary parameters on the \ac{GW} strain}
\label{sss:effect_of_params}

\paragraph{Masses and mass-ratio}
A change in the total mass $\Mtot$ of a binary while holding all other parameters fixed results in a reciprocal change in the
binary's frequency evolution, so that the dimensionless product $\Mtot \omega(t)$ or $\Mtot f$ is unchanged. Therefore, an increase
of $\Mtot$ is equivalent to shifting the \ac{GW} strain in the Fourier domain to lower frequencies by the same factor, which makes
the observable signal in the sensitive frequency band of a \ac{GW} detector shorter. % At the same time the \ac{GW} signal amplitude at fixed distance grows proportional to $\Mtot$. 

For fixed total mass the symmetric mass-ratio enters the phasing at 0\ac{PN}\ order (see Eq.~\eqref{eq:TF2_phase}). A binary becomes
more and more asymmetric as the mass-ratio $q$ decreases from its maximum $q=1$ (equal mass) and this also decreases the symmetric
mass-ratio $\eta$ from its maximum $1/4$ to lower values. The main effects of decreasing $\eta$ for a \ac{GW} signal with constant total mass starting from a fixed initial frequency are: (i) a steep increase in the time it takes for the binary to merge; as in
Eq.~\eqref{eq:time_to_coalescence} the so-called \emph{chirp time}~\cite{Sathyaprakash:2009xs} is inversely proportional to $\eta$,
and (ii) an increase in the number of \ac{GW} orbits or cycles until merger, as the 0\ac{PN}\ phase term is inversely proportional
to $\eta$.

The chirp mass depends both on total mass and symmetric mass-ratio. An increase of the chirp mass has two main effects: (i) the \ac{GW} signal from a binary becomes brighter as its Fourier domain amplitude in Eq.~\eqref{eq:TF2_restricted_amplitude} scales as $\Mc^{5/6}$, and (ii) the number of \ac{GW} orbits or cycles from a fixed initial frequency until merger are reduced because the 0\ac{PN}\ phase term in Eq.~\eqref{eq:TF2_phase} goes as $\Mc^{-5/3}$.

% Note on measurability of parameters
We have seen in Sec.~\ref{ss:PN} that the chirp mass $\Mc$ enters the phase evolution of Eq.~\eqref{eq:TF2_phase}
at leading order, while the mass-ratio $q$ and effective aligned spin $\chi_\mathrm{eff}$ appear in the 
phasing at higher orders. It would not be unreasonable to expect that terms at lower \ac{PN}\ order which make
larger contributions to the phasing will be easier to measure than terms entering at higher \ac{PN}\ order, although
this is not always true. See Sec.~\ref{ss:mass_measurement} for a discussion.
Similarly, we expect that the two component masses $m_1, m_2$ will be harder to constrain than the 
chirp mass and symmetric mass-ratio because they will involve correlations between chirp mass and mass-ratio.

\paragraph{Aligned spins}
The dominant effect for spins aligned with the orbital angular momentum of the binary is captured by
an effective aligned spin parameter. The parameter $\chi_\mathrm{eff}$ is a simplified version of the
combination which appears in the spin-orbit term of the \ac{PN}\ phasing in Eq.~\eqref{eq:TF2_phase}. The full term
is
$\chi_\mathrm{eff} - \frac{76}{226}\eta (\chi_1 + \chi_2)$~\cite{Poisson:1995ef,Ajith:2011ec}.
%\kc{This does not look right. The 1.5PN terms reduces to $\chi_\mathrm{eff}$ for equal masses, so there should be some factors of $m_1-m_2$ here.}. 
% MP: The asymmetric (m1 - m2) contribution is contained in chi_eff. There are additional symmetric contributions proportional to  (chi1 + chi2) which are not part of chi_eff. See https://arxiv.org/pdf/1107.1267.pdf.
The fact that the phasing depends on
such a combination implies that it is in general difficult to measure the aligned spins on the 
individual \ac{BH}s; for very unequal mass binaries only the spin on the primary will be measurable~\cite{Purrer:2015nkh,Chatziioannou:2018wqx}.
In a similar way as for the symmetric mass-ratio, increasing the aligned spin relative to a non-spinning 
compact binary (i.e. positive $\chi_i$) starting from a fixed initial frequency, and holding other parameters
constant, increases the time and the number of \ac{GW} cycles until merger. If the spins are anti-aligned with
the orbital angular momentum (i.e. negative $\chi_i$) the time to merger is decreased relative to the spinless
case. The slower merger process of binaries with positive aligned spins is due to additional angular momentum that
needs to be shed before the binary can settle down into a stationary Kerr black hole. This effect has been called
``orbital hang up''~\cite{Campanelli:2006uy}. % https://arxiv.org/pdf/1007.4789.pdf p. 11

\paragraph{Luminosity distance}
We saw in Eq.~\eqref{eq:hoft_simple} that the \ac{GW} strain in Minkowski space
depends inversely on the distance: \ac{GW} signals from closer binaries are stronger, as well as ``louder'' in a given detector.
% emitted from a binary will
%become  the closer it is in terms of distance.
We also saw in Sec.~\ref{sss:redshift} that cosmological effects can be expressed by using the 
Minkowski space signal model but defining `redshifted' masses larger by a factor $(1+z)$, and 
% 
%if we take the mass to refer to the `detector frame'
%mass, then we should also take the distance to refer to the \emph{luminosity distance} $d_L$,
%in which case we can continue to use the Minkowski space signal model 
with $d_L$ taking the place of $d$. %in our signal model.

The phase evolution of a coalescing compact binary which radiates \acp{GW} determines the redshifted masses, and, together with the measurement of the \ac{GW} signal amplitude, the source luminosity distance can thus be deduced~\cite{1986Natur.323..310S,Sathyaprakash:2009xs,Singer:2016eax}; 
however the effects of angular parameters, particularly inclination as discussed below, on signal amplitude can induce large uncertainties in this measurement.

\paragraph{Sky location and polarization angle}
As will be discussed in Sec.~\ref{ss:detector_response}, the \ac{GW} strain recorded by a \ac{GW} detector is a 
weighted linear combination of the \ac{GW} polarizations $h = F_+ h_+ + F_\times h_\times$,
where the (real) detector pattern functions $F_+, F_\times$ depend on the sky position $(\alpha, \delta)$ and 
the polarization angle $\psi$. The \ac{GW} polarizations are defined with respect to a system of axes
in the plane perpendicular to the propagation direction of the wave. The polarization angle
effectively rotates this system of axes (see Fig.~\ref{fig:cbc-coords-source}). The triple $(\alpha, \delta, \psi)$ defines the location
of the source with respect to the frame of a \ac{GW} detector on Earth.

\paragraph{Spherical harmonic modes and inclination}
% Mode expansion and inclination angle
In the wave-zone, far away from the coalescing binary, it is customary to expand the complex combination of gravitational-wave polarizations $H = h_+ - i h_\times$ on a space-like 2-sphere centered on the source in terms of
appropriate basis functions for metric perturbations around Minkowski space.
These basis functions are spin-weighted spherical harmonics~\cite{Goldberg:1966uu}
${}_{-s}Y^{\ell m}$ (of spin weight $s=-2$) which depend on the usual spherical polar angles $(\theta, \varphi)$
\begin{equation}\label{eq:SpHarmWaveform}
    h_+ - i h_\times =
    H(t; \theta, \varphi) = \sum_{\ell=2}^\infty \sum_{m=-\ell}^\ell  h^{\ell m}(t) \, {}_{-2}Y^{\ell m}(\theta, \varphi)\,,
\end{equation}
where $h^{\ell m}(t)$ are the \ac{GW} \emph{modes} or \emph{harmonics}.
% Note: together with our definition of the forward FT with a minus in the exponential, this implies that the (2,2) mode mainly has power in the negative frequencies, but this shouldn't be a problem.
%
At leading order, for the dominant $(\ell, m)=(2, \pm 2)$ modes in the \ac{GW} signal, we have
\begin{equation}
    H(t; \theta, \varphi) = h^{22}(t) \, {}_{-2}Y^{22}(\theta, \varphi)
                          + h^{2,-2}(t) \, {}_{-2}Y^{2,-2}(\theta, \varphi)\,,
\end{equation}
where
$${}_{-2}Y^{2, \pm 2}(\theta, \varphi) = \sqrt{\frac{5}{64\pi}} (1 \pm \cos\theta)^2 e^{\pm 2 i \varphi}$$
with $h^{22}(t) = A^{22}(t) e^{-i \phi_\mathrm{GW}(t)}$, and, assuming equatorial symmetry of the binary, 
$h^{2,-2} = (h^{22})^*$.
This can be simplified to
\begin{multline}
%    H(t; \theta, \varphi) \propto A^{22} \frac{1 + \cos^2 \theta}{2} \cos(\phi_\mathrm{GW} - 2 \varphi)
%    - i \, A^{22} \cos\theta \sin(\phi_\mathrm{GW} - 2 \varphi)\,.
    H(t; \theta, \varphi) = \sqrt{\frac{5}{4\pi}} A^{22}(t)
    \biggl[ \frac{1 + \cos^2 \theta}{2} \cos\left(\phi_\mathrm{GW}(t) - 2 \varphi\right) \\
    - i \, \cos\theta \sin\left(\phi_\mathrm{GW}(t) - 2 \varphi\right) \biggr]\,.
\end{multline}
The angles $\theta$ and $\varphi$ are related to the direction under which the observer sees the orbital plane
defined in the source frame of the binary. In general the orbital plane of the binary will be inclined with respect to
the plane perpendicular to the direction of wave propagation $\hat N$,
which is described by an inclination angle $\iota$ ($\theta_\mathrm{JN}$ for precessing binaries). This 
inclination angle takes the place of the angle $\theta$ in the spherical harmonics.
The angle $\varphi$ can be substituted by the phase of the binary at coalescence $\phi_c$  at time $t_c$.
The effect of changing these parameters $\phi_c, t_c$ is a simple offset in phase and shift in the time
of the \ac{GW} strain, say the time at which the signal reaches its maximum amplitude.
An expansion into spin-weighted spherical harmonic modes allows us to separate the dependence of the waveform 
on inclination and a phase angle so that the modes $ h^{\ell m}(t)$ only depend on the remaining binary parameters.

As can be seen from the above expression for the \ac{GW} strain (substituting $\iota$ for $\theta$), 
if the source is seen ``face-on'' ($\iota=0$) 
or ``face-off'' ($\iota=\pi$) the \ac{GW} signal leads to a much stronger response than if the inclination 
tends towards ``edge-on'' ($\iota=\pi/2$) where the response is weakest.
Moreover, for $\iota=0$ the strain is a linear combination 
$h_+ - i h_\times \propto A^{22} \exp [ - i(\phi_\mathrm{GW} - 2 \varphi) ]$ and the \ac{GW} is circularly polarized, 
whereas for $\iota=\pi/2$ only $h_+ \propto (A^{22}/2) \cos(\phi_\mathrm{GW} - 2 \varphi)$ is non-zero and
the \ac{GW} is linearly polarized.
% Add a figure like 1.1 in https://arxiv.org/abs/2201.03428?

\begin{figure}[t]
  \centering
    \includegraphics[width=\textwidth]{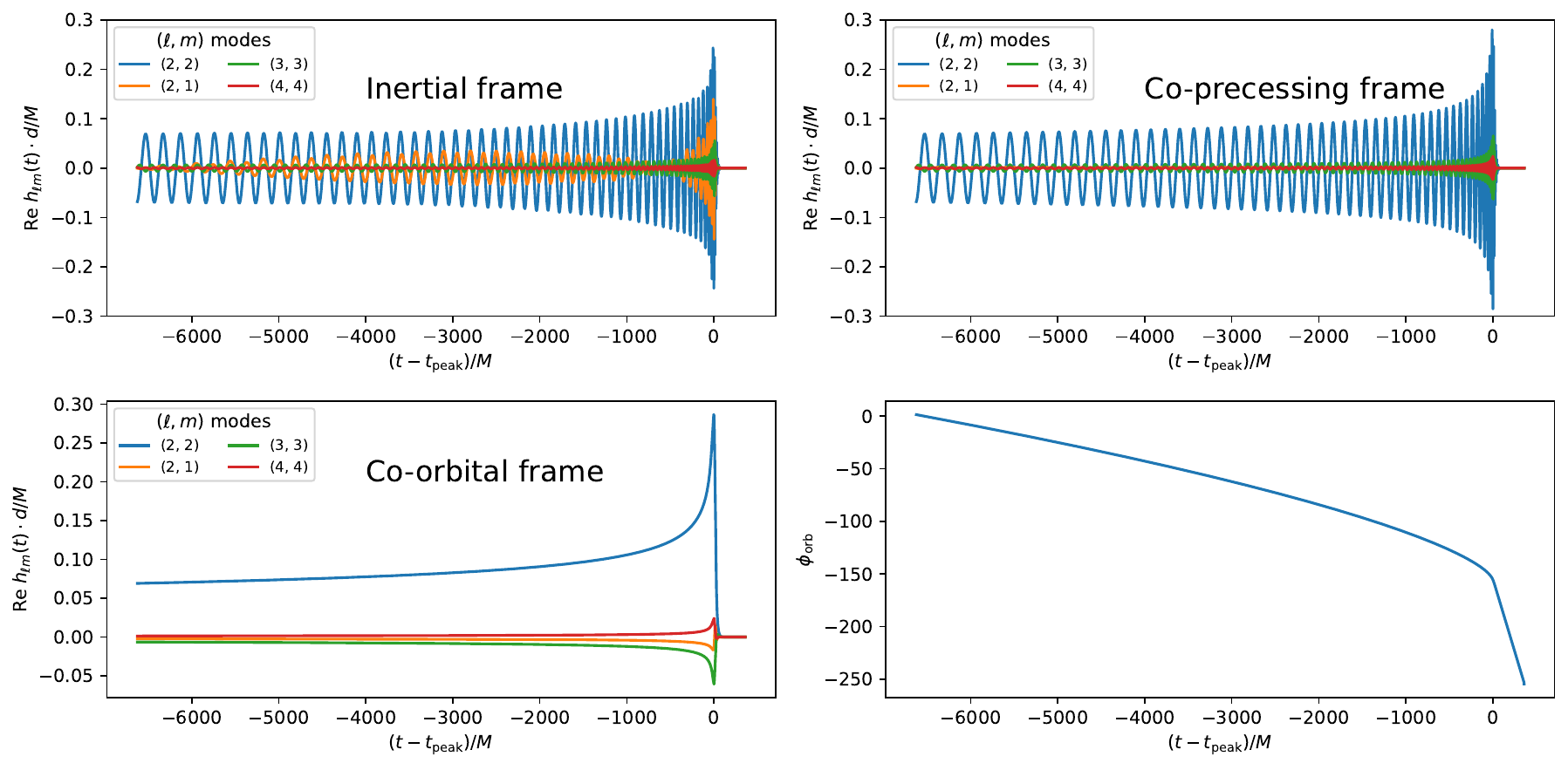}
  \caption{
      Selected spherical harmonic modes ($(2,2), (2,1), (3,3), (4,4)$) of the \ac{GW} signal emitted 
      by a precessing black hole binary. Shown is the real part of the modes in three reference frames.
      \emph{Top left:} inertial frame, \emph{top right:} co-precessing frame,
      \emph{bottom left:} co-orbital frame, \emph{bottom right:} one half the averaged phase of 
      the co-precessing frame $(2, \pm 2)$ modes.
      The binary has mass-ratio $q=1/4$, and spin vectors $\chi_1 = (0.2, 0.3, 0.35)$, 
      $\chi_2 = (0.5, 0.05, 0.42)$ defined at a reference frequency of $20\, \mathrm{Hz}$ for a total
      mass of $50 \,M_\odot$. The waveform model used is \texttt{SEOBNRv5PHM}~\cite{Ramos-Buades:2023ehm}.
  }
  \label{fig:precessing_waveform_frames_phase}
\end{figure}

\paragraph{Adapted reference frames for precessing binaries}
% similar to Fig 1 of https://arxiv.org/pdf/1905.09300.pdf
Figure~\ref{fig:precessing_waveform_frames_phase} illustrates how waveform modes of a precessing binary
exhibit modulations when analyzed in the inertial reference frame of the binary. The modes simplify
greatly when one switches to a non-inertial, rotating reference frame which tracks the time-dependent
motion of the precession: the co-precessing frame.
There, the $(2, 2)$ mode is clearly dominant over the weaker higher order modes
shown. Mathematically, the transformation law for spin-weighted spherical harmonics ${}_{-s}Y^{\ell m}$
under arbitrary rotations can be used to find that the inertial \ac{GW} modes $h^I_{\ell m}$ transform 
to the co-precessing modes $h_{\ell m}^{copr}$ as follows~\cite{Schmidt:2010it,Boyle:2013nka,Hannam:2013oca}
\begin{equation}
\label{eq:transf_coprecessing}
    h_{\ell m}^{copr}(t) = e^{i m\alpha(t)} \sum_{\ell=2}^{\ell_\mathrm{max}} \sum_{m' = -\ell}^\ell
                        e^{i m'\gamma(t)} \, d^\ell_{m', m}(-\beta(t)) h^I_{\ell, m'}(t)\,,
\end{equation}
where $d^\ell_{m', m}$ denote Wigner d-matrices, and the time dependent Euler angles 
$\alpha(t), \beta(t), \gamma(t)$ parametrize the evolution of $\vec \hat L_N$ in a frame with $ \vec \hat z =\vec \hat J$.
An arguably superior representation of time-dependent frame rotations is given by quaternions~\cite{Boyle:2013nka}.

In the co-precessing frame the precessing waveform is fairly well approximated by the waveform
which has spin components in the orbital plane set to zero, a binary with spins aligned with the orbital
angular momentum. However, in general $h_{\ell, m}(t) \neq (-1)^\ell h^*_{\ell, -m}(t)$ and this 
approximation therefore discards asymmetry in the waveform modes~\cite{Boyle:2014ioa,Varma:2019csw,Thompson:2023ase,Kolitsidou:2024vub}.
A further simplification can be achieved by taking out the orbital phase times the
$m$ number of each mode, 
$h^\mathrm{coorb}_{\ell m}(t) = h^\mathrm{copr}_{\ell m}(t) e^{im\phi_\mathrm{orb}}$,
where a quantity that approximates the orbital phase is defined as the average of the dominant mode
phases in the co-precessing frame
$\phi_\mathrm{orb} = (\arg [h^\mathrm{copr}_{2,-2}(t)] - \arg [h^\mathrm{copr}_{2,2}(t)]) / 4$.

\begin{figure}[t]
  \centering
    \includegraphics[width=\textwidth]{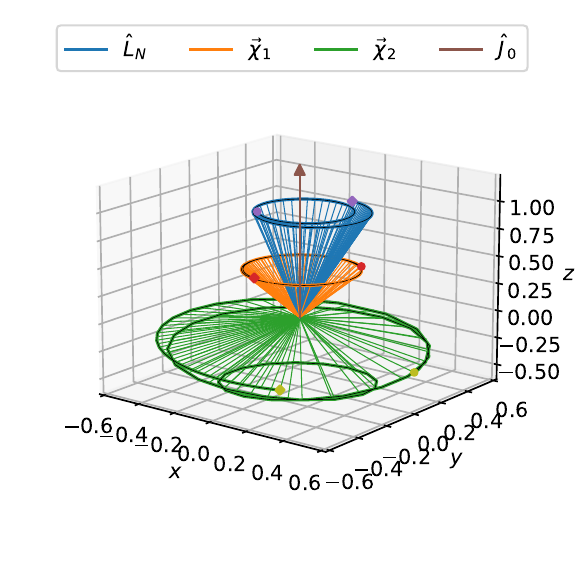}
  \caption{
      Evolution of Newtonian orbital angular momentum $\hat L_N$ and spin vectors $\vec\chi_i$
      over time for the precessing binary with waveform shown in
      Fig.~\ref{fig:precessing_waveform_frames_phase}.
      The initial point is indicated by a filled circle and the end point (merger) by a diamond.
      The evolution of each vector traces out a separate precession cone about the direction 
      of the total angular momentum $\hat J_0$ at the initial point (slightly scaled).
  }
  \label{fig:figs_precession_cone}
\end{figure}

An example of the waveform emitted by a strongly precessing binary decomposed into modes is shown in
Fig.~\ref{fig:precessing_waveform_frames_phase} and the precession cones traced by the orbital
angular momentum and spin vectors for the same binary are displayed in Fig.~\ref{fig:figs_precession_cone}.
In this configuration both spin vectors have non-zero projections in the orbital plane, and the binary
has unequal masses which amplifies the effect of precession on the waveform. 
% See Ossokine paper: https://inspirehep.net/literature/1343311

The impact of sizable precessing spins as subsumed in the effective
precession parameter $\chi_p$ is mainly visible in strong modulations in the inertial frame waveform
as seen in the modes shown in the top left panel of Fig.~\ref{fig:precessing_waveform_frames_phase}.
The strength of these modulations increases as the binary becomes more unequal in mass-ratio and
becomes more prominent for an inclination angle tending towards edge-on rather than face-on inclination. 
Precession also has more subtle effects on the phasing which are not visible in
Fig.~\ref{fig:precessing_waveform_frames_phase}.

\paragraph{Eccentricity}
% An eccentric binary radiates \acp{GW} at all higher harmonics with frequencies being integer multiples of
% the dominant \ac{GW} frequency for a non-eccentric binary. This strong enhancement of the \ac{GW} radiation 
% shortens the time to coalescence in a major way for highly eccentric binaries with initial 
% eccentricities close to unity. % But see Fig 9 of https://arxiv.org/pdf/2112.06952.pdf
% Fig 1 of https://arxiv.org/pdf/2209.03390.pdf indicates that ecc shortens the time to coalescence
The waveform emitted from a highly eccentric binary is highly deformed and resembles bursts of radiation
compared to a quasi-circular inspiral. The binary alternates orbital passages by the periastron (where the two 
bodies have their closest approach) and the apastron (where they are the furthest apart within one orbit).
Contrary to quasi-circular binaries, waveforms emitted from eccentric binaries (even if they have aligned spins)
have oscillatory amplitudes and frequencies. The periastron of the orbit is also no longer fixed and instead 
precesses or advances during the inspiral. The emitted gravitational radiation is enhanced for highly eccentric 
binaries and radiates peak power in the higher harmonics in the inspiral regime. The increased \ac{GW} emission also 
radiates away eccentricity during the inspiral of an eccentric binary so that the orbit eventually circularizes~\cite{Peters:1964zz}. Therefore, as the binary approaches the merger, the higher 
harmonics become subdominant. % see e.g. Fig 4 of https://arxiv.org/pdf/1807.07163.pdf
The time to coalescence for eccentric binaries is shortened compared to quasi-circular binaries. 
A radial phase parameter $l$ fixes where in the orbit a binary finds itself at a reference frequency and so 
this parameter will shift the oscillation pattern in the waveform in time.
% perhaps cite PoissonWillGravity

% see scripts/seobnrv5phm_dynamics_and_modes_plots.py

% see scripts/seobnrv5phm_dynamics_and_modes_plots.py

\subsection{Numerical relativity simulations}
\label{ss:NR}

As a coalescing binary approaches merger, the orbital velocity reaches a sizable fraction of the 
speed of light. In this regime the \ac{PN}\ expansion breaks down and the solution of Einstein's 
equations can only be approximated by full numerical simulations. Simulations of the merger of \ac{BBH}s
have become routine since the breakthroughs by several independent groups in 2005~\cite{Pretorius:2005gq,Campanelli:2005dd,Baker:2005vv}.
The Einstein equations formally depend on the ten degrees of freedom in the metric tensor $g_{\alpha\beta}$
\footnote{Here we do not assume a weak perturbation around Minkowski space as in Sec.~\ref{sec:Newtonian_binary}, but a general metric.}.
The introduction of a coordinate system (usually one time-like and three space-like coordinates) allows us to separate
Einstein's equations into six dynamical and four constraint equations of second order.
The dynamical equations describe the time evolution of the spatial metric tensor and its time derivative
(or, equivalently, the extrinsic curvature variable) given some initial data on an initial hypersurface
which in turn needs to satisfy the
elliptic constraint equations. To stably evolve the Einstein equations numerically it is important to choose 
a formulation of Einstein's equations which makes the dynamical equations hyperbolic of a strong enough degree
so that the system is well-posed~\footnote{Well-posedness requires that a unique solution exists and the solution 
depends continuously on the initial conditions.}.
This is satisfied by the BSSN and the generalized harmonic formulations~\cite{baumgarte_shapiro_2010,Alcubierre_2008}.
To fix the coordinate or gauge freedom four variables need to be specified. For instance, they can
describe the proper time of an observer's normal to the current spatial hypersurface (lapse function)
and how the timeline of an observer deviates from the normal to the hypersurface (shift vector).

Starting from a numerical solution of the constraint equations for a \ac{BBH} system,
one can evolve the system in time until the merger with the help of special techniques to deal with the curvature
singularity inside each BH: either the BH interior inside the \emph{apparent horizon} (a quasi-local definition 
of the BH horizon which is based on the spatial locations where outgoing light rays cannot escape the BH~\footnote{
Apparent horizons, when they exist, always lie in the interior of the globally-defined event horizon. NR simulations
rely on quasi-local quantities for the positions and properties of \acp{BH} in each spatial hypersurface.}) is
excised from the computational domain and the excision surface carefully follows the BH motion, or one can use
the so-called ``moving puncture'' method which places the singularity in-between points of the numerical grid. 
Special prescriptions for the gauge conditions are also used to avoid singularities.
The spatial derivative operators in the dynamical equations are usually discretized with high order finite
difference stencils or spectral methods. The spatial resolution is set by the size of the apparent horizons
of the BHs and therefore scales with the size of the smaller BH if the masses are unequal, while the total mass 
can be scaled out. Time evolution is performed by methods for numerically solving ordinary differential equations.
The spatial discretization scale then determines the size of the largest time step for which the system can be 
stably evolved. Furthermore, BBH simulations need to resolve both the small scales of the BH horizons, and
extract the GWs in the wave zone, sufficiently far away from the binary. Covering these disparate scales with a 
single uniform grid is infeasible; instead, practical simulations require the use of fixed or adaptive mesh refinement
methods. Errors due to extraction of the waves at finite distances from the source can be mitigated by
extrapolation methods or, ideally, by transporting the radiation to future null infinity.

A non-eccentric binary black hole coalescence can be divided into three regimes: (i) \emph{inspiral} where the orbital
velocity is low, $v \ll c$, and the system undergoes quasi-circular inspiral, (ii) \emph{merger} where the
two separate BH horizons combine to form a single distorted potato-like horizon and
orbital velocities reach $v \simeq 0.5 c$ or more in this highly nonlinear regime, and (iii) \emph{ringdown}, 
where the final black hole radiates away excess energy in terms of quasinormal modes (QNMs) and settles down 
to a stationary Kerr black hole solution. This last stage can be well described by BH perturbation theory. 
The emitted waveform is a chirp signal that peaks in amplitude and frequency at the merger, before decaying 
exponentially in the ringdown. The mass and spin (angular momentum) of the BHs can be determined numerically 
from their apparent horizon properties.
% The \ac{GW} frequency reaches the value of the frequency of the fundamental QNM of the final Kerr BH

For an equal-mass non-spinning BH binary, about 5\% of the total mass of the system is radiated away in GWs and
therefore the mass of the final Kerr black hole is about $M_f / M \simeq 95\%$ of the initial mass of the binary.
The final Kerr black hole of an equal-mass non-spinning binary turns out to have a spin of about $\chi_f \simeq 0.7$
acquired from remaining angular momentum of the binary that has not been radiated away during the coalescence.
For unequal mass binaries the radiated energy is lower than for the equal mass case.
% Maybe show graphically how Mf, af depend on q or eta, and spins?
In general, the dependence of the final mass and the final spin is a function of the initial binary masses 
and spins which is needed for waveform models that include a description of the merger and ringdown parts of the signal.
This information can be fitted from numerical relativity (NR) simulation data available over the binary parameter space.

NR simulations can produce highly accurate waveforms (with phase errors a fraction of a radian
required by data analysis applications), but the high computational cost limits the number of
simulations and their length in time which can be carried out in practice. Moreover, the cost of simulations also
varies strongly over the binary parameter space and, in particular, increases steeply the more unequal the mass-ratio 
and the higher the aligned spin are, based on the induced size of the apparent horizons to be resolved and the allowed time step.
A single simulation can take several months on a supercomputer while running in parallel on hundreds of CPU cores.
Over 4000 publicly available BBH NR simulations have been performed so far and have been collected in several catalogs 
for use in modeling and \ac{GW} data analysis~\cite{SXS:catalog,Mroue:2013xna,Boyle:2019kee,RIT:catalog,Healy:2017psd,Healy:2019jyf,Healy:2020vre,Healy:2022wdn,MAYA:catalog,Jani:2016wkt,BAM:catalog,Hamilton:2023qkv}.

\subsection{Quasi-normal modes}
\label{ss:QNMs}
% Maggiore 12.3

After the apparent horizons of the two individual \acp{BH} in a binary have merged, the \ac{BH} can be treated 
as a deformed Kerr \ac{BH} in linear perturbation theory.
It is therefore natural to seek to describe the gravitational radiation emitted by this object by metric 
perturbations of an analytical BH solution. The linear perturbations around a Schwarzschild black hole are 
described by the Regge-Wheeler (RW) and Zerilli equations and perturbations of a Kerr black hole by the 
Teukolsky equation~\cite{Maggiore:2018sht}. 
Perturbations are in general decomposed into spheroidal harmonics of spin-weight $-2$ 
which depend on quantum numbers $\ell, m$ and the (complex) mode frequency $\omega$.
In the Schwarzschild case, for a fixed $\ell$ mode, these equations are formally of the form
\begin{equation}
    \frac{d^2\phi}{dx^2}(\omega, x) + \left[ \omega^2 - V(x) \right] \, \phi(\omega, x) = 0\,,
\end{equation}
with $x \in \mathbb{R}$ a rescaled version of the Schwarzschild tortoise coordinate, $\phi(\omega, x)$ the Fourier transform with respect to time of the RW or Zerilli function, and $V(x)$ the RW or Zerilli potential. These potentials
have a peak in the vicinity of small and positive $x$ and fall off smoothly to zero as $x \to \pm \infty$.

Consider for a moment perturbations of a one-dimensional string with fixed endpoints. In this case, a general
perturbation can be expressed as a sum over normal modes $\phi_n(t,x) = e^{i\omega_n t}\psi_n(x)$, which form
a complete set. In contrast to the fixed (Dirichlet) boundary conditions used in the string example,
the metric perturbations of a BH are not describing a stationary system, but will either escape into the BH horizon
or towards infinity with radiation boundary conditions $\phi(\omega, x) \propto e^{i\omega |x|}$ as $x \to \pm\infty$.
We will obtain a discrete set of complex frequencies $\omega_\mathrm{QNM} = \omega_R + i \omega_I$ belonging 
to the so-called \emph{quasi-normal mode} (QNM) solutions of the system. 
% A rigorous treatment of QNMs relies on the Laplace transform in time instead of the Fourier transform.
The QNMs vanish in time as damped oscillations of the form
\begin{equation}
    e^{\omega_I t} \left[ a \sin(\omega_R t) + b \cos(\omega_R t) \right]\,,
\end{equation}
where $\omega_I < 0$.
For a Schwarzschild black hole, QNM solutions depend on the azimuthal quantum number $\ell$ which scales the frequency $\omega_R$, and the overtone index $n$ which increases $\omega_I$ and therefore determines how quickly the mode is damped in time. 
QNMs for Kerr also depend on the (magnetic) quantum number $m$. The Kerr QNM modes can co-rotate ($m>0$) 
or counter-rotate ($m<0$) with the Kerr \ac{BH}.
We are most interested in the least-damped QNM modes which will dominate the emitted \ac{GW} signal.
The least-damped mode for a non-spinning black hole has $\omega_{n=1,\ell=2} \approx (0.75 - i 0.18) c/R_S$,
with the Schwarzschild radius
$R_S = 2GM/c^2$. This yields a frequency $f \simeq 12\, \mathrm{kHz} (M_\odot/M)$ and a damping time 
$\tau = 1/|\omega_I| \simeq 55 \,\mu \mathrm{s} (M/M_\odot)$.

Perturbations of the Kerr metric have been analyzed in terms of curvature perturbations in the Newman-Penrose
null tetrad formalism which leads to the Teukolsky equation. The QNMs of Kerr BHs are denoted as 
$\omega_{n \ell m}$, with $\omega_{122}$ being the least-damped mode, now depend on the Kerr parameter 
of the background BH around which we are perturbing.

\subsection{Modeling the signal from inspiral to ringdown}
\label{sec:LEOBandPhenom}

% \emph{Describe EOB and Phenom: the basic strategy, and tricks used to include more and more effects.
% Also say a few words about quasinormal modes and the ringdown.
% Mention that binaries containing NS have no ringdown but a very complex post-merger}

% look at Buonanno & Sathyaprakash review for EOB: \cite{Buonanno:2014aza}
% look at Maggiore Vol II
% look at my text from Sec 3 of https://arxiv.org/pdf/2005.03745.pdf
% check refs given in Gadre:2022sed

% Already said in the NR section
% The signal of coalescing black hole binaries consists of an inspiral, a merger, and a ringdown stage.
% The early inspiral is well described by the \ac{PN}\ expansion while the merger signal requires full
% numerical relativity simulations. The ringdown can be described by black hole perturbation theory where
% the final black hole radiates away excess energy through the emission of quasi-normal modes

We next discuss two approaches that are used to construct accurate waveform models covering all
regimes of \acp{GW} emitted in a black hole binary coalescence: the phenomenological (Phenom) and
effective-one-body (EOB) frameworks.  Both combine theoretical understanding from \ac{PN}\ and 
perturbation theory with NR simulations, and both 
have also been modified to model waveforms from binaries which include NSs.
Another approach that builds a direct surrogate to numerical simulations is discussed in Sec.~\ref{sec:surrogate}.

In the data-driven Phenom and surrogate approaches, and to a lesser degree for EOB, 
the waveform is usually decomposed to simplify modeling.
This is informed by physical intuition about the class of binaries to be modeled, 
in particular taking into account the effect of certain parameters on the signal as described in 
Sec.~\ref{ss:signal_characteristics}.
For non-precessing binaries the dominant $(\ell, m) = (2, \pm 2)$ spherical harmonic mode of the 
GW signal can be well represented by its amplitude and phase $h(t) = A(t) e^{i \phi(t)}$.
If higher harmonics are included in the model the amplitude - phase decomposition can lead to 
zero-crossings which are hard to model and it is advantageous to extract the orbital motion of 
the two bodies by multiplying the complex modes with an exponential of half the phase of the 
$(2, 2)$ mode. In the resulting ``co-orbital frame''~\cite{Varma:2019csw}
the real parts of the modes become non-oscillatory and amplitude-like, while the
imaginary parts add some fine structure.
If the signals include precession effects, one can transform into a 
time-dependent reference frame in which the orbital plane of the binary is fixed. In this
``co-precessing frame''~\cite{Schmidt:2010it,OShaughnessy:2011pmr,Boyle:2011gg}, 
precession-induced modulation effects are significantly reduced. 
Both the motion of the reference frame and the signal in the precession-adapted frame need 
to be modeled.

\paragraph{Effective-one-body framework}
\label{par:EOB}

In the late inspiral the \ac{PN}\ approximation deteriorates and eventually breaks down as the orbital 
velocity becomes a sizable fraction of the speed of light. This difficulty can be addressed by the 
effective-one-body (EOB) framework~\cite{Buonanno:1998gg,Buonanno:2000ef}. In the following we qualitatively describe the 
main idea of EOB following Ref.~\cite{Maggiore:2018sht}.
% 14.2.1
We start by considering the equations of motion of a two-body system. Up to 2\ac{PN}\ order and assuming zero spins, no
radiation reaction force is present and the system is conservative. After introducing a particular set of coordinates (ADM coordinates)
we can write the equations of motion in terms of a Hamiltonian in the center-of-mass frame of the binary, using
relative position and conjugate momentum coordinates $(\vec q, \vec p)$. Due to the invariance of the Hamiltonian
under time translations and rotations, energy and angular momentum are conserved. We write energy as a function of
the angular momentum and a radial action variable which is an adiabatic invariant. % def in mechanics

The basic idea of EOB is to find a one-body problem in an external spacetime that reproduces the solution of the given two-body problem (written in a gauge invariant form). % eq 14.50
We can consider a test particle of reduced mass $\mu = m_1 m_2/(m_1 + m_2) = \eta M$ which moves along the geodesic of a
static spherically symmetric metric $g_\mathrm{eff}^{\mu\nu}$ written in terms of some effective coordinates. At the
Newtonian level this is the Schwarzschild metric which describes the gravitational potential of a mass $M = m_1 + m_2$.
The EOB line element is
\begin{equation}
    ds_\mathrm{eff}^2 = g_\mathrm{eff}^{\mu\nu} dx_\mathrm{eff}^{\mu} dx_\mathrm{eff}^{\nu}
                      = - A(R) dt_0^2 + B(R) dR^2 + R^2 (d\theta^2 + \sin^2\theta d\phi^2),
\end{equation}
where the functions $A(R)$ and $B(R)$ need to be determined.
The EOB approach then defines a natural mapping between the real two-body problem and the effective one-body problem. 
This can be motivated by a quantum mechanical analogy (the Bohr-Sommerfeld quantization condition of the hydrogen atom).
Adiabatic invariants corresponding to the principal and angular momentum quantum numbers can be directly related 
between the two descriptions, whereas the mapping of the energy is nontrivial. One arrives at the following result at 
2\ac{PN}\ order: % eq 14.64
\begin{align}
    A(R) &= 1 - \frac{2 M}{R} + 2\eta \left(\frac{M}{R}\right)^3\,,\\
    B(R) &= 1 + \frac{2 M}{R} + (4 - 6\eta) \left(\frac{M}{R}\right)^2\,,
\end{align}
along with an energy mapping. % $E_0 = \frac{E^2 - m_1^2 - m_2^2}{2M}$.
Therefore, within this approximation we can substitute the original two-body problem by an easier to solve effective
one-body problem and solve it for $\eta > 0$.

Considering the conservative dynamics of the system one finds that, as is the case for the Schwarzschild metric, the EOB metric has a horizon (because $A(R)$ has a simple zero at $R > 0$ for all $\eta$) and an \emph{innermost stable circular
orbit} (ISCO). Similarly, one can define the \emph{light ring}: the smallest possible (unstable) circular orbit for a massless
particle. For very unequal masses a small mass inspirals through a sequence of quasi-circular orbits until it reaches 
the ISCO and suddenly starts to plunge toward the central BH. In contrast, for comparable masses, there is no sharp
transition between inspiral and plunge.

The effect of the emission of GWs (first entering at 2.5\ac{PN}\ order) can be expressed as a radiation reaction force
$\hat {\mathcal{F}_\varphi}$ which can be added into the Hamiltonian equations of motion for the conservative dynamics. % Eqs 14.96 - 14.99
Written in polar coordinates in the orbital plane these equations read
\begin{align}
    \frac{dr}{d\hat t} &= \frac{\partial \hat H}{\partial p_r}\,,\\
    \frac{d\varphi}{d\hat t} &= \frac{\partial \hat H}{\partial p_\varphi}\,,\\
    \frac{d p_r}{d\hat t} &= - \frac{\partial \hat H}{\partial r}\,,\\
    \frac{d p_\varphi}{d\hat t} &= \hat {\mathcal{F}_\varphi}(r, p_r, p_\varphi)\,,
\end{align}
where
\begin{equation}
    \hat H(r, p_r, p_\varphi) = \frac{1}{\eta} \sqrt{1 + 2\eta \left[ \sqrt{A(R) \left( 1 + \frac{p_r^2}{B(r)} + \frac{p_\varphi^2}{r^2} \right)} - 1 \right]}\,,
\end{equation}
and $r = R/M$, $\hat t = t/M$.

After solving these equations of motions numerically to obtain the inspiral and plunge dynamics, we can compute the 
emitted GWs via the quadrupole formula. To complete the waveform, a merger part needs to be smoothly attached after 
the plunge. This ringdown waveform consists of a superposition of damped sinusoids which results from the oscillation 
of the final Kerr BH in its quasi-normal mode oscillations. This yields a qualitatively correct waveform of the 
inspiral, merger and ringdown.
The above EOB construction can be extended up to higher \ac{PN}\ orders (4\ac{PN}\ is used is practical models) and can be generalized
to include spin effects. In practice, EOB models also add corrections for neglected non quasi-circular effects and
introduce unknown higher order \ac{PN}\ terms in the Hamiltonian which are then tuned or calibrated to NR simulations to 
increase the accuracy of the model.

EOB models from two different groups have been used for LVK analyses. This includes the ``SEOBNR''-type models
for aligned spin and the dominant mode~\cite{Taracchini:2012ig,Taracchini:2013rva,Bohe:2016gbl}, models including 
higher harmonics~\cite{Cotesta:2018fcv,Pompili:2023tna} and precession~\cite{Pan:2013rra,Babak:2016tgq,Ossokine:2020kjp,Ramos-Buades:2023ehm},
as well as prescriptions for tidal models~\cite{Steinhoff:2016rfi,Matas:2020wab}, and models incorporating further optimizations~\cite{Devine:2016ovp,Knowles:2018hqq}.
Of these, the most complete recent model for BBHs is ``SEOBNRv5PHM''~\cite{Ramos-Buades:2023ehm}
for generically precessing binaries which includes higher harmonics.
``TEOBResum''-type models have been constructed for non-precessing binaries including tidal effects~\cite{Nagar:2018zoe,Nagar:2018plt,Akcay:2018yyh}, and for precessing binaries~\cite{Akcay:2020qrj,Gamba:2021ydi}.

% EOB AEI
% ===
%
% v1, v2, Taracchini:2012ig,Taracchini:2013rva
% v3 (v3 prec?) Pan:2013rra,Babak:2016tgq,Steinhoff:2016rfi
% v4 (v4, v4HM, v4P, v4PHM, v4_ROM, v4_ROM_NRTidal/v2, v4_ROM_NRTidalv2_NSBH, v4HM_ROM, v4T, v4T_surrogate)
%   Bohe:2016gbl,Cotesta:2018fcv
% [v5]
%
% NSBH: Matas:2020wab
%
% which TEOBResumS models to cite? https://inspirehep.net/literature/1777194 ?
%  https://docs.ligo.org/lscsoft/lalsuite/lalsimulation/group___l_a_l_sim_i_m_r_t_e_o_b_resum_s__c.html
% TEOBResum_ROM
% TEOBResumS
% Check LAL wf table for references; ask Sergei / Vijay if I'm missing anything important
%
%
% Phenom
% ======
%
% ancient (A, B, C)
%
% IMRPhenomD, Pv2, Pv3, HM, XAS, XHM, XPHM, NRTidal / v2, NSBH
% IMRPhenomT, THM, TP, TPHM
%  -- just the families:
% old (D, HM, Pv2, Pv3); NRTidal
% current (X: XAS, XHM, XP, XPHM); NRTidal
% TD (T, THM, TP, TPHM)

\paragraph{Phenomenological waveform model framework}
\label{par:Phenom}

The class of phenomenological models takes a very practical approach at constructing models of the GW
signal emitted from compact binaries. This approach matches a post-Newtonian inspiral to a phenomenological
description of the merger and ringdown which aims at encapsulating the waveform morphology from
numerical-relativity simulations. The method posits ansatz functions in the Fourier domain for different
regimes of the coalescence along with a polynomial dependence on mass-ratio and spins.
Hybrid inspiral - NR waveforms are created by stitching together NR
simulations with inspiral waveforms produced by an uncalibrated EOB model (evaluated at the same binary parameters
as the respective NR simulation). These ``hybrids'' are created for all mass-ratios and aligned spins where
accurate NR simulations are available and are used to calibrate the model.
The resulting phenomenological models focus on computational efficiency by avoiding the numerical solution
of systems of complicated ODEs (as used in EOB) and instead provide analytical expressions that can be directly evaluated.

%The main characteristics of the approach are to write the waveform as a combination of parametrized ansatz functions for inspiral, late inspiral  and merger, and ringdown pieces, 

Traditionally, phenomenological models have been constructed in the Fourier domain~\cite{Ajith:2009bn}, so that inner 
products between waveforms used in data analysis can be directly computed. Therefore, modeling starts by 
considering the morphology of waveforms in the Fourier domain in contrast to the EOB time domain method.
Ansatz functions are picked in three frequency regimes: inspiral, an intermediate late-inspiral regime,
and merger-ringdown. In the following we describe some elements of the ``IMRPhenomXAS'' waveform
model~\cite{Pratten:2020fqn}.

%%% example IMRPhenomXAS
In the inspiral regime the phase model is written as the 3.5\ac{PN}\ TaylorF2 approximant plus higher order
``pseudo-\ac{PN}'' terms
\begin{equation}
  \varphi_\mathrm{Ins}(f) - \varphi_\mathrm{TF2}(f) = 
  %1/\eta \sum_{i=0}^5 \sigma_i f^{(i+3)/3}\,,
  \frac{1}{\eta} \left[ \sigma_0 + \sum_{i=0}^4 \sigma_{i+1} f^{(i+3)/3}  \right]\,,
\end{equation}
where the $\sigma_i$ are unknown \ac{PN}\ coefficients that have not yet been calculated. % by \ac{PN}\ theory.
The $\sigma_i$ are determined by fitting a rescaled difference between frequency derivatives of
hybrid waveform and TaylorF2 phases $f^{-8/3} (\varphi'_\mathrm{Hybrid}(f) - \varphi'_\mathrm{TF2}(f))$.
A similar \ac{PN}-based ansatz is made for the inspiral amplitude
\begin{equation}
    A_\mathrm{Ins}(f) = A_\mathrm{PN} + A_0 \sum_{i=1}^3 \rho_i (\pi f)^{(6+i)/3}\,,
\end{equation}
with ``pseudo-\ac{PN}'' coefficients $\rho_i$, again determined from the hybrids.

As discussed in Sec.~\ref{ss:QNMs} the ringdown can be described by BH perturbations in the time domain.
It is therefore instructive to consider the Fourier transform of a damped oscillation
$h(t) = \Theta(t) \, e^{2\pi t (i f_\mathrm{RD} - f_\mathrm{damp})}$, where $\Theta(t)$ is the Heaviside
step function. This leads to a Lorentzian function (Cauchy distribution) for the phase derivative
%$\frac{d \mathrm{arg} \tilde h(f)}{df} \propto \frac{f_\mathrm{damp}}{(f - f_\mathrm{RD})^2 + f_\mathrm{damp}^2}$,
and for the square of the Fourier domain amplitude
$|\tilde h(f)|^2 \propto 1 / [(f - f_\mathrm{RD})^2 + f_\mathrm{damp}^2]$. 
Since the physical waveform should fall off faster than $1/f$ due to its smoothness, the
ansatz for the merger-ringdown amplitude multiplies the Lorentzian by a decaying exponential
\begin{equation}
  A_\mathrm{MR} = \left[ \frac{a_R (f_\mathrm{damp}\sigma)}{(f - f_\mathrm{RD})^2 + (f_\mathrm{damp}\sigma)^2} \right]
                  e^{-\lambda (f - f_\mathrm{RD}) / (f_\mathrm{damp}\sigma)}\,,
\end{equation}
with fitting parameters $a_R, \sigma, \lambda$.
The merger-ringdown model for the phase derivative consists of a Lorentzian plus terms with negative powers
in frequency to capture to capture steep gradients
\begin{equation}
    \eta \varphi'_\mathrm{RD} = c_\mathrm{RD} + \sum_i^n c_i^{-p_i} + \frac{c_0 a_\varphi}{(f - f^\mathrm{RD})^2 + f^2_\mathrm{damp}}\,,
\end{equation}
again introducing a number of new parameters $c_\mathrm{RD}, c_i, a_\varphi$.
Moreover, the ringdown model depends on a fitting formula that predicts the final BH's mass and spin from the
progenitor BHs. From these one can determine the ringdown and damping frequencies $f_\mathrm{RD}$ and $f_\mathrm{damp}$.

Lastly, an intermediate regime bridges the gap between the inspiral and merger-ringdown parts of the model and
uses ansatz functions inspired by the neighboring regimes. For a description of the model in the intermediate regime
and details about where in frequency the models in the three frequency regions are situated and how they are matched
together, see~\cite{Pratten:2020fqn}.

With the amplitude and phase ansatz functions defined, one can fix the model coefficients for any given point
in parameter space where an accurate NR simulation is available and a hybrid waveform has been computed.
To obtain a practical model one posits a polynomial or rational dependence of the model coefficients
over parameter space, motivated by the mass-ratio and spin dependence of \ac{PN}\ approximants. The model uses
hierarchical fits in symmetric mass-ratio, an effective aligned spin parameter, and the difference between
the BH's aligned spins.
%%% end example IMRPhenomXAS

%%%
This approach leads to fast, closed form models that are convenient to use for data analysis.
Examples include the aligned-spin BBH models up to ``IMRPhenomD''~\cite{Ajith:2009bn,Santamaria:2010yb,Khan:2015jqa}
and the recent ``IMRPhenomXAS''~\cite{Pratten:2020fqn}. Newer models are available in versions 
that include higher-order modes~\cite{London:2017bcn,Garcia-Quiros:2020qpx}.
Models for precessing binaries require additional modeling of the time-dependent rotation of the
reference frame, in a similar way as for EOB models. To use this prescription in the Fourier
domain, additional approximations are introduced via the stationary phase approximation~\cite{Hannam:2013oca}.
An extension to two-spin dynamics in given in~\cite{Khan:2018fmp}.
A more accurate solution of the frame dynamics for generically precessing binaries is used in ``IMRPhenomXPHM''~\cite{Pratten:2020ceb}. A recent model includes tuning against numerical waveforms in the
precessing sector~\cite{Hamilton:2021pkf}.
Further models incorporate tidal effects for BNSs~\cite{Dietrich:2017aum,Dietrich:2019kaq}
or NS - BH binaries~\cite{Thompson:2020nei}.
Recently, a family of phenomenological models has been created in the time domain~\cite{Estelles:2020osj,Estelles:2020twz,Estelles:2021gvs}
in an effort to avoid some of the approximations used by Fourier domain phenomenological models.

% Construction:
% * Look at morphology of waveform data pieces in Fourier domain
% * Pick ansatz functions in frequency (motivated by theory) to describe inspiral, merger and ringdown morphologies
% * Posit a polynomial dependence of coefficients over parameter space
% * Add higher order unknown terms in inspiral (EOB or \ac{PN}) for calibration
% * Fit coefficients to hybrid \ac{PN}\-NR waveforms over the parameter space
% * Obtain a fast, closed form model
% Analytical descriptions used:
% * \ac{PN}\ inspiral phasing, and amplitude (SPA)
% * \ac{PN}\ frame Euler angles in IMRPhenomPv2

% [Ajith+07, 08, Santamaria+10, Hannam+…MP13,
% Khan+…MP15, Khan+18, London+18] + PhenomX models

% IMRPhenomXAS~\cite{Pratten:2020fqn}: https://arxiv.org/pdf/2001.11412.pdf
% IMRPhenomXHM~\cite{Garcia-Quiros:2020qpx}: https://arxiv.org/pdf/2001.10914.pdf
% IMRPhenomXPHM~\cite{Pratten:2020ceb}: https://arxiv.org/pdf/2004.06503.pdf

% Validity of common modeling approximations for precessing binary black holes with higher-order modes: https://arxiv.org/pdf/2001.10936.pdf \cite{Ramos-Buades:2020noq}
% assesses the accuracy of two main approximations commonly used to construct phenomenological IMR waveforms from precessing BBHs: (i) identification between aligned-spin and co-precessing waveforms, (ii) inclusion of higher-order aligned-spin modes in the construction of ap- proxmiate precessing modes
% need to take into account mode asymmetries and mode-mixing to significantly improve accuracy of prec BBHs where HMs contrubute

\subsection{Reduced-order and surrogate models}
\label{sec:surrogate}

% \textbf{Possible figures:}
% \begin{itemize}
%   \item EIM sketch
%   \item data piece over time (and parameter) for the two surrogate methods
% \end{itemize}

% \textbf{Topics to discuss:}
% \begin{itemize}
%   \item [x] orthonormal bases: Gram-Schmidt, SVD, greedy basis
%   \item [x] empirical interpolation
%   \item [x] fitting and interpolation methods (greedy polynomial fits, GPR, splines, ML); cross-validation
%   \item [x] time-domain vs Fourier domain surrogates
%   \item [x] mention some classes of widely used models: NR surrogates, EOB ROMs
%     - construction, typical accuracy and speed
%   \item [ ] mention / forward reference ROQs if discussed in PE section (mentioned, but not yet discussed in PE section)
% \end{itemize}

Over the past decade, data driven models of GWs have come to prominence in \ac{GW} 
data analysis with the promise of delivering high accuracy and fast evaluation speeds.
Reduced-order and surrogate modeling aims at creating an efficient and accurate \emph{surrogate}~\footnote{
``Surrogate model'' is the more general term, while ``reduced-order model'' implies the use of truncated
basis expansions in the model construction.} or substitute model for a given set of gravitational waveform data, 
with the help of numerical methods for decomposing, compressing, and fitting the original data.

\paragraph{Waveform data set and data pieces.}
A data set $\{h(\vec t; \vec\lambda_i)\}_{i=1}^N$ can be generated from
a \ac{GW} model $h(t; \vec\theta)$\footnote{
  In the following we assume time domain waveforms, but surrogates can also be constructed in the frequency domain.
} 
(or a set of NR waveforms)
at parameter values $\mathcal{T} = \{\vec\lambda_i\}_{i=1}^N$ on a time grid $\vec t = \{ t_j \}_{j=1}^m$.
The input data is then a set of waveforms given over the space of relevant intrinsic binary parameters
$\vec\lambda$ (a subset of the set of all physical parameters $\vec\theta$). The values $\mathcal{T}$
can form a grid or, more generally, can be scattered data points in the domain of the original model.
% \paragraph{Data pieces.}
Each waveform in this input data set is decomposed into a number of \emph{data pieces} $f_k(t; \vec\lambda)$
chosen in such a way as to obtain as simple a function of time and binary parameters as possible
as discussed in Sec.~\ref{sec:LEOBandPhenom}.

\paragraph{Reduced basis.}
Each waveform data piece $f(t; \vec\lambda)$ in a surrogate model aims to approximate its 
discrete training data set $\mathcal{H} = \{f(\vec t; \vec\lambda_i)\}_{i=1}^N$ as closely as possible.
%$\mathcal{H} = \{f(\vec t; \vec\lambda_i)\}_{i=1}^N$ at corresponding parameter values $\mathcal{T} = \{\vec\lambda_i\}_{i=1}^N$
%with the training data given on a (usually equally spaced) time grid $\vec t = \{ t_j \}_{j=1}^m$.
Usually, one first computes a discrete orthonormal basis $\{ \vec B_i \}_{i=1}^N$,
$\langle \vec B_i, \vec B_j \rangle = \delta_{ij}$
with elements $\vec B_i = B_i(\vec t) = (B_i(t_1), \dots, B_i(t_m))^T$ from the training data,
with $\langle \cdot, \cdot \rangle$ the Euclidean inner product on $\mathbb{R}^m$.
% Usually one assumes the Euclidean inner product on $\mathbb{R}^m$,
% $\langle f(\cdot; \vec\lambda_1), f(\cdot; \vec\lambda_2) \rangle =
% \sum_{j=1}^m f(t_j; \vec\lambda_1) f(t_j; \vec\lambda_2) \, \Delta t$ if the data piece is real valued.
Often a \emph{reduced basis} with $\mathrm{dim} \, \vec B = n < N$ is computed which amounts to a
truncated basis expansion 
\begin{equation}
	%\vec f(\vec\lambda) = f(\vec t; \vec\lambda) \approx \sum_{i=1}^{n} c_i(\vec\lambda) \vec B_i,	
  \vec f(\vec\lambda) = f(\vec t; \vec\lambda) \approx \sum_{i=1}^{n} c_i(\vec\lambda) B_i(\vec t)\,,	
\end{equation}
with $n$ chosen according to a desired accuracy threshold. The basis expansion coefficients are given by
\begin{equation}
	c_i(\vec\lambda) = \langle \vec f(\vec\lambda), \, \vec B_i \rangle\,.	
\end{equation}
Two methods have been used in the literature to build a reduced basis for a data piece $f$:
(i) a greedy algorithm which at step $k$ adds waveform data at the point in parameter space where the
data piece vector has the highest projection error $\lVert f(\cdot; \vec\lambda) - \mathcal{P}_k f(\cdot; \vec\lambda) \rVert$; here, $\mathcal{P}_k f(\cdot; \vec\lambda)$ is the orthogonal projection of the data piece
onto the span of the basis $\{ \vec B_i \}_{i=1}^k$ constructed in the first $k$ steps of the algorithm
(see Appendix A of \cite{Field:2013cfa} for details),
or (ii) a method where the waveform data is suitably downsampled in the independent variable, arranged 
in a matrix with elements $\mathcal{F}_{ij} = f(t_j; \vec\lambda_i)$ and the (truncated) singular value decomposition~\cite{GolubVanLoan} $\mathcal{F} = U^T \Sigma V$ is used to compute the basis~\cite{Purrer:2014fza,Purrer:2015tud}.
If the training set is sufficiently dense and the data are smooth, waveforms inside the domain covered 
by the training set, but not part of the training set itself will be well approximated by the reduced basis.

% The first approach requires some care to ensure that the Gram-Schmidt orthonormalization process is carried
% out in a numerically stable way, but has the added advantage that it selects a subset of the most disparate
% waveforms across parameter space. One would typically expect these selected parameter locations to lie as far
% away from each other as possible, near the boundary of the parameter space domain. If, instead clusters of
% points are observed in the interior of the domain this may indicate that the data is lacking smoothness and
% may be hard to model.
%
% On the other hand, the SVD is a standard matrix decomposition with readily available implementations,
% but is fairly expensive for large data sets. Therefore, it is crucial to first suitably downsample the training set in time, to reduce the rank of the matrix $\mathcal{F}$ (see~\cite{Purrer:2014fza} for a discussion
% of computational cost of the SVD and greedy basis methods).

% Comment on the computational complexity of these algorithms:
% - count for greedy alg
% - SVD (n^3) -- look up CQG paper and refs therein
% refer to Field et al PRX, my CQG, another Field paper on modified GS?
% -- this is discussed in Sec. 7 of Purrer:2014fza

%This can be achieved by Gram-Schmidt orthonormalization.

% EIM p. 6ff PRX, App B PRX

\paragraph{Surrogate construction.}
Given a reduced orthonormal basis $B_i(\vec t)$ for a waveform data piece $f(t; \vec\lambda)$ 
two contrasting approaches have been used to build a surrogate for this piece by fitting or 
interpolating coefficients over parameter space:
(i) one may directly fit or interpolate the coefficients $c_i(\vec\lambda)$\footnote{
Note that, in contrast to method (ii), sparseness in time has already been imposed before basis construction.} 
as in Ref.~\cite{Purrer:2014fza,Purrer:2015tud}, or
(ii) the problem can be recast as interpolation in time by selecting a ``good'' set of sparse \emph{empirical}
time nodes $\{ T_1, \dots, T_n \}$ from the full time grid such that the resulting interpolant recovers the 
waveform data piece exactly at the $T_i$ while minimizing the interpolation error. 
As discussed in Ref.~\cite{Field:2013cfa} this is achieved by the so-called 
\emph{empirical interpolant}~\cite{maday2007general,chaturantabut2010nonlinear} (EI)
\begin{equation}
  \mathcal{I}_n [f] (\vec t; \vec \lambda) = \sum_{j=1}^n \mathcal{B}_j(\vec t) \, f(T_j; \vec\lambda)\,,	
\end{equation}
where
$\mathcal{B}_j(\vec t) = \sum_{i=1}^n B_i(\vec t) \left(V^{-1}\right)_{ij}$
and
$V_{ij} = B_j(T_i)$, $V \in \mathbb{R}^{n \times n}$.
The EI time nodes are chosen such that $V$ has a small condition number.
This can be accomplished with an algorithmic complexity of $\mathcal{O}(n^3)$
as shown in App. A of Ref.~\cite{antil2013two}.
%The original EIM method accomplishes this with an algorithmic complexity of $\mathcal{O}(n^4)$,
%but can be improved to $\mathcal{O}(n^3)$ using the algorithm described in Appendix A of Ref.~\cite{antil2013two}.
With this transformation in hand, the data piece $f(\cdot; \vec\lambda)$ then needs to be fitted at each of
the EI time nodes $T_j$.

% fits (poly / splines / GPR / ML)
% Fit vs interpolate
% briefly summarize Tp-splines - https://github.com/mpuerrer/TPI/blob/master/doc/doc.pdf
% Look at refs in https://arxiv.org/pdf/1909.10986.pdf
\paragraph{Fitting and interpolation methods.}
A variety of methods have been used to predict expansion coefficients $c_i(\vec \lambda)$ or 
the waveform data piece at a fixed EI node $f(T_j; \vec\lambda)$ at unknown parameter space values 
$\vec\lambda$ given training data. See Ref.~\cite{Setyawati:2019xzw} for an introductory review. 
Interpolation methods enforce that the interpolant coincides with the given data at the training set 
points, while a fitting method provides no such guarantee and minimizes the residual between the 
given data and e.g.\ a polynomial ansatz made by the method.
In low dimensional parameter spaces interpolation by tensor-product splines~\cite{deBoorSplines}
is straightforward and reliable~\cite{Purrer:2014fza,Purrer:2015tud} whenever a large enough training set 
can be generated on a regular grid. A combination of radial basis functions with monomials has been used in Ref.~\cite{Pathak:2024zgo} which does not require a grid.
% This amounts to determining a (usually cubic) spline in each parameter space dimension
% threading through the points of the Cartesian product of one-dimensional grids which define the training set.
For moderately large parameter spaces with dimensions $\gtrsim 5$ %regular grids are no longer feasible and 
methods for fitting scattered data must be used. This includes polynomial regression with feature selection such as the greedy forward stepwise regression method with cross-validation error estimates described in Appendix A of Ref.~\cite{Blackman:2017dfb},
Gaussian process regression (GPR)~\cite{rasmussen:williams:2006} which can also provide a measure of uncertainty
for each fit~\cite{Doctor:2017csx,Lackey:2018zvw,Varma:2018mmi,Varma:2019csw}, but is costly for large training sets ($\mathcal{O}(N^3)$), and, more recently, neural networks~\cite{goodfellow2016deep} which have been shown to handle problems of moderate dimensionality with good accuracy and are fast to evaluate~\cite{Khan:2020fso,Schmidt:2020yuu,Thomas:2022rmc}.

% SF: the paper the describes/introduces the forward stepwise greedy (with hyperparameter tuning using cross-validation):
% https://journals.aps.org/prd/pdf/10.1103/PhysRevD.95.104023  = Blackman:2017dfb
% all the precessing models use the forward stepwise greedy.  The aligned-spin hybrid model uses GPR
% The greedy forward stepwise method is especially useful for larger parameter spaces, such as for precessing
% binaries~\cite{Blackman:2017dfb,Blackman:2017pcm,Varma:2019csw,Gadre:2022sed}. There, the intrinsic parameter space is 7-dimensional, comprised by the mass-ratio and three
% Cartesian components of each spin vector. For instance, a multi-variate cubic polynomial in 7 dimensions leads to
% a large number of $\binom{7 + 3}{3} = 120$ terms and the greedy method iteratively picks the features which decrease the residual the most at each step and stops when the fitting error is deemed sufficiently small.
% Notably, this method does not pick terms according to t-tests on the coefficients as in the standard forward stepwise method which has been criticized in the literature.
% check which NRsur use GPR w SF: aligned

% **CHECK:** do we want to list publicly available codes implementing some of these methods: rompy, greedycpp, romgw -- ask SF for refs

\paragraph{Available surrogate models.}
% Discuss these and cite some important papers
Surrogate models have been built for dominant mode aligned-spin binaries~\cite{Field:2013cfa,Purrer:2014fza,Purrer:2015tud,Lackey:2018zvw},
models including higher harmonics~\cite{Varma:2018mmi,Cotesta:2020qhw} and precession~\cite{Blackman:2017dfb,Blackman:2017pcm,Varma:2019csw,Gadre:2022sed}.
% % Repeats what was discussed earlier
% For dominant mode aligned-spin binaries a decomposition in terms of amplitude or phase leads to efficient and
% accurate surrogate models~\cite{Field:2013cfa,Purrer:2014fza,Purrer:2015tud,Lackey:2018zvw}. If higher harmonics
% are included it has been advantageous to extract the binary's orbital phase from the spherical harmonic modes~\cite{Varma:2018mmi,Cotesta:2020qhw}.
% Precessing binaries require transformation into special adapted reference frames to simplify the
% structure of the waveform~\cite{Blackman:2017dfb,Blackman:2017pcm,Varma:2019csw,Gadre:2022sed}.
%
% EOB; TD vs FD surrogates
Many surrogate models (or ROMs) have been built for EOB models, see Sec.~\ref{par:EOB}, which are in general significantly 
slower to evaluate than phenomenological models, see Sec.~\ref{par:Phenom}, limiting their applicability for data analysis purposes.
Similar to phenomenological models EOB surrogates tend to be built directly
in the frequency domain~\cite{Purrer:2014fza,Purrer:2015tud,Lackey:2018zvw,Cotesta:2020qhw}.
%  to avoid having
% to deal with conditioning of the time domain signal and subsequent discrete Fourier transform.
% Frequency domain surrogates are therefore usually a bit faster and easier to use than their time domain
% equivalents, but some physical effects (precession, eccentricity) are harder to model in the frequency domain without further approximations and thus frequency domain surrogates tend to be restricted to binaries with aligned spins.
%
% NR surrogates
Over the past several years surrogate models of NR simulations have been successfully
constructed including effects of precession, higher harmonics, memory, and eccentricity~\cite{Blackman:2017dfb,Blackman:2017pcm,Varma:2019csw,Varma:2018mmi,yoo2023numerical,Islam:2021mha}. 
As NR simulations are extremely costly and sparse, these studies have worked with carefully designed 
training sets to keep the number of simulations small without sacrificing high accuracy. Naturally, 
their parameter space coverage is more restrictive than for semi-analytic models, but where they are 
available, NR surrogates are arguably more accurate than models based on semi-analytical techniques.
Surrogates have also been constructed for remnant properties of BH mergers (final mass and spin, 
as well as recoil velocity)~\cite{Varma:2018aht}.

% Speed and accuracy
The evaluation speed of surrogate models is typically several orders of magnitude faster than 
it took to generate one of the training space waveforms; the speedup is especially high for
NR surrogates. Surrogate models usually achieve accuracies of $10^{-3}$ (or better) 
in the \emph{mismatch}, a normalized and maximized inner product between waveforms,
(see Eq.~\eqref{eq:match} and the following discussion in Sec.~\ref{ss:overlap_and_match})
against original data which is in general sufficient for most \ac{GW} data analysis tasks for current
ground based interferometers, but see~\cite{Purrer:2019jcp} for estimated accuracy requirements 
for future ground-based \ac{GW} detectors.

\subsection{Comparison of waveform accuracy and efficiency} % (fold)
\label{sub:comparison_of_waveform_accuracy_and_efficiency}

\begin{figure}[tb]
  \centering
    % Plots generated in compare_waveforms.ipynb
    \includegraphics[width=.45\textwidth]{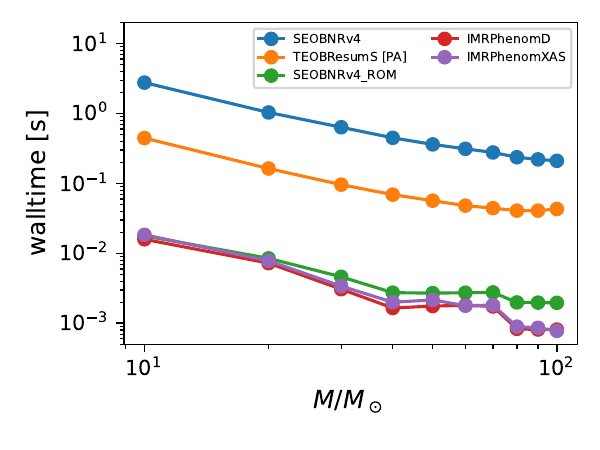}
    \includegraphics[width=.46\textwidth]{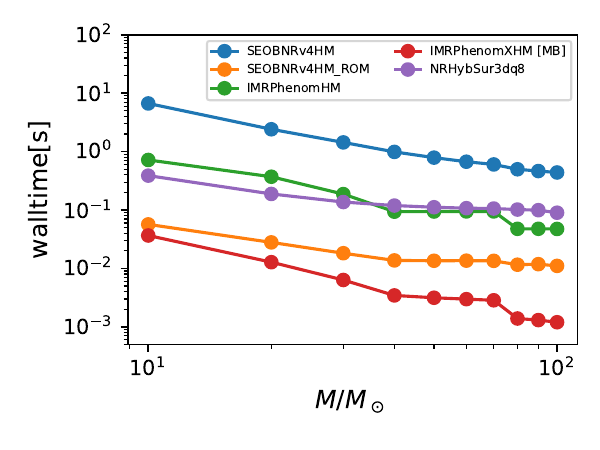}\\ % TD - Figures don't seem quite the same size.  Also there is quite a bit of padding around them, use tight layout? 
    \includegraphics[width=.45\textwidth]{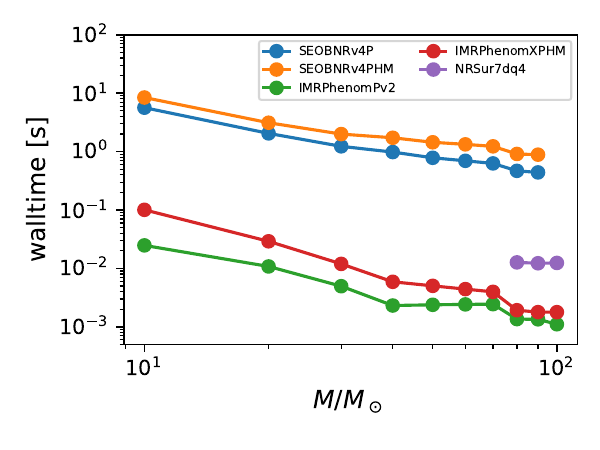}
  \caption{
      Evaluation times for Fourier domain polarizations of a number of BBH waveform models
      as a function of the total mass of the binary.
      \emph{Top Left:} aligned-spin models for the dominant \ensuremath{(2, 2)} mode,
      \emph{Top Right:} aligned-spin models including higher harmonics,
      \emph{Bottom:} models for precessing binaries. 
      The total mass was varied from 10\,\Msun\ to 100\,\Msun, while all
      other parameters were kept fixed (mass-ratio 1:3, all spin components set to zero
      except \ensuremath{\chi_{1z} = \chi_{2z} = 0.5} and, for models supporting precession, \ensuremath{\chi_{1x} = 0.5};
      the starting and maximum frequencies were set to 20 and 16384\,Hz, with the frequency spacing
      automatically determined by LALSuite's~\cite{lalsuite} \texttt{SimInspiralFD} function.)
      PA refers to the post-adiabatic approximation and MB to multi-banding techniques.
      The NR surrogate model in the precession panel is limited by the number of cycles available, which translates to
      a lower total mass for which it can be evaluated. In contrast, the NR surrogate in the middle panel 
      has been hybridized with \ac{PN}\ waveforms in the inspiral, allowing it to produce longer waveforms.
  % \kc{is this adapted from some published paper?}, MP: No, these plots come from some scripts I had written.
  % \kc{How can the surrogate in the middle panel go down to 10\,\Msun? Is this the hybrid version?} MP: Yes, indeed.
  % \kc{Use \ensuremath{M} of the total mass in the labels and also use consistent fonts with other plots.} Fixed total mass; leave font for later
  }
  \label{fig:IMR_walltimes}
\end{figure}
Waveform models aim to provide accurate approximations of the true underlying waveform (as a solution to
Einstein's equations in GR). Given that the most accurate merger waveforms come from a sparse set of
expensive NR simulations, but are too short to cover the frequency sensitivity of ground
based \ac{GW} detectors, practical models invariably need to combine analytical and numerical waveforms in addition
to some interpolation or fitting. Therefore, waveform models compound various sources of errors from their
constituent parts and model assumptions. Given these imperfections and the undesirable consequences of waveform
systematics in data analysis applications (missed signals, bias in binary parameters) it is crucial to study
how these models compare among themselves and against NR simulations in an effort to
further improve their fidelity.

% Virtually any paper which discusses a waveform model includes some accuracy comparisons.
A detailed comparison of the accuracy of current waveform models is beyond the scope of this review.
The following articles provide comparisons between a number of models in terms of mismatch or in
terms of a shift (\emph{bias}) in recovered parameters compared to the parameters of a synthetic signal
in parameter estimation. For comparisons of aligned spin models see Refs.~\cite{Khan:2015jqa,Bohe:2016gbl,Pratten:2020fqn,Varma:2018mmi}, for
models including precessing see Refs.~\cite{Khan:2018fmp,Varma:2019csw,Ossokine:2020kjp,Pratten:2020ceb,Hamilton:2021pkf}.
Tidal models are being compared in Refs.~\cite{Dietrich:2019kaq,Thompson:2020nei}.

% Waveform model papers including comparisons
%
% Purrer:2015tud: SEOBNRv2_ROM vs SEOBNRv2
% Cotesta:2020qhw: SEOBNRv4HM_ROM vs SEOBNRv4HM
% Varma:2018mmi  NRHybSur3dq8, SEOBNRv4HM vs NR hybrid waveforms Fig 6
% Varma:2019csw: NRSur7sd4, SEOBNRv3 vs NR
%
% Bohe:2016gbl: SEOBNRv4 vs IMRPhenomD, SEOBNRv2 and SEOBNRv4_ROM vs SEOBNRv4
% Cotesta:2018fcv: SEOBNRv4HM vs NR
% Babak:2016tgq: SEOBNRv3 vs NR
% Ossokine:2020kjp: SEOBNRv4P, SEOBNRv4PHM, IMRPhenomPv3 vs NR
% Nagar:2018zoe: TEOBResumS vs NR
%
% Khan:2015jqa: IMRPhenomD vs SEOBNRv2_ROM
% Khan:2018fmp: IMRPhenomPv2, IMRPhenomPv3, SEOBNRv3 vs NR
% Dietrich:2019kaq: IMRPhenomD, IMRPhenomD_NRTidal, IMRPhenomD_NRTidalv2, IMRPhenomPv2_NRTidal, IMRPhenomPv2_NRTidalv2, SEOBNRv4_ROM_NRTidalv2 vs TEOBResumS-NR hybrids
% Thompson:2020nei: PhenomNSBH, PhenomD, PhenomDNRT, SEOBNRv4NRT, SEOBNRv4T, LEA+ vs NR
% Pratten:2020fqn: IMRPhenomXAS, IMRPhenomD, SEOBNRv4 against NRHybSur3dq8 Fig 17
% (Garcia-Quiros:2020qpx  IMRPhenomXHM vs NR Fig 16)
% Pratten:2020ceb: IMRPhenomD, IMRPhenomXAS, IMRPhenomHM, IMRPhenomXHM, SEOBNRv4HM_ROM and IMRPhenomXPHM and NRSur7dq4 models with NRHybSur3dq8 Fig 7
% Hamilton:2021pkf: Phenom\ac{PN}\R, IMRPhenomXP, SEOBNRv4P, NRSur7dq4 vs NR Fig 22

For data analysis applications the efficiency of waveform models is very important. A comparison of the computational cost
of BBH models implemented in the LSC's Algorithm Library Suite~\cite{lalsuite} is shown in Fig.~\ref{fig:IMR_walltimes}.
The evaluation time increases steeply for low-mass binaries driven by the waveform length. Models including higher
harmonics and/or precession are in general slower that simpler models. Phenomenological and surrogate / ROM models 
are several orders of magnitude faster than time domain EOB models. Further acceleration for frequency domain models
is possible with multi-banding techniques as used by IMRPhenomXPHM~\cite{Vinciguerra:2017ngf,Garcia-Quiros:2020qlt},
while time domain EOB models can trade efficiency for accuracy through the post-adiabatic approximation~\cite{Nagar:2018gnk,Mihaylov:2021bpf}.

\newpage
\section{Statistical framework}
\label{sec:structure}

In this section we develop a statistical model of the \ac{GW} signal
and detector noise, showing how the noise curves from Sec.~\ref{sec:intro}
and the parameter-dependence of the waveforms from Sec.~\ref{sec:cbc}
combine to produce a natural geometric structure for the parameter space.
This structure gives rise to the typical correlations in the parameter
estimation results discussed in Sec.~\ref{sec:pe}, and gives us a method
for covering the parameter space with templates for a matched-filter search,
discussed in Sec.~\ref{sec:detection}.

Detector noise is generally treated as a stationary Gaussian process with a
correlation structure described by a given \ac{PSD}, resulting
in the Whittle likelihood function~\cite{Whittle:1957}.
The noise \ac{PSD} used may be estimated from data or predicted via the
physics of the detector(s) considered.
This noise model was initially applied to GW data analysis %by Finn\kc{Do we want to call him out by name? We haven't done that for anyone else so far},
studying
both detectability and parameter estimation for inspiral signals in~\cite{Finn:1992wt}.
The likelihood ratio of a signal under the Gaussian noise model
also gives rise to the classical matched filter for signal detection~\cite{Turin:1960}.

More recent works have examined relaxing some of the assumptions
%made below when deriving
behind the Whittle likelihood: for example, the \ac{PSD} may not be
known exactly, so its uncertainty may be marginalised over~\cite{Rover:2008yp,Littenberg:2014oda,Chatziioannou:2019zvs,Talbot:2020auc,Biscoveanu:2020kat},
or the \ac{PSD} may be jointly estimated along with the signal parameters~\cite{Chatziioannou:2021ezd,Plunkett:2022zmx}.

\subsection{Likelihood function for detector noise}
\label{ss:noise_likelihood}

The calibrated output of a \ac{gw} detector $d(t)=h(t)+n(t)$ contains the \ac{GW}
signal $h(t)$ and additive noise $n(t)$. The %stochastic part of the
noise is a
combination of fundamental, environmental and technical noise sources for the
particular detector; references in Sec.~\ref{ss:detectors} provide more details.
Mathematically, we can say that the noise is drawn from a stochastic process,
$n\sim N$ %which we can
%\kc{I do not understand this notation},
% TD : the notation means no more than what the text says, the noise is a realization of a process which we label N
described by a statistical model $H_N$ that gives us
the probability of observing a particular noise realisation $p(n(t)|H_N)$.

The noise model is chosen in the light of available information,
%In order to choose a %specific
% %noise model we need to consider the information available, %to us,
which we can quantify in terms of the moments of the process, %noise distribution,
written as expectation values integrating over the space of noise
realisations.\footnote{Thus, the moments are \emph{not} defined as
time-averages for a specific realisation $n(t)$, although knowledge of surrounding data may be used to inform the model, assuming ergodicity of the noise generating process.}
%\kc{are we drawing here a distinction between how they are defined and how they are in practice computed?} \td{Yes! The moments are mathematical objects as part of the noise model, numerical values can be estimated in various ways involving actual time series ..}
The first moment, %of the stochastic process,
the mean, may for simplicity be set to zero by redefining the zero of detector strain output:
\begin{equation}
E_N[n(t)] = \int n(t) p(n(t)|H_N) \, dn(t) = 0\,.
\end{equation}
The second moment of the stochastic process
\begin{equation}
E_N[n(t)n(t')] = \int n(t)n(t') p(n(t)|H_N) \, dn(t) \equiv C_{N}(t,t')\,,
\end{equation}
specifies a two-point kernel $C_{N}(t,t')$ known as the auto-covariance function.
%The expectation values above are taken by integrating over the
%space of noise realisations, and thus reflect our knowledge of the stochastic process that generates
%the noise.

We may continue in this fashion to higher-order moments to completely specify
the noise stochastic process.  % using the method of moments.
However, the common case where we only have constraints for the first and second
moments, i.e., the mean and variance, the probability distribution that maximises
the entropy, therefore representing the least informed distribution,
is the \emph{multivariate Gaussian} distribution. This motivates the use of
the Gaussian likelihood function in our analyses, although it is possible to derive
a likelihood with higher order information.
%However, at this point we want to make a concrete connection to the case at hand, and introduce some terminology.

\paragraph{Stationarity}
A stochastic process is said to be stationary if it is invariant under a translation
in time,
\begin{equation}
 p(n(t)|H_N) = p(n(t + \tau)|H_N)\,,
\end{equation}
which is satisfied if all its moments are time-invariant.
A less restrictive requirement is that of \emph{weak-sense stationarity}, where
only the first two moments are required to be invariant.  The mean may
be set to zero at all times, but for the second moment stationarity implies that
\begin{equation}
C_N(t, t') = C_N(0, t'-t) \equiv C_N(\tau)\,,
\end{equation}
i.e., the auto-covariance is a function only of the
time lag $\tau \equiv t'-t$.

\paragraph{Discretization}
The noise considered above as a continuous process $n(t)$ is a mathematical idealization
of actual detector output, which is a set of discrete samples from the generating process.
For data sampled at a finite set of $J$ times, $t_j$, separated by uniform
intervals $\Delta t$, the sampled values can be thought of as a vector $\vec{n}$
in a $J$-dimensional space, and the covariance function becomes a symmetric matrix
\begin{equation}
C_{jk} \equiv E[n_j n_k] = \int n_j n_k p(\vec{n}|H_N) \, d^J\vec{n}\,.
\end{equation}

Neglecting information from higher moments, the probability %distribution
of a particular noise realisation $\vec{n}$ is given by the multivariate Gaussian
distribution
\begin{align}
\label{eq:TimeDomainLikelihood}
 p(\vec{n}|C_N,H_N) = \frac{1}{\sqrt{\det 2\pi \mat{C_N}}}
 \exp\left[ -\frac{1}{2} \sum_j \sum_k  n_j (C_N^{-1})_{jk} n_k\right]\,.
\end{align}
This time-domain likelihood function requires a double sum over the data, thus
its computational cost scales as $J^2$. %as the square of the length of the data considered.
The likelihood may be much more efficiently evaluated %is much more efficient to express this
in the frequency domain using the discrete Fourier transform (DFT): %which can be
the frequency-domain noise realization is expressed via a matrix operation as
$\tilde{\vec{n}} = \mat{F}\vec{n}\Delta t$, where $F_{jk}=J^{-1/2}\exp(-2\pi ijk/J)$
is a (unitary) discrete Fourier transformation matrix.
Since $\mat{F}\mat{F^*}=\mat{I}$,
the time-domain likelihood can be rewritten as %in terms of $\tilde{\vec{n}}$ as
\begin{align}\label{eq:FreqDomainLikelihood}
 p(\vec{n}|C_N,H_N) &= \frac{1}{\sqrt{\det 2\pi \mat{C_N}}}
 \exp\left[ -\frac{1}{2} (\vec{n}^T\mat{F})(\mat{F^{-1}}\mat{C_N^{-1}}\mat{F})(\mat{F^{-1}}\vec{n})\right]\nonumber\\
 &= \frac{1}{\sqrt{\det 2\pi \mat{C_N}}} \exp\left[ -\frac{1}{2\Delta t^2} \vec{\tilde{n}}^T\mat{\tilde{C}_N^{-1}}\vec{\tilde{n}}^{*}\right]\,,
\end{align}
where $\mat{\tilde{C}_N}=\mat{F^{-1}}\mat{C_N}\mat{F}$. 
% \MP{should we not have $\mat{\tilde{C}_N}=\mat{F}^{-1}\mat{C_N}\mat{F}$ so that $\mat{\tilde{C}_N}\mat{\tilde{C}_N^{-1}} = I$?}.

If the noise is stationary, then $\mat{C_N}$ is a circulant matrix, and
$\mat{\tilde{C}_N}$ is diagonal, allowing us to replace
the double sum with a single sum over the diagonal elements.
We define the \emph{one-sided} noise \ac{PSD} $S_N(f)$ via the Wiener-Khinchin theorem
as the Fourier transform of the noise auto-covariance function,
%\jv{add citation for WK}
\begin{align}
 \frac{1}{2} S_N(f) &= \int_{-\infty}^{\infty} C_N(\tau) \exp(-2\pi i f\tau)\,d\tau\,,
\end{align}
allowing the diagonal of $\mat{\tilde{C}_N}$ to be written as $\tilde{C}_{ii} = \frac{J}{2\Delta t} S_N(f_i)$.
Therefore, the frequency domain likelihood can be written as
\begin{align}
 p(\vec{\tilde{n}}|S_N,H_N) &= \det\left(2\pi \mat{C_N}\right)^{-1/2}
  \exp\left[ - \frac{1}{2J\Delta t} \sum_i \frac{\tilde{n}_i {\tilde{n}}_i^*}{\frac{1}{2}S_N(f_i)} \right]
  \nonumber \\
  &= \det\left(2\pi \mat{C_N}\right)^{-1/2}
  \exp\left[ - \frac{1}{T} \Re \sum_{i>0} \frac{\tilde{n}_i {\tilde{n}}_i^*}{\frac{1}{2}S_N(f_i)} \right]\,,
\end{align}
using the fact that the original time-series $\vec{n}$ is real, implying
$\tilde{n}_{-i}=\tilde{n}_i^*$, and we have introduced $T = J \Delta t$.
%, and we can write the likelihood as a sum over positive frequency components as
%\begin{align}
% p(\vec{\tilde{n}}|S_N,H_N)
%\end{align}
%
If we further define the noise-weighted inner product
\begin{align}\label{eq:inner_product}
 \braket{\vec{a}|\vec{b}} \equiv 2 \Re \int_0^\infty \frac{\tilde{a}^*(f)\tilde{b}(f)}{\frac{1}{2}S_N(f)} \,df\,,
\end{align}
the noise likelihood can be written in terms of the frequency-domain noise realisation $\tilde n(f)$ as
\begin{equation}\label{eq:noise_likelihood_inner_product}
 p(\tilde n(f)|S_N,H_N) = \det\left(2\pi \mat{C_N}\right)^{-1/2}
 \exp \left( -\frac{1}{2}\braket{\vec{n}|\vec{n}} \right)\,.
\end{equation}
With this definition the space of data is formulated as an \textit{inner product
vector space}, which we will later use to access various (differential)-geometric
methods, for example considering distances between neighbouring signals.
See~\cite{Romano:2016dpx,LIGOScientific:2019hgc} for further discussion of the
derivation of the Gaussian likelihood and its assumptions.

%\jv{Say something about timescales for stationarity? Cite IAS work?}

\subsection{Likelihood for a signal model}
\label{ss:signal_likelihood}

We now return to the case where the data contains both noise
and a signal, which we denote %For brevity
as $H_S$ for signal hypothesis.  This hypothesis includes our previous assumptions
about the noise $\vec{n}$ given by the noise model $H_N$ with \ac{PSD} $S_N$, but adds the
assumption that the mean of the data is non-zero due to the presence of a signal $\vec{h}$,
such that $\vec{d} = \vec{h} + \vec{n}$.
Calculating the precise form of the signal requires assuming a particular waveform model;
%\jv{The logic of this does not depend on which waveform model in particular is used.}
%Here $h$ is the signal, which depends
%on a set of parameters $\pvec$, whch in turn can be considered as a point in
%parameter space.
for a given set of source parameters $\pvec$, the waveform model yields the signal time series
$\vec{h}$, or \textit{signal model},
and, from \ref{eq:noise_likelihood_inner_product}, the likelihood of the residuals $\vec{h}-\vec{d}$ is
% TD : Have replaced '\vec{h}, \pvec' by '\vec{h}(\pvec)' as otherwise it doesn't seem to make sense (you don't independently condition on h and theta)
\begin{align}\label{eq:LikelihoodSignal}
 p(\vec{d}|\vec{h}(\pvec), H_S) =
 \det(2\pi \mat{C_N})^{-1/2}
 \exp\left(-\frac{1}{2} \braket{\vec{d} - \vec{h}|\vec{d} - \vec{h}} \right)\,.
\end{align}
%For every point in the parameter space $\pvec$ that allows us to compute $\vec{h}$, we
%can evaluate Eq.~\eqref{eq:LikelihoodSignal}.
The shape of this function across the parameter space
$\pvec$ will determine the precision with which we can estimate the
physical parameters of the source, see Sec.~\ref{sec:pe}.
The likelihood function over $\pvec$ is also of importance when searching for a
signal, as values of the likelihood close to those at any global or local maximum
are generally only obtained over a tiny fraction of the possible parameter space.
We therefore need to know how finely to sample the space
to avoid missing possible signals, i.e.\ how to lay out a \emph{template bank},
as discussed in Sec.~\ref{ss:detection_banks}.

The inner product in Eq.~\eqref{eq:LikelihoodSignal} can be expanded as
$
 \braket{\vec{d}-\vec{h}|\vec{d}-\vec{h}} =
 \braket{\vec{d}|\vec{d}} + \braket{\vec{h}|\vec{h}} - 2 \braket{\vec{d}|\vec{h}}.
$
The quantity $\OptimalSNR^2=\braket{\vec{h}|\vec{h}}$ is the square of the optimal \ac{SNR},
a useful measure of the detectability of a signal in a detector with given noise
\ac{PSD}.
In the picture where data of fixed length is described as a $J$-dimensional
vector space with the inner product Eq.~\eqref{eq:inner_product},
%We already considered the space of data of a fixed number of samples as a $J$-dimensional
%vector space, and have defined an inner product in this space via Eq.~\eqref{eq:inner_product}.
%We can also think of
the optimal SNR is the length of the vector representing a particular
signal. % in the space of possible realisations of the data.
%\td{Not sure about discussing a time or frequency series as a data \emph{point}.
%Is this standard nomenclature?  Data realizations?  Data streams?}
%In this picture
We may consider the space of data streams $\vec{d}$ to have an associated metric $\mat\Sigma$:
for data %are represented
in the frequency domain, the metric is a diagonal matrix with nonzero elements given by %the inverse of the noise \ac{PSD},
$\Sigma_{ii} = (T/2) S_N(f_i)^{-1}$, indicating that the Fourier frequency bins are orthogonal dimensions of
this space. The operation of \emph{whitening} the data -- achieved by multiplying a data
stream $\vec{\tilde{d}}$ by $S_N(f)^{-1/2}$ -- can then be seen as the normalisation of these
frequency bins, such that the metric becomes the identity matrix.

\subsection{The Fisher information metric}\label{ss:FisherMatrix}

The space of possible signals $\vec{h}$ can be considered as a
sub-manifold of the full vector space of possible data streams $\vec{d}$.
This sub-manifold of signals, with dimensionality equal to the number of
independent parameters used to
describe the signal, is not itself a vector space, since a linear
combination of signals in general is not another valid signal.
Still, provided the waveform is a smooth function of these parameters
we may consider it as a continuous manifold.

We can then use the signal parameters as a coordinate system
to identify points on the signal manifold; the flat-space metric
restricted to the signal manifold will induce a metric $\mathbf{\Gamma}$ in the parameter
space, determined by the ``closeness'' of neighbouring signals.
Formally, for an infinitesimal change in parameters $\d\pvec$ we have a change in the waveform $\d\pvec
%\frac{\partial \tilde{h}}{\partial \pvec}$,
%(\partial \tilde{\vec{h}}/\partial \pvec)$  TD - don't need a tilde??
(\partial \vec{h}/\partial \pvec)$
and a squared interval
\begin{align}
{\d s}^2 = \|\!\d\pvec\|^2 &= \left(\d\pvec \frac{\partial \vec{\tilde{h}}}{\partial \pvec}\right)^T \mat{\tilde{C}_{N}}^{-1} \frac{\partial \vec{\tilde{h}}}{\partial \pvec} \d\pvec \nonumber\\
&= \d \pvec^T \mat{\Gamma} \d \pvec\,,
\end{align}
where the metric in parameter space is given by
\begin{align}\label{eq:deFishernition}
% do the bra-ket thing by treating the | as a left bracket, requiring one extra null \right
\Gamma_{ij} &= \left< \frac{\partial \vec{h}}{\partial \pvec_i}
 \left| \frac{\partial \vec{h}}{\partial \pvec_j} \right\rangle \right. \,.
\end{align}
The matrix $\mathbf{\Gamma}$ is known as the Fisher information matrix, or %the Fisher information
metric in this context~\cite{Amari}.

The Fisher matrix formalism is widely used to determine parameter estimation
precision under a Gaussian approximation of the likelihood function~\cite{Finn:1992wt,Cutler:1994ys,Vallisneri:2007ev}.
If we have identified the parameter values $\pvec_0$ which maximise the
log-likelihood at $l_0=\log L(\pvec_0)$, the expansion of the log-likelihood
to second order %as a function of $\pvec$
around $\pvec_0$ is given by
\begin{equation}
 \log L(\pvec_0 + \delta\pvec) \approx l_0 - \frac{1}{2}\delta\pvec_i\Gamma_{ij}\delta\pvec_j \equiv l_0 - \frac{1}{2} \|\delta\pvec_i\|^2 \,.
\end{equation}
%where is the maximum log-likelihood value:
Neglecting higher order terms results in a Gaussian function on parameter space with
covariance matrix $\mat \Gamma^{-1}$.
Assuming a uniform prior distribution, the
resulting posterior probability distribution for the parameters is also a multi-variate
Gaussian.  The diagonal elements of $\mat \Gamma^{-1}$ will then give the variance
of the marginal posterior, i.e.\ the precision at which a given parameter may be
measured.  In particular, the variance is proportional to $\langle h | h \rangle^{-1}$,
i.e.\ to $\OptimalSNR^{-2}$.

In practice, this predicted precision may be inaccurate if the prior distribution
is not uniform over the scale of the likelihood peak,
%(including the case of a hard boundary near the peak)
or if the Gaussian approximation of the likelihood
fails, e.g.\ if there are multiple maxima of comparable height due to symmetries
of the likelihood. These limitations restrict the applicability of the Fisher
matrix formalism for parameter estimation precision to cases with high \acp{SNR}
and a single, well-defined peak~\cite{Rodriguez:2013mla} -- though
even in the low SNR case where we cannot neglect higher-order terms, it provides
the Cramer-Rao bound on the covariance of an unbiased point estimator~\cite{Vallisneri:2007ev}.
One can alleviate these restrictions using a higher order expansion of the likelihood
to predict parameter estimation precision~\cite{Vitale:2011zx}, but in
general the parameters of a signal must be estimated by sampling the posterior
probability distribution, as described in Sec.~\ref{sec:pe}.

\subsection{Detector response}\label{ss:detector_response}

%\kc{Why are these here and not in Sec 2?} \td{because Section 2 did not include anything relevant to the detectors: here is the first place where we consider any detector output.}
Having described the generic likelihood function for a transient
GW signal, we now introduce some specifics for the \ac{CBC}
case, with the ultimate aim of understanding the design of
searches and parameter estimation methods.

The strain from a generic frequency-domain compact binary signal can be written as
\begin{align}
\tilde{h}_+(f) &= A_+(f)\exp(-i\phi_+(f)+\phi_0)\,, \\
\tilde{h}_\times(f) &= A_\times(f)\exp(-i\phi_\times(f)+\phi_0)\,,
\end{align}
for the $+$
and $\times$ polarisations respectively, see Sec.~\ref{ss:signal_characteristics}. A
detector, labelled by $I$, responds linearly to a weighted combination of these,
\begin{equation}
\tilde h_I(f) = F_+ \tilde{h}_+(f) + F_\times \tilde{h}_\times(f)\,,
\end{equation}
where $F_+$ and $F_\times$ are the \emph{detector response functions} (or
\emph{pattern functions}).  These are independent of signal frequency in the
long-wavelength approximation, but
depend on the direction of propagation of the waves with respect to the detector, as
parameterised by the right ascension $\alpha$ and declination $\delta$
of the source and the \ac{GW} polarisation angle $\psi$, %of the radiation,
as in Sec.~\ref{sss:binary_params} and Table~\ref{tab:parameters}.
For an interferometric detector with arms %at right angles
along the $x$- and $y$-axes, the response functions are given by, e.g.~\cite{Maggiore:2007ulw},
\begin{align}
 F_+(\theta, \phi;\psi =0) &= \frac{1}{2} (1+\cos^2\theta) \cos 2\phi\,, \nonumber \\
 F_\times(\theta, \phi;\psi =0) &= \cos \theta \sin 2\phi\,,
\end{align}
where $\theta$ and $\phi$ are the polar and azimuthal angles in the detector
frame; for nonzero polarization angle $\psi$ the responses follow from
considering a rotation about the direction of propagation.

In the simplified case of emission in the $l=m=2$ mode of a
binary whose component spins are parallel to the orbital
angular momentum vector, the orbital plane will not precess over time and the
two polarization phases are related by $\phi_+ = \phi_\times + \pi/2$, thus we
have
\begin{align}
A_+(f) &= \frac{1}{2}(1+\cos^2 \theta_{JN}) \frac{A_0(f)}{d_L}\,,\\
A_\times(f) &= \cos\theta_{JN} \frac{A_0(f)}{d_L}\,,
\end{align}
where $A_0(f)$ is a common frequency-dependent amplitude, and other
parameters are defined in Table~\ref{tab:parameters}.
%$\theta_{JN}$ is the angle between the line of sight vector $N$ and the total angular momentum vector $J$, $D$ is the luminosity distance to the source,

The response in detector $I$ for a signal whose arrival time and coalescence phase are $t_{c,I}$ and $\phi_{c,I}$ respectively may then be written
\begin{equation} \label{eq:signal_fd1}
 \tilde{h}_I(f) = A_I(f) \exp\left(i (\Phi(f) - 2\pi f t_{c,I} - \phi_{c,I}) \right)\,,
\end{equation}
where the amplitude $A_I(f) = A_0(f)/D_{\mathrm{eff},I}(\alpha, \delta, \psi, \theta_{JN}, t_c)$,
with $D_{\mathrm{eff},I}$ being the \emph{effective distance} to the source for detector $I$,
which folds together the dependence on position and orientation angles~\cite{Allen:2005fk}.
Equation~\eqref{eq:signal_fd1} implicitly defines $\Phi(f)$; see Eq.~\eqref{eq:TF2_phase}
as an example of the phase of the TaylorF2 approximant.

\subsection{Measurement precision of signal parameters}
\label{ss:precision_of_parameters}

Now that we have a definition of the notion of ``closeness'' of two signals or
templates via the Fisher matrix, we consider the case of actual compact binary
signals. This will let us understand which quantities make the largest
contributions to measurable waveform differences, and therefore, about which ones 
observed signals are expected to yield the most information.

In Sec.~\ref{ss:FisherMatrix} we saw that the definition of the Fisher
matrix, Eq.~\eqref{eq:deFishernition}, requires waveform derivatives $\partial \tilde{\vec{h}}/\partial \pvec$.
In the frequency domain, considering a single detector, we can write
the waveform as Eq.~\eqref{eq:signal_fd1}, i.e., a product of amplitude and
phase factors, to obtain the derivatives with respect to $\vec{\theta}$:
\begin{align}
 \tilde h(f)&=A(f)\exp i\Phi(f)\,,\\
 \frac{\partial \tilde h(f)}{\partial \pvec} &= \frac{\partial A(f)}{\partial \pvec}\frac{\tilde h(f)}{A(f)} + i \frac{\partial \Phi(f)}{\partial \pvec}\tilde{h}(f)\nonumber\\
&=\left(\frac{\partial \log A(f)}{\partial \pvec} + i \frac{\partial \Phi(f)}{\partial \pvec}\right)\tilde{h}(f)\,.
\end{align}
From this we can see the relative importance of the signal amplitude and
phase to the measurement of parameters: roughly speaking,
a change in the phase by the \emph{addition} of one radian (counted over
the entire observed bandwidth of the signal) is equivalent to a change
in the amplitude by a \emph{factor} of $e$.  Since the waveform may spend
many thousands of cycles in the detector's sensitive frequency band, the
signal phase evolution is of primary importance: % in measuring source parameters:
parameters which affect the phase are typically much more
precisely measured than those that vary only the amplitude.
The derivative is also proportional to the signal itself: as expected,
louder signals provide more information about their parameters.
% TD - not sure if it makes sense to bring in searches here or if it's even
% correct that searches basically ignore amplitude ...
% As the phase evolution is so much more important in computing the likelihood, search algorithms commonly consider only this part by using normalised templates, leaving estimation of amplitude and other extrinsic parameters to full parameter estimation in follow-up.

We can take the post-Newtonian waveform expansion as a starting point to
investigate the influence of some key parameters.  As detailed in Sec.~\ref{ss:PN}
and Eq.~\eqref{eq:TF2_phase}, the phase is written as a sum over
powers of frequency $f^{(k-5)/3}$:
\begin{equation}
 \Phi(f) = 2\pi f t_c - \phi_c -\frac{\pi}{4} + \sum_k \psi_k(\Mc,q,\vec{a_1},\vec{a_2}) f^{(k-5)/3} + \cdots\,,
\end{equation}
omitting terms in $\log f$ for simplicity.\footnote{See e.g.~\cite{Arun:2008kb} for
closed form expressions to 3.5\ac{PN} order.}
The derivative with respect to the reference phase $\phi_c$ is trivial; %; which is simply $1$,
for the other parameters we have
\begin{align}
  \frac{\partial \Phi(f)}{\partial t_c} &= 2\pi f\,,\\
  \frac{\partial \Phi(f)}{\partial \pvec} &= \sum_k f^{(k-5)/3} \frac{\partial \psi_k}{\partial \pvec}\,.
\end{align}
In Table~\ref{tab:PN_terms} we show the order at which different parameters appear. At lowest order only $\Mc$ contributes: we have $\psi_0(\Mc)=(3/4)(8\pi\Mc)^{-5/3}$, such that $\partial \psi_0(f)/\partial \Mc \propto \Mc^{-8/3}$. We therefore expect that only the (redshifted) chirp mass, not the individual masses $m_1,m_2$, will be well-measured for inspiral-dominated waveforms. Only by continuing the expansion to higher orders in $k$ will a second combination of masses, the symmetric mass ratio $\eta$, enter the expansion; contributing less to the integral of Eq.~\eqref{eq:deFishernition}, it will be less well measured.  Therefore we should expect the measurement uncertainties in the two component masses to show a high degree of correlation, as for instance in~\cite{LIGOScientific:2018hze}.
%\jv{CITES results papers}

At higher order still, the spin vectors $\vec{a_1},\vec{a_2}$ enter the expansion, and are accordingly less well determined. As with the individual masses, certain combinations of spins appear earlier than others, the most important being the effective aligned spin $\chi_\text{eff}$, and then the effective precessing spin $\chi_p$. Since the spin orientations are defined on the sphere, the corresponding parameters have nontrivial prior boundaries and/or distributions; for relatively low \acp{SNR}, the Fisher matrix expansion is thus less useful in predicting their measurability, and the full apparatus of Bayesian inference should be used, as in Sec.~\ref{sec:pe}.
%, which takes into account the bounded priors. 
Tidal deformability, finally, enters for systems containing one or more \acp{NS} only from 5\ac{PN} order, as in Sec.~\ref{sss:tidal}, thus requires much higher \acp{SNR} for a nontrivial measurement, although the high order is partly counteracted by a large numerical coefficient, compare~\cite{LIGOScientific:2018hze,Wade:2014vqa}.

\begin{figure}[tb]
\centering
\includegraphics[width=0.6\textwidth]{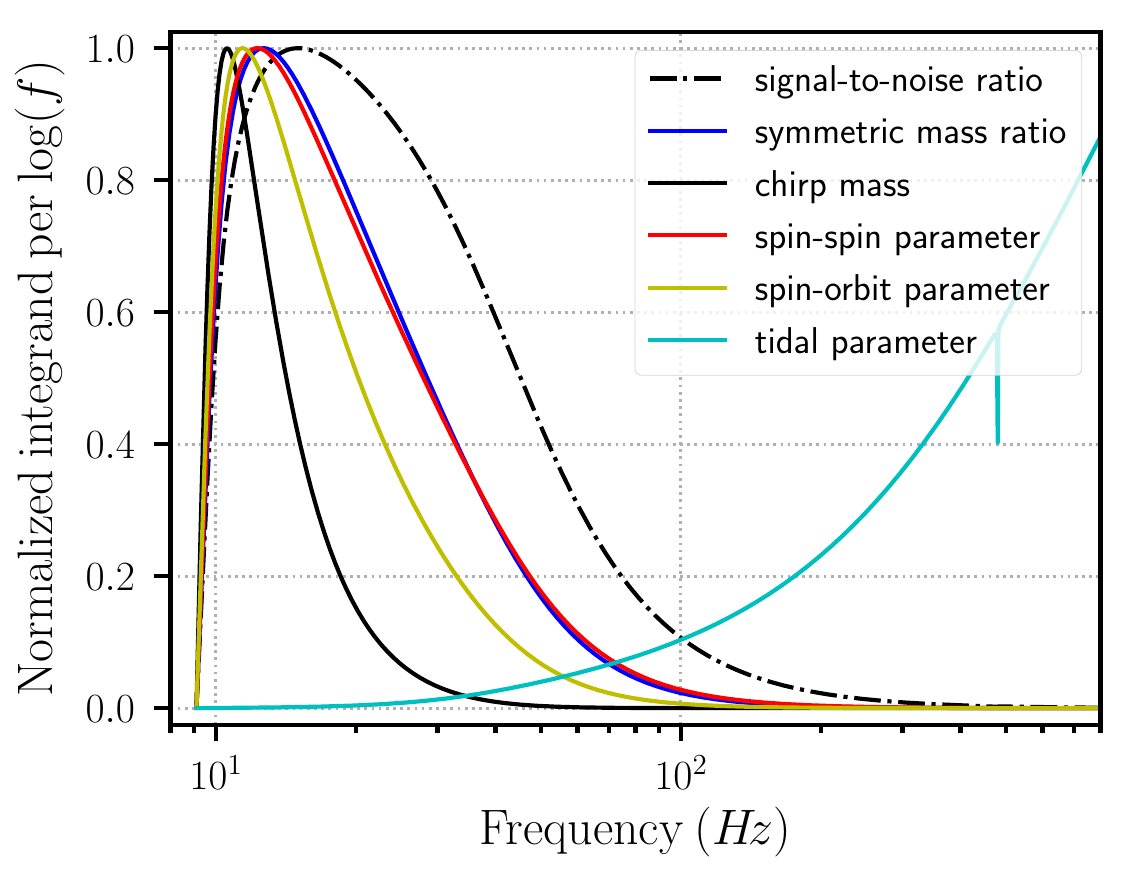}
\caption{Frequency dependence of Fisher matrix element integrands, reproduced from~\cite{Harry:2018hke} (see also~\cite{Damour:2012yf}).  The quantity plotted versus frequency is the density of information about each binary parameter over log(frequency), $|\partial \tilde{\vec{h}}/\partial\theta|^2 /(f S_n(f))$ where $S_n$ is taken to be the zero-detuned high power Advanced LIGO model \ac{PSD}; each curve is 
normalized to peak at unity, except the tidal parameter, normalized to unity at $1\,$kHz.  
Although the tidal parameter information appears to increases monotonically towards high frequency, this calculation will become inapplicable by the BNS merger phase, at frequencies beyond the scale of this plot. 
%\kc{Should we explain why the tidal one appears to be monotonically increasing?}
\label{fig:hh_fisher_integrands}
}
\end{figure}
Further insights %into the interplay between measurability of different \ac{PN} parameters and the influence of detector sensitivity
arise by considering the \emph{integrand} of Fisher matrix elements: for a parameter $\theta$, the diagonal matrix element is the frequency integral of $|\partial \tilde{\vec{h}}/\partial\theta|^2 /S_n(f)$,  This may be thought of as a spectral density of information or parameter precision; Fig.~\ref{fig:hh_fisher_integrands} illustrates the density over \emph{log} frequency for various parameters of \ac{BNS} systems.
%\footnote{Here, as the frequency scale is logarithmic the density per \emph{log} frequency is shown.}
Moreover, for parameters having similar frequency dependence of $\partial \tilde{\vec{h}}/\partial\theta$, the off-diagonal matrix elements are expected to be relatively large, implying strongly correlated measurement errors, for instance between orbit-aligned spins and mass ratio~\cite{Cutler:1994ys,Baird:2012cu}.

\paragraph{Post-Newtonian coefficient space}  As the coefficients of many \ac{PN} expansion terms $\psi_k$ are complicated non-linear functions of physical binary parameters, the Fisher matrix varies strongly over parameter space, leading to additional pitfalls for low- or moderate-\ac{SNR} signals.  While the nonlinearity can be partly addressed by careful choice of parameter basis, e.g.~\cite{Veitch:2014wba},
%~\cite{Lee:2022jpn} - Recent paper about PE makes use of a `less correlated' parameter set - Lee et al. https://arxiv.org/abs/2203.05216
% TD - Not sure if that is best discussed here as it's really about PE not FM forecasts ..
an alternative strategy considers the \emph{space of \ac{PN} coefficient values}, over which the Fisher matrix is (near)-constant~\cite{Tanaka:2000xy,Pai:2012mv,Brown:2012qf,Ohme:2013nsa} for low-mass binaries.  Linear combinations of $\psi_k$ may then be found for which the matrix is diagonal, making tasks such as template placement in a multi-dimensional space (Sec.~\ref{ss:detection_banks}) significantly simpler.

\subsection{Multiple detectors}\label{sec:multi_detector}

We can extend the single-detector analysis to the case of multiple detectors by applying the chain rule of probability. For example, if we are analysing data from Hanford (H), Livingston (L) and Virgo (V), the joint likelihood is
\begin{align}
  p(\vec{d}_H,\vec{d}_L,\vec{d}_V|H_S)&=p(\vec{d}_H|\vec{d}_L,\vec{d}_V,H_S)p(\vec{d}_L|\vec{d}_V,H_S)p(\vec{d}_V|H_S) \nonumber \\
  &= \prod_{I\in\{H,L,V\}} p(\vec{d}_I|H_S)\,,
\end{align}
where the expression reduces to a product over the detectors if the noise in each detector is assumed to be statistically independent.\footnote{Schumann resonances could cause correlated magnetic noise in distant detectors, potentially impacting searches for the stochastic background, e.g.~\cite{Thrane:2013npa}.}
The multi-detector likelihood may then be written as
%Since Eq.~\eqref{eq:LikelihoodSignal} takes the form of an exponential, combining the likelihoods amounts to summing the inner products,
\begin{align}\label{eq:NetworkLikelihood}
  p_\mathrm{net}(\{\vec{d}_I\}|H_S) & \propto \exp\left( -\frac{1}{2}\sum_{I}
  \braket{\vec{d}_I-\vec{h}_I | \vec{d}_I-\vec{h}_I}_I \right)\,,
\end{align}
where $\braket{a|b}_I$ denotes an inner product using the PSD of detector $I$ and $\vec{h}_I$ is the signal as measured at detector $I$, including the detector responses and time shifts due to propagation time between different sites. % in the network.
This {\it coherent} likelihood function gives a complete model of the detector network response as a whole: the signals at each detector $\vec{h}_I\equiv\vec{h}_I(\pvec)$ are dependent on a {\it single} parameter vector $\pvec$. This form is used in coherent analyses, including for parameter estimation of the signal in Sec.~\ref{sec:pe}. %, where the emphasis is placed on estimating the single set of physical parameters.  TD - seems a bit redundant to say this

If we allowed the signal parameters to (unphysically) vary independently in each detector $I$, which we may write as $\pvec_I$, we could use this additional freedom to find a higher maximum of the product of likelihoods: $\prod_{I} \max_{\pvec_I} p(\vec{d}_I|\pvec_{I},H_S) \geq \max_{\pvec} p_\mathrm{net}(\{\vec{d}_I\}|\pvec,H_S)$.  Thus we can simplify the problem of maximising the likelihood, for the purposes of detection, by maximising the \emph{separate} per-detector likelihoods, at the cost of some inaccuracy if unphysical combinations of parameters are allowed.  In practice, constraints are imposed upon maxima to ensure compatible signals in all detectors, for example by enforcing consistent relative arrival times.  This approach is referred to as a {\it coincident} likelihood, as opposed to the fully {\it coherent} likelihood of Eq.~\eqref{eq:NetworkLikelihood}, and is commonly applied in matched filter searches, as discussed further in Sec.~\ref{ss:multi_search}.

\subsection{Normalized signals and similarity measures}
\label{ss:overlap_and_match}

Since \ac{CBC} sources occur at a wide range of \emph{a priori} unknown
distances, the absolute amplitudes of \acp{GW} at a detector also have a wide range of uncertainty.
%ing widely over different sources.
We may discuss the properties %form
of signals %and their statistical properties
without specifying an amplitude or distance %.  This may be achieved
via \emph{normalized} signals with an optimal \ac{SNR} (for some notional or real detector sensitivity) of unity,
\begin{align}
 \hat{\vec{h}} &= \frac{\vec{h}}{\braket{\vec h|\vec h}^{1/2}}  \implies \vec h = \OptimalSNR \hat{\vec h}\,.
\end{align}
The Fisher matrix is then
\begin{align}
% do the bra-ket thing by treating the | as a left bracket, requiring one extra null \right
\Gamma_{ij} &= \OptimalSNR^2 \left< \frac{\partial \hat{\vec h}}{\partial \pvec_i}\left|\frac{\partial \hat{\vec h}}{\partial \pvec_j} \right\rangle\right. \,,
\end{align}
giving an expected inverse relation of parameter estimation accuracy with optimal
\ac{SNR}.  For data containing a signal $\OptimalSNR \hat{\vec{h}}$ plus noise $n$, if we
consider a model that is identical to the signal except for an unknown amplitude
$\mathcal{A}\hat{\vec h}$, the likelihood is maximized for
$\mathcal{A}_\mathrm{ML} = \OptimalSNR + \braket{\hat{\vec h}|\vec n}$,
where the noise term averages to zero over data realizations.
% and has unit variance. TD .. does it ?

%\td{Overlap eg defined in Damour,I,S gr-qc/9708034 (as 'ambiguity function'), also 'effectualness' and 'faithfulness'}
Various \emph{measures of similarity} between different signal models $\vec h(\pvec)$
are linked to aspects of parameter estimation and detection.
While the amplitude of a predicted signal has an influence on astrophysical
interpretation, similarity of signals is generally assessed independently
of amplitude: the most basic measure is the \emph{overlap} of two signals $\vec h_1$, $\vec h_2$
\begin{equation}
  \mathcal{O}(\vec h_1, \vec h_2) = \frac{\braket{\vec h_1|\vec h_2}}{(\braket{\vec h_1|\vec h_1} \braket{\vec h_2|\vec h_2})^{1/2}} \equiv \braket{\hat{\vec h}_1|\hat{\vec h}_2}\,.
\end{equation}
The overlap can be shown to be maximized at unity, when $\vec h_1=\vec h_2$.

Furthermore, the absolute values of the orbital phase of a binary or its time of merger
are rarely of astrophysical interest (in the absence of non-\ac{GW} observations %beyond \ac{GW}
that may depend on these quantities), and detection algorithms do
not generally make use of them.  Allowing arbitrary differences in phase and
coalescence/merger time between signal models, we arrive at a less restrictive
measure called the \emph{match} %of similarity
\begin{equation}
\label{eq:match}
%\mathbf{M}(\pvec_1,\pvec_2) \equiv
    \mathbf{M}(\vec h_1(t_{c1}, \phi_{c1}, \vec\theta_1), \vec h_2(t_{c2}, \phi_{c2}, \vec\theta_2)) =
    \max_{t_{c1}-t_{c2}, \phi_{c1}, \phi_{c2}} \mathcal{O}(\vec h_1, \vec h_2)\,,
\end{equation}
where $\vec\theta$ represents source parameters excluding time and phase.\footnote{In
the approximation that the dependence of each signal on orbital phase is via a complex
phase $\sim e^{i\phi_{ci}}$, it is only necessary to maximize over the phase difference.}
The \emph{mismatch} is simply $1 - \mathbf{M}$, which like $\mathbf{M}$ ranges between
$0$ and $1$.

If, conversely, we are comparing different signal models, for example two approximants discussed in Sec.~\ref{sec:LEOBandPhenom},
our goal is to assess systematic differences. The comparison between signals for binaries with identical
parameter values $\vec\theta'$ (including intrinsic parameters), %of component masses and spins,
$\mathbf{M}(\vec h_1(t_{c1}, \phi_{c1}, \vec\theta'),
\vec h_2(t_{c2}, \phi_{c2}, \vec\theta'))$, is then called the \emph{faithfulness}.  This quantity
informs how far we expect the models to give the same inferences on such parameters; differences between models will contribute to systematic uncertainties and/or biases in measured source properties (see Sec.~\ref{sub:comparison_of_waveform_accuracy_and_efficiency}).
%
%\td Add cross-ref to PE ?
% PE doesn't seem to discuss waveform bias .. :/

\newpage
\section{Detection}
\label{sec:detection}

%\emph{Responsible:} Tom

In this chapter we present the problem of detection for compact binary merger signals in the data of a \ac{GW} detector network, %one or more gravitational-wave detectors, 
and describe solutions deployed or proposed for real %LIGO-Virgo 
data as well as the principles underlying them. 

In the ground-based detector network, sensitive at %restricted to %observations at 
frequencies of a few Hz and above, \ac{CBC} signals are comparatively rare and weak relative to the noise.  By contrast, future space-based detectors are expected to face the opposite case where signals (not necessarily all from compact binaries) dominate the output stream, see e.g.~\cite{Amaro-Seoane:2012aqc}.  Even for 3rd generation detectors 
%(Einstein Telescope / Cosmic Explorer) case, 
with a sensitivity to \ac{BBH} mergers covering nearly the whole of the observable Universe, the rate of signals is expected to be at most one per tens or hundreds of seconds; while these ``chirp'' signals will overlap in time within the detector's %observational 
frequency band, they will not overlap significantly in the time-frequency plane~\cite{Regimbau:2012ir,Meacher:2015rex,Johnson:2024foj}.
%i.e.\ at any given time, two merger signals will almost always have well separated and clearly distinguishable frequencies 

%for ground-based detectors, 
Detection thus implies \emph{selecting} the small fraction of data containing distinguishable signals, 
%and from which astrophysical information can be extracted by further analysis, out of much larger volumes of 
as opposed to noise or indistinguishably weak signals.  %Note that 
This problem is close, but not identical, to classic detection theory %where there is 
with a well-defined ``null hypothesis'' denoting the lack of signal.
% a signal is either present or absent and there is thus a well-defined `null hypothesis'.  Nevertheless much of the machinery of classical hypothesis testing can be adapted to the gravitational wave search case. 
% 
In realistic cases, there are important differences from idealized %detection 
theory.
% of detection in realistic compact binary detection scenarios.  Some such caveats concern comparisons of the efficiency or optimality of different search methods and algorithms. 

For one, possible binary signals cover a large parameter space: %before many detections are made 
we do not \emph{a priori} know 
the true distribution of signals over that space.  A theoretically optimal search maximizes the number of
detected signals for a given rate of false alarms, but we are unable to evaluate this for an unknown population. 
We also might choose a figure of merit that differs from simply the expected number of detections. 
%, since for example detecting a large number of near identical binary systems may be of relatively small scientific interest. 

Second, even for a known signal distribution, the optimal search method
% -- assuming that we have decided on an appropriate signal distribution -- 
depends on the noise content of the data.  Since real GW detector data contains %non-ideal
transient excess noise events 
%which can mimic some features of binary merger signals, 
whose rate and 
morphology %and rate of occurrence  %these transients 
cannot be predicted, search optimization requires 
%cannot be %determined or implemented without %gathering 
empirical information on such noise properties, and 
thus %in realistic cases only a finite sample of noise can exist, 
%optimization %of a search method 
is necessarily limited by the finite statistics of real noise.
%limits to the %statistical precision to which we can determine the efficiency or optimality of a method for identifying signals within such noise. 

Furthermore, searches precede %are %employed as a first stage before 
a much more computationally intensive exploration of the %parameters and 
properties of candidate signals, see Sec.~\ref{sec:pe}. 
Thus, beyond 
%The search algorithm is thus expected to indicate more than 
the mere presence %or absence 
of a signal, sufficiently accurate point estimates of arrival time and of  
%(redshifted) 
binary component masses provided by the search algorithm %of the source binary 
will significantly reduce the cost of subsequent analysis. 
%ensuing parameter estimation analysis.  
Searches also provide detailed timing %time series 
information 
%on the time of arrival at each detector in order 
%which is 
used to estimate 
%approximately reconstruct a full 
%likelihood as a function of
the source's location for low-latency followup~\cite{Singer:2015ema,Singer:2016eax}. 
%; thus, the capability to provide some degree of parameter estimation is already inherent in the search analysis. 

%In addition to these differences of principle from classic %detection 
%theory, 
\ac{CBC} detection methods are also subject to technical and computational limits.  The technical challenges of setting up a \mbox{(near-)optimal}
search are currently solved only for signals in the four-dimensional parameter space $\{m_1, m_2,$ $\chi_{1z}, \chi_{2z}\}$
of quasi-circular, non-precessing coalescences %without in-plane component spins, 
neglecting subdominant \ac{GW} multipoles, see Sec~\ref{sec:cbc}. %[CITES]. 
%Detection methods for 
Methods for detection of more complex (and thus more realistic) signals 
%with nontrivial contributions from subdominant multipoles, orbital precession or eccentricity 
are under active investigation, as discussed below.

In the rest of this chapter we revise standard detection theory, before discussing alterations or extensions for real \ac{GW} data and reviewing the challenges of more complex \ac{CBC} signals, computational issues and different astrophysical search applications.

\subsection{Theory of transient signal detection}
\label{ss:detection_theory}

%In the standard case, %detection 
Given a stretch of data $d(t)$, we want to distinguish between the hypothesis $H_S$ that 
%we have for which we want to answer the question whether 
it contains a signal of known form, and the hypothesis $H_N$ that it contains %-- considered as a hypothesis, ; or 
no signal.\footnote{We sometimes designate the no-signal case as `noise' although noise is present in both hypotheses.} 
For %In realistic cases, we may have 
GW data spanning several months, %over which 
more than one signal may arrive; to begin we consider a short period %of data 
containing no more than one signal\footnote{Signals
from compact binary mergers arrive at Earth at a rate of one per $\sim$tens of seconds, 
although the great majority have an amplitude well below that currently considered detectable~\cite{LIGOScientific:2017zlf}.}, see Sec.~\ref{sec:combine} for joint analysis of multiple signals. 
The data are supposed to contain noise with known statistical properties, usually described via the correlation function, which in the simplest case is assumed stationary, see Sec.~\ref{ss:noise_likelihood}.  The detection algorithm calculates a scalar %function of the data, a 
`test statistic' $\varrho(d)$ which %if the algorithm is optimal, 
will as far as possible take a higher value under the $H_S$ hypothesis than under $H_N$.
%\footnote{For definiteness we take the mean value over noise realizations to be higher for the $H_S$ case than for $H_N$.}
% in the presence as opposed to absence of the signal. 

\paragraph{Efficiency and false alarm rate} 
The performance of a test statistic %considered as a function of the data, 
may be evaluated by imposing a threshold value $\varrho^\ast$: 
%and notionally treating all data stretches with $\varrho>\varrho^\ast$ as containing a signal, and all stretches with $\varrho<\varrho^\ast$ as containing no signal. 
the \textit{efficiency} $\epsilon(\varrho^\ast)$ is the expected fraction of cases actually containing a signal for which $\varrho>\varrho^\ast$, \textit{i.e.}\ $P(\varrho>\varrho^\ast|H_S$).  The \textit{false alarm rate}\footnote{`Rate' here denotes a probability, not a Poisson rate parameter.} $F(\varrho^\ast)$ is the expected fraction of cases containing no signal for which $\varrho>\varrho^\ast$, \textit{i.e.}\ $P(\varrho>\varrho^\ast|H_N)$.  
In terms of probability density over the data realization $d$, 
% under each hypothesis, we have
\begin{align}
\label{eq:effdef}
  \epsilon(\varrho^\ast) &= \int \mathrm\Theta(\varrho(d) - \varrho^\ast) p(d|H_S)\, \mathrm{d}d\,, \\
  F(\varrho^\ast) &= \int \mathrm\Theta(\varrho(d) - \varrho^\ast) p(d|H_N)\, \mathrm{d}d\,.
\end{align}

\paragraph{Receiver operating curve and likelihood ratio}
We may calculate $\epsilon(\varrho^\ast)$ and $F(\varrho^\ast)$ over a range of $\varrho^\ast$ values, 
%-- assuming perfect knowledge of the probabilities $p(d|\{\mathbf{N},\mathbf{S}\})$ -- 
%we calculate . 
%Both the efficiency and false alarm rate are constant or decreasing functions of $\varrho^\ast$, thus 
%The threshold $\varrho^\ast$ thus 
parameterizing a curve in the ($F$, $\epsilon$) plane called the ``receiver operating curve'' (ROC).
%\footnote{The name stems from 
%%development of 
%detection theory for radio signals.} 
%with $\epsilon$ being a constant or increasing function of $F$. 
%It is fairly obvious that the statistic $\varrho(d)$ that yields a given ROC is not unique, since replacing it with any monotonically increasing function of $\varrho(d)$ will yield the same ROC: the new statistic partitions the space of all possible data realizations into above-threshold and below-threshold in the same way as the old, the only difference being the numerical value corresponding to any given threshold.  In this context $\varrho(d)$ is also referred to as a `ranking statistic' as its job is to \emph{order} possible data realizations from least to most likely to contain a signal. 
Any monotonic increasing function of $\varrho(d)$ will yield the same ROC, giving the same ranking of data from noise-like to signal-like; hence $\varrho(d)$ is called a ``ranking statistic''.

%\paragraph{Likelihood ratio as optimal statistic}
A ``Neyman-Pearson'' optimal statistic is one that yields the largest possible $\epsilon$ for a given value of $F$.  It may be shown, e.g.~\cite{Biswas:2012tv} that the likelihood ratio $\Lambda_\mathrm{SN}$ is an optimal statistic, where 
\begin{equation}
 \Lambda_\mathrm{SN}(d) \equiv \frac{p(d|H_S)}{p(d|H_N)}\,. 
\end{equation}
If this likelihood ratio can be efficiently and accurately computed for the true GW signal population under all possible data realizations, the problem is solved; 
% and \kc{I do not understand this, why would they skip ahead if we are saying that this is not the answer?}the reader may skip to the end of the chapter after reviewing the classic solution,
%for idealized noise processes, 
%the matched filter.  
however it will not be surprising that in real applications the classic solution, the matched filter, is only the first %step and is one
of many tools required. 

\paragraph{The idealized matched filter}
For a signal of known form $h(t)$, %and a Gaussian noise model of known spectrum $S_n$ in a single detector, 
the likelihood of a data set $d(t)$ in a single detector %in the presence of the signal 
as defined in Sec.~\ref{ss:signal_likelihood} is 
\begin{equation}
 p(d|H_S,h(t)) = p(n(t)\equiv d(t)-h(t)|H_N) = \mathrm{const}\cdot e^{-\braket{d-h|d-h}/2}\,,
\end{equation}
where %$\mathbf{S}[h]$ denotes the signal hypothesis for a signal $h(t)$ and 
the inner product $\braket{a|b}$ is given by Eq.~\eqref{eq:inner_product} for Gaussian noise with a known spectrum $S_N$. 
%In words, the probability of the data, given a specific signal, is the probability of the noise being equal to the data minus the signal.  
The likelihood in the absence of signal is just
\begin{equation}
 p(d|H_N) = p(n(t)\equiv d(t)|H_N) = \mathrm{const}\cdot e^{-\braket{d|d}/2}\,,
\end{equation}
thus the likelihood ratio is $\Lambda_\mathrm{SN}(d) = e^{\braket{d|h} - \braket{h|h}/2}$.  For a fixed signal, %$h$, 
$\braket{h|h}$ is constant, 
%the second term in the exponential is a constant, 
thus %the 
%; since the exponential is a monotonic function, the 
%inner product 
$\braket{d|h}$ %linear in $d$ 
is an equivalent ranking statistic to $\Lambda_\mathrm{SN}$.  As a product of the (whitened) data with a whitened template in the frequency domain, $\braket{d|h}$ is called a ``matched filter''. 
%, i.e.\ has the same ROC curve as $\Lambda_\mathrm{SN}(d)$. 

\subsubsection{Signal with unknown parameters: composite hypotheses}
In astrophysical applications, some properties of %signal hypothesis where %the form of 
the signal are unknown; we parameterise them by a set of values $\pvec$, for instance the amplitude or time of arrival at the detector.  Hence, we have a ``composite signal hypothesis''.  
We write the signal hypothesis for specific parameter values as $h(\pvec),H_S$ or just $\pvec,H_S$. 
%We then wish to determine whether data contains a signal with \emph{any} possible parameter values, or no signal.  
In order to determine an optimal detection method, we require the \emph{distribution} over %parameters 
$\pvec$, notated as $p(\pvec),H_S$; this distribution is analogous to a prior %PDF 
in parameter estimation, see Sec.~\ref{ss:peprior}. 
%we notate the signal parameter distribution as 
% given %for the signal hypothesis with 
%a specific distribution.  %The parameter PDF arises from 
Considering transient signals as %independent events in 
an inhomogeneous Poisson process with %parameter-dependent 
rate $\lambda(\pvec)$, the PDF of a single merger's parameters is proportional to $\lambda(\pvec)$. 

For some parameters, $p(\pvec)$ is trivial:
the distributions over arrival time $t_c$ and coalescence phase $\phi_c$ are uniform.  Since we believe the
Earth is not in a special location in the Universe, the distribution of source locations %relative to Earth 
is uniform over spatial volume, up to cosmological effects at high redshift: hence we expect a distribution 
of luminosity distances $p(d_L) \propto d_L^2$, and %thus 
of signal amplitudes $p(A) \propto A^{-4}$~\cite{Schutz:2011tw}.  Similarly %the distribution of 
the orbital axis direction is isotropic, %relative to the line of sight is uniform on the surface of a sphere,
thus for the inclination %we have 
$p(\iota) \propto \sin \iota$. 

For other parameters, in particular binary %component 
masses and spins, there is no clear expectation. % for the signal distribution.
%Until recently, 
Published LIGO-Virgo searches generally specify only the boundaries of the region to be searched: for instance in~\cite{LIGOScientific:2016dsl} a range of (redshifted) component masses $[1,99]\,M_\odot$ and binary total masses $[2,100]\,M_\odot$ was considered, with a maximum component spin magnitude $\chi < 0.9895$. 
%TD : (CHECK!).  
However, search methods were not designed for optimal sensitivity to a specific distribution over this space. 
%The resulting search method may be optimized for \emph{some} signal distribution covering this region, but in general this ``optimal target distribution'' is not straightforward to determine.  
%We will revisit this point after discussing template placement methods over a target space, which are closely connected to optimizing searches for a (partly) unknown signal distribution. 

For a signal distribution $p(\pvec)$, the optimal likelihood ratio statistic is 
\begin{equation}  \label{eq:marglr}
 \Lambda_\mathrm{SN}(d;p(\pvec)) = \frac{\int%_\Theta 
 p(\pvec) p(d|\pvec,H_S)\,\mathrm{d}^n\pvec}{p(d|H_N)}\,. 
\end{equation}
The numerator here is mathematically identical to the \emph{evidence} for a signal model specified by parameters $\pvec$ with prior $p(\pvec)$, see Sec.~\ref{sec:pe}.
%If the signal model $\mathbf{S}(\pvec)$ is a smooth function of $\pvec$, then the integrand will also be one  
% TD - Not sure why that was mentioned explicitly - we don't need to differentiate it AFAIK

Current search methods treat the signal likelihood $p(d|h(\pvec),H_S)$ as a strong\-ly peaked function of $\pvec$, and approximate the integral over a finite parameter space $\Theta$ by a small region around the global maximum.  For a signal of optimal \ac{SNR} $\OptimalSNR$ as in Sec.~\ref{ss:signal_likelihood}, %proportional to the signal amplitude, %$A$ [REFER BACK?], 
the peak likelihood is higher by a factor $\sim$$\exp(\OptimalSNR^2/2)$ than at parameters far from the peak; %is of order $1$, 
thus, such an approximation may be useful for $\OptimalSNR\sim (5-10)$  and above. 
%That this may be a reasonable approximation can be seen by considering a data set which contains \emph{two} signals of different amplitudes $A_{1,2}$ and well separated parameter values $\pvec_{1,2}$: the likelihood for $\pvec$ values in the neighbourhood of each signal will be peaked at values scaling as $\exp(A_{1,2}^2/2)$ resp., thus almost regardless of the form of each peak or the relative values of the `prior' distribution $p(\pvec_{1,2})$ around each, the integral is dominated by the higher amplitude peak, if higher than all other peaks due to noise (cf.\ Laplace's method).  For this not to be the case, the relative numbers of weak (low-amplitude) signals/peaks should be exponentially large compared to high-amplitude [CITE?]. 
%
%This approximation leads to the idea of a \emph{trigger} (terminology borrowed from high-energy physics): in this language, 
Approximate evaluation of $\Lambda_\mathrm{SN}$ is also possible if one records only the maxima of $p(d|\pvec,H_S)$, or better of $p(\pvec)$$p(d|\pvec,H_S)$, that exceed a predetermined threshold: these maxima are called \emph{triggers}, a term borrowed from high-energy physics.  The threshold value for real (non-ideal) data results from a tradeoff between search sensitivity vs.\ computational and data storage limits.  
%Namely if the value of the marginalized signal likelihood can be predicted approximately from the maxima of the likelihood over parameter space, it is sufficient to record those maxima as and when they exceed a predetermined threshold.  The threshold may be determined by imposing that any lower maximum would make a negligibly small contribution to the signal evidence; 
%
%\td{Add pointer to further discussion of optimized population search later??}

\subsection{Matched filtering for single-mode CBC signals}
%Next we consider the application of the formalism above to the CBC signal parameter space. 
%Considering this formalism applied to a
For a general merging binary system, the signal has a complicated structure if including effects of precessing component spins and GW emission in non-dominant modes; yet more so if the orbit has non-negligible eccentricity, i.e.\ is not quasi-circular.  %Optimizing searches for such signals is an area of active research [REFER FORWARD/CITES]. 
We begin, though, with the simple case of dominant mode GW emission from quasicircular binaries with orbit-aligned component spins, where the signal at a given detector has the form of a sinusoid with frequency-dependent amplitude and phase factors, Eq.~\eqref{eq:signal_fd1}.  The likelihood may easily be \emph{maximized} over the %overall 
signal amplitude parameter and coalescence phase~\cite{Sathyaprakash:1991mt}. 
%For a binary with known masses and (orbit-aligned) spins, the distance, direction and inclination of the source, time of merger and the phase of the GW oscillation at a given binary separation (or ``coalescence phase'') are all unknown and we typically maximize likelihood over these parameters.  For the signal at a given detector $I$ we have [CHECKME]
\emph{Marginalizing} over these parameters yields corrections, which at least for high-\ac{SNR} signals are small~\cite{BrownThesis,Dent:2013cva}.

The maximized likelihood ratio is %proportional to? 
$\exp(\rho(t_c,\pvec^-)^2/2)$, where $\rho$ is the phase- and amplitude-maximized matched filter \ac{SNR} for a signal with coalescence time $t_c$ and template parameters $\pvec^-$ (excluding amplitude, coalescence time and phase).  We find~\cite{Allen:2005fk}
\begin{gather}
 \rho(t_c,\pvec^-)^2 = \braket{d|\hat{h}(t;t_c,\phi_c=0,\pvec^-)}^2 + \braket{d|\hat{h}(t;t_c,\phi_c=\tfrac{\pi}{2},\pvec^-)}^2  \label{eq:complex_snr1}
  %= \left| 2 \int_0^\infty \frac{\tilde{d}(f)\hat{\tilde{h}}^*(f;\pvec',t_c)}{\frac{1}{2}S_N(f)} df \right|^2
  \\ = \left| 2 \int_0^\infty \frac{\tilde{d}(f)\hat{\tilde{h}}^*(f;\pvec^-)}{\frac{1}{2}S_N(f)} e^{2\pi i f t_c} df \right|^2, \label{eq:complex_snr2}
\end{gather}
where $\hat{h}(\pvec^-)$ is a normalized template waveform with $\phi_c$, $t_c$ set to 0.  This ``complex matched filter'' uses the fact that $\tilde{h}(t_c;\phi_c=\tfrac{\pi}{2})$ = $i\tilde{h}(t_c;\phi_c=0)$.  
The DFT~\cite{Allen:2005fk} allows us to straightforwardly evaluate $\rho(t_c,\pvec^-)$
%Furthermore, it may be %straightforwardly evaluated 
as a series over $t_c$, 
although time-domain implementations have also been considered~\cite{Cannon:2011vi,Luan:2011qx}. %particularly for low latency search 
It is appropriate to call $\rho$ a \ac{SNR}, as each scalar product of the form $\braket{d|\hat{h}}$ 
%on the RHS of Eq.~\eqref{eq:complex_snr1} 
has zero mean and unit variance in stationary Gaussian noise; thus in the absence of signal we have the distribution $p(\rho|H_N) = \rho \exp(-\rho^2/2)$, peaking at unity. 

For data containing a signal $\mathcal{A}\hat{h}(t;t_c,\phi_c,\pvec^-)$ that exactly matches a template, the most probable value of $\rho$ is the optimal \ac{SNR} $\OptimalSNR = \mathcal{A}$, see Sec.~\ref{ss:overlap_and_match}.
%\braket{h(\pvec^-)|h(\pvec^-)}^{1/2}$.  
Heuristically, with some fairly large number of independent templates (see discussion below in Sec.~\ref{ss:detection_banks}), the rate of noise triggers will fall to a sufficiently low level at high \acp{SNR} %at $\rho\gtrsim 8$
such that signals with $\OptimalSNR>8$ have a high probability of being distinguishable from noise.  The \emph{horizon distance} $D_\mathrm{h}$ of a detector for a binary with given intrinsic parameters is the largest distance for which %at which the merger signal has 
$\OptimalSNR>8$: this is attained for a system located directly overhead the detector and with zero orbital inclination to the line of sight.  Often, $D_\mathrm{h}$ for a \ac{BNS} system with non-spinning components of mass $1.4\,M_\odot$ is considered as a ``standard candle'' to quantify detector sensitivity. 

By contrast, the sensitive \emph{range} of a detector %for a standard coalescing binary 
is an averaged distance, considering the distribution of angular parameters specifying the binary position and orientation.  It can be defined by considering a distribution of mergers with a homogeneous rate density $\mathcal{R}_\ast$. %the total rate within a sphere of radius $D_\mathrm{h}$ is then $(4\pi/3)\mathcal{R}_\ast D_\mathrm{h}^3$. 
The expected rate of mergers having $\OptimalSNR>8$, $\mathcal{R}_\mathrm{det}$, may then be calculated~\cite{Finn:1992xs}; %defining this as, 
the range $D_\mathrm{det}$ is the radius of a sphere such that $(4\pi/3)D_\mathrm{det}^3 \mathcal{R}_\ast = \mathcal{R}_\mathrm{det}$~\cite{Allen:2005fk,Sutton:2013ooa}.  Equivalently, drawing a merger with random position and orientation within a distance $D_\mathrm{h}$, the probability that it has $\OptimalSNR>8$ is $(D_\mathrm{det}/D_\mathrm{h})^3 \simeq 2.26^{-3}$. 

These definitions neglect cosmological expansion of space, which affects both the signal seen at the detectors~\cite{Krolak:1987ofj}, see Sec.~\ref{sss:redshift}, %via a constant factor in the \ac{GW} frequency,
and the geometric relation between luminosity distance and volume of space.  While for Advanced detectors the redshift $z$ of detectable \ac{BNS} is small compared to unity, for more massive and distant source 
% to obtain an accurate estimate of sensitivity to more distant sources 
redshift effects must be included~\cite{Chen:2017wpg}.
The actual sensitivity of a search over real data will be determined by the detector \emph{network}, and as we discuss in Secs.~\ref{ss:detection_nonideal} and \ref{ss:detection_stats}, does not correspond to a fixed \ac{SNR} threshold.  The rate of mergers is also likely not constant even in comoving cosmological units~\cite{Fishbach:2018edt,KAGRA:2021duu}, implying further corrections to the $D_\mathrm{det}$ calculation. %%beyond spacetime geometry must %come into play.  
Sensitive range is, though, still a useful standard measure of broadband detector performance.

\subsection{Signal manifolds and template banks}
\label{ss:detection_banks}

In searching for signals from binaries with unknown parameters (component masses, spins, etc.), in addition to the scientific problem of determining the correct range and distribution of signal parameters, 
we face a technical challenge of computational cost to evaluate the likelihood at a ``densely spaced'' set of points in the parameter space. 
%may be computationally very expensive.  
%Even if we do not completely solve the scientific problem of the signal distribution, 
This challenge has a well-defined solution via the construction of template banks, the topic of this section. 

%The reasoning used is as follows: Regardless of the actual distribution of signals, we decide to search for possible signals within a specific parameter domain or region.  Once the size of this region has been fixed (in a sense that we will better define soon), for data with a given PSD we expect there will be a well-defined minimum rate of noise events at or above a given SNR threshold.  Thus, for a statistically confident detection any signal must have a matched filter SNR above a given threshold. 
%(the actual threshold value is immaterial, the relevant point is its existence).

Given a desired parameter search region, for data with a given \ac{PSD} there will be some rate of noise events (maxima of likelihood or matched filter \ac{SNR}) inside this region.  As the rate density of noise events is a rapidly falling function of \ac{SNR}, for a statistically confident detection a signal must have \ac{SNR} above a specific threshold. %depending on the \ac{PSD} and size of the search region. %and true signal rate). 

The matched filter \ac{SNR} obtained for a %(normalized) 
template %$\hat{h}(\pvec)$ 
exactly matching a signal with parameters $\pvec$ %(neglecting coalescence time and phase) 
has a non-central $\chi^2$ distribution with mode equal to \OptimalSNR, as above. 
%the ``optimal'' SNR $\rho_{\rm opt} = \sqrt{\braket{h[\theta]|h[\theta]}}$ as above. 
For a template $\hat{h}(\pvec')$ differing from the signal, %we consider 
the ``match'' of Eq.~\eqref{eq:match}, 
%between the template $\hat{h}[\theta']$ and the signal $h[\theta]$, 
i.e.\ the normalized overlap maximized over arrival time and orbital phase, notated here as
%, assuming that the template is a modulated sinusoid [as above], 
\begin{equation}
    \mathbf{M}(\pvec,\pvec') \equiv \mathbf{M}(h(t_1, \phi_1, \pvec'), h(t_2, \phi_2, \pvec)),
 %\max_{bla} \frac{ \braket{h[\theta]|\hat{h}[\theta']} }
 %{ \braket{h[\theta]|h[\theta]} },
\end{equation}
%where templates notated as $\hat{h}$ are normalized. 
is bounded to be $\leq 1$.  The most likely value of \ac{SNR} recovered by $\hat{h}(\pvec')$ is then $\mathbf{M}\OptimalSNR$; thus for a given signal %of given $\OptimalSNR$ 
the probability of detection drops progressively with decreasing $\mathbf{M}$.  For a bank, 
% templates $\hat{h}(\pvec'_p)$, %any signal has a match $\mathbf{M}[\theta,\theta'_p]$ with each of them, however 
we consider only the maximum of $\mathbf{M}(\pvec,\pvec'_p)$ over templates $\hat{h}(\pvec'_p)$.  For a given probability of detection, the optimal \ac{SNR} required %for a given detection probability 
scales as $1/\mathbf{M}$, which for fixed intrinsic parameters $\pvec$ implies a maximum source distance %at which a signal is detectable with given probability 
scaling as $\mathbf{M}$.  The expected number of detectable signals %with parameters $\pvec$ detected 
then scales as the volume of space $\propto\mathbf{M}^3$ (neglecting cosmological effects).  If all signals had the same maximum match $\mathbf{M}$ we would lose a fraction $1-\mathbf{M}^3$ of signals relative to the ideal case of an infinitely dense bank. 

In general the match ranges between $1$ and the smallest value in the target parameter region, the `minimal match' (MM), i.e.\ the largest `hole' in the bank.  Thus a minimax strategy places templates to obtain a sufficiently high MM value, losing a fraction $1-\mathrm{MM}^3$ of signals in the worst case. 
%since we cannot rely on signals having matches above MM.  %in the worst case we `lose' .  
Typical target values are $\mathrm{MM}=0.965$ ($0.97$, $0.98$), losing at worst $\sim$$10\%$ ($\sim$$9\%$, $\sim$$6\%$).  %As pointed out for instance in
For a broad and smooth distribution of signal parameter values, the \emph{average} match will be significantly higher than $\mathrm{MM}$~\cite{Allen:2021yuy}. 

For a template $\hat{h}(\pvec'_p)$, the region of parameter space ``covered'' with match $\geq \mathrm{MM}$ is that within a given distance of $\pvec'_p$ using the information (Fisher) metric of Sec.~\ref{ss:FisherMatrix}, after projecting out the coalescence time and phase dimensions; the covered region is an approximate ellipsoid in suitable coordinates~\cite{Owen:1995tm,Owen:1998dk}.  To completely cover an extended region, we require overlapping ellipsoids; the larger such overlaps are the more templates are required in total, increasing computational cost.  If coordinates can be found with an approximately constant metric,
%where the ellipsoids are of (nearly) constant form, implying an approximately constant metric, 
geometrical grid methods will minimize the %overlaps and cover the space with fewest templates
overlaps~\cite{Owen:1998dk,Cokelaer:2007kx,Brown:2012qf,Harry:2013tca,Roulet:2019hzy}: this is the case where the \ac{PN} parts of the inspiral waveform dominate the \ac{SNR}.  In general for \ac{IMR} signals there is no convenient coordinate basis with a near-constant metric, and stochastic or random methods must be used~\cite{Babak:2008rb,Harry:2009ea,Manca:2009xw}, with a modest increase in computational cost relative to theoretically optimal placement. 

Random placement using a uniform distribution over specific coordinates is likely to be inefficient, %in the sense of 
requiring a large number of templates to cover a given fraction of the space, as first, the ideal density of templates indicated by the metric may vary significantly over the space, and second, due to random fluctuations many templates may be much closer %to one another 
than required for coverage. 
Stochastic methods address both issues by sequentially proposing new random templates, where a template is added to the bank only if its highest match over the current template set is below some threshold. %[CITES].  
%Under-coverage due to random fluctuations may also be addressed by ``nudging'' an initially random or stochastic set of templates to locally improve coverage [CITE Indik?].  
The match may either be evaluated via an approximate metric (e.g.~\cite{Ajith:2012mn,Kalaghatgi:2015nia}) or by directly computing the maximized overlap with existing templates.  For both methods the computational cost of building stochastic banks can become large: there exist several strategies to limit this cost while maintaining %the resulting bank's 
efficiency of coverage, e.g.~\cite{Capano:2016dsf}, including combinations of geometric and stochastic placement~\cite{Roy:2017qgg,DalCanton:2017ala,Roy:2017oul}.  High-frequency effects arising from tidal deformation of \ac{NS} components may invalidate the geometric placement framework, as a quadratic approximation of the overlap is insufficient, however stochastic methods are still applicable~\cite{Harry:2021hls}. 

For target signals considered as the dominant multipole emission from quasi-circular \acp{BBH}\ with non-precessing spins, which may be written as a single sinusoidal term, the template bank ranges over two masses and two aligned spin components, though in practice only the combination $\chi_\mathrm{eff}$ has a large enough effect on the signal to affect placement~\cite{Brown:2012qf,Ajith:2012mn,Privitera:2013xza,DalCanton:2014hxh}.  This parameter space is essentially completely covered by banks in use for LVK searches, up to limits imposed by the waveform models.  The case of searches for low-mass eccentric binaries may also be treated via this standard route, considering the effect of eccentricity as a correction to the dominant mode \ac{PN} evolution~\cite{Nitz:2019spj,Dhurkunde:2023qoe}, though the larger parameter space leads to severe computational limitations. 

\paragraph{Hierarchical search and alternative templated methods}
The above discussion assumes the use of a single, fixed bank for a given data set; however, alternative schemes may yield computational savings.  Hierarchical search first applies a ``coarse'' bank with lower minimal match, then selectively filters an additional higher-density bank around times and parameter points with high \ac{SNR}, e.g.~\cite{Gadre:2018wsb,Soni:2021vls,Dhurkunde:2021csz}.  Fixed banks may also be dispensed with altogether, in favour of time-dependent stochastic selection of template parameters, given a suitable optimization method~\cite{Hanna:2021luk}; thus, particle swarm optimization (PSO) promises to reduce computational cost in cases where a very large fixed bank is required~\cite{Weerathunga:2017qce,Srivastava:2018wvy}, as for instance eccentric inspirals~\cite{Pal:2023dyg}. 

\paragraph{Effectualness and bank robustness}
We do not expect any real \ac{CBC} signal to be exactly represented by the %dominant mode quasi-circular aligned spin model
template waveforms -- real signals will be ``outside'' the submanifold in signal space over which a bank is placed, even if \ac{NR} waveforms are used directly as templates~\cite{Kumar:2013gwa}.  The bank's usefulness is measured by its \emph{effectualness} or \emph{fitting factor}~\cite{Apostolatos:1995pj} for a given signal $h_\mathrm{test}(\pvec)$, defined as $\max_p \mathbf{M}(h_\mathrm{test}(\pvec),h(\pvec'_p))$, where the phase and coalescence time of the template $h(\pvec'_p)$ are maximized over when calculating $\mathbf{M}$; for multi-modal signals, this is not equivalent to maximizing over the signal phase.  In practice, %with the exact form of true signals unknown, 
various strategies exist to evaluate the robustness of banks: using $h_\mathrm{test}$ from different waveform models or approximants, or from \ac{NR}; taking the test signal's parameters $\pvec$ to lie outside the region covered by the bank; and including the effects of non-dominant multipoles, spin-induced orbital precession or orbital eccentricity (beyond the leading \ac{PN} effect).  We discuss %the changes needed in search 
methods for improving coverage of such general signals below in Sec.~\ref{ss:complex_search}. 

\subsubsection{Optimizing searches for a target signal population}
\label{sss:opt_pop_search}
We now consider implementation of the theoretically optimal search method for a signal of unknown parameters, i.e.\ the marginalized likelihood ratio of Eq.~\eqref{eq:marglr}.  If sufficient computational resources are available, the likelihood integral may in principle be evaluated by directly sampling the entire search parameter space, using methods adapted to source parameter estimation in place of template filtering, Sec.~\ref{ss:pe_sampling}; see also~\cite{Smith:2017vfk}.  In practice, the integral over parameter space can be approximated by a sum over templates (marginalizing over phase and coalescence time for each template)~\cite{Dent:2013cva}.  In such a sum, the density or distribution of signals over the search parameter space $p(\pvec)$ is replaced by a signal probability \emph{per template} (at given \ac{SNR}); this probability will then scale inversely with the density of templates.

The expected behaviour of the likelihood in the neighbourhood of a local maximum is given by the Fisher metric,
see Sec.~\ref{ss:FisherMatrix}, thus the marginalized likelihood over a region around the template with maximum \ac{SNR} may also be approximated analytically.  The resulting ranking statistic contains a term for %representing the
%ratio of signal density to the Fisher metric density
the ratio of the astrophysical signal distribution to the template bank density induced by the Fisher metric~\cite{Dent:2013cva} (assuming a  %near-ideal bank placement with 
near constant correlation/match between neighbouring templates).  The bank or metric density may be thought of as the \emph{density of independent noise events} over parameter space, yielding an interesting connection between parameter measurability and detection: the more precisely the parameters of a signal can be measured at a given \ac{SNR}, the higher is the rate of noise events in a matched filter search. 
%\kc{I did not understand this Are we saying that parameters can be measured better in regions of high template density? I do not understand the connection here to the ``noise", or is this not detector noise?}.  
% TD: Yes, correct, higher template density implies larger Fisher metric components which implies better measurability at a given SNR. 
% By 'noise' I mean noise events (maxima of SNR above some threshold).  So, the response of the search to (Gaussian) detector noise. 
Template or noise density is very highly concentrated in the low component mass regions of \ac{CBC} parameter space, in contrast to signal density which is concentrated mainly over component masses $10$--$50$\,\Msun~\cite{KAGRA:2021duu}.  For a signal density that varies rapidly over the search space, it may also be important to account for the effect of noise, and/or of biases in the search templates relative to expected signal waveforms, on the position of likelihood maxima~\cite{Fong:2018elx}.

While the detection rate is maximized by a ranking where the signal model corresponds to the best current estimate of the astrophysical merger distribution~\cite{Kumar:2024bfe}, other figures of merit may be optimized by suitable model choices. %of signal density.
Specific subpopulations of astrophysical interest may be targeted via a strongly peaked signal density~\cite{Magee:2019vmb,Li:2023zdl}; conversely, search sensitivity may be maintained over a broad parameter space, accounting for template density effects, via a simple monotonic model, for instance a merger rate uniform over $\log m_1$, $\log m_2$~\cite{Fong:2018elx,LIGOScientific:2020ibl,Phukon:2021cus}.

\subsection{Computational optimization and low latency search}
\label{ss:low_latency}

The evaluation of matched filter time series via DFT~\cite{Allen:2005fk} is realized by instruction sets 
%for various computing architectures 
that optimize %the 
throughput %of templates 
per CPU or GPU core when processing large batches of data.  However, DFT is not immediately adapted to all search applications: %specifically, 
as one requires a segment's worth of samples to be convolved with the template, %at once,
waiting for a complete input data segment may introduce tens or hundreds of seconds additional latency.  In practice, DFT optimization allows for the use of redundant, overlapping data segments to reduce latency to $\leq \mathcal{O}(10\,\mathrm{s})$~\cite{Nitz:2018rgo,DalCanton:2020vpm}.

Significant computational savings may also be realized by exploiting the time-depen\-dence of the inspiral-merger signal: the cost of filtering scales with the sample rate, but the highest rate is only required for the last few template cycles (considering only the dominant sinusoidal mode), the previous cycles being well represented at a lower rate.  Thus, separate filter banks operating with different bandwidths may be combined to reconstruct the \ac{SNR} of the full template at much lower total cost~\cite{Buskulic:2010zz,Adams:2015ulm,Aubin:2020goo}. 

An alternative strategy for streaming low latency search is the time domain filter~\cite{Cannon:2011vi,Luan:2011qx,Hooper:2011rb} which introduces negligible latency, at the cost of significantly higher basic computational cost.  (Time-domain whitenening or overwhitening filters still require a nontrivial data buffer, unless carefully designed~\cite{Tsukada:2017cuf}.)  In practice, the number of template filters must be minimized.  Given a bank with high minimal match, the outputs of neighbouring templates 
%must have a high match with each other, thus their outputs 
are highly correlated: one template may be written as a sum of neighbouring templates plus a small (few \%) additional term.  The number of matched filtering operations can be reduced by \emph{decomposition} of a bank into a smaller number of (near) orthogonal filters having a negligible match with each other, for instance singular value decomposition~\cite{Cannon:2010qh}.  However, reconstructing the 
%entire 
\ac{SNR} time series of the original bank from a smaller basis requires a large number of matrix multiplications: careful tuning is required to realize a computationally workable search~\cite{Messick:2016aqy,Chu:2020pjv}. 

If our figure of merit for the detection algorithm gives enough weight to how early a signal may be identified, as against sensitivity (probability of detection) at fixed $F(\varrho^\ast)$, then a low bandwidth filter which can produce an output \emph{before the coalescence time}, even at the cost of underestimating the likelihood ratio, may form part of an optimized strategy~\cite{Sachdev:2020lfd,Nitz:2020vym,Kovalam:2021bgg}.  While such ``early warning'' filters also imply reduced parameter precision, even a rough indication of sky position and distance may be valuable for guiding initial followup observations.  %by non-\ac{GW} facilities.

\subsection{Searches beyond a single GW emission mode} 
\label{ss:complex_search}

The machinery of complex matched filters and matches (overlaps) maximized over phase, which underlies current \ac{CBC} searches, requires templates with a single sinusoidal component whose phase and amplitude are functions of either time or frequency.  This ``dominant mode'' assumption is a good approximation over a significant range of parameter space, as the effects of orbital precession or non-dominant emission multipoles on signals seen at the detectors are often small.  Furthermore, the actual detected population to date lies predominantly in this range, with mostly near-equal binary masses and small in-plane spins.\footnote{These trends in the merging binary population persist after accounting for search selection biases: see Secs.~\ref{subsec:sel} and \ref{ss:populations_cosmo}.}  Nevertheless, the single-component banks have low effectualness for signals whose (optimal) \ac{SNR} receives large contributions from non-dominant modes: specifically, unequal-mass ($q \lesssim 1/4$) binaries, especially when viewed edge-on~\cite{Capano:2013raa,CalderonBustillo:2015lrt}; signals with strong amplitude modulation during the inspiral phase due to orbital precession, which likewise implies total angular momentum nearly perpendicular to the line of sight; signals for binaries undergoing transitional precession; and very high mass binaries, for which dominant mode emission is at frequencies below the detector sensitive band.  Non-negligible eccentricity within the detectors' sensitive frequency range will also decrease effectualness and sensitivity, given that the template bank assumes quasi-circular orbits~\cite{Brown:2009ng,Divyajyoti:2023rht,Gadre:2024ndy}.

Low effectualness does not imply such signals \emph{cannot} be detected, but significant losses in sensitivity 
%relative to dominant-mode signals with the same $\OptimalSNR$ 
are expected.  Methods to suppress instrumental artefacts in non-ideal data by penalizing signals that deviate from single-component templates, see Sec.~\ref{sss:glitchy}, will in general further reduce sensitivity to such signals.  
As their sources, though rare, are of high interest for both astrophysics and fundamental gravitational physics, there is a strong motivation to improve detection prospects.  

Methods to do so fall broadly into two categories.  One is to supplant the matched filter by a ``weakly modelled'' algorithm that places only generic constraints on transient signal morphology -- for instance on \ac{GW} signal duration, frequency range and evolution, and polarization -- rather than matching to exact templates. 
%aims to reconstruct (transient) signal power present consistently over the detector network, while . 
The details of weakly modelled methods are beyond the scope of this review: see for instance~\cite{Klimenko:2015ypf,LIGOScientific:2016fbo}.  Such methods are expected to approach the detection efficiency of an optimal matched filter for high-\ac{SNR} signals, regardless of their match to any given template model; their sensitivity to quieter signals will depend on implementation and on the parameter range targeted.

The second basic approach to detection of multi-component signals is to extend the standard matched filtering machinery in order to account for the higher signal complexity, by either maximizing or marginalizing over the extrinsic parameters which control the relative amplitudes and phases of the component modes.  These parameters comprise the sky location of the source and its spatial orientation (i.e.\ the inclination, orbital phase and polarization angle); Ref.~\cite{Dhurandhar:2017rlr} gives a general discussion of such marginalization. 

\paragraph{Precessing binaries}
For the inspiral of single-spin precessing binaries, Ref.~\cite{Pan:2003qt} gives a semi-analytic maximization over these angular parameters, considering the %with a single spinning component, 
$l=2$ \ac{GW} emission formulated as a sum of 5 ``harmonic'' components.  %The implementation of 
A search for precessing binaries along these lines requires a template bank covering spin magnitude and orientation parameters, which so far has only been achieved by imposing strong restrictions on other source parameters~\cite{Indik:2016qky} in order to reduce the dimensionality of the problem: even then, the size of the bank is significantly larger than the non-precessing case. 
%, even prohibitively so.  
To circumvent the complexity of signals from generic spinning binaries, approximate search strategies have considered a simpler space of templates that also contains non-physical signals~\cite{Buonanno:2002fy}, or have maximized over extrinsic parameters without applying all physical constraints~\cite{Pan:2003qt,Buonanno:2004yd}.  However, the resulting gain in SNR %due to better matching
for precessing signals is largely offset by a higher noise level relative to a standard single-component search~\cite{VanDenBroeck:2009gd}.  Restricting to the physical signal space, \cite{Harry:2016ijz} demonstrates a method for maximizing likelihood over extrinsic parameters, with the exception of inclination, which significantly affects how precession appears in the detected signal.  The resulting search is in principle more sensitive for some precessing systems, particularly unequal-mass \ac{BBH} and \ac{NSBH}, despite the higher noise level; however a prohibitively large number of templates is still required.% to cover a full range of masses and spins. %~\cite{Harry:2016ijz}. 

More recently, the multi-component model of~\cite{Pan:2003qt} was reformulated as a power law in $\tan(\beta/2)$, where $\beta$ is the precession cone opening angle (Eq.~\eqref{eq:transf_coprecessing}), equivalent to $\theta_{JL}$ in the notation of Sec.~\ref{sss:binary_params}.  Over most of parameter space where $\tan(\beta/2) \ll 1$, the signal
%  it was noted that for a substantial part of the precessing signal parameter space 
%for small or moderate precession opening angle $\beta \lesssim 44^\circ$ [CHECK NOTATION] 
is well approximated by a subset of harmonics, often 
%covered by a simplified model summing only two components, each corresponding to a non-precessing binary
as few as two~\cite{Fairhurst:2019vut}.  
%As a consequence, a search using a straightforward extension of existing template banks, where the amplitudes and phases of the two non-precessing components are freely maximized over, is expected to identify signals with $\beta \gtrsim 18^\circ$ with higher significance than a non-precessing search. 
Using this hierarchy, \cite{McIsaac:2023ijd} constructed a bank with effectualness over $0.9$ for %essentially 
a full range of \ac{NSBH} spins, without excessive increases in computational cost and noise background, to yield substantially increased sensitivity to highly precessing signals. 

\paragraph{Non-dominant multipoles}
A multi-component search involving multipoles of different $l$ (emission ``modes'') is conceptually similar to the precessing inspiral $l=2$ case, but the component \ac{GW} frequencies are widely separated and cannot be approximated as a single sinusoid with ``sidebands''.  The additional complexity of multi-mode templates is expected to increase the noise level in searches \cite{Capano:2013raa}, requiring targeted strategies to achieve improved sensitivity. 
In~\cite{Harry:2017weg} an approach similar to~\cite{Harry:2016ijz}, but without analytic maximization over the source orbital rotation angle (corresponding to the coalescence phase) was given, %maximizing over a parameter controlling the relative amplitudes of the two \ac{GW} polarizations, 
%at the detector, 
%thus accounting for both the sky location and the polarization angle; 
treating both this angle and the inclination as parameters to be covered by a bank of multi-mode templates, in addition to binary masses and spins.  The two additional angular parameters control the relative amplitudes and phases of different multipoles.  Despite the larger bank, this strategy is more sensitive than the standard dominant mode search for binaries with significantly unequal masses and/or high (redshifted) total mass, and with nearly edge-on inclination~\cite{Chandra:2022ixv}.  Recently, a general numerical method of likelihood marginalization over extrinsic parameters~\cite{Roulet:2024hwz} was used to effectively reduce the multi-mode search to a combination of separate matched filters for each mode, potentially yielding significant efficiency gains~\cite{Wadekar:2023kym}. 
For searches using ``higher mode'' templates, as for standard single-mode templates, in addition to false alarms caused by Gaussian noise, sensitivity is also impacted by transient noise artefacts~\cite{Chandra:2022ixv,Wadekar:2023gea}; the next section presents the general problem of non-ideal detector data, before discussing relevant strategies and methods to maintain search sensitivity. 

%\kc{We should also discuss the recent IAS search with higher order modes}
%  TD - mention eccentric search strategy ? 

\subsection{Non-ideal data}
\label{ss:detection_nonideal}

\subsubsection{Noise spectrum drift and calibration} 
The theoretically ideal matched filter assumes stationary noise with a known \ac{PSD}, but the mere fact that real \ac{GW} detectors do not observe for indefinitely long times, and are subject to %long-term changes
variations in sensitivity, breaks this assumption~\cite{LIGOScientific:2019hgc}.  Even within segment lengths analyzed in practice, i.e.\ $10^2$--$10^3$\,s, \acp{PSD} can disagree significantly between shorter %($\mathcal{O}(10)\,s$) 
stretches of data, 
%indicating a short-term drift or instability 
typically by a few percent, e.g.~\cite{Venumadhav:2019tad}.  Such drift is not necessarily surprising, as %in contrast to, for instance, particle detectors, 
\ac{GW} interferometers have myriad moving parts, subject to complex and sometimes nonlinear passive isolation and active control systems~\cite{Abbott:2016xvh,aLIGO:2020wna} to suppress low-frequency ground motion; 
%, with the objective of suppressing low-frequency ground motion by many orders of magnitude [CITE].  
alignment degrees of freedom affecting the strain noise spectrum may then vary on few-second timescales.  Mild short-term \ac{PSD} variations may be accounted for in detection algorithms by suitable moving averages or ad-hoc correction factors~\cite{Messick:2016aqy,Zackay:2019kkv,Mozzon:2020gwa}. 

Search sensitivity may also be affected by inaccuracies in detector calibration: specifically, large %enough
calibration errors may increase the mismatch of templates with \ac{GW} signals in the calibrated strain. %time series 
A constant amplitude or phase calibration error has no effect on search sensitivity in a single detector, though if errors are different between detectors a multi-detector search may be affected, see Sec.~\ref{sss:coinc}. 
The effect of frequency-dependent errors on single-detector sensitivity scales quadratically with the size of the miscalibration -- considered as a fractional amplitude error, or phase error in radians~\cite{AllenCalib,Buskulic:2001vh} -- implying a negligible loss given the scale of current errors~\cite{LIGOScientific:2016xax,Sun:2020wke,Virgo:2018gxa,VIRGO:2021kfv}. 

\subsubsection{Transient excess noise: glitch suppression}
\label{sss:glitchy}

While the great majority of existing data can be well described as quasi-stationary or locally stationary Gaussian noise~\cite{LIGOScientific:2019hgc}, it has been evident since the first LIGO observing run~\cite{LIGOScientific:2003qos} that multiple, short ($\lesssim\!1$\,s) stretches of data exist that are highly inconsistent with any Gaussian noise model.\footnote{Actual \ac{GW} signals are also ``non-Gaussian'', but in addition are correlated over the detector network, correspond to astrophysical sources, and would not be interpretable as instrumental artifacts.}  Significant excess power is observed for short times over some (narrow or wide) frequency range: such noise transients are termed \emph{glitches}, e.g.~\cite{LIGOScientific:2016gtq}.  These times are in general visible as large deviations in the whitened strain time series.  Without a method for subtracting them from the data, %set searched, %[CITES?], 
glitches invalidate the assumptions of the optimal matched filter and can generate arbitrarily high spurious matched filter \ac{SNR} and $\Lambda_\mathrm{SN}$ values. 

Implementing an optimal search in the presence of glitches would require an adequate model of their morphology and amplitude and time distributions.  For some types of glitch, these parameters may in part be predicted or inferred from detector auxiliary data or environmental monitor data monitoring the physical effects, internal or external to the instrument, that cause excess measured strain power~\cite{LIGO:2021ppb,Virgo:2022ysc}.  However, many glitch types remain partly or completely undiagnosed, e.g. ``blips''~\cite{Cabero:2019orq}.  Hence, glitch suppression strategies have so far been based on more heuristic or empirical methods.  These essentially assume that glitch morphology is significantly different from \ac{CBC} signal morphology, and thus aim to find discriminator statistics which optimally separate the two populations of transient signals.  In principle, glitches in a given detector might replicate the form of \ac{CBC} signals, though this is highly implausible, except for very high-mass systems with only a few signal cycles at detectable frequencies.  In this ``worst case'', no suppression is possible except within a multi-detector analysis, where a single-detector glitch will be inconsistent with a signal present in the network, see Sec.~\ref{ss:multi_search}.  Discriminators based on multi-detector signal properties have also been proposed using methods from parameter estimation~\cite{Veitch:2009hd,Isi:2018vst}.

\paragraph{Loud glitch removal: gating and inpainting}

\begin{figure}[b]
\begin{minipage}{0.6\textwidth}
 \includegraphics[width=\linewidth]{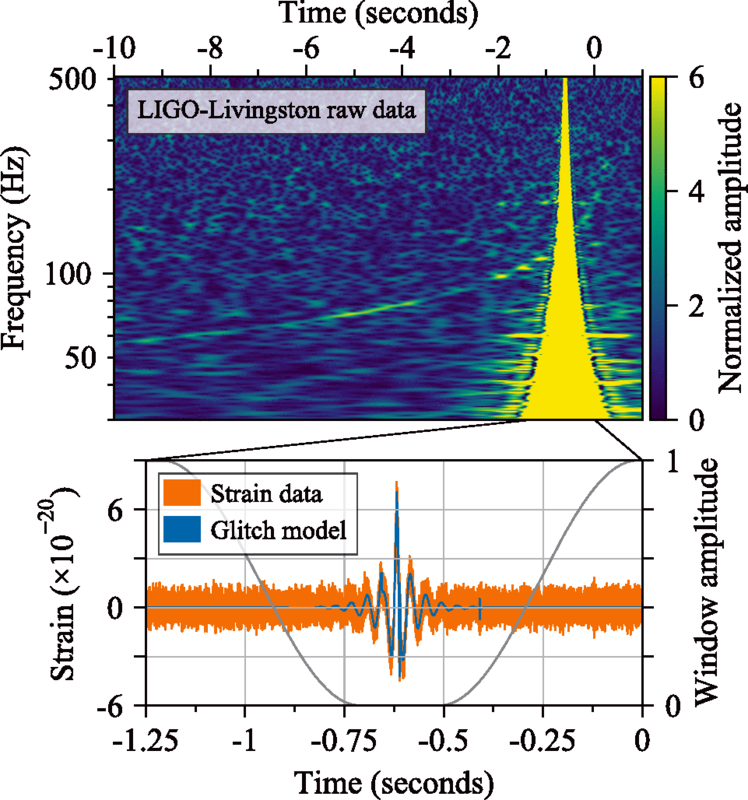}
\end{minipage}
\hspace{0.2cm}
\begin{minipage}{0.4\textwidth}
 \vspace*{0.3cm}
 \includegraphics[trim=0 0 10 0,clip,width=0.94\linewidth]{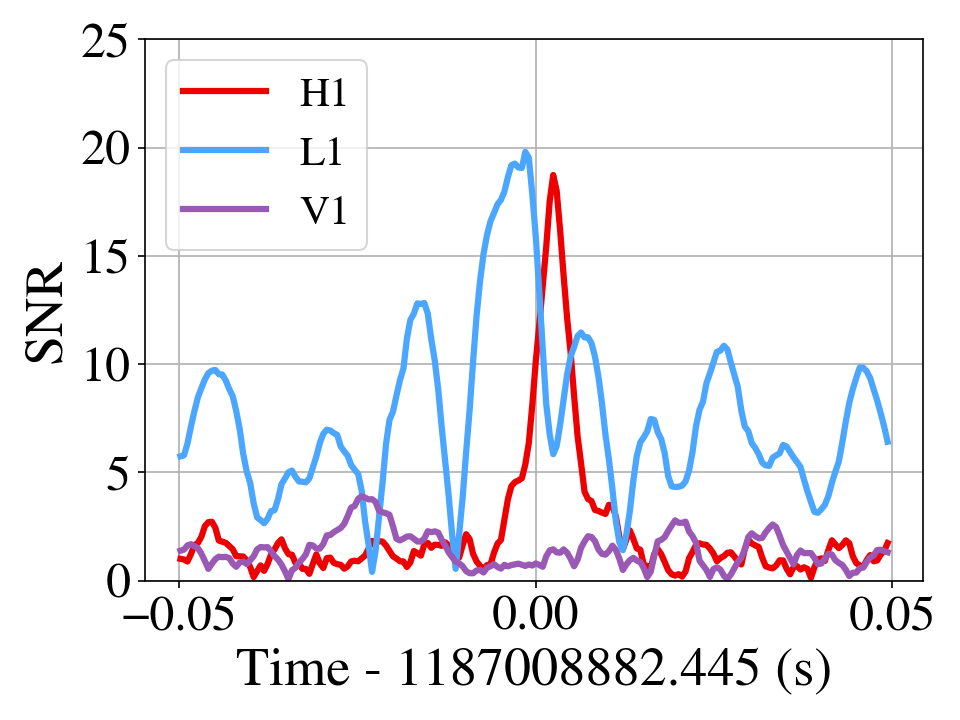}
 \vspace{0.1cm}\\ 
 \includegraphics[trim=0 0 10 0,clip,width=0.94\linewidth]{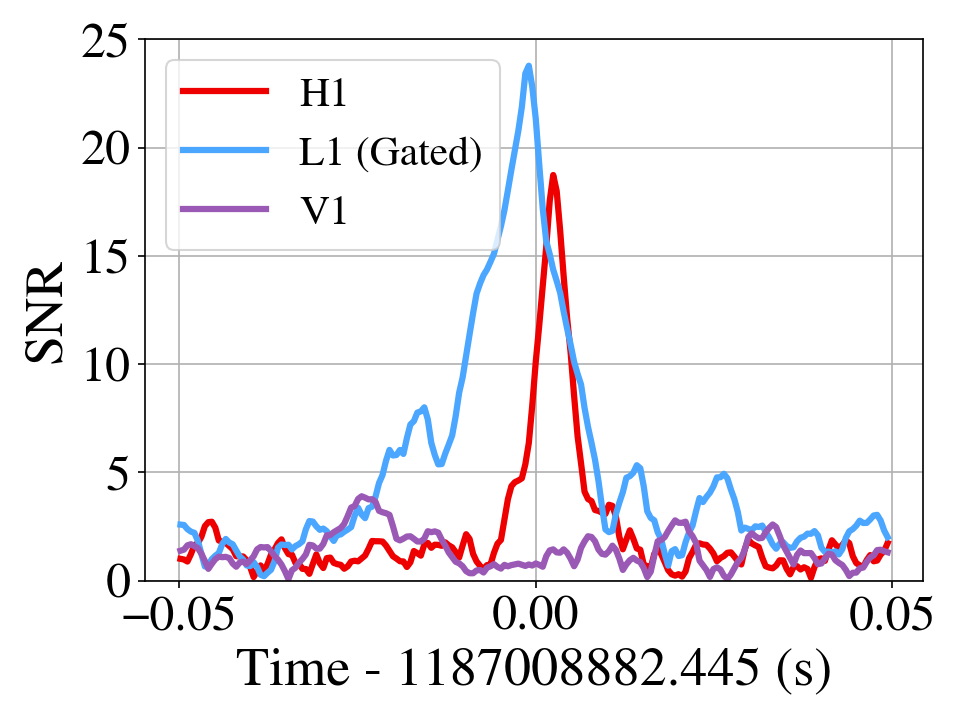}
\end{minipage}
\caption{Mitigation of the glitch in LIGO-Livingston data during the GW170817 signal. 
 %Times are shown relative to August 17, 2017 12?41:04 UTC. 
 Top left: A time-frequency representation of the raw LIGO-Livingston data; 
 % used in the initial identification [76]. 
 %The coalescence time reported by the search is at time 0.4 s in this figure and the glitch occurs 1.1 s before this time. T
 the time-frequency track of GW170817 is clearly visible despite the presence of the glitch. Bottom left: Band-passed LLO strain data (orange curve),
 %bandpassed between 30\,Hz and 2\,kHz, 
showing the glitch in the time domain. To mitigate the glitch in rapid reanalysis to produce a 3-detector sky map, 
 %shown in Fig. 3 [77], 
the raw detector data were multiplied by an inverse Tukey window (gray curve, right axis) that zeroed out (``gated'') the data around the glitch.  The blue curve shows a wavelet-based reconstruction of the glitch~\cite{Cornish:2014kda} which was subtracted from the data for subsequent analysis of the source properties.  
%The time-series data visualized in this figure are bandpassed between 30\,Hz and 2\,kHz.
 % so that the detectorÕs sensitive band is emphasized
 Figure reproduced from~\cite{LIGOScientific:2017vwq}.  Right: SNR time series from the PyCBC search, without gating (top) and with gating applied to LLO data (bottom).
}
 \label{fig:bns_gating}
\end{figure}
Very loud glitches that are not suppressed or removed may cause severe artefacts in any frequency domain (i.e.\ Fourier-transform based) analysis, and will also mask real signals if the glitch has even a small overlap with the relevant \ac{CBC} template.  Very high amplitude outliers in whitened strain ($10^2$--$10^3$ standard deviations) are sufficiently unlikely to be caused by \ac{GW} signals that the affected data can safely be ignored.  \emph{Gating} achieves this by zeroing (windowing) out short data segments ($\lesssim\!1$\,s) around these outliers, at the possible cost of introducing edge artifacts \cite{Usman:2015kfa,Messick:2016aqy,Sachdev:2019vvd}.  Searches using gating successfully detected and localized the GW170817 signal, a \ac{BNS} where one detector's data contained a loud glitch $\sim\!1\,$s before merger~\cite{LIGOScientific:2017vwq}, see Fig.~\ref{fig:bns_gating}.  Gating is restricted to short segments, as zeroing significant fractions of the data would introduce significant biases in \ac{PSD} estimation which  
%both this bias and gating a significant fraction of a template length 
may invalidate the matched filter output (on top of reducing expected signal \ac{SNR}). 

A theoretically better justified but computationally more expensive technique is \emph{inpainting}: rather than zeroing a short segment, it is replaced by a time sequence determined by the condition that the resulting likelihood or matched filter is completely insensitive to template values inside the ``hole''~\cite{Zackay:2019kkv}; i.e.\ the inpainted data contributes identically zero to the likelihood.  While inpainting removes the issue of gating edge artefacts, it still introduces some \ac{SNR} bias which requires correction.  In practice, the matched filter is not valid or useful if a significant fraction of the template duration is affected by gating or inpainting, thus short duration templates are typically vetoed in times surrounding gates/`holes' \cite{Zackay:2019kkv,Chandra:2021wbw}.  While these techniques prevent very loud glitches from raising the overall search noise distribution, they cannot avoid a loss of sensitivity around the glitch time, as they necessarily remove a portion of any \ac{GW} signal. %implies a reduction in \ac{SNR}. 

\paragraph{Signal consistency: chi-squared and friends}
Glitches that are not \emph{a priori} too loud to be confused with \ac{CBC} signals can still produce significant ($\rho \gtrsim 10$) matched filter \acp{SNR}.  If the morphology of a glitch differs significantly from the template that best matches it, the data will remain inconsistent with Gaussian noise even after subtracting the corresponding signal: i.e.\ there is excess power orthogonal to the template.  In addition, properties of the matched filter apart from the \ac{SNR} itself are unlikely to be typical of an astrophysical signal.  These considerations may be exploited to %effectively 
exclude or suppress triggers caused by glitches.  

A classic example is the time-frequency chi-squared test~\cite{Allen:2004gu}: the template is divided into disjoint frequency bins and the distribution of \ac{SNR} over the bins (at the supposed merger time of the trigger) is employed to calculate a measure of fit with the expected signal behaviour.  Typically, while the matched filter integrand of a true signal is supported over the entire template bandwidth, that of glitches is concentrated in a narrow frequency band; this chi-squared test is thus typically most effective for long duration or high bandwidth templates.
%; it does not, however, consider excess power orthogonal to the template.  
Comparable signal consistency tests use the expected time dependence of \ac{SNR} based on the filter template autocorrelation~\cite{Harry:2010fr,Messick:2016aqy,HannaThesis}, or the distribution of trigger \acp{SNR} over different templates~\cite{Harry:2010fr}, the so-called ``bank chi-squared'' test~\cite{Harry:2010fr,Tsukada:2023edh,HannaThesis}. 

The above tests consider only the glitch power that matches (is contained within) the search templates.  The ability to discriminate glitches can be improved by also measuring power orthogonal to the templates, see~\cite{Dhurandhar:2017aan} for a general discussion.  For instance, the ``sine-gaussian'' veto of~\cite{Nitz:2017lco} deploys additional filters at frequencies higher than the end (ringdown) frequency of the binary template, in order to exclude or suppress %glitches of a specific morphology:
very short duration transients with excess power over a wide frequency range~\cite{Cabero:2019orq}.  If some assumption is made on the morphology of glitches, or if their morphology can be inferred or ``learned'' from bulk data, e.g.~\cite{Soni:2021cjy,Glanzer:2022avx}, then near-optimal strategies may be designed to distinguish them from binary signals, e.g.~\cite{Joshi:2020eds,Merritt:2021xwh}: the idea is to calculate or approximate the likelihood ratio between \ac{CBC} signal and glitch hypotheses.  The feasibility of such methods depends on the number of parameters or principal components required to model glitches accurately.  
%It should not be necessary to obtain a very high match with glitch waveforms, unless we want to optimally suppress low-\ac{SNR} glitches, or unless the glitches themselves have a high match with binary templates. 

Care must be taken when applying signal-based vetoes and discriminators, as true \ac{CBC} signals do not necessarily match the search templates closely. 
For 
%A degree of mismatch can be expected with 
current searches that only model the dominant emission mode from non-precessing binaries, \ac{GW} power orthogonal to the best-matched template may arise from precession and/or ``higher modes'', posing a risk of false positives in identifying glitches (see e.g.~\cite{Nitz:2017lco}).  Any reduction in sensitivity to signals with such physical effects can be checked via analysis of simulations (as in~\cite{Capano:2016dsf}) and may be mitigated by the design of signal-based tests. 

\paragraph{Extended and/or broad-band excess noise}

\begin{figure}[tb]
 \centering
 \includegraphics[width=0.6\linewidth]{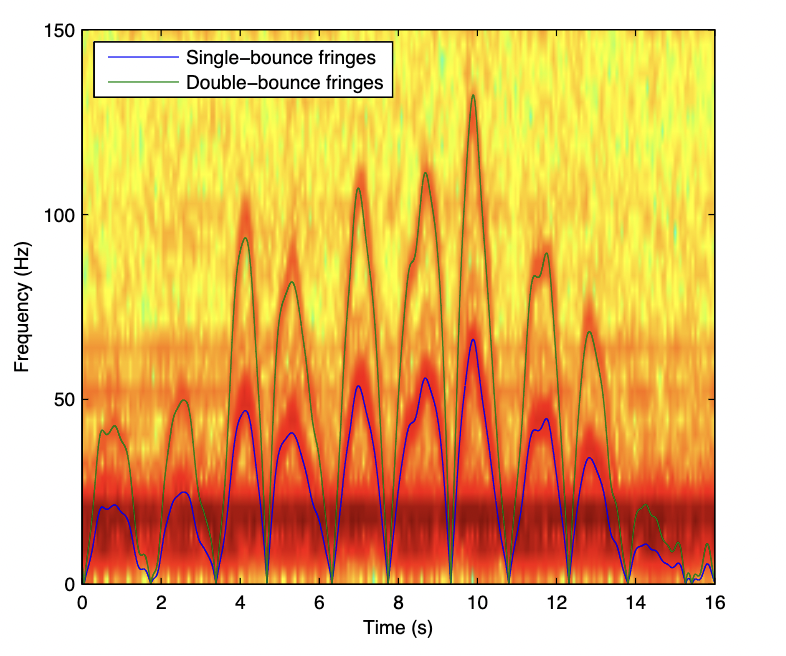}
\caption{
Spectrogram of the Virgo interferometer output signal during intense microseism activity (logarithmic color scale), showing extended broad-band artefacts due to light scattering.  Overlaid are curves predicting the frequency of excess power from the measured velocity of the source of unwanted beam reflection, an optical bench.  Figure reproduced from~\cite{Accadia:2010zzb}. 
%[REPLACE FROM JOURNAL WEBSITE] - I don't have the ability to read CQG apparently :/ (TD)
% Permission has been ordered (as of May 13)
}
 \label{fig:virgo_scatter}
\end{figure}

Some types of non-ideal detector behaviour violate the assumption that departures from Gaussian-distributed noise take the form of isolated, short glitches.  
%The preceding discussion was based on an assumption that \ac{GW} data are , apart from isolated/ transient excess power which is due either to a \ac{GW} signal or a glitch.  However, 
Longer periods of excess noise, for instance ``clusters'' of glitches correlated with a temporary instrumental problem, may be diagnosed and suppressed by monitoring the time variation of an averaged search trigger rate~\cite{Aubin:2020goo}. 
More serious noise contamination over extended periods can arise from light scattering off moving optical components into the main beam, % can cause excess power 
lasting several seconds to tens of seconds over specific frequency bands~\cite{Accadia:2010zzb,LIGO:2020zwl,Was:2020ziy}, see Fig.~\ref{fig:virgo_scatter}. 
The matched filters for templates intersecting the affected time-frequency region may be invalidated: short of ignoring such data %segments 
altogether and effectively zeroing search sensitivity, there is at present no clear strategy for tackling these artefacts.  They are an increasing concern as detector low-frequency sensitivity improves, and hence the duration of signals in the sensitive frequency band increases: the longer the signal duration, the higher the probability it will intersect an excess noise region.  Gating/inpainting may still be applied over small fractions of the template duration, but in general more complex and potentially costly methods are required, for instance explicitly modelling and subtracting the excess noise time series~\cite{Davis:2022ird,Udall:2022vkv,Tolley:2023umc}, currently not adapted for automatic use in searches.

\subsection{Multi-detector searches: coherent, coincident \& externally triggered}
\label{ss:multi_search}

The discussion has, so far, been restricted to data from a single detector, but we now consider strategies %and methods 
for searching global network data.  While noise time series in different observatories can be considered as statistically independent (with minor possible exceptions~\cite{Thrane:2013npa}), binary signals will appear consistently in each detector with predictable variations in amplitude, time of arrival and other relevant parameters over the network.  Network search methods account for these facts, while respecting constraints from computational cost. 

\subsubsection{Co-located detectors and composite data streams}
The most trivial network consists of detectors with the same location and orientation: in the past, this described the two LIGO Hanford interferometers~\cite{Lazzarini:2004hk,Fotopoulos:2008yq} and it may be relevant for a future Einstein Telescope config\-uration\cite{Freise:2008dk,Punturo:2010zz}.\footnote{In addition, for a broad range of source directions, the LIGO Hanford and Livingston detectors behave approximately as co-aligned, up to a sign flip of $h(t)$ and a time offset.}  The joint likelihood is then a product over detectors, taking the same signal in each.  
%The resulting statistic is \emph{not} a (quadrature) sum over single-detector \acp{SNR}, as maximization or marginalization over amplitude, coalescence phase \emph{etc.}\ is performed jointly over the detectors.  %Consider 
For two co-located detectors with independent noise processes we have %: 
%\begin{equation}
%	p(d_1,d_2|h(t),H_S) = const.\cdot e^{-\braket{d_1-h|d_1-h}_1/2} e^{-\braket{d_2-h|d_2-h}_2/2}, 
%\end{equation}
a log likelihood ratio 
\begin{equation} \label{logl_twoifo}
	\log \Lambda_\mathrm{SN} = \braket{d_1|h}_1 + \braket{d_2|h}_2 - \frac{1}{2}\left(\braket{h|h}_1 + \braket{h|h}_2\right)\,, 
\end{equation}
where $\braket{a|b}_{\!A}$ denotes an inner product using the \ac{PSD} of detector $A$.  The first two terms may be rewritten as $\braket{d_\mathrm{co}|h}_\mathrm{co}$ via a composite or coherent data stream $\tilde{d}_\mathrm{co} = S_\mathrm{co}(f)(\tilde{d}_1 S_{n,1}^{-1}(f) + \tilde{d}_2 S_{n,2}^{-1}(f))$, where $S_\mathrm{co}(f)$ is an initially arbitrary function %of frequency 
defining the inner product $\braket{|}_\mathrm{co}$.  Requiring that $S_\mathrm{co}(f)$ in fact be the \ac{PSD} of the data stream $\tilde{d}_\mathrm{co}$, we obtain $S_\mathrm{co}(f)^{-1} = \sum_{A} S_{n,A}(f)^{-1}$, and the terms in brackets may be rewritten as $-(1/2) \braket{h|h}_\mathrm{co}$.  Thus, co-located and -aligned detectors behave as a single composite detector with a lower noise floor than any individual one.  This remains the case if there are significant correlations between the detectors' noise streams, %as for the Initial LIGO Hanford pair, 
although the resulting expressions are more complicated~\cite{Lazzarini:2004hk}.   In addition to the composite detector, a ``null'' stream may be derived for which the \ac{GW} signal vanishes: while under ideal conditions the null stream is not useful for detection, it may be used to diagnose %a variety of 
non-ideal behaviour such as glitches and/or calibration systematic error~\cite{Ajith:2006qk,Schutz:2020hyz,Goncharov:2022dgl}. 

%Writing $h=\mathcal{A}\hat{h}$ with $\mathcal{A}$ a scalar amplitude parameter, we may maximize %over $\mathcal{A}$ 
%to obtain 
%\begin{equation}
%    \log \Lambda_\mathrm{SN} = \frac{1}{2}\left(\frac{\braket{d_1|\hat{h}}_1 + \braket{d_2|\hat{h}}_2}{\sqrt{\braket{\hat{h}|%\hat{h}}_1 + \braket{\hat{h}|\hat{h}}_2}} \right)^2
%\end{equation}
%The quantity in brackets can be thought of as a \emph{coherent} \ac{SNR}: the numerator depends on the combination $(d_1/S_{n,1}) + (d_2/S_{n,2})$ of the two data streams. 
\subsubsection{General networks}
For %a network including 
non--co-aligned detectors, %that are not co-aligned, rather than a single signal stream $h$
we consider two \emph{a priori} independent signals $h_{+,\times}(t)$ with different amplitudes and phases.  The signals seen at different %spatially separated or misaligned 
detectors are then controlled by the extrinsic parameters of source direction, inclination, and polarization angle, which for the single-detector dominant mode case discussed earlier collapse into a single amplitude factor.\footnote{For co-located but not co-aligned detectors, the likelihood may be simplified considerably~\cite{Bose:1999bp}.} 
%degenerate with luminosity distance. 
% 
In general, an optimal search is obtained by marginalizing the joint multi-detector likelihood over these extrinsic parameters given their known priors, as well as over the intrinsic mass and spin parameters.  For short transient signals the dependences on extrinsic (``sky'') and intrinsic parameters can be approximately separated~\cite{Keppel:2013uma} such that component %search may be implemented by covering the 
masses and (orbit-aligned) spins may be covered by a fixed template bank, as in the single-detector case.  The intrinsic space metric is determined by a composite \ac{PSD} which, as in the co-located case, is the harmonic sum of individual detector \acp{PSD}~\cite{Keppel:2013uma}. 
 
Currently implemented approaches treat the extrinsic parameter dependence approximately via maximization. 
For a %general network and a %non-precessing quasicircular 
binary of given masses and spins, the two %$+$ and $\times$ 
signals in the radiation frame can be written as, e.g.~\cite{Harry:2010fr},
\begin{equation}
 h_{+,\times} = \mathcal{A}^{1,2} h_0(t) + \mathcal{A}^{3,4} h_{\pi/2}(t)\,, 
\end{equation}
where $h_{0,\pi/2}(t)$ are the two waveform phases for a source at a fixed fiducial distance %(e.g.\ 1\,Mpc) 
such that $\braket{h_0|h_0}_{\!A} = \braket{h_{\pi/2}|h_{\pi/2}}_{\!A} \equiv \sigma_A^2$, and the $\mathcal{A}^\mu$ ($\mu = 1,\ldots,4$) are amplitudes dependent on the %extrinsic parameters of 
source's distance and orientation relative to the given frame.\footnote{ 
%orientation angles $\iota, \psi, \phi_0$ [CHECK NOTATION]; 
For transient signals lasting up to several minutes, 
%up to several minutes in the detectors' frequency band, 
$\mathcal{A}^\mu$ may be taken as constant.}
The signal seen by detector $A$ is decomposed as%~\cite{Harry:2010fr}
\begin{equation}
 h^A(t) = \sum_{1,2} \mathcal{A}^{1,2} F^A_{+,\times} h_0(t^A) + \sum_{3,4} \mathcal{A}^{3,4} F^A_{+,\times} h_{\pi/2}(t^A)
  \equiv \sum_{\mu=1}^4 \mathcal{A}^\mu h^A_\mu(t^A)\,,
\end{equation}
defining the four components $h^A_\mu(t^A)$, where $t^A$ is the signal arrival time at the detector.  The multi-detector likelihood ratio generalises from Eq.~\eqref{logl_twoifo} as
\begin{multline}
 \log \Lambda_\mathrm{SN} = \sum_A \left( \braket{s^A|h^A}_{\!A} - \frac{1}{2} \braket{h^A|h^A}_{\!A} \right) \\
  \equiv \sum_A \left( \mathcal{A}^\mu \braket{s^A|h^A_\mu}_{\!A} -
   \frac{1}{2} \mathcal{A}^\mu \mathcal{M}_{\mu\nu} \mathcal{A}^\nu \right)\,,
\end{multline}
where $\mathcal{M}_{\mu\nu} = \sum_A \braket{h^A_\mu|h^A_\nu}_{\!A}$~\cite{Harry:2010fr}.  Since $\braket{h_0|h_{\pi/2}} \simeq 0$, we take the matrix $\mathcal{M}_{\mu\nu}$ to be block-diagonal.  Maximizing $\Lambda_\mathrm{SN}$ freely over $\mathcal{A}^\mu$ we find a coherent \ac{SNR} 
\begin{equation}
 \rho^2_\mathrm{coh} \equiv 2 \log \max_{\mathcal{A}} \Lambda_\mathrm{SN} = 
  \sum_A \braket{s^A|h^A_\mu}_A (\mathcal{M}_{\mu\nu})^{-1} \sum_B \braket{s^B|h^B_\nu}_B\,.
\end{equation}
This coherent \ac{SNR}$^2$, comparable to the $\mathcal{F}$-statistic for continuous wave detection~\cite{Jaranowski:1998qm} is chi-squared distributed with 4 degrees of freedom~\cite{Cutler:2005hc}.  Further redefining the radiation frame via a rotation,\footnote{This redefinition will depend on the source sky direction via $F^A_{+,\times}$ and on the detector sensitivities and template waveform via $\sigma_A$.} the cross-terms between $+$ and $\times$ elements in $\mathcal{M}_{\mu\nu}$ can be eliminated, rendering the matrix diagonal.  The resulting ``dominant polarization'' \ac{SNR}$^2$, e.g.\ Eq.~(2.34) of~\cite{Harry:2010fr}, is a sum of independent power terms over the two waveform phases and two polarizations.  

For a template of given intrinsic parameters, the signal arrival times in each detector remain to be searched over.  As these times are determined by the direction to the source, i.e.\ sky position, this aspect can be handled %analogously to the intrinsic parameter space template bank, 
by setting up a discrete grid or bank of sky points with spacing determined by a condition on minimal match, based on expected loss of \ac{SNR} from slightly inaccurate relative arrival times~\cite{Keppel:2013uma,DalCanton:2015fnd,Macleod:2015jsa}; this sky mismatch contributes in addition to intrinsic parameter mismatch.  The computational cost of constructing the coherent \ac{SNR} from individual matched filters scales directly with the number of sky points, which grows rapidly for larger networks of widely separated detectors (see Chapter~4 of~\cite{DalCanton:2015fnd}), as expected given the localization accuracy of such networks~\cite{Fairhurst:2010is,Wen:2010cr}.  It is then the dominant cost except for networks with relatively poor localization, e.g.\ the initial LIGO configuration with two co-located Hanford interferometers H1 and H2~\cite{Macleod:2015jsa}, or the LHO-LLO-Virgo network before reaching Advanced design sensitivity. 

%In most cases of interest, 
Coherent search computational cost is thus 
%generally scales as the product of the number of intrinsic templates and number of sky points; both are 
strongly dependent on the maximal mismatch (i.e.\ expected loss in \ac{SNR}) used to construct both the intrinsic and sky banks.  At realistic fixed computational cost, %to complete the search 
one is forced to mismatches higher than the typically considered $\sim\,$3\%, implying a nontrivial loss of sensitivity~\cite{DalCanton:2015fnd}.  (This limitation only becomes stronger considering %the need to perform 
additional searches of time-shifted data for estimating the noise background, see Sec.~\ref{sss:significance}.)  Therefore for covering a complete range of sky directions, rather than coherent search, %theoretical optimality of the 
\emph{coincident search} has been preferred, where triggers are first obtained as peaks in \ac{SNR} for each separate detector's data, then combined over detectors in a subsequent stage, see Sec.~\ref{sss:coinc}.

\paragraph{Externally triggered searches}
If, however, the signal arrival time and source direction are more or less exactly known, such as 
%when searching for \ac{GW} counterparts to 
for sources detected via \ac{EM} transient emission, %(notably \acp{GRB}), 
a coherent search is viable at realistic computational cost with negligible loss of sensitivity~\cite{Harry:2010fr,Williamson:2014wma}.  Such \emph{externally triggered} searches have been employed for short \ac{GRB} counterparts~\cite{LIGOScientific:2010wgu,LIGOScientific:2012fcp,LIGOScientific:2017zic,LIGOScientific:2021iyk}, %over several Advanced observing runs, 
and have also been considered for fast radio bursts (FRBs)~\cite{LIGOScientific:2022jpr}.  As these searches cover small volumes of parameter space, particularly coalescence time, significantly lower \acp{SNR} are required to obtain a high evidence in favour of $H_S$ than in ``all-sky'' searches, where all source parameters are unknown.  Beyond optimizing sensitivity in the ideal Gaussian noise case, the coherent search offers specialized methods for suppressing glitches, %see Sec.~\ref{sss:glitchy}, 
for example the presence of excess power or \ac{SNR} in a null stream; these tests become more powerful if the sky direction, and thus the detector antenna factors, are known.  It is believed that \ac{GRB} are emitted within a narrow cone around the binary orbital axis (so-called ``beaming''), thus the range of orientations to be maximized over is significantly reduced, essentially assuming circular \ac{GW} polarization, which further increases sensitivity relative to the all-sky case~\cite{Williamson:2014wma}. 

\subsubsection{Coincident search and multi-detector consistency}
\label{sss:coinc}
Published searches of LIGO-Virgo data using \ac{CBC} templates have employed the coincident method, which conducts a single-detector search for each detector and then combines candidate signals over the detector network, e.g.~\cite{Babak:2012zx}.  While coincident search is in general less sensitive than coherent, it is computationally feasible and simpler to implement.  The product of the single-detector stage is a set of triggers above a low \ac{SNR} threshold, typically 4-5; %given their number 
for large template banks and long stretches of data, typically only a subset are stored.  For instance, ``clustering'' may be applied to discard triggers with significantly lower \ac{SNR} than others in nearby templates within a small time window. 

Then, candidate events for the observing network are derived from combinations of single-detector triggers which are sufficiently likely to be generated by the same \ac{GW} signal~\cite{Robinson:2008un}.  (Depending on detector sensitivities and pattern functions, a signal may also give rise to only one single-detector trigger~\cite{Callister:2017urp,Sachdev:2019vvd,Nitz:2020naa}.)  Typically, ``coincidence'' between triggers in different detectors is determined in two stages: initially, triggers are enforced to be generated in the same template and to have consistent merger times, allowing for \ac{GW} propagation time between observatories plus a small buffer. 
%few$\times 10^{-3}$\,s buffer).  
For multi-detector candidates surviving these cuts, the likelihood of the remaining trigger parameters -- \acp{SNR}, relative merger times and relative \ac{GW} phases -- is estimated based on the expected signal distribution~\cite{Cannon:2015gha,Nitz:2017svb,Hanna:2019ezx}.  As these trigger properties are determined by the putative source's extrinsic parameters, this step is analogous to maximization over source direction and orientation in the coherent search.  However, as the coincident search only considers maxima of single-detector \ac{SNR} above threshold, it will not in general be able to reproduce the maximum signal likelihood of the coherent search.  %Rather than approximating the likelihood of the \emph{data} given a signal as in coherent search, 
Instead, the coincident search estimates the likelihood of the \emph{set of single-detector triggers}. 

The single-detector \ac{SNR} threshold, arising from computational limitations, is the main limitation on the sensitivity of coincident search, as compared to an ideal (and computationally infeasible) coherent search.  
For a 2-detector network the two approaches are expected to be nearly equivalent, but
%The sensitivity of an optimized 2-detector coincident search is expected to be very similar to coherent search; 
the loss relative to optimal sensitivity increases rapidly with the number of observatories of similar sensitivity in the network from 3 upwards (see e.g.~\cite{DalCanton:2015fnd}).  For networks with many detectors, a signal may have moderate or low \ac{SNR} in several detectors, thus thresholding may cause an inaccurate estimate of its likelihood ratio. 

\paragraph{Hierarchical coherent search}
A more complicated scheme to obtain some benefits of coherent search with moderate computational cost involves making an initial selection of trigger times using a coincident search, then running the full coherent search only around these trigger times.  With careful selection of thresholds, such a hierarchical approach may approach the coherent search sensitivity for large networks at manageable cost~\cite{Bose:2011km,DalCanton:2015fnd} although the need for accurate background estimation, to be discussed in the next section, places computational limitations.  The complexity of coherent hierarchical search, as well as the status of the Advanced detector network to date, have restricted its implementation in practice, though see~\cite{Chu:2020pjv} for an exception with large scale GPU use to cover a 3-detector sky grid, and~\cite{Olsen:2022pin} for a recent application of hierarchical coherent search to the LHO-LLO network. 

%\kc{Maybe worth a mention here that cWB (sure, not a CBC search) is coherent.}
% TD - not sure really how to work it into the text, I added a reference to cwb in the 'weakly modelled search is beyond the scope of this review' remark earlier. 

%\kc{Should we again mention the IAS search here? Isn't this coherent?}
%\td{Yeah, well is it?  I'll have to try and dig through all the papers ..}
% It seems they have a coherent followup stage in the O3 search, but I can so far only find it having been applied to H-L data, which is not such of a challenge for coherent methods as there is only 1 time delay dimension. 

\subsection{Statistical methods for significance and search sensitivity}
\label{ss:detection_stats}

Having searched a stretch of data, the remaining task is to interpret the output: have signals actually been detected, and which possible signals \emph{could} have been detected?  
Given the unpredictable real detector noise and unknown signal distribution, regardless which function of trigger parameters is chosen as detection statistic, the statistic value cannot be directly interpreted in terms of odds, i.e.\
% Bayes factor\kc{it can be interpreted as a BF, I think what you mean is that we cannot easily get the actual odds ratio as the prior odds are highly non trivial} or 
probability of signal \emph{vs.} noise origin; instead the relation must be estimated empirically.  This section will describe statistical methods for interpreting search results, both concerning individual candidate detection and the overall rate of detections, and ultimately the rate of mergers for an astrophysical population of sources.  Inference on population properties beyond the rate requires more sophisticated inference, which is treated in Sec.~\ref{sec:combine}.

%\begin{itemize}
 %\item \textit{Estimating significance - timeslides, offsources etc.}
 %\item \textit{Estimating search sensitivity - injections, Monte Carlo etc.}
%\end{itemize}
 
\subsubsection{Detection significance in non-ideal data}
\label{sss:significance}
The initial objective is to measure the (frequentist) significance of a candidate event, in terms of the probability that noise would generate a comparable event.  The \emph{false alarm probability} of an event with ranking statistic $\varrho^\ast$, in a search of data with duration $T$, %[NOTATION?], 
is the probability at least one event with $\varrho \geq \varrho^\ast$ would occur if searching this duration of instrumental noise, i.e.\ in the absence of \ac{GW} signals.  Considering noise events as an inhomogeneous Poisson process with differential rate $\mu_N(\varrho)$, we have 
\begin{multline} \label{eq:fap1}
 p(\geq\hbox{1 noise events with } \varrho \geq \varrho^\ast|T) = 
 1 - p(\hbox{0 noise events with } \varrho \geq \varrho^\ast|T) \\ = 
 1 - \exp\left( -T \int_{\varrho^\ast}^\infty \mu_N(\varrho)\, \mathrm{d}\varrho \right). 
\end{multline}
This may be written in terms of \ac{FAR} at a threshold $\varrho^\ast$, i.e.\ $\int_{\varrho^\ast}^\infty \mu_N(\varrho)\, \mathrm{d}\varrho$.  

Considering a duration $\sim\,$$T$ of noise data, the \ac{FAR} may only be directly measured with precision $\sim\,$$1/T$.  However, the noise rate density $\mu_N(\varrho)$ is estimated in practice via bootstrap resampling of the actual data (although this data may contain signals~\cite{Cannon:2012zt,Capano:2017fia}): 
% KC: Doesn't that contradict some of our quoted FARs that are more than a Hubble time? I know there is some disagreement on whether these FARs should be quoted, but maybe we can soften the text a bit.
% TD : I mean you can only go this far by direct measurement without using resampling/time slides.
in order to reach very small false alarm probabilities, resampling schemes must generate much larger durations of notional noise data~\cite{Cannon:2015gha,Usman:2015kfa}.  The most conceptually straightforward method is to apply a large range of relative time-shifts to data or triggers from different detectors, obtaining samples comparable with the actual search but without coincident signals~\cite{Was:2009vh,Babak:2012zx}; the time-shifts must thus be larger than signal arrival time differences over the network.  Additional steps may be taken to remove potential signals for a mean-unbiased estimate of $\mu_N(\varrho)$, e.g.~\cite{Tsukada:2023edh,Joshi:2023ltf}; however, to ensure the correct rate of false positive outcomes, i.e.\ candidates with a very small estimated false alarm probability in the absence of signals, the time-shifted analysis must use the same data selection as the actual search~\cite{Capano:2017fia}.  For events observed only in a single detector, 
%the background time is limited to the total data duration, 
the precision of background estimates is severely limited unless extrapolation beyond the limit $\sim\,$$1/T$ is considered~\cite{Callister:2017urp,Nitz:2020naa,Davies:2022thw}. 

Typically, estimates of $\mu_N(\varrho)$ and candidate \ac{FAR} are produced by resampling hours up to a few weeks of data; on such timescales, detector behaviour (e.g.\ average rate of glitches) is assumed stable.  \ac{FAR} may be interpreted by comparison with an expected rate of astrophysical signals, as discussed further in Sec.~\ref{sss:sensitivity}, or by finding the $p$-value, i.e.\ the false alarm probability over a total duration of data $T$, via $p = 1 - \exp(T \cdot \mathrm{FAR})$ \cite{Usman:2015kfa,LIGOScientific:2016vbw}. 

\subsubsection{Search sensitivity, merger rate and probability of astrophysical origin}
\label{sss:sensitivity}

To connect the results of a search with astrophysical processes, its response to \ac{GW} signals must be estimated: this is achieved by analyzing large sets of simulated signals (``injections''), added to the data searched, or at least data with the same noise properties.  The search output for any individual injection should be the same as for an equivalent astrophysical signal, up to possible waveform systematic errors.  Injected signal parameters are drawn randomly from a population distribution which may be intended as astrophysically realistic, or which may be reweighted via importance sampling~\cite{Tiwari:2017ndi} to represent other distributions of interest. 

For relatively low-mass systems in the Initial detector network, the potential sources within a detectable range could not be approximated as homogeneous, thus search results were interpreted using an estimate of the distribution of blue light luminosity over the local Universe, an observable which is assumed to trace compact binary formation~\cite{Fairhurst:2007qj}.  For more recent observations, sources are assumed uniformly distributed in (comoving) volume and time, with random orientations~\cite{LIGOScientific:2011hqo}: thus, search sensitivity is described by the fraction of signals $\varepsilon_{\rm det}$ up to a maximum distance $d_{\rm max}$ that are detected with \ac{IFAR}\footnote{\ac{IFAR} is defined as \ac{FAR}$^{-1}$ so that higher ranking corresponds to a higher numerical value.} or ranking statistic above a threshold value. 
If a total time $T$ is searched, the search sensitive volume-time is given by~\cite{LIGOScientific:2016kwr,LIGOScientific:2016ebi}
\begin{equation}
 VT = \int_0^{d_{\rm max}}\! \frac{T}{1+z(d_L)} \epsilon_{\rm det}(d_L) \frac{dV_c}{d\,d_L} \, d\,d_L \equiv \mathcal{V}_{\rm max} \varepsilon_{\rm det}\,,
\end{equation}
%where  is the fraction of detected signals; 
where $\epsilon_{\rm det}(d_L)$ is the average probability of detection at a given distance, $V_c(d_L)$ is the comoving volume within a sphere out to distance $d_L$, and the factor $1/(1+z)$ accounts for time dilation; the total comoving volume-time $\mathcal{V}_{\rm max}$ %out to $d_{\rm max}$ 
is defined by evaluating the integral with $\epsilon_{\rm det}$ set to 1. 
If a total of $N_{\rm inj}$ injections are performed with the desired distribution and $N_{\rm det}$ are recovered above threshold, we estimate a detection fraction $\varepsilon_{\rm det} \simeq N_{\rm det} / N_{\rm inj}$.  The sensitivity $VT$ may be evaluated for an arbitrary distribution of masses and spins: in particular, for a delta function in (source frame) masses and spins, $VT$ is a function of these intrinsic parameters, proportional to the detection probability for a merger with randomly drawn extrinsic parameters within $d_{\rm max}$. 

For a constant (in comoving units) astrophysical merger rate $R$, the expected number of detections is then $R\cdot VT$.  The result of the search is then considered as a counting experiment: how many signals above threshold were actually seen?  The likelihood of this outcome given a value of $R$ allows us to infer statistical rate limits or estimates.  For this procedure to be consistent, the detection threshold should be set such that, with high probability, any event above it is a true signal; in practice, the threshold choice is somewhat arbitrary, and may bias rate estimates~\cite{Biswas:2007ni}.  In addition, the value of $V$ is strongly dependent on the assumed binary mass and spin distribution, causing, in general, large unknown systematic errors in astrophysical rate estimates. 

While this dependence on the target distribution will be addressed by astrophysical population inference in Sec.~\ref{subsec:sel}, the number of detections may be inferred more precisely, and avoiding arbitrary bias, by including potential noise events in the inference~\cite{Farr:2013yna} (cf.~\cite{Messenger:2012jy}).  We consider both the rate density of noise events $\mu_N(\varrho) \equiv r_N b(\varrho)$ and that of signals $\mu_S(\varrho) \equiv r_S f(\varrho)$, with \acp{pdf} over detection statistic $b(\varrho), f(\varrho)$ respectively.  We assume that these rates and \acp{pdf} are measured with negligible uncertainty, except for $r_S$; 
%are the respective for triggers due to noise or due to signals, where %with some assumed population; 
$f(\varrho)$ may be estimated via injections. 
In the resulting Poisson mixture model, the likelihood that a single search trigger has statistic $\varrho$ is $p(\varrho|r_S) \propto r_N b(\varrho) + r_S f(\varrho)$, 
and the likelihood of the entire set of events obtained in the search $\{\varrho_p \}$ given a value of $r_S$ is~\cite{Farr:2013yna,LIGOScientific:2016ebi}
% Laredo / Wasserman ?
\begin{equation}
 P(\{\varrho_p \}|r_S) \propto \exp [ -T (r_N + r_S) ] \prod_p \left(r_S f(\varrho_p) + r_N b(\varrho_p)\right)\,.
\end{equation}
Choosing a prior over $r_S$ (e.g.\ of ``Poisson-Jeffreys'' form $p(r_S) \propto r_S^{-1/2}$) we obtain a posterior distribution $p(r_S|\{\varrho_p \})$.  
%over $r_S$ is analytically integrable (see also [CITE Jolien]).  
If $r_S$ were known exactly, the probability that a given event is of astrophysical rather than terrestrial (noise) origin would be $p_{\rm astro}(\varrho) = r_S f(\varrho) / (r_S f(\varrho) + r_N b(\varrho))$; accounting for rate uncertainty, we find~\cite{LIGOScientific:2016kwr}
\begin{equation}
 p_{\rm astro}(\varrho) \equiv 1 - p_{\rm terr}(\varrho) = \int p(r_S|\{\varrho_p \})\, \frac{r_S f(\varrho)}{r_S f(\varrho) + r_N b(\varrho)}\, \mathrm{d} r_S\,. 
\end{equation}
The signal and noise \acp{pdf} may be extended to a multi-dimensional space of search outputs beyond the ranking statistic $\varrho$, notably template masses and spins; 
%, giving rate densities over a multi-dimensional space.  
this extension allows for signal model features (for instance a specific mass distribution) which are not included in the ranking $\varrho$~\cite{Andres:2021vew}.  The method has also been extended to include multiple distinct signal components~\cite{Kapadia:2019uut}. 

The values of $p_{\rm astro}$ obtained in catalogues of \ac{CBC} events, e.g.~\cite{LIGOScientific:2021djp}, %via this formalism [CITES] 
are mainly determined by the strongly falling dependence of the background distribution $b(\varrho)$: significant variations appear between different search pipelines for some events with odds $p_{\rm astro}/p_{\rm terr}$ $\sim \mathcal{O}(1)$, reflecting both fluctuations in event parameters, e.g.\ in \ac{SNR}, and differences in methods used to suppress noise events.  The $p_{\rm astro}$ values of past events may also be revisited when new search results enable a more precise determination of the rate and distribution of true signals. 

%having \ac{FAR} ($\varrho$) below (above) threshold. 

%\subsection{Weakly modelled (`burst') searches}
% \begin{itemize}
% \item \textit{Motivation: precession, HM, eccentricity, non-GR??}
% \item \textit{Overview of methods}
% \end{itemize}

\subsection{Machine learning applications}
% \begin{itemize}
% \item \textit{Motivation}
% \item \textit{Methods and state of the art}
% \end{itemize}

Since \ac{CBC} searches still face complex issues of implementation, machine learning  methods have seen significant interest in recent years, ranging from random forest and similar binary classifiers %~\cite{Baker:2014eba,Kim:2014nba} 
to convolutional deep neural networks. %[CITES].  
Among the motivations for such methods are the unpredictable nature of glitches, motivating algorithms which can ``learn'' glitch properties empirically from data rather than using hand-tuned vetoes.  A related problem is finding a near-optimal detection statistic over the multi-dimensional parameter space of multi-detector events, where %the noise trigger distribution similarly cannot be derived from first principles and 
the ``curse of dimensionality'' may impede accurate empirical estimation of the noise and signal distributions. %from a trigger set. 

A second broad motivation for applying %machine learning or 
deep learning techniques is computational speed.  %In many applications 
Search sensitivity or latency are typically still limited by the computational cost of matched filtering, or of reconstructing template outputs from a smaller set of filters;
% is typically still limiting on search sensitivity or latency: %of obtaining results: 
in principle, classifiers such as neural networks may be evaluated with very little latency, depending on the compute architecture and constraints such as memory. 

Classical machine-learning methods, notably maximum likelihood fitting~\cite{Nitz:2017svb} and kernel density estimation~\cite{Cannon:2012zt}, are already employed to characterize noise distributions in LVK searches.  Multivariate classifiers %developed in particle physics 
such as random forest and simple neural networks have been considered since the Initial detector era~\cite{Baker:2014eba,Kim:2014nba,Kapadia:2017fhb,Kim:2019ktw}, though finding limited application due to the complexity of incorporating them
%machine learning stages 
into existing analysis pipeline architecture. 

More recently, image recognition methods have been applied to time-frequency representations of short data segments, in order to identify and suppress glitches~\cite{Zevin:2016qwy} or to directly classify signal vs.\ noise segments, bypassing matched filtering, e.g.~\cite{Alvares:2020bjg}.  While such ``visual'' methods can easily identify contaminated data, they discard the signal phase information and so cannot approach the sensitivity of matched filtering for longer duration signals. 

A more fundamental approach that has dominated recent research applies the \ac{CNN} directly to (whitened) strain data streams, thus preserving their information content~\cite{George:2016hay}. % with in principle negligible latency.  
With suitable training, a \ac{CNN} may reproduce the matched filter's detection efficiency~\cite{Gabbard:2017lja} and could address further search issues, %discussed in this section, 
for instance optimizing sensitivity for a given signal population, or in non-ideal data, e.g.~\cite{Gebhard:2019ldz}.  Deep learning training typically optimizes sensitivity at operating \ac{FAR} thresholds of 1 per $\sim$few hours, leaving a challenge of performance at lower (false) event rates, which requires more complicated schemes~\cite{Huerta:2020xyq,Schafer:2021cml}. 

When comparing \ac{CNN} performance to ``classic'' templated searches, the large-scale compute resources required for training, as well as hardware for evaluation (typically high-grade GPUs) should be borne in mind -- see e.g.~\cite{Marx:2024wjt} for a recent practical application.  There is an intriguing similarity between the compute operations in templated (time domain) low latency searches, i.e.\ fast convolution of filters with the data and reconstruction of physical template outputs via large-scale matrix multi\-plic\-ation, and the steps required to train and evaluate CNNs~\cite{Yan:2021wml}.   Application to longer duration (low mass) signals also appears to be impeded by memory constraints given the large number of waveform cycles, although see~\cite{Krastev:2019koe}.  A hybrid approach using aspects of existing templated searches within a deep learning compute architecture may yield further advances.

\newpage
\section{Parameter Estimation}
\label{sec:pe}

\subsection{Introduction to Bayesian Inference}
\label{ss:introBayes}

As an observational science, GW astronomy is %particularly 
well suited to the Bayesian probability interpretation,
whereby uncertainties resulting from noisy observations are quantified through a probability distribution. %function.
This statistical approach stands in contrast to the frequentist methods %that are 
usually employed in the detection problem, where the significance of a potential signal is estimated through comparison to a large number of noise-generated events. Bayesian inference has been widely employed in astronomy and cosmology where one is often dealing with a small amount of data, or a single realisation of a phenomenon. Historically, its use for anything but simple problems was hindered by computational difficulty, %in using Bayesian inference  has been prohibitive, 
but with the advent of modern computers and the development of algorithms such as Markov Chain Monte Carlo the method has seen a resurgence. We will introduce the building blocks necessary to understand Bayesian analysis of \ac{GW} signals; many textbooks are available for a more in-depth understanding of relevant techniques~\cite{SiviaBook,GelmanBook} and their historical and theoretical underpinnings~\cite{jeffreys1998theory,jaynes2002probability}.

Since the binary mergers giving rise to \ac{GW} signals are unique events rather than repeated trials of the same experiment, statements about their parameters have evidential probability, i.e.\ corresponding to degrees of belief based on the available evidence. In the usual case where the parameters are continuous (masses, angles, etc.), rather than finite, discrete probabilities, we must use a \ac{PDF} defined on the parameter space: integrating the \ac{PDF} over (part of) its domain then gives a finite probability.
As a density, the \ac{PDF} depends on the specific parameterisation used, thus a Jacobian must be applied to transform between different parameterisations %of the model 
(see Sec.~\ref{ss:pe_degeneracies}).

In Bayesian inference, we update our state of knowledge about a model by calculating the posterior \ac{PDF} over the model parameters $\pvec$ using Bayes' theorem:
\begin{align}\label{eq:Bayes}
p(\pvec|d,H_S) &= \frac{p(\pvec|H_S)p(d|\pvec,H_S)}{p(d|H_S)}.
\end{align}
The posterior \ac{PDF} is computed from the prior \ac{PDF}
%in the parameters of interest 
$p(\pvec|H_S)$, the likelihood function $p(d|\pvec,H_S)$, and the normalisation constant $p(d|H_S)$. 
The prior \ac{PDF} describes our prior knowledge about the parameters, before we consider the observations.
For example, without fore-knowledge of the binary's location we use a distribution uniform on the celestial sphere (see Sec.~\ref{ss:peprior}).
The likelihood function is the probability of observing the data $d$ if the parameters are $\pvec$, see Eq.~\eqref{eq:LikelihoodSignal}.
$H_S$ denotes the ``signal hypothesis'', a model of the data that defines what parameters are contained in $\vec\theta$, along with models of the waveform and the residual noise, and any additional information required to completely specify the likelihood function.
The constant $p(d|H_S)$ is the `marginal likelihood' of the data given $H_S$, it is computed by marginalising the likelihood over all parameters, i.e.\ integration weighted by the prior density:
\begin{align}\label{eq:evidence_integral}
p(d|H_S) &= \int p(\pvec|H_S)p(d|\vec\theta,H_S)\,d\pvec \equiv Z_{S}.
\end{align}
This term is also often called the `Bayesian evidence' for a model, and is commonly denoted by the symbol $Z$ in analogy to the partition function of thermodynamics. It serves as a normalisation constant for the posterior probability distribution, being independent of the parameters $\pvec$, though varying between one model and another.
%as such does not depend on the parameters of the model (although it does vary from model to model).
For a given model, %Since the evidence is a constant, 
the posterior distribution is often defined only up to a constant of proportionality:
\begin{align}
p(\pvec|d,H_S) &\propto p(\pvec|H_S)p(d|\pvec,H_S).
\end{align}

The marginal likelihood is also the expectation value of the likelihood of the data, taken over the prior for a given model: thus, as it quantifies the mean likelihood for a particular model, it can also be used to compare the posterior probability of two models.
%From Eq.~\eqref{eq:evidence_integral} we can also see that the evidence is the . In other words it .
The ``posterior odds'' $O_{1,2}$ between two models $H_1$ and $H_2$ is given by the ratio of their posterior probabilities
\begin{align}\label{eq:Bayes_model}
\frac{P(H_1|d)}{P(H_2|d)} &= \frac{P(H_1)}{P(H_2)}\frac{p(d|H_1)}{p(d|H_2)}\equiv\frac{P(H_1)}{P(H_2)}B_{1,2} \,,
\end{align}
where the Bayes factor $B_{1,2}$ is the ratio of the marginal likelihoods for hypotheses 1 and 2.
The prior odds $P(H_1)/P(H_2)$ is the ratio of model probabilities prior to considering an observation, for example the two models might be ``the source is a binary black hole'' versus ``the source is a binary neutron star''. 
Since in this example we might have different prior \acp{PDF} for the two models, based on assuming the presence or absence of tidal deformability, 
we see that the models do not need to have the same number of parameters; in general they do not even need to have any common parameters. Nevertheless, the Bayesian framework allows one to compare different assumptions quantitatively via the posterior odds ratio.  To go even further, we can allow the prior \ac{PDF} to be dependent on model `hyper-parameters', which then may themselves be estimated in hierarchical inference, for which see Sec.~\ref{sec:combine}.

%Relevant examples of model comparison are calculating the Bayes' factor for $H_1=$(General Relativity is correct) and $H_2=$(some deviation from General Relativity), or $H_1=$(compact objects' spins are always exactly aligned with their orbital angular momentum) versus $H_2=$(compact objects' spins can be misaligned). In this later case, formally the prior on $M_1$ is null, but \sout{there is still very important information to be gained by computing the Bayes' factor}\kc{the result is of astrophysical interest as it assesses the presence of spin-precession in the signal.} (CITATION, examples of precessing PE analyses).

In practical cases, while one seeks to compute a ratio of marginal likelihoods for generic models (e.g.\ General Relativity versus a particular modified gravity theory), typically the actual computation requires much more precisely specified models, which may define
%actually computes a ratio between much more specific models, which may specify 
a particular \ac{PSD}, waveform model, and additional signal processing parameters as required to unambiguously evaluate the likelihood.
Since the marginal likelihood is integrated over the model's priors, it also requires that we fully specify both the parameters of the model, and the form of the prior distribution over these parameters.
This makes a Bayesian hypothesis test sensitive to very specific model assumptions, which is desirable when relevant information is available a-priori; such assumptions should be made explicit when reporting such a result.
%TD - removed repetition of 'specific' !

\subsection{Priors}
\label{ss:peprior}

%As mentioned in Sec.~\ref{sec:detection}, Bayesian statistics are particularly well suited to gravitational-wave astronomy. 
Bayesian inference makes explicit the need for a prior probability distribution over the model parameters, which
can be chosen to incorporate any prior information or constraints. %that one may have when performing inference. 
For some parameters the prior is determined unambiguously by symmetry arguments: this is the case for the direction of the source on the sky, its spatial orientation relative to Earth, and the polarization angle, 
%(rotation around the line of sight), 
assuming that no other information is available. 
However a model assuming an electromagnetic counterpart to a binary coalescence, such as a gamma-ray burst, may use a prior encoding its (known) sky direction rather than the default prior \ac{PDF} uniform on the celestial sphere. 
All parameters present in the model (see Table~\ref{tab:parameters}) must have a prior specified before inference can be done; in practice, priors are set using a combination of such symmetry arguments, astrophysical foreknowledge or modelling, and convenience in interpreting the resulting posteriors.
%\kc{Can we also add that the prior needs to be normalizable for the marginal likelihood to make sense?}
%\td{In practice all priors have to be normalizable, because one cannot sample over a really infinite range...?}

\subsubsection{Prior from astrophysics}
The prior may be chosen as a functional form derived from studies of independent observations or other physics considerations.
% can be used to specify the prior. 
A standard example is the a-priori spatial distribution of GW sources. Following the cosmological principle and the Copernican principle, given the very large distances 
of %at which 
typical \ac{GW} sources, %are expected to be detected, 
a prior uniform in co-moving volume is most commonly used (Eq.\eqref{eq:cosmoprior} in Sec.~\ref{ss:populations_cosmo}).  While this forces the analysis to assume a cosmology, the choice can be amended in post-processing (see Sec.~\ref{sss:reweight} below).  A prior uniform in co-moving volume also has the convenient advantage of being normalisable, as the integrated volume is finite. 
%However, in some use-cases, a prior including knowledge of galaxy catalogues can be useful, and in practice this knowledge can also be included in post-processing. 

In some cases \ac{PE} results are obtained for different choices of prior representing different astrophysical assumptions, for example allowing wider or smaller ranges of \ac{NS} spin magnitude~\cite{LIGOScientific:2018hze}.  Another example would involve deriving a prior over the masses of \acp{BH} using previous electromagnetic or GW observations (in conjunction with other model assumptions), e.g. \cite{LIGOScientific:2018hze}. 
% TD - Citation? If it's an example, has anyone actually done it - eg Farr/Ozel?
%, to explicitly describe the prior knowledge of their mass distribution.

\subsubsection{Computational considerations}
The computational cost of most parameter-estimation analyses is dominated by calculating the likelihood, not the prior. However, it should still be possible to evaluate the prior quickly, %it remains important to have a prior form able to be quickly evaluated, 
as some sampling strategies (in particular nested sampling based on MCMC, see Sec.~\ref{sss:nesty}) %, e.g.\ LALInferenceNest) 
require many evaluations. %of the prior. 
Furthermore as noted above, the posterior and the prior being probability densities, any change of parametrisation $x_i \rightarrow y_i$ (using generic notation) requires computing a Jacobian,
\begin{equation}
p(\vec{y}|H) = p(\vec{x}(\vec{y})|H)\left|\frac{\partial x_i}{\partial y_j}\right|.
\end{equation}
This computation is much more convenient for priors of closed (analytic) form.

\subsubsection{Change of priors via reweighting}\label{sss:reweight}
From Eq.~\eqref{eq:Bayes}, it is clear that having calculated the posterior \ac{PDF} $p_1(\pvec|d,H_S)$ for a prior $p_1(\pvec|H_S)$ with given data, it is possible to calculate the posterior $p_2(\pvec|d,H_S)$ for a different prior $p_2(\pvec|H_S)$ over the same parameter space:
%TD - how about this
%\footnote{Strictly, we should consider different signal hypotheses in the two cases}
% A different prior implies a different model / signal hypothesis .. no?
\begin{align}
    p_2(\pvec|d,H_S) %&= \frac{p_2(\pvec|M)p(d|\pvec,M)}{p(d|M)}  \nonumber\\
                          %&= \frac{p_2(\pvec|M)}{p_1(\pvec|
                          %M)}\frac{p_1(\pvec|M)p(d|\pvec,M)}{p(d|M)} \\
                          &= \frac{p_2(\pvec|H_S)}{p_1(\pvec|H_S)} p_1(\pvec|d,H_S).
\end{align}
Thus, provided that the \emph{weight} $p_2(\pvec|H_S)p_1(\pvec|H_S)^{-1}$ can be computed, it is possible to \emph{reweight} the posterior to accommodate a different prior assumption (even a parametrised one --- see Sec.~\ref{sec:combine} for details). 
To obtain an accurate result, the estimation of the original posterior $p_1(\pvec|d,H_S)$ must have delivered enough information (usually in the form of posterior samples, see Sec.\ref{ss:pe_sampling}, or an accurate interpolant) where the weight is large, so where $p_2(\pvec|H_S)$ is large and $p_1(\pvec|H_S)$ small. 
% TD : a reader won't know yet what are posterior samples or interpolants.  Refer forward or rephrase?
This is made usually more difficult as most \ac{PE} schemes are designed to spend most computing resources in regions where $p_1(\pvec|d,H_S)$ (and $p_1(\pvec|H_S)$) is large. 
% TD : large prior does not imply large posterior .. 

To enable flexible use of posterior \ac{PDF} computations, it is thus usually better to avoid priors that exclude part of the parameter-space, and instead be very conservative for the computationally expensive first estimation of the posterior, relying on reweighting for subsequent analyses with different priors. 
This often leads in practice to adopting priors that are uniform over parameters which are of particular interest for further studies: this choice allows a wide range of reweighting options, and also makes summary statistics easier to interpret.

\subsubsection{Other considerations on priors}
Another approach involves uninformative priors, or the so-called Jeffreys prior, which is a density proportional to the square root of the determinant of the Fisher information matrix. It is invariant under a change of parametrisation, but can require extra computation to determine its density at points of interest, since the Fisher information matrix depends on the signal and noise models for a particular observation.

As mentioned above, a prior \ac{PDF} represents a distribution of possible signals over parameter space, thus prior choices are necessarily connected with issues of astrophysical or cosmological population modelling.  A systematic treatment of such population models will require Bayesian hierarchical inference, where the goal is to determine properties of the global population, rather than of individual sources; this will be presented in Section~\ref{sec:combine}. 
% Further prior considerations can arise from the \textit{Malmquist Bias}: as with most fields of observational astronomy, GW astronomy can be impacted by the Malmquist bias. Weak signals might not be detected and not trigger generated, see Sec.~\ref{sec:detection}. 
%The samples of signals on which \ac{PE} is run thus is not the same as an astrophysical sample of all possible signals. However in practice this effect is handled at population-level analyses, see Sec~\ref{sec:combine}, and one can use astrophysical prior to answer the well-posed question of the probability of parameters assuming said astrophysical priors.

\subsection{Likelihood function}
%\emph{Responsible:} John

In the context of Bayesian \ac{PE}, the statistical model describes the entire process by which the data is %could be 
created, including both signal and noise models and all relevant parameters required to predict the signal. In Sec.~\ref{sec:structure} we modeled the
noise as a stationary Gaussian process, giving us a noise likelihood function.
%for instance the assumption of stationary Gaussian noise (see section~\ref{sec:detection} for non-Gaussian, non-stationary noise and PSD). However, often in the context of CBC analyses, Models refers the models of the gravitational-wave signal, parametrised by the parameters of astrophysical interest \pvec.
We have seen in Sec.~\ref{sec:cbc} how the signal model is defined using a parameter vector $\pvec$; together with a stationary Gaussian noise model, this yields the likelihood function in Eq.~\eqref{eq:LikelihoodSignal}, defined in terms of the data $d$, signal model $h(\pvec)$, and the signal hypothesis $H_S$, which includes the statistical description of the noise via its 
%the noise 
power spectral density $S_N(f)$: 
\begin{align}
 p(\vec d|\vec{h}(\pvec), H_S) \propto \exp\left(-\frac{1}{2}\braket{\vec{d}-\vec{h}|\vec{d}-\vec{h}} \right)\,.
\end{align}
%\td{discuss selection effects, or their apparent absence, here?}
% I.e. remark that PE is usually only run on data for which a search algorithm indicates detection, meaning some measure of significance above threshold.  So the likelihood ought to be p(d, det|h(theta)). 
% But if we assume the search is deterministic, we know that p(det|d) = 1 for such data, so p(d, det|anything) is identically equal to p(d|anything).  So we can effectively ignore this when doing PE. 
% Refer forward to population selection subsection for more discussion. 
% Previous text from the prior section is pasted here .. 
%as with most fields of observational astronomy, GW astronomy can be impacted by the Malmquist bias. Weak signals might not be detected and not trigger generated, see Sec.~\ref{sec:detection}. 
%The samples of signals on which \ac{PE} is run thus is not the same as an astrophysical sample of all possible signals.
The inner product inside the exponential is in practice evaluated with a finite amount of data sampled over time $T$, with a sampling rate $f_s$, yielding a discrete sum over $N=T/f_s$ points in the time domain
instead of an integral.
In this case the maximum frequency that can be represented in this data is the Nyquist frequency $f_s/2$,
and since the data time series is real, we have in the frequency domain
%the conjugation relation 
$\tilde{d}(-f)=\tilde{d}(f)^\ast$. 
The (log) likelihood for a \ac{GW} signal template $\tilde{h}$ is then 
%defined by the frequency-domain 
a sum over a finite range of positive frequencies,
\begin{align}\label{eq:likelihoodsum}
 p(\vec d|\vec{h}(\pvec), H_S) \propto \exp\left(-\frac{2}{T}\sum_{i=0}^{N/2}\frac{|\tilde{d}_i-\tilde{h}_i|^2}{S_N(f_i)}\right)\,.
\end{align}
Multiple detectors can be incorporated %in the analysis 
by simply multiplying their individual likelihood functions, assuming their noise is statistically independent,
\begin{equation}
 p(\vec d|\vec{h}(\pvec), H_S) \propto \prod_{A} p(\vec d_A|\vec{h}_A(\pvec), H_S, A),
\end{equation}
where the index $A$ runs over available detectors and its use as conditioning information denotes that the corresponding noise \ac{PSD} is to be used. 

In cases where the waveform is calculated in the time domain, care needs to be taken to transform into the frequency domain. To avoid aliasing, $f_s$ must be chosen so that the maximum frequency of the signal is no greater than the Nyquist frequency $f_s/2$, and a window function is applied before applying a discrete Fourier transform.
%In particular, 
The window function is used to taper the start and end of the waveform, reducing the spurious high frequency content introduced by the discontinuity between the start and end time if no window is applied.
This windowing may lead to an overall loss of the signal's power: 
the Fourier transform of the data $\tilde{d}$, after applying a window whose Fourier transform is $\tilde{w}$, is $\tilde{d}_w=\tilde{d}*\tilde{w}$, and the likelihood needs to be modified to take this into account~\cite{Talbot:2021igi}.

For signal models generated natively in the frequency domain, particular care needs to be taken of the waveform's abrupt start and end: such unphysical features may match against data and lead to artificially precise and biased results~\cite{Mandel:2014tca}. In general with CBC templates including merger and ringdown, the signal model naturally tapers at its end, while the start is in general at low enough frequencies so as not to contribute to the likelihood integral.
% TD - it's still not clear to me how we do actually do PE for BNS signals to avoid effects from the unphysical upper frequency cutoff ..  I guess it could be assumed to be at high enough frequency that the effect on likelihood is small?

%Note that 
The likelihood of Eq.~\eqref{eq:likelihoodsum} is the most commonly used form, but relies on assumptions of a stationary Gaussian noise process with exactly known \ac{PSD} and no windowing, damping or correlation effects. Various works have looked at relaxing these assumptions, e.g.
\cite{Rover:2008yp,Littenberg:2014oda,Chatziioannou:2021ezd,Chatziioannou:2019zvs,Talbot:2020auc,Hourihane:2022doe,Ashton:2022ztk}. %\td{explain or omit damping/correlation?}

 %\begin{itemize}
 %\item Range of frequencies, windowing, abrupt ends in signal models.
 %\item See section~\ref{sec:detection} for non-Gaussian, non-stationary noise and PSD
 %\item Marginalisation over glitches, detector calibration uncertainties, PSD uncertainty.
 %\end{itemize}

\subsection{Exploring parameter space}
\label{ss:pe_sampling}

%Stochastic sampling (MCMC etc), grid (rapid skyloc) and hybrid methods

The posterior \ac{PDF} contains all the information about the model parameters that can be extracted from the data.
However, its domain is the entire prior, which for compact binaries is a high-dimensional space containing in practice 15 parameters or more for a generic quasi-circular coalescing black hole binary.
While we can evaluate the density at any point in parameter space using Eq.~\eqref{eq:Bayes}, it is not practical to evaluate the posterior entirely on a grid due to the size of the parameter space.
%This can be quantified by the information gained between the prior and posterior probability density functions, given by the Kullbeck-Liebler (K-L) divergence~\cite{}.
One is often concerned with the \emph{marginal} distributions, given by integrating the posterior over ``nuisance'' parameters, for example we may only be interested in the chirp mass $\Mc$, in which case all the other parameters are nuisance parameters denoted $\pvec'$. The marginal posterior is
\[
p(\Mc|d,H_S)=\int  p(\Mc,\pvec'|d,H_S) d\pvec' = \int  \frac{p(d|\Mc,\pvec',H_S)p(\Mc,\pvec'|H_S)}{p(d|H_S)} d\pvec'\,.
\]
Evaluating this integral using standard quadrature would be very difficult in high dimensional spaces.
Instead, in practice, Monte Carlo methods are often used to compute quantities of interest. These rely on a \emph{sampling} of
the posterior distribution at various points, chosen with a particular application in mind.

Final \ac{PE} results are typically provided as samples from the posterior distribution itself, i.e.\ a set of points in parameter space, where the density of the samples is proportional to the posterior \ac{PDF}.
With such a sampling, integrals weighted by the posterior probability are easily evaluated by Monte Carlo methods, with each sample
given an equal weight in the integrand. For example the marginal distribution above can be approximated by a histogram,
where the number of samples in each $\Mc$ bin would be proportional to the probability density in that bin.
%\jv{Add in a diagram of histogram - could be used to explain how to read results plots?}
Using posterior samples also makes it trivial to produce \emph{credible intervals} for individual parameters, defined to be an interval of one-dimensional parameter space $(\theta_\mathrm{min},\theta_\mathrm{max})$ that contains a chosen fraction $X$ of the probability, and therefore posterior samples, i.e.
\begin{align}
    \int_{\theta_\mathrm{min}}^{\theta_\mathrm{max}} p(\theta|d) d\theta &= X\,.
\end{align}
There are however many choices of $\theta_\mathrm{min}$ and $\theta_\mathrm{max}$ that satisfy this equation. A common choice of interval to quote is the \emph{minimum credible interval}, which spans the region of highest probability density that contains $X$.

A number of methods have been developed to draw the posterior samples efficiently, based on a variety of statistical methods, with different degrees of specialisation. We give an overview of the main methods and their historical development below.

\subsubsection{Markov Chain Monte Carlo}
% inspiral_mcmc, SpinSpiral, LALInference, BayesWave, Bilby-MCMC...

%In practice, this involve numerical exploration of the parameter space, using the resulting samples to measure credible intervals, and numerical integration of the posterior to obtain the evidence and perform model selection. Various samplers have been used over the years in gravitational-wave astronomy, including:

Markov-Chain Monte-Carlo (MCMC) methods are a class of algorithms that enable efficient exploration of high-dimensional probability distributions. First introduced in 1953 in the context of simulating the behavior of particles in a gas~\cite{10.1063/1.1699114}, they are often a building block of other sampling algorithms.

MCMC methods work by constructing a Markov chain whose stationary distribution is the desired target distribution: the chain is thus designed to converge to the target distribution after a sufficient number of steps. At each step of the chain, the algorithm proposes a new state, which is accepted or rejected based on a probabilistic criterion that preserves the desired stationary distribution.

One of the most commonly used such methods %MCMC algorithms 
is the Metropolis-Hastings algorithm
%, which was 
introduced %by Nicholas Metropolis 
in 1953~\cite{osti_4390578} and extended %by W. Keith Hastings 
in 1970~\cite{10.1093/biomet/57.1.97}. 
In order to sample from a target probability distribution $\pi(\theta)$, the algorithm consists of the following steps:
\begin{itemize}
    \item Start with an initial parameter vector $\theta_0$, often drawn from the prior.
    \item At each iteration $t$, propose a new parameter vector $\theta'$ from a proposal distribution $q(\theta'|\theta_{t-1})$, where $q$ is a proposal distribution that depends on the current parameter value $\theta_{t-1}$ (also knowns as a ``jump proposal distribution"). A common choice for $q$ is a Gaussian distribution with mean $\theta_{t-1}$ and some fixed covariance matrix.
    \item Calculate the acceptance ratio 
    \begin{equation}
    A = \frac{\pi(\theta')}{\pi(\theta_{t-1})} \frac{q(\theta_{t-1}|\theta')}{q(\theta'|\theta_{t-1})}\,,
    \end{equation}
     where $\pi(\theta)$ is the target distribution and $q(\theta'|\theta_{t-1})$ is the probability of proposing $\theta'$ given the current parameter value $\theta_{t-1}$. The ratio of the proposal distributions $q$ cancels out if $q$ is symmetric, but if not it is required to maintain detailed balance.
    \item Set $\theta_t = \begin{cases} \theta' & \text{with probability } \min(1,A) \\ \theta_{t-1} & \text{with probability } 1 - \min(1,A) \end{cases}$\,.
    \item Repeat steps 2-4 for a sufficient number of iterations to obtain a stationary distribution.
\end{itemize}

In step 3, the acceptance ratio $A$ determines whether the proposed parameter vector $\theta'$ is accepted or rejected. If $A > 1$, then $\theta'$ is accepted with probability 1, since the new value has a higher probability density than the current value. If $A < 1$, then $\theta'$ is accepted with probability $A$, which is proportional to the likelihood of the proposed value relative to the current value. This ensures that the chain eventually converges to the target distribution $\pi(\theta)$.

PTMCMC (Parallel Tempering Markov Chain Monte Carlo) is an advanced variant of MCMC that incorporates parallel tempering to improve the exploration of complex probability distributions (\cite{PhysRevLett.57.2607,keramidas1991computing,B509983H}, and see~\cite{vanderSluys:2008qx,Veitch:2014wba,Farr:2013tia} for examples of GW-specific applications). In PTMCMC, multiple chains, each associated with a specific temperature, are run simultaneously. These different temperatures act as scaling factors, controlling the distribution's ``flatness''. By having chains at higher temperatures explore the distribution more broadly and chains at lower temperatures focus on regions of higher density, PTMCMC enables a more efficient exploration of the entire parameter space.
To achieve this, the PTMCMC chains periodically swap parameters between neighboring temperatures, so that information about the tails of the distribution from the higher temperature chains can percolate to the lower temperature ones. This involves a special jump proposal that exchanges states between adjacent chains, which is  accepted or rejected based on the temperatures and the likelihoods of the current states in the chains. This process promotes better chain mixing and facilitates the movement of states between different regions of the distribution. Since different temperature chains can run in parallel when a swap is not taking place, this aspect of PTMCMC can be efficiently parallelised.

There are several other MCMC methods that are commonly used in addition to the Metropolis-Hastings algorithm. One popular method is the Gibbs sampler (\cite{Geman:1984,Gelfand:1990}, and see \cite{Veitch:2014wba} for examples of GW-specific applications), which is efficient when the joint distribution of the parameters can be factorized into conditional distributions. The Gibbs sampler updates one parameter at a time, by drawing from its conditional distribution given the current values of the other parameters. This can be more efficient than the Metropolis-Hastings algorithm when the conditional distributions are easier to sample from than the full joint distribution.

Another commonly used MCMC method is the Hamiltonian Monte Carlo (HMC) algorithm (\cite{1987PhLB..195..216D,PhysRevD.75.083525}, and see \cite{Porter_2014,PhysRevD.107.043013} for examples of GW-specific applications), which is designed to move more efficiently through high-dimensional parameter spaces. HMC generates proposals by simulating the Hamiltonian dynamics of the system, which involves using the gradient of the log-posterior distribution to simulate the motion of a particle in a higher-dimensional space. HMC can be particularly useful for sampling from target distributions with complex geometries or correlations between parameters.

There are also variations of the Metropolis-Hastings algorithm, such as the adaptive Metropolis-Hastings algorithm \cite{bj/1080222083,Andrieu2008}, which adjusts the proposal distribution during the sampling process to improve efficiency. The slice sampling algorithm \cite{10.1111/1467-9868.00179,10.1214/aos/1056562461} is another variation, which generates proposals by slicing the target distribution at a random point and using a bounded interval to select a new value. Slice sampling can be particularly useful when the target distribution is difficult to sample from using other methods.

In general, MCMC methods tend not to suffer as much from the curse of dimensionality as other sampling algorithms, but %despite their usefulness 
they can be computationally expensive, and require careful tuning of the proposal distribution to ensure efficient exploration of the parameter space. Problem-specific proposal distributions are almost always required, and in the context of GW astronomy depend on the detector network configuration and the signal properties.

\subsubsection{Nested Sampling}\label{sss:nesty}
% Inspnest, LALInference, Bilby

Nested sampling is a more recently developed Bayesian inference algorithm, proposed initially in~\cite{Skilling:2006gxv}. 
Its aim %of the nested sampling algorithm 
is to evaluate the marginal likelihood integral of Eq.~\eqref{eq:evidence_integral} through a stochastic sampling of the parameter space. As a by-product, it can also produce samples from the posterior \ac{PDF}. 
The method relies on a mapping of the $D$-dimensional integral to a one-dimensional integral
\begin{align}
    Z = \int p(d|\pvec, H)p(\pvec|H) d\pvec = \int_0^1 L(X) dX\,,
\end{align}
where $X$ is defined as the fraction of the prior mass enclosed by a $(D-1)$-dimensional surface at constant likelihood $L(X)$:
\begin{align}
 X(L') = \int_{L(\pvec)>L'}p(\pvec|H)d\pvec\,.
\end{align}
As $X$ decreases from 1 to 0, $L(X)$ becomes progressively higher, so that $L(0)$ is the maximum of the likelihood function.
Finding the actual surface that encloses any particular likelihood (i.e.\ finding $X(L)$) is prohibitively difficult in general, so the method instead uses a statistical estimate of $X$ obtained through nested sampling of the prior distribution.
This relies on the fact that the prior is by definition normalised over the entire parameter space, so we can initiate the algorithm with $X_0=1$. The algorithm then draws $\Nlive$ samples $\theta_i$ from the prior, and evaluates their likelihoods to get $L_i$. The samples are then ordered by likelihood value, such that $L_i<L_{i+1}$. 
Since the samples are drawn from the prior, the fraction of the prior mass $t$ contained within a likelihood contour that passes through sample $j$ is given by the beta distribution $p(t_j)=\beta(t_j; \Nlive, 1)$.

At each iteration of the algorithm the lowest-likelihood sample is removed, and replaced by another drawn from the prior but with a higher likelihood.
The number of samples above the current likelihood contour, the \emph{live points}, is then kept constant, and at each iteration the prior mass enclosed by the contour shrinks by a factor $t$, so $X_{i+1} = t_i X_i$.
By repeatedly drawing nested samples, $X$ shrinks geometrically toward $0$, and we force the samples into a smaller volume of higher likelihood. Working with logarithms we have the expected shrinkage $\langle \log t \rangle = -\Nlive^{-1}$, with variance $\sigma^2_{\log t} = \Nlive^{-2}$, so after $i$ iterations $\log X \approx -i/\Nlive$.

The marginal likelihood integral can be approximated using the trapezoid rule,
\begin{align}
Z \approx \frac{1}{2}\sum_{i} (L_i + L_{i+1}) w_i\,,
\end{align}
where $w_i = X_{i+1}-X_{i}$.
The algorithm is typically run until a termination condition is met, which is often chosen via an upper bound on the remaining evidence $L_\mathrm{max} X_i / Z_i < \exp (0.1)$. The set of samples visited during the run is referred to as the nested sampling chain, and provides a scan of how the parameter values are constrained as the likelihood increases.
Once the algorithm is complete, samples from the posterior can be produced by re-sampling the nested sampling chain with weight $L_i w_i / Z$, in a procedure similar to that described in Sec.~\ref{sss:reweight}.

The challenge in nested sampling lies in being able to draw samples efficiently from the likelihood-limited prior distribution. In this regard it shares many common features with the challenge of efficiently sampling the posterior distribution that is encountered in \ac{MCMC} methods, and can benefit from many of the same techniques to improve efficiency.
In initial applications of nested sampling to \ac{CBC} data analysis, the samples were drawn using a short \ac{MCMC} chain~\cite{Veitch:2009hd} with customised jump proposals to explore the symmetries of the likelihood function. This was developed further with \texttt{LALInferenceNest}~\cite{Veitch:2014wba} which combined the custom jump proposal with adaptive \ac{MCMC} steps based on the affine-invariant sampling method of~\cite{10.2140/camcos.2010.5.65} (also used in the popular Emcee sampler~\cite{foreman2013emcee}).
An alternative method 
%of sampling the limited prior 
requires modelling the constrained region in such a way that it can be sampled exactly. This has the advantage 
%over \ac{MCMC} 
that the samples can be independently drawn, eliminating the need for 
%to draw 
a mini-MCMC chain at each iteration of nested sampling.  %This method was pioneered by 
The popular MultiNest sampler thus fits the target regions with a set of ellipsoids that can be sampled exactly~\cite{Feroz:2008xx}. However, if the likelihood contour does not match the ellipsoids exactly, rejection sampling will be needed, introducing an inefficiency that can be as bad as running an MCMC chain.

As nested samplers have become more sophisticated, the need for custom \ac{CBC}-specific jump proposals has reduced somewhat, and many general-purpose sampling algorithms have been applied. % to the problem. 
The \texttt{PyCBC}~\cite{Biwer:2018osg} and \texttt{Bilby}~\cite{Ashton:2018jfp} libraries provide user-friendly python interfaces for Bayesian inference of \ac{CBC} signals, and can inter-operate with many different sampling algorithms, including MultiNest and Dynesty~\cite{Speagle:2019ivv}.
%kc{ bilby has custom proposal.}

% Peregrine \cite{Bhardwaj:2023xph}

\subsubsection{Hybrid grid-sampler exploration}
% RIFT
An alternative to the purely stochastic sampling methods described above is a mixed approach, 
% to the problem, 
where different parameters are tackled with different methods. This is particularly relevant for the division between intrinsic parameters, which affect the phase or amplitude evolution of the signal, and extrinsic parameters which only result in an overall amplitude or phase shift. Since exploration of the intrinsic parameter space requires recomputing a potentially costly waveform, there is a strong motivation to use a directed search method for that subspace, while the less costly extrinsic space can still be sampled in a pseudo-random fashion. 
The RIFT pipeline has adopted such a hybrid grid plus stochastic sampling scheme~\cite{Pankow:2015cra,Lange:2018pyp}, which has enabled the use of otherwise prohibitively expensive waveform models for \ac{PE}, and even the direct use of \ac{NR} waveforms~\cite{Lange:2017wki,LIGOScientific:2016kms}.

% Say more here about the treatment of multiple modes in RIFT which I think is particularly nice

\subsubsection{Machine Learning methods}
\label{sss:pe-ml}
% Why do we need ML?
Recently, the rapid development of machine learning (ML) techniques based on large neural networks has enabled novel approaches to gravitational wave analysis (see e.g.~\cite{Cuoco:2020ogp} for a review). There are a wide array of methods that have been proposed, which can be broadly categorised as enhancements to existing sampling algorithms, or being complete replacements of more traditional methods. Generally these have the goal of decreasing the time required to obtain a \ac{PE} result.

% Enhancements
% BAMBI, NESSAI, ...
Early applications of ML were based around improving a particular aspect of the sampling algorithm. For instance, the difficulty in drawing samples from the likelihood-constrained prior distribution in the nested sampling algorithm. One approach that has been explored by the BAMBI sampler was to enhance the efficiency of these draws by training a neural network to approximate the likelihood surface~\cite{graff2012bambi}, which was available within the LALInference software as far back as the first observing run. More recently, the technique of \emph{normalizing flows} has been used to accelerate both nested sampling~\cite{Williams:2021qyt} and \ac{MCMC}~\cite{Ashton:2021anp} within the Bilby framework, and can offer at least an order of magnitude speedup over previous samplers.

Normalizing flows are a technique that is particularly well-suited to Bayesian inference problems, see~\cite{kobyzev2020normalizing} for a recent review. Their goal is to learn an invertible transformation  that maps a complicated distribution in physical parameter space $p_\Theta(\pvec)$ to a simple distribution $p_Z(\vec z)$ (often a multivariate normal - hence the name normalizing flow) in a \emph{latent space}. The transformation function is defined by $z = f(\pvec; \vec\Phi)$ where $\vec\Phi$ is the output of a neural network, and $f$ is typically built up from multiple layers of simple transformations (for example an affine transformation). Using ML optimization methods such as stochastic gradient descent it is possible to find the optimum $\Phi$ that results in the output of the flow $p_\Theta(\pvec)=p_Z(\vec z)|J(\pvec,\vec z)|$ approximating a desired target distribution, where $J_{ij}(\pvec,\vec z) = \frac{\partial z_i}{\partial \theta_j}$ is the Jacobian matrix of the transformation.
The output of the flow can then be used to produce a trainable proposal distribution for use in MCMC and nested sampling algorithms.

Another hybrid application of normalizing flows was explored with a combination of
normalising flow jump proposals and Hamiltonian Monte Carlo, using the auto-differentiation tools
of modern machine learning packages to accelerate the computation of gradients~\cite{Wong:2023lgb}.

% Replacements
A particularly appealing application of machine learning methods is the possibility of completely replacing the stochastic sampling algorithms which require repeated draws, and explicit likelihood calculations which are computationally costly. \emph{Likelihood-free inference} techniques have been applied which completely replace the likelihood and prior in Bayesian inference with a joint distribution of the data and parameters, which is then later conditioned on a specific piece of observed data.
% Vitamin, DINGO
Two techniques which have been used for this are conditional variational auto-encoders~\cite{Gabbard:2019rde}, and normalising flows~\cite{Dax:2021tsq} (in a different configuration from the ``enhancements'' mentioned above).
Both these methods have the advantage of being very low-latency compared to stochastic sampling,
as they rely on a neural network that is trained in advance on a large number of simulated signals
and noise realisations. This moves the dominant cost of inference up-front, so that when an actual
observation is made the network simply conditions its output on observed data.
Such methods can be slow to train, especially on large datasets, and so their application to
longer \ac{BNS} signals is still to be fully demonstrated. However there have been recent works
showing promising combinations of amortised inference with classical principal component analysis,
a technique used elsewhere \cite{Cannon:2010qh} for reducing data volumes~\cite{Langendorff:2022fzq}.

Another simulation-based inference technique that has been developed for computationally expensive problems is neural ratio estimation~\cite{Miller:2022shs}, which has been shown to reduce the number of likelihood evaluations required to perform full parameter estimation by a factor of $\sim 50$\cite{Bhardwaj:2023xph}.

%% JV: This section seemed incomplete so commented for now.
% \subsubsection{Test of samplers}
% One of the strongest checks used to ensure that those posterior samples are indeed randomly drawn from the posterior is to compare one set with another, obtained independently. This is done in practice both with the same sampling code (running the same analysis multiple times) and with different sampling code to increase the robustness of the results.
%
% Furthermore, some sanity checks are also performed. From section~\ref{sec:detection} (check reference), we can expect that the maximum likelihood obtained by the sampler is close to $SNR^2/2$ (define close, using dimension/2 or SNR depending on the SNR definition).

\subsection{Degeneracies of the likelihood}\label{ss:pe_degeneracies}

Parameter estimation results for \ac{CBC} signals, and of gravitational waves in general,
often exhibit certain characteristic features in the posterior \ac{PDF}, a classic example
being the `V'-shaped degeneracy between distance and inclination angle.
%\jv{[FIG]}.
These degeneracies
may result in correlations between parameters, or even in multi-modal posteriors, corresponding
to similar waveforms that cannot be distinguished from each other 
due either to an exact symmetry in the waveform model, or because the noise level is too high
to resolve their differences.
Understanding these degeneracies is essential to the interpretation of \ac{PE} results,
but it can also help us in the design of \ac{PE} methods, since these structures
can be difficult to sample with a naive approach (for example \ac{MCMC} with fixed jump scales
leads to poor mixing of chains). Several works have proposed specific \ac{MCMC} jump proposals or
transformations of the parameter space to reduce or remove these degeneracies~\cite{Raymond:2014uha,Veitch:2014wba,Farr:2014qka,Roulet:2022kot}.

\subsubsection{Distance - inclination}
Expanding the logarithm of the likelihood function defined in Eq.~\eqref{eq:LikelihoodSignal}
results in a quadratic in the waveform $\vec{h}$,
\begin{align}
\log L = -\frac{1}{2}\left(\braket{\vec{d}|\vec{d}} + \braket{\vec{h}|\vec{h}} - 2\braket{\vec{d}|\vec{h}}\right) + k
\label{eq:likelihood_expansion}
\end{align}
where $k=-\frac{1}{2}\log \det 2\pi \mat{C_N}$ is a normalisation constant that depends
only on the (usually fixed) \ac{PSD}. (If the noise model is fixed then $\braket{\vec d|\vec d}$ is
also a constant.) %as the waveform does not enter into this expression.

From Eq.~\eqref{eq:signal_fd1} we see that the response of a single detector labelled $I$
to the dominant $l=m=2$ mode of a non-precessing signal can be separated into frequency-dependent
and non-frequency dependent parts as
\begin{equation}
    \vec h^I = \frac{A_0(f)}{d_L}\left(\frac{F_+^I}{2} (1+\cos^2\theta_{JN}) + F_\times^I \cos\theta_{JN}\right)\exp i \left[\Phi(f)-2\pi f t_{c,I} - \phi_{c,I}\right].
\end{equation}
Since the inner product is an integral over frequency, this allows us to move the common non-frequency-dependent factors $\alpha_I$ out of the inner product, to obtain
%so that the log-likelihood is
\begin{align}
    \log L &= -\frac{1}{2}(\alpha_I)^2 |A_0(f)|^2 + \alpha_I \braket{A_0(f)\exp i\left[\Phi(f)-2\pi f t_{c,I}\right]| d}\cos \phi_{c,I} + \mathrm{const.}\,, \\
    \alpha_I&= \frac{1}{2d_L}\left(F_+^I (1+\cos^2\theta_{JN}) + 2 F_\times^I \cos\theta_{JN}\right)\,,
\end{align}
where $|A_0|^2=\braket{A_0(f)|A_0(f)}$. Since the single parameter $\alpha_I$ incorporates
the effect of both $d_L$ and $\theta_{JN}$, as well as the antenna response for a particular
detector $I$, there exists a degeneracy among these parameters where $\alpha$ is held
constant.
The transformation $\theta_{JN}\rightarrow \pi-\theta_{JN}$,  $F_\times^I\rightarrow -F_\times^I$,
results in the same signal appearing in the detector, and corresponds to a flip between
left-handed and right-handed elliptical polarisations of the gravitational wave, with
a change in the polarisation angle. Intuitively we can understand this perfect
symmetry as a single detector only sees one of the two polarisation modes of a
general waveform.
The detector response functions themselves depend on three parameters
$\alpha$, $\delta$ and $\psi$, and it is always possible for a single detector to
find a new sky position and polarisation angle that maps $F_\times\rightarrow-F_\times$
while keeping $F_+$. This can be achieved by changing the sky location,
or by mapping $\psi\rightarrow\pi-\psi$.
For a two-detector network, there are two $\alpha_I$'s, and the sky location can be
inferred from the time delay between detectors (see below). However the degeneracy
in $\psi$ still exists. A third detector in the network is required to fully break
this degeneracy, and therefore infer the polarisation state of the GW, although
in the case where detectors are of unequal sensitivity, or where a signal originates
near the null of one detector's response function, it may be impossible in practice
to do so due to the limited signal-to-noise ratio.

Even in a multiple-detector case, the measurement of the amplitudes $\alpha_I$ is still
subject to a statistical uncertainty,
%with variance $\sigma^2_\alpha=1$ \jv{(check this calculation)},
therefore there is a anti-correlation between $d_L^{-1}$ and $\cos\theta_{JN}$ when $0<\theta_{JN}<\pi/2$,
which reduces the accuracy to which distance can be determined, unless the inclination angle
can be measured in another way (for example by multi-messenger observation). 
%TD Seems like references are needed here ..
This correlation limits the possible precision of inference on the Hubble constant 
%to be a limiting factor when 
using \ac{GW} distance estimates~\cite{LIGOScientific:2017adf}.

\subsubsection{Sky localization}
The localisation of compact binary sources is a highly important output of
parameter estimation %that is 
used to enable multi-messenger followup observations and
correlations with other astrophysical transients, e.g.\ \acp{GRB}.
The location of the source on the celestial sphere is typically described
by the right ascension $\alpha$ and declination $\delta$ coordinates.
As a single gravitational wave detector is approximately omnidirectional, a
network of two or more detectors is required to achieve any reasonable precision. 
%for localisation. 
As we have seen, %already, 
the antenna response functions %can 
determine the amplitude of the signal at %as it appears in 
each detector, and so provide weak information about the probable location; far more
information, though, is obtained from differences in signal arrival times between observatories.

\begin{figure}
 \centering
 \includegraphics[width=0.75\textwidth]{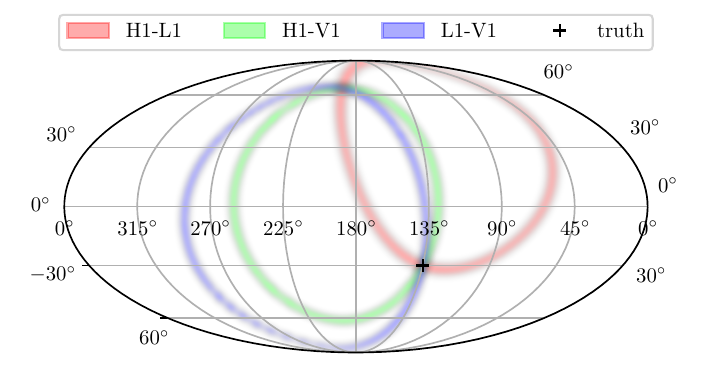}
 \includegraphics[width=0.75\textwidth]{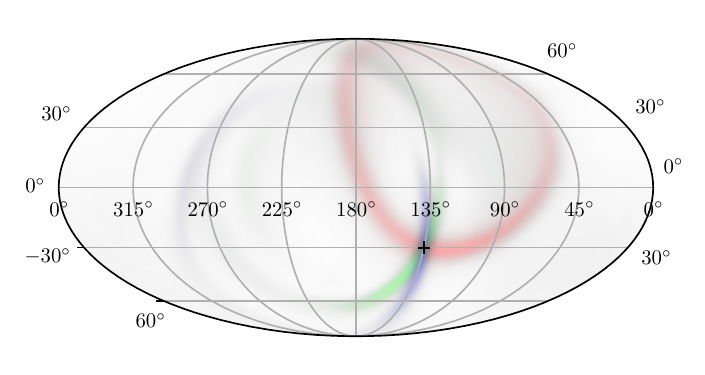}
    \caption{\label{fig:localisation} Localisation of a gravitational wave source by a detector network using (top) only timing information and (bottom) timing and antenna response information. Colours indicate which network is used to perform localisation, and the true location is marked with a $+$.
}
\end{figure}
Fig.~\ref{fig:localisation} shows the localisation of a $\Mc=10\,\Msun$ source by different two-detector networks, illustrating the difference between localisation with timing alone versus that including the true antenna response function. From timing alone, each pair of detectors located at $\vec{x}$ is able to localise a source with propagation direction $\vec{k}$ to a ring on the sky consistent with a constant time offset between them $\Delta t_{IJ} = (\vec{x}_J - \vec{x}_I)\cdot\vec{k}/c$.
The width of the ring is produced by the uncertainty in the difference in the time of arrivals at the two sites, which is inversely proportional to the effective bandwidth
of the signal, weighted by the noise \acp{PSD}~\cite{Fairhurst:2010is}.

If we consider the three pairs of detectors possible from a three detector network,
the three resulting rings intersect at the true location of the source, but also at the point which is the reflection of the source in the plane containing the detectors, which generates the same times of arrival (see left panel of Fig.~\ref{fig:localisation}). There are therefore always two locations on the sky consistent with any observed time of arrival in a three detector network.
With the inclusion of the antenna response function, the symmetry of the two modes is broken, since in general only one has a consistent set of antenna response functions
that lead to the observed signal in all three detectors. However if one detector has
significantly poorer sensitivity than the others, then the signal to noise ratio may be low in one detector, and the degeneracy only weakly broken.

Sampling the sky location posterior can be challenging for algorithms because of
the spherical topology of the parameter space and the ring-like structures which even
when broken must be explored at lower likelihood levels for nested sampling
algorithms, or at higher temperatures for annealed MCMC methods. Local Gaussian
jump proposals that are not adaptive to the structure will have difficulty achieving
good acceptance rates at all points in the space. One solution is to perform
a reparameterisation that rotates the sky so the north pole is aligned with the
vector separating the detectors. If the dominant pair of detectors is chosen then
the posterior ring will align with a constant latitude in the new coordinate system,
decoupling the two angular parameters so that they can be sampled more effectively.
The time of arrival of the signal can also be remapped as part of this transformation:
the time of arrival is typically well measured in each detector individually,
but the time of arrival at the geocentre, relative to the detectors,
is dependent on the direction of propagation of the signal.
Therefore using the common reference position of the geocenter leads to degeneracy between the sky location and the reference time. When performing the sky rotation it is
also useful to remap the reference time to that of arrival in the most sensitive
detector.

\subsubsection{Phase and polarisation angle}
The orientation of a quasi-circular binary is given by the three angular parameters,
$\theta_{JN}$, $\psi$ and $\phi_c$, which completely determine the polarisation
content of the signal through the decomposition into spin-weighted spherical
harmonics, as in Eq.~\eqref{eq:SpHarmWaveform}. For the dominant $\ell=m=2$ mode,
the signal can be written as
\[
h_+-ih_\times = h^{22}{}_{-2}Y^{22}(\iota, \phi_c) + h^{2-2}{}_{-2}Y^{2-2}(\theta_{JN}, \phi_c)
\]
where ${}_{-2}Y^{\ell m}(\theta_{JN}, \phi_c) = f^{lm}(\theta_{JN})e^{i m \phi_c}$,
and $f^{lm}(\theta_{JN})$ is the inclination-dependent emission pattern determined
by the spin-weighted spherical harmonics~\cite{Mathews:1962}.
The effect of the phase parameter on the $m=2$ mode is therefore to rotate
between the $+$ and $\times$ parts of the waveform when written in this complex
format. Recall that the effect of the $\psi$ parameter is also to rotate
between $+$ and $\times$ polarisations with angle $2\psi$.
When the source is viewed directly face-on or face-off $(\theta_{JN}=0,\pi)$, 
varying the $\phi_c$ and $\psi$ parameters produces equivalent effects on the $m=2$
mode, causing a positive or negative 
degeneracy, $\phi + \psi = \mathrm{const}$ or $\phi - \psi = \mathrm{const}$.
Each condition describes a characteristic stripe in the $\phi,\psi$ %two-dimensional 
parameter space %of $\phi,\psi$ 
that can be difficult to sample, being periodic in both
parameters and potentially quite narrow. For a generic orientation $\theta_{JN}$,
the posterior exhibits a combination of both positive and negative degeneracies.
%\td{Hard to visualize what this means!!}
%\jv{Illustration to be produced}

\subsubsection{Masses and spins}
\label{ss:mass_measurement}
As discussed in Sec.~\ref{ss:precision_of_parameters}, those parameters which
most influence the phase evolution of the signal
tend to be the best measured.  
These are referred to as intrinsic parameters, including both the masses and
spins of the binary components, combinations of which enter into the phasing
formulae at different \ac{PN} orders 
%in the post-Newtonian expansion 
for the inspiral (see Tab.~\ref{tab:PN_terms}).
Generally, the lower the order, the more effect the parameter has on the phasing,
and the easier it will be to measure it for lower-mass signals. Unfortunately the parameter combinations
that enter into the \ac{PN} expansion are not necessarily those of the greatest
astrophysical interest. 
%Rather than the component masses, 
The chirp mass, entering the \ac{PN} expansion at lowest order, is by far the best
determined intrinsic parameter for inspiral-dominated signals.  As a consequence,
the posterior on component masses is constrained to lie on a line of approximately
constant chirp mass. The length of the line in the direction perpendicular to
chirp mass is determined by the uncertainty in the second mass combination, which
can be taken to be the symmetric mass ratio $\eta$ which enters at 1PN order. 
%\jv{Add figure showing degeneracy for BNS, BBH, IMBH}

High-mass signals, though can have a significant or even dominant contribution from
the merger and ringdown parts. Since the quasi-normal mode frequencies of
the ringdown signal scale inversely with the final mass, in these cases
the final mass measurement may be as effective as the chirp mass measurement~\cite{Veitch:2015ela}.

Sampling the highly correlated $(m_1,m_2)$ joint distribution can be challenging, especially
for binary neutron star signals where the chirp mass is extremely well measured.
It is therefore common to use the $(\Mc,q)$ parameterisation of the mass space,
where $q=m_2/m_1 < 1$ provides a naturally bounded domain that is still easily
interpretable.

Turning to the spin parameters, the first combination that enters the \ac{PN} expansion
is the $\chieff$ parameter, a mass-weighted combination of the aligned spin components
that determines the spin-orbit coupling and has a dominant effect on the rate of inspiral in the near-equal-mass regime~\cite{Ajith:2009bn}. It is conventionally defined as
\begin{align}
\chi_\mathrm{eff} \equiv \frac{(m_1 \vec{\chi}_1 + m_2 \vec{\chi}_2)\cdot\hat{\vec{L}}}{m_1 + m_2}.
\end{align}
%Even though this is the dominant spin combination, 
As $\chieff$ enters at 1.5PN order, it is typically poorly measured for near-threshold
signals. Furthermore, the prior on this parameter is not trivial: if one assumes that the spin vectors
$\vec{\chi}_1,\vec{\chi}_2$ have a uniform prior within the $3$-ball $|\vec{\chi}|<1$, the marginal prior
on the $z$-components of the spin vectors contains an integrable singularity at $\chi_z = 0$,
$p(\chi_z)= \frac{1}{2\chi_\mathrm{max}}(-\ln|\chi_z/\chi_\mathrm{max}|)$~\cite{Lange:2018pyp}.
This in turn creates a singularity in the marginal prior at $\chieff=0$ since it is
the weighted sum of the $\chi_z$'s.
This makes interpreting the results of parameter estimation %on the effective aligned spin 
difficult: typically one is looking primarily for evidence that the
posterior is significantly shifted compared to the prior (see for example the $\chieff$ analysis in \cite{Mandel:2020lhv}).

Further spin measurements enter at even higher \ac{PN} order in the phasing, although
the non-aligned parts of the spin vectors can be measured by another means: through
the precession of the orbital plane that occurs due to spin-orbit coupling, introduced  in Sec.~\ref{sss:precession}.
This effect produces an amplitude and a phase modulation of the signal, as the system's orbital angular momentum vector tips toward and away from the observer.
The measurability of this effect is dependent on the rate of change of the emission
pattern ${}_{-2}Y_{lm}(\iota,\phi_c)$ with respect to the instantaneous inclination
angle $\iota = \tan^{-1}\vec{\hat{L}}\cdot\vec{\hat{N}}$, which determines the size of the amplitude
modulation seen by an observer viewing the system along the $N$ vector.
This time-dependence of $\iota$ for a precessing binary makes the choice of angular
parameter $\theta_{JN}$ a more useful choice in parameter estimation, since
the total angular momentum of the source, $J$ is more closely conserved during the
merger process.
To encode the dominant information about the precession, the $\chip$ parameter
has been introduced, which measures the relevant in-plane spin
combination~\cite{Hannam:2013oca,Schmidt:2014iyl}. This too has a non-uniform prior induced
by the uniform component spins, which complicates the interpretation of its posterior distribution.
See \cite{Callister:2021gxf} for a detailed comparison of spin priors under different constraints.

%In terms of parameterisation, 
The use of these two dominant parameters is also complicated in practice
by the need to specify 4 additional spin degrees of freedom to fully determine
the source properties. It is therefore more common to use an angular coordinate
system determined by the spin magnitudes $a_{1,2}$, tilts $\theta_{1,2}$, and
the azimuthal angles $\phi_{JL}$ and $\phi_{12}$ which measure the difference
between $J$ and $L$ an between the component spin vectors respectively~\cite{Farr:2014qka}.

\begin{figure}[htb]
\begin{center}
\includegraphics[width=0.5\textwidth]{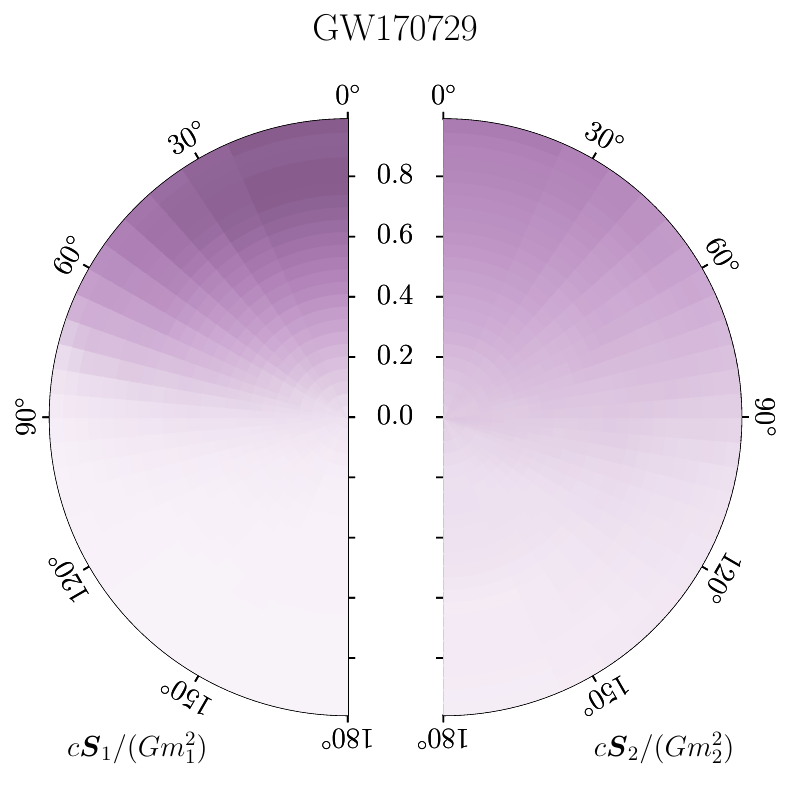}
\end{center}
\caption{\label{fig:spin_disk} An example `spin disk' plot for GW170729, reproduced from \cite{LIGOScientific:2018mvr}. The plot shows the posterior on the spin vectors' magnitudes and tilts, marginalised over the spin azimuthal angles. Bin sizes are adjusted to contain equal prior probability, and the colour density represents the posterior probability in each bin. Left and right panels show the primary and secondary spin, respectively.}
\end{figure}

Results on the spin angular can be visualised by the marginal posteriors on
the various parameters of interest, particularly $\chieff$, $\chip$, $\theta_{1,2}$,
$a_1$, $a_2$. To examine the 4D parameter space of component spin magnitudes
and tilts a `spin disk' plot, (e.g.\ Fig.~\ref{fig:spin_disk}), can be useful.

\subsection{Optimization}

Parameter estimation usually involves millions of likelihood evaluations, dominating the total computational cost. Numerous schemes have therefore been developed to reduce the cost of a likelihood evaluation.
One can reduce the number of summation elements in the likelihood sum from Eq.~\eqref{eq:likelihoodsum}
 with various techiques. Multi-banding~\cite{Cannon:2011vi,Vinciguerra:2017ngf,Garcia-Quiros:2020qlt,Morisaki:2021ngj}
 involves downsampling the data at different sampling rates depending on the signal's frequency content. As \ac{CBC}s increase in frequency as a 
function of time, one can compute the overlap for earlier parts of the signal with a much lower 
sampling rate than later parts of the signal, without loss of accuracy.

Another technique that can reduce the signal bandwidth to even greater extremes involves computing the waveform relative to a reference signal. The resulting likelihood can be written in terms of the difference between the signals, which occupies a much smaller bandwidth than the full signal, provided they are sufficiently similar. This technique is employed in various forms as the heterodyned likelihood~\cite{Cornish:2010kf}, or relative binning~\cite{Zackay:2018qdy} methods.
It can result in a speed-up of up to four orders of magnitude in the likelihood computation~\cite{Cornish:2021lje}. This technique has also been shown to apply to signals containing multiple modes, which can otherwise prove a challenge to acceleration schemes~\cite{Leslie:2021ssu} since the modes interfere with each other, and the signal cannot be described as having an instantaneous frequency.

Furthermore, using the reduced order basis approach described in section~\ref{sec:surrogate}, it is 
possible to precompute a reduced set of quadrature weights that accurately approximate the 
waveform likelihood integrals, called a \ac{ROQ} rule \cite{Antil:2012wf,Canizares:2013ywa,Canizares:2014fya,Smith:2016qas,Morisaki:2020oqk,Qi:2020lfr,Morisaki:2023kuq,Morras:2023pug}. Instead of evaluating the likelihood integral 
at each frequency sample, \ac{ROQ} exploits the precomputed quadrature weights to efficiently 
compute the likelihood by taking a linear combination of the reduced basis waveforms. As show in section~\ref{ss:pe_degeneracies}, one can expand the log-likelihood of Eq.~\eqref{eq:LikelihoodSignal} into a linear $\braket{\vec{d}|\vec{h}}$ and a quadratic $\braket{\vec{h}|\vec{h}}$ term, and a constant. The quadratic term is the same for all waveforms, and can be precomputed. The linear term can be written as a sum over the quadrature weights $w_i$ and the inner product of the reduced basis waveforms $\vec{h}_i$ with the data $\vec{d}$, $\braket{\vec{d}|\vec{h}} = \sum_i w_i \braket{\vec{d}|\vec{h}_i}$. The inner product of the data with the reduced basis waveforms can be precomputed and stored in a matrix $\mat{R}$, so that the likelihood can be written as $\log L = -\frac{1}{2}\braket{\vec{d}|\vec{d}} + \vec{w}^T\mat{R}\vec{d} + \mathrm{const.}$, where $\vec{w}$ is the vector of quadrature weights. The \ac{ROQ} rule is then the set of quadrature weights $\vec{w}$ that minimise the error in the likelihood integral. This can be computed by solving the linear system $\mat{R}\vec{w} = \vec{1}$, where $\vec{1}$ is a vector of ones. The \ac{ROQ} rule can be used to compute the likelihood for any waveform in the reduced basis, and so can be used to compute the likelihood for any signal in the parameter space.
This approach significantly reduces the computational burden, as the number of reduced order bases is typically much smaller than the number of frequency points in the likelihood integral. Integrals over more points, typically those for longer, lower-mass signals, will see a greater speed-up.

An effective optimization strategy %for parameter estimation 
is to marginalise (either analytically or numerically) over as many parameters as possible when evaluating the likelihood. While this usually makes each likelihood evaluation more computationally expensive, the reduced dimensionality almost always yields an overall speed-up.  Marginalization has been applied to the orbital phase, coalescence time and distance parameters~\cite{Veitch:2014wba,Dhurandhar:2017rlr,Ashton:2018jfp}. It is possible to recover the probability density function of the marginalised parameters in post-processing, at the cost of evaluating more likelihood values; however, as this post-processing 
step is done only on samples after thinning, %post-sampling, 
the extra cost is small. 

\newpage
\section{Combining Information}
\label{sec:combine}

\subsection{From one to many: Hierarchical inference}
\label{sec:hier}

Section~\ref{sec:pe} discussed the inference of the parameters $\pvec$ describing an individual source. The posterior PDF $p(\pvec | d)$ with given data $d$ depends on a choice of prior $p(\pvec | H_S)$.
%\td{Unfortunately the PE section had two different notations for the prior information, $H_S$ is more consistent with the rest of the review.}
Section~\ref{sec:pe} also explained how to select between two models $H_1$ and $H_2$, each with its corresponding prior distribution, using Bayes factors and posterior odds. The goal of hierarchical inference, discussed in this section, is to find the ``best'' prior, or equivalently population model that is common to all events. This can be thought of as an extension of the model selection presented in Sec.~\ref{sec:pe}.

In general we consider a family of models for the prior PDF parametrized by $\vec{\Lambda}$, a vector of population hyper-parameters, writing the source prior as $p(\pvec | \vec{\Lambda})$. 
%\td{I am dropping $H_S$ here as it describes all the assumptions that are required to calculate the likelihood for a given source, eg the waveform and the noise model.}
A hierarchical inference aims to use the data from all events $\left\{ d_i \right\}_{i = 1}^N$ to infer the hyper-parameters $\vec{\Lambda}$, in addition to the source parameters $\left\{ \pvec_i \right\}_{i = 1}^N$ for every event $i$.
Such an analysis stacks information across events by jointly inferring parameters specific to each event as well as (hyper-)parameters common to all events. Thus, when additional events are observed, we update not only the population from which they are drawn, but also the parameters of {previously seen events}. 

We can use Bayes' theorem to write the joint posterior PDF over the source parameters and population hyper-parameters as~\cite{2004AIPC..735..195L,2010PhRvD..81h4029M,2011ApJ...731..120M}:
\begin{equation}
 p(\left\{ \pvec_i \right\}_{i = 1}^N, \vec{\Lambda} | \left\{ d_i \right\}_{i = 1}^N, H_S ) = \
  \frac{p( \left\{ \pvec_i \right\}_{i = 1}^N | \vec{\Lambda} ) 
        p( \left\{ d_i \right\}_{i = 1}^N | \left\{ \pvec_i \right\}_{i = 1}^N, H_S ) 
        p(\vec{\Lambda}) }
  {p( \left\{ d_i \right\}_{i = 1}^N | H_S )}\,,
\end{equation}
where $p(\vec{\Lambda})$ is a prior distribution of hyper-parameters (or hyper-prior); 
note that this is a simplified expression that neglects observational selection effects, which are treated in the following Section~\ref{subsec:sel}. 
Assuming that each observation $i$ is independent, we have
\begin{equation}
\label{eq:joint-hier-likelihood-p-nosel}
 p(\left\{ \pvec_i \right\}_{i = 1}^N, \vec{\Lambda} | \left\{d_i \right\}_{i = 1}^N, H_S ) = 
  \frac{p(\vec{\Lambda}) \prod_{i=1}^N p( d_i | \pvec_i, H_S ) p( \pvec_i | \vec{\Lambda} ) }{ p( \left\{ d_i \right\}_{i = 1}^N | H_S) }\,.
\end{equation}
In many situations, we are only interested in inferring the population hyper-parameters $\vec{\Lambda}$, so we can marginalize over the single-event parameters $\left\{ \pvec_i \right\}_{i = 1}^N$:
\begin{equation}
 p(\vec{\Lambda} | \left\{d_i \right\}_{i = 1}^N, H_S ) = 
  \frac{p(\vec{\Lambda}) \prod_{i=1}^N \int p( d_i | \pvec_i, H_S ) p( \pvec_i | \vec{\Lambda} )  d\vec\theta_i }{ p( \left\{ d_i \right\}_{i = 1}^N | H_S) }\label{eq:hierlike}\,.
\end{equation}
The numerator of Eq.~\eqref{eq:hierlike} is the %prior on the hyper-parameters $\vec\Lambda$, referred to as the 
hyper-prior multiplied by the ``single-event evidences'' $\int p( d_i | \pvec_i, H_S ) p( \pvec_i | \vec{\Lambda} ) d\vec\theta_i$. This makes intuitive sense because the ``best'' (maximum-likelihood) $\vec\Lambda$ corresponds to the choice of prior on $\vec\theta$ that maximizes the product of single-event evidences.

Single-event evidences may be evaluated by various methods.  Assuming that parameter estimation has already been performed on each event using some ``interim prior'' $\pi_\mathrm{PE}(\pvec)$, resulting in  $J$ \ac{PE} samples $\pvec_i^j$ for each event $i$, importance sampling is often convenient to apply.  The \ac{PE} samples are drawn from a posterior PDF given by the likelihood $p( d_i | \pvec_i)$ with the ``PE'' prior,
\begin{equation}
 \pvec_i^j \sim p( d_i | \pvec_i)\pi_\mathrm{PE}(\pvec)\,.
\end{equation}
The single-event evidence can then be estimated as the following importance sampling, or Monte Carlo, approximation {(in what follows we will drop the explicit dependence on the signal hypothesis $H_S$)}:
\begin{equation} \label{eq:evidence_importance_sample}
 \int p( d_i | \pvec_i) p( \pvec_i | \vec{\Lambda} )  d\theta \approx \frac{1}{J} \sum_{j = 1}^J  \frac{p ( \pvec_i^j | \vec{\Lambda} )}{\pi_\mathrm{PE}(\pvec_i^j)}\,.
\end{equation}
The error in this Monte Carlo approximation is small when the number of \ac{PE} samples is large and the interim prior is not too different from the population model $p(\pvec | \vec{\Lambda})$. %In other words, in order to keep the Monte Carlo error sufficiently small compared to the uncertainty of the target population posterior, 
For this error to be small compared to the uncertainty in the population posterior, the effective number of \ac{PE}, which depends on the chosen population model(s) $p(\pvec | \vec{\Lambda})$ for all choices of $\vec{\Lambda}$ in the hyper-prior, must be large enough~\cite{2022arXiv220400461E}.
Certain population models, especially those that allow for rapid variation of the population \ac{PDF} $p(\pvec | \vec{\Lambda})$ over parameter space, may require an especially large number of \ac{PE} samples for each event, with a possibly prohibitive computational cost.

Instead of estimating the single-event evidences via importance sampling, one could first obtain a {continuous functional approximation} to the single-event likelihood $p( d_i | \pvec_i)$ using  %density estimation 
techniques such as kernel density estimates, Gaussian mixture models or Gaussian processes, based on the \ac{PE} samples~\cite{Landry:2018prl,2020arXiv200101747W,2021MNRAS.508.2090D,2022ApJ...926...79G,2022arXiv220514154D}. Equipped with an estimate for $p( d_i | \pvec_i)$, it is straightforward to evaluate the joint posterior over the single-event parameters and population hyper-parameters from Eq.~\eqref{eq:joint-hier-likelihood-p-nosel}. The density estimate still carries an uncertainty that scales inversely with the number of \ac{PE} samples. Nevertheless, this approach can be better-behaved than importance sampling, for instance it correctly assigns regions of $\pvec_i$ parameter space without discrete samples a nonzero (but small) likelihood.

The hierarchical Bayesian likelihood may also be evaluated without the use of \ac{PE} samples by directly sampling source parameters ($\pvec_i$) and population hyper-parameters ($\vec{\Lambda}$) from the joint likelihood (Eq.~\eqref{eq:joint-hier-likelihood-p-nosel}). Thus, rather than sampling from the \ac{PE} posterior for each event separately with a fixed `interim' prior, 
%(a fixed population model), 
\ac{PE} samples are drawn jointly with the population hyper-parameters across {all} events simultaneously. 
This may be desirable when we are interested in the population-informed parameters for individual events.
However, directly evaluating the joint likelihood may require many more evaluations of the \ac{PE} likelihood $p(d_i | \pvec_i)$ in total, {in part due to the much higher dimensionality of the joint parameter space}, requiring in turn many more waveform evaluations and higher computational cost.
Therefore, 
%if all we are interested in is
in order to update our inference of individual events' parameters with our knowledge of the population, we can reweight pre-existing \ac{PE} samples (drawn with an interim prior) in post-processing, once we have sampled from the marginalized likelihood of Eq.~\eqref{eq:hierlike}~\cite{2020ApJ...891L..31F,2020PhRvD.102h3026G,2020ApJ...895..128M,Essick:2021,2021PhRvD.104h3008M,2023MNRAS.525.3986M}. 
This remains true 
%for the hierarchical Bayesian likelihood that 
when including selection effects (discussed in the following section).

\subsubsection{Astrophysical events as a Poisson process}
So far we have written the source prior, or ``population model'', as a probability density $p(\pvec | \vec{\Lambda})$. For any fixed $\vec{\Lambda}$, such a PDF, integrating to unity across the $\pvec$ parameter space, describes the \emph{relative} proportion of systems with different source properties. %$\pvec$. Integrating across all possible $\pvec$, the PDF normalizes to unity.
However, if considering \emph{absolute} numbers of systems, for instance the astrophysical rate of {binary mergers within some volume of space}, we prefer to consider number densities $dN / d\pvec$ instead of PDFs $p(\pvec)$, the difference being that $dN / d\pvec$ integrates to the total expected number of systems $\mathcal{N}$ within the considered range of $\pvec$, 
\begin{equation}
 \frac{dN}{d\pvec} (\pvec | \vec{\Lambda}) = \mathcal{N} p(\pvec | \vec{\Lambda})\,.
\end{equation}
We often consider % number density 
$dN / d\pvec$ as the intensity function of an inhomogeneous Poisson process -- ``inhomogeneous'' because the intensity %number of systems $dN / d\pvec$ 
is not constant across the space of $\pvec$. The corresponding likelihood is given by 
\begin{equation}
\label{eq:inhom-pos-likelihood-nosel}
 p(\left\{ \pvec_i \right\}, \left\{d_i \right\} | \vec{\Lambda}, \mathcal{N}) \propto \mathcal{N}^N e^{-\mathcal{N}} \prod_{i=1}^N p( d_i | \pvec_i ) p( \pvec_i | \vec{\Lambda} )\,
\end{equation}
where $\left\{ \ldots \right\}$ denotes the set over $N$ events. 
Using this likelihood, we can then write the posterior over the source parameters and population hyper-parameters (which now include the total number of systems $\mathcal{N}$):
\begin{equation}
\label{eq:inhom-pos-posterior-nosel}
 p(\left\{ \pvec_i \right\}, \vec{\Lambda}, \mathcal{N} | \left\{d_i \right\} ) \propto
 p(\vec{\Lambda}, \mathcal{N}) \mathcal{N}^N e^{-\mathcal{N}} \prod_{i=1}^N p( d_i | \pvec_i ) p( \pvec_i | \vec{\Lambda} )\,.
\end{equation}
We have hidden the $p(\left\{d_i \right\})$ factors in the proportionality because they are independent of $\vec{\Lambda}$ and $\mathcal{N}$.
We discuss the relation between Eq.~\eqref{eq:inhom-pos-posterior-nosel} and Eq.~\eqref{eq:joint-hier-likelihood-p-nosel} in the following subsection.
Again, these expressions only hold in the absence of observational selection effects, where the expected number of \emph{observed} events $\mathcal{N}$ is the same as the number of events occurring in the astrophysical population.

\subsection{Accounting for selection biases}
\label{subsec:sel}

\ac{GW} observations, like most astrophysical observations, are subject to selection effects: most \acp{CBC} that occur in the Universe produce \acp{GW} that are too quiet to be confidently distinguished from noise in current detectors, thus only a small subset {of mergers on our past light-cone}
%of the Universe's \acp{CBC} 
make it into detection catalogs. Nearby, face-on binaries with high masses and large spins aligned with the orbital angular momentum %axis 
are over-represented among GW detections, because they emit louder signals. To learn about the underlying astrophysical population of \acp{CBC}, we must account for such observational selection effects in the hierarchical Bayesian inference.

Each \ac{CBC}, with source properties $\theta$, produces a signal lying within data $d$ and is either detected, {meaning that its estimated statistical significance, in terms of false alarm rate or probability of astrophysical origin, satisfies some numerical threshold}, in which case we label it ``det'' and record its corresponding data $d_i$; or non-detected, in which case we do not record it.
% as it fails to pass our detection threshold. 
Detection is therefore a binary property of the data resulting from a \ac{CBC} signal within the detector noise at a particular time. 
%\td{But the data is produced by the detectors, not by the astrophysical system ...} 
We can write the probability that some hypothetical data is detected as $P(\mathrm{det} | d)$. Typically, $P(\mathrm{det} | d)$ is deterministically 0 or 1 for any piece of data $d$, based on whether the data passes some threshold in, for example, \ac{FAR} or \ac{SNR}~\cite{2019MNRAS.486.1086M}, but this need not be the case~\cite{2014arXiv1412.4849F}. 

In order to correct for selection effects, we must account for the fact that our analysis only includes data from detected sources. Given $N$ detections, we explicitly include the label $\left\{ \mathrm{det}_i \right\}_{i = 1}^N$ among $\left\{ d_i \right\}_{i = 1}^N$, $\left\{ \pvec_i \right\}_{i = 1}^N$ and $\vec{\Lambda}$ when describing joint probabilities:
\begin{equation}
p( \mathrm{det},  d, \pvec, \Lambda ) = P( \mathrm{det} | d ) p( d | \pvec ) p( \pvec |\vec{\Lambda} )  p(\vec{\Lambda})\,.
\end{equation}
Note the conditional structure implied by the above equation: given data $d$, detectability $\mathrm{det}$ is conditionally independent of the source properties $\vec\theta$~\cite{2004AIPC..735..195L,2019MNRAS.486.1086M,2023arXiv231002017E}. 
Marginalizing the above expression over %all possible 
data realizations %sets 
$d$ and then over the source parameters $\pvec$, we obtain $P(\mathrm{det} | \vec{\Lambda})$, which also appears in the GW literature with the notation $\beta(\vec{\Lambda})$~\cite{LIGOScientific:2016dsl} or $\alpha(\vec{\Lambda})$~\cite{2019MNRAS.486.1086M}.  
This is the fraction of sources that are detected (out of the total number $\mathcal{N}$ of astrophysical sources) for a hypothetical \ac{CBC} population described by hyper-parameters $\vec{\Lambda}$:
\begin{equation}
\label{eq:beta}
P(\mathrm{det} | \vec{\Lambda})  \equiv \iint P(\mathrm{det} | d) p(d | \pvec) p(\pvec | \vec{\Lambda})\, \mathrm{d}d\, \mathrm{d}\pvec\,.
\end{equation}

%To include the $P(\mathrm{det} | d)$ term, 
The inhomogeneous Poisson process likelihood of Eq.~\eqref{eq:inhom-pos-likelihood-nosel} is then modified as follows. 
The expected number of observed events in the exponent term is now $\mu(\vec{\Lambda}) \equiv \mathcal{N}P(\mathrm{det} | \vec{\Lambda})$, while each factor $i$ in the product over events is multiplied by $P( \mathrm{det}_i | d_i)$~\cite{2004AIPC..735..195L,2019MNRAS.486.1086M,2019PASA...36...10T,2022hgwa.bookE..45V,2023arXiv231002017E}: %\td{I am changing small $p$ to big $P$ for probability of detection!}
\begin{multline}
\label{eq:inhom-pos-likelihood-sel}
 p( \left\{ \pvec_i \right\},  \left\{d_i \right\},  \left\{\mathrm{det}_i \right\}  |  \vec{\Lambda}, \mathcal{N} ) \\ 
  \propto \mathcal{N}^N e^{-\mu(\vec{\Lambda})} \prod_{i=1}^N P(\mathrm{det}_i | d_i ) p( d_i | \pvec_i )  p( \pvec_i | \vec{\Lambda} ) \\ 
  \propto \frac{ \mu(\vec{\Lambda})^N e^{-\mu(\vec{\Lambda})} }{ P(\mathrm{det} | \vec{\Lambda})^N } \prod_{i=1}^N p( d_i | \pvec_i )  p( \pvec_i | \vec{\Lambda} ).
\end{multline}
In the last line, we substitute for $\mathcal{N}$ and hide the $P(\mathrm{det}_i | d_i )$ terms in the proportionality because they are independent of the population hyperparameters $ \vec{\Lambda}$ and $\mathcal{N}$ \cite{2023arXiv231002017E}. 

Now we can write the joint posterior over $ \left\{ \pvec_i \right\}$, $\vec{\Lambda}$, and $\mu$ (or alternatively $\mathcal{N}$) as 
\begin{equation}
\label{eq:inhom-pos-posterior-sel}
 p(\left\{ \pvec_i \right\}, \vec{\Lambda}, \mu | \left\{d_i \right\}, \left\{\mathrm{det}_i \right\} ) \propto 
  p(\vec{\Lambda}, \mu) \frac{ \mu^N e^{-\mu} }{P(\mathrm{det} | \vec{\Lambda})^N} \prod_{i=1}^N p( d_i | \pvec_i) p( \pvec_i | \vec{\Lambda} )\,.
\end{equation}

If we are only interested in inferring the shape of the population rather than the number of sources, we can marginalize Eq.~\eqref{eq:inhom-pos-posterior-sel} over $\mu$. 
If we adopt a separable prior on $\mu$ and $\vec{\Lambda}$, we recover the following posterior PDF for $\left\{ \pvec_i \right\}$ and $\vec{\Lambda}$:
\begin{equation}
\label{eq:joint-hier-likelihood-p-sel}
p(\left\{ \pvec_i \right\}, \vec{\Lambda} | \left\{d_i \right\} \left\{\mathrm{det}_i \right\} ) \propto
\frac{p(\vec{\Lambda})}{ P(\mathrm{det} | \vec{\Lambda})^N } \prod_{i=1}^N  p( d_i | \pvec_i) p( \pvec_i | \vec{\Lambda} )\,,
\end{equation}
which reduces to Eq.~\eqref{eq:joint-hier-likelihood-p-nosel} if $P(\mathrm{det} | \vec{\Lambda}) = 1$, implying there are no selection effects.

If we choose to work with the parameter $\mathcal{N}$ rather than $\mu$, an arbitrary prior on $\mathcal{N}$ may not generally lead to a separable induced prior on $\mu$ and $\vec{\Lambda}$.
However, the choice of a uniform prior over $\ln\mathcal{N}$ will always induce a uniform prior on $\ln\mu$, independently of $\vec{\Lambda}$: with this choice, marginalization over $\mathcal{N}$  
%Therefore, it is convenient to choose a flat-in-log prior on , so that when we marginalize over $\mathcal{N}$ 
recovers Eq.~\eqref{eq:joint-hier-likelihood-p-sel}.
This allows us to sample from the joint posterior over the shape of the population $\vec{\Lambda}$ and its amplitude $\mathcal{N}$ or $\mu$ in two steps.
First we can sample $\vec{\Lambda}$ from the marginalized posterior of Eq.~\eqref{eq:joint-hier-likelihood-p-sel}.
Then we can draw $\mu$ samples from the distribution $p(\mu | N) \propto p(N | \mu) p(\mu)$, where the observed number of events follows a Poisson distribution, $p(N | \mu) \propto \mu^N e^{-\mu}$, and $p(\mu)$ is the prior on $\mu$. 

We now discuss how to evaluate the {detection fraction} $P(\mathrm{det} | \vec{\Lambda})$. 
%that appears in the above equations. 
In Eq.~\eqref{eq:beta}, it %$P(\mathrm{det} | \vec{\Lambda})$ 
is defined as a double integral over data and single-event parameters: this may be estimated
% this integral is 
by Monte Carlo sampling. 
We draw $\pvec_j$ samples from $p(\pvec | \vec{\Lambda})$, simulate the GW signal for each sample, add it to a realization of the noise distribution to get the corresponding mock detector data $d_j$, and evaluate $P(\mathrm{det} | d_j)$ by processing the mock data with search pipelines. 
The integral $P(\mathrm{det} | \vec{\Lambda})$ can then be approximated as the average over samples $j$:
\begin{equation}
P(\mathrm{det} | \vec{\Lambda}) \approx \langle P(\mathrm{det} | d_j) \rangle\,.
\end{equation}
If $P(\mathrm{det} | d_j)$ is deterministic (either 0 or 1 for each simulated data $d_j$), this is simply the fraction of simulated signals that pass the detection threshold.
As discussed in \ref{sss:sensitivity}, we refer to these simulated signals as ``injections,'' and the injections that pass the detection threshold are known as ``found injections.''

To save computational cost, we often evaluate $P(\mathrm{det} | d_j)$ for just one set of $\pvec_j$ samples, drawn from some reference PDF $p_\mathrm{draw}(\pvec)$ (see the discussion in  \ref{sss:sensitivity}). 
We then estimate $P(\mathrm{det} | \vec{\Lambda})$ by reweighting these injections:
\begin{equation}
P(\mathrm{det} | \vec{\Lambda}) = \left\langle \frac{P(\mathrm{det} | d_j) p(\pvec_j | \vec{\Lambda})}{p_\mathrm{draw}(\pvec_j)} \right\rangle\,.
\end{equation}
The use of importance sampling here is analogous to estimating single-event evidences using \ac{PE} samples %for each event 
drawn under a reference prior distribution, Eq.~\eqref{eq:evidence_importance_sample}; in both cases, the uncertainty in the Monte Carlo integral must be under control in order to obtain valid inferences~\cite{2022arXiv220400461E}.\footnote{\cite{Farr:2019rap} gives a simple initial treatment for the selection function case.}

For estimating $\beta(\vec{\Lambda})$, the effective number of injections required to avoid biasing the likelihood increases in proportion to the number of detected events. In principle, the injected distribution $p_\mathrm{draw}(\pvec)$ may be adjusted to better match the inferred astrophysical distribution $p(\pvec | \vec{\Lambda})$, so that fewer injections are required; 
%to reach the required precision; 
however this requires \emph{a priori} knowledge of the true population, which is necessarily incomplete if derived from previous observations. 
%in the Monte Carlo estimate of $P(\mathrm{det}|\vec{\Lambda})$. 
As an alternative to importance sampling, we can compute the integral $P(\mathrm{det} \mid \vec{\Lambda})$ by first estimating a smooth approximation to $P(\mathrm{det} \mid d)$~\cite{Wysocki:2018mpo,2020PhRvD.102j3020G,2022ApJ...927...76T}.
This is analogous to calculating the single-event evidence terms with a smooth approximation to the \ac{PE} likelihood instead of averaging over \ac{PE} samples.

\subsection{Inference on astrophysical populations}
\label{ss:populations_cosmo}

A major goal of \ac{GW} astrophysics is to understand the origin of merging binary compact objects. The formation and evolutionary histories of neutron stars and black holes in merging binaries is an open problem %that likely 
involving many astrophysical processes {that are currently poorly understood}. 
%Despite being poorly understood, 
Many such processes also play a fundamental role in a variety of astrophysical contexts outside the production of \ac{CBC} sources, including stellar evolution, galaxy assembly, and the production of heavy elements. The population of \ac{CBC} sources also has implications for cosmology, as \acp{GW} from cosmological distances encode information about the cosmic expansion history of the Universe.

Binary compact object mergers likely represent extremely rare outcomes of massive stellar evolution. %
Stars more massive than $\sim8\,M_\odot$ end their lives as \acp{NS} or \acp{BH}~\cite{2023arXiv230409350H}; in rare circumstances, such a \ac{NS} or \ac{BH} will be close enough to another compact object that they eventually merge within the age of the Universe. 
There are, broadly, two distinct channels to achieve this: either the two objects started off in a binary star system that remained gravitationally bound throughout its evolution~\cite{1998ApJ...506..780B,2002ApJ...572..407B,2007PhR...442...75K}, or they started off in a dense star cluster, interacting dynamically with other objects before assembling a tight binary~\cite{1993Natur.364..421K,1993Natur.364..423S}; see Refs.~\cite{2021hgwa.bookE..16M,2022PhR...955....1M} for recent reviews. 
In either case, the two compact objects must reach a sufficiently tight orbit that will shrink through gravitational radiation to the point of merger. 
Alternatively, it has been proposed that primordial BHs produced shortly after the Big Bang could contribute to the CBC population~\cite{2016PhRvL.116t1301B}. 
The existence of a subpopulation of primordial origin, if discovered (for example, by finding BHs at subsolar masses or at redshifts higher than the first star formation), would have dramatic implications for early Universe cosmology and the composition of dark matter~\cite{2020ARNPS..70..355C}.

The hierarchical Bayesian approach discussed in the previous subsections allows us to infer properties of the compact binary population, such as their merger rate and how BHs and NSs are distributed in mass, spin and merger redshifts. 
These population distributions depend on the evolutionary pathways of BHs and NSs in merging binaries. 
By measuring the population properties of GW sources, we ultimately hope to reverse engineer the histories of CBC progenitors. 
The population of CBCs also depends on the cosmological parameters, particularly the relation
between luminosity distance, which controls the \ac{GW} signal amplitude, and redshift, which affects the signal frequency in the same way as a change in component masses (see Sec.~\ref{sss:redshift}). 
Any inference on the (source-frame) mass distribution thus depends on %some estimate of 
the events' redshifts, derived from their luminosity distances via a cosmological model. 
The cosmological redshift--distance relation may also be simultaneously inferred together with the astrophysical population parameters~\cite{1986Natur.323..310S,2012PhRvD..86b3502T,2019ApJ...883L..42F,2021PhRvD.104f2009M,2022PhRvL.129f1102E}.

In general, a \ac{BBH} population model describes the distribution over source-frame primary and secondary masses (or alternatively primary mass and mass ratio), component spin magnitudes and orientations (or alternatively, ``effective'' spins like $\chi_\mathrm{eff}$ and $\chi_p$; {see Section~\ref{sss:binary_params}}), merger redshift (plus background cosmology relating luminosity distance to redshift), and the binary position and orientation. %(including sky position and inclination). 
Populations with \ac{NS} components also have hyperparameters describing their tidal deformabilities, which are discussed in the following subsection.

Several different strategies for population modeling have then been followed.
The first approach aims to infer physical parameters by directly comparing observations to simulations.
The formation of \acp{CBC} is subject to several uncertainties, ranging from unknown initial conditions (for example, the initial mass distribution of stars) to uncertain physical quantities (for example, nuclear reaction rates in stellar cores), to evolutionary stages where direct modelling is too computationally difficult and we instead use approximate prescriptions (for example, the common envelope in binary stellar evolution).  These uncertain elements may all leave some imprint on the \ac{CBC} population, so one could in principle model the CBC population in terms of the corresponding physical parameters, and directly infer their values via hierarchical Bayesian analysis. 
We refer to such population models as \emph{physics-driven models}.

At the opposite pole of methodology, %to} physics-driven models, 
\emph{data-driven models}, also known as non-parametric models, aim to be flexible enough to recover an arbitrary CBC population distribution.  
For example, the ``parameters'' of these population models might simply be the merger rate density at each point in $\pvec$-space, and the number of parameters approaches infinity as one considers arbitrarily high resolution in $\pvec$ space (e.g.\ a histogram in the limit of infinitely many, closely-spaced bins).  In practice, the range of distributions covered is limited both by computational and statistical considerations, particularly in the limit of densities which vary rapidly over a small range of parameter space. 
Data-driven models give the most reliable description of the population in that they are unbiased by the systematic uncertainties that can plague (incomplete) physics-driven models.  However, for the same number of observations, the statistical uncertainties in the inferred population tend to be larger because of the high model flexibility.
Furthermore, unlike physics-driven models, the parameter inferences often cannot be immediately read off as physical or astrophysical implications, usually requiring further interpretation. 

Falling in between physics-driven and data-driven models, \emph{parametric phenomenological models} are designed to consist of relatively simple functional forms, such as Gaussians or (truncated) power laws, but contain 
% These models may contain 
features that are motivated by astrophysical theory: e.g.\ a gap in the \ac{BH} mass spectrum or a local excess of density for \ac{BH} spin magnitudes close to zero. 
The main advantage of parametric phenomenological models is simplicity in writing them down and drawing initial astrophysical conclusions.
Oftentimes, these models are also cheaper to fit to data than data-driven or physics-driven models, although this stops being the case if the parametric model is over-prescriptive and struggles to fit the data. % performing the analysis and drawing some preliminary astrophysical conclusions.  %Thus, they are often fit to the data first before more computationally demanding or sophisticated data-driven or physics-driven models are applied. 

In the simplest parametric and data-driven population models, the probability distribution over source parameters is assumed to be separable, and thus can be written as a product,
% In other words, masses, spins, redshifts and orientations are assumed to be independently distributed, e.g.,
\begin{equation}
 p(\pvec | \mathbf{\Lambda}) = p_m(m_1, m_2 | \mathbf{\Lambda}_m)p_\chi(\mathbf{\chi}_1, \mathbf{\chi}_2 |  \mathbf{\Lambda}_\chi) p_z(z |  \mathbf{\Lambda}_z) p_\Omega(\mathbf{\Omega} |  \mathbf{\Lambda}_\Omega)p_i(\boldsymbol{\iota} | \mathbf{\Lambda}_i)\,.
\end{equation}
% TD - note that mathbf doesn't work for lower case Greek letters !!
Here $\mathbf{\Omega}$ and $\boldsymbol{\iota}$ are the angular parameters describing the position on the sky and the binary orientation, respectively; the corresponding distributions are usually assumed to be isotropic, 
% Often, the distributions over sky positions $\mathbf{\Omega}$ and orientation angles $\mathbf{\iota}$ are assumed to be isotropic, 
but see for instance~\cite{2023PhRvD.107d3016E,2023arXiv230413254I,2022arXiv220400968V} for counter-examples.

In general, model parameters include cosmological parameters that set the redshift--distance relation and also the distribution $p_z$, which may be written
\begin{equation}\label{eq:cosmoprior}
 p_z(z) \propto \frac{1}{1+z}\frac{dV_c}{dz}f(z),
\end{equation}
with $V_c$ the comoving volume, such that the merger rate density per comoving volume per source-frame time evolves with redshift as
\begin{equation}
 \frac{dN}{dV_c\, dt} \propto f(z), 
\end{equation}
and %the differential comoving volume 
$dV_c/dz$ also depends on the cosmological parameters. 
Common phenomenological models for $f(z)$ include a power law in $(1 + z)$ or functional forms inspired by the star formation rate~\cite{Fishbach:2018edt,2020ApJ...896L..32C}.

The two-dimensional mass distribution is often factorized as~\cite{LIGOScientific:2016dsl,2017ApJ...851L..25F,2017PhRvD..95j3010K}:
\begin{equation}
 p_m(m_1, m_2) = p_1(m_1) p_2(m_2 | m_1),
\end{equation} 
or similarly using the mass ratio $q = m_2/m_1$ in place of $m_2$.  Alternative factorizations may, though, be more natural in some contexts; for example, if we believe that both component \ac{BH} masses are drawn from the same distribution, such that they both share features like local overdensities in the same locations, we may consider a ``pairing function'' parametrization~\cite{2020ApJ...891L..27F,2023arXiv230805102F}:
\begin{equation}
 p_m(m_1, m_2) \propto p(m_1) p(m_2) f(q, M_\mathrm{tot}),
\end{equation}
where $p$ is the component mass distribution and $f$ is a pairing function that depends on the mass ratio and/or total mass; two BH masses are randomly paired in a binary if $f \equiv 1$. 

Typical phenomenological parametric models for the primary mass distribution $p_1(m_1)$ or the component mass distribution $p(m)$ include mixtures of (broken) power laws and Gaussians, with some smoothing applied at the minimum and maximum mass ends to avoid sharp cutoffs, and possible notch filters to represent dips or gaps~\cite{2017ApJ...851L..25F,2017PhRvD..95j3010K,2018ApJ...856..173T,2019PhRvD.100d3012W,2020ApJ...899L...8F,2021ApJ...913L..23E,2021ApJ...913L...7A}. 
The conditional secondary mass distribution, conditional mass ratio distribution, or pairing function is most commonly taken as a power law in mass ratio~\cite{2017ApJ...851L..25F,2017PhRvD..95j3010K}. 
However, there are already indications of additional structure in the mass ratio distribution; for example there are hints that at higher masses, the mass ratio distribution more strongly prefers symmetric ($q \approx 1$) systems compared to low masses~\cite{2022ApJ...933L..14L,2022ApJ...931..108F,Sadiq:2023zee}.

The spin distribution can be parameterized in terms of ``effective'' spin parameters $\chi_\mathrm{eff}$ and $\chi_p$. Past work has modeled these distributions as truncated Gaussians~\cite{2019MNRAS.484.4216R,2020ApJ...895..128M} or histograms with 3-5 bins~\cite{2017Natur.548..426F,2018ApJ...854L...9F}. 
Alternatively, one can consider all six spin degrees of freedom; in spherical coordinates, the dimensionless spin magnitude and spin tilt angle and azimuthal angle of each component.\footnote{Because these spin parameters evolve during the binary inspiral, a reference frequency must also be specified~\cite{2022PhRvD.105b4045V,2022PhRvD.105b4076M}.}
A common parameterization for the spin magnitude distribution is a beta distribution, because it is naturally truncated over the physical range between 0 and 1~\cite{2019PhRvD.100d3012W}.  To avoid statistical sampling issues, the prior is typically restricted to non-singular beta distributions; however, this choice excludes spin magnitude distributions that peak sharply at zero, 
%a priori, 
which can be undesirable if we wish to measure a possible excess of nonspinning BHs, as some theories suggest~\cite{2019ApJ...881L...1F,2021PhRvD.104h3010R,2021ApJ...921L..15G,2022ApJ...937L..13C}. 
The cosine of the spin tilt distribution is often modeled as a truncated Gaussian; Ref.~\cite{2017PhRvD..96b3012T} introduced a mixture model between such a truncated Gaussian 
%in cosine tilt 
centered at zero (aligned spin) 
% with uncertain standard deviation 
% TD - not sure why this needs to be said explicitly
and a flat distribution in cosine tilt (isotropic spins); motivated by   
%This model aims to capture 
expectations from a mixture between isolated binary evolution formation channels, which are thought to produce binaries with nearly-aligned spins (\cite{2000ApJ...541..319K} ( although the double pulsar system may provide a counterexample~\cite{2011ApJ...742...81F})) and dynamical assembly, which, at least in gas-free environments like globular clusters, is thought to produce binaries with isotropic spins~\cite{2000ApJ...528L..17P,2016ApJ...832L...2R}. 
The distributions of azimuthal spin angles are typically fixed as uniform, but theoretical predictions for spin-orbit resonances may also motivate explicit inference over these parameters~\cite{2014PhRvD..89l4025G,2022PhRvD.105b4045V,2022PhRvL.128c1101V}.
The simplest models assume that the two compact object spins in a binary are independently drawn from the same distribution, but this contradicts some predictions 
%for the origin of BH spin, 
in which, for example, first-born and second-born BHs preferentially follow different spin distributions~\cite{2018MNRAS.473.4174Z,2020A&A...635A..97B}. 
Recent work has thus relaxed this assumption, allowing primary and secondary BH spins to be drawn (independently or not) from different distributions~\cite{2022ApJ...929L..26F,2023arXiv231105182A}.

Each phenomenological parametrization of the mass, spin or merger redshift distributions can be replaced by a data-driven or non-parametric model. %So far, such data-driven models have been applied in only one or two dimensions. For example, the primary mass distribution may be modeled with a flexible, data-driven model, while the secondary mass distribution, spin distributions, and redshift distributions are all modeled with a simple parametric function. 
Data-driven models commonly used in the \ac{GW} literature include Gaussian mixture models~\cite{Tiwari:2020vym,2021ApJ...913L..19T}, splines~\cite{2022ApJ...924..101E,2023ApJ...946...16E,2023PhRvD.108j3009G}, Gaussian processes (e.g.~\cite{2024arXiv240402210F}, including Gaussian-process regularized histograms~\cite{2017MNRAS.465.3254M,Li:2021ukd,2023ApJ...957...37R} and autoregressive processes~\cite{2023arXiv230207289C}), Dirichlet processes~\cite{2022MNRAS.509.5454R}, and optimized kernel density estimates~\cite{2022PhRvD.105l3014S,Sadiq:2023zee}. 
Many such methods achieve flexibility by
%such as splines or Gaussian-process regularized histograms, 
setting up an underlying parametric model, such as a power law, and allowing more or less general forms of deviation from it. 
Data-driven models typically impose some structure on the correlations between neighboring points in $\pvec$-space~\cite{Heinzel:2024jlc}, in order to control the (lack of) smoothness of the population \ac{PDF}, and thus limit the effective number of independent degrees of freedom to be fit. 
Such strategies to regularize data-driven models will become less necessary in the limit of large catalogs. 

Once data-driven models are fit to the data, there is usually a second ``feature extraction'' step, where interesting features, such as local maxima, minima or trends, are identified. 
Although the model is not parameterized in terms of such features, posteriors on, e.g., the location of local maxima in the mass distribution may be calculated in post-processing~\cite{2023ApJ...955..107F,2023arXiv230207289C}.   
The significance of any features can also be evaluated by comparing the posterior against the prior of the data-driven model, noting that 
although such models are designed to be flexible they can induce nontrivial priors on the population distribution. 

In both the parametric and data-driven models, the next level of complexity in modeling the \ac{CBC} population
%joint mass-spin-redshift distribution 
is to allow parameter correlations %between various parameters, 
rather than assuming that masses, redshifts and spins are independent.  Two-parameter correlations investigated up to now include evolution of the mass or spin distribution with redshift~\cite{2019ApJ...883L..24S,2021ApJ...912...98F,2022ApJ...932L..19B,2022A&A...665A..59B,2022ApJ...932L..19B}, or the spin or mass ratio distributions varying over primary mass~\cite{2020ApJ...894..129S,2021MNRAS.507.3362T,2022PhRvD.105l3024F,2022ApJ...935L..26F,Wang:2022gnx}.  Empirically, a statistically significant correlation between the effective inspiral spin and the mass ratio has been identified~\cite{2021ApJ...922L...5C}, which may indicate that primary and secondary BH spins are drawn from different distributions, as discussed above.

The simplest such model allows for only a linear correlation, although quadratic and higher order polynomial terms may straightforwardly be included. 
For instance one can %express the evolution of the primary mass distribution with redshift by 
allow the location of a Gaussian peak $\mu$ in the primary mass distribution to vary linearly with redshift, $\mu(z) = \mu_0 + z\cdot d\mu/dz|_{z = 0}$, with parameters $\mu_0$ 
%(the location of the peak at redshift 0) 
and $d\mu/dz|_{z = 0}$. 
%(the linear slope of $\mu(z)$) are free parameters. 
Another simple parametric model {for two-parameter correlations}
%used to search for correlations between any two variables (e.g. mass and redshift) 
is a switch point analysis: in such a model %with a switch point, 
the parameters describing the distribution in one variable (e.g.\ mass) are allowed to take on different values depending if the second variable is above or below the switch point (e.g.\ at low versus high redshifts). 
Although a discontinuous jump in the distribution may not be physically well-motivated, switch point analysis can be a quick way to check if the data prefer correlations between two variables~\cite{2021ApJ...912...98F}. 
Correlations can also be encoded via mixture models, in which the population consists of different components (subpopulations) with different properties; for example, allowing distinct parameterized power law or Gaussian peak components in the mass distribution to take on different spin distributions~\cite{Wang:2022gnx,2023arXiv230401288G,2023arXiv230302973L}.
Copulas also naturally allow the analyst to fix the form of the one-dimensional marginal distributions, while exploring different two-dimensional correlations specified by a copula density function~\cite{2022MNRAS.517.3928A}. 
Data-driven models can also be extended to include correlations between variables, either by directly modeling multiple dimensions non-parametrically (e.g., a multi-dimensional Gaussian mixture~\cite{Tiwari:2020vym,Tiwari:2021yvr}, a multi-dimensional Gaussian-process regulated histogram~\cite{2023ApJ...957...37R,Ray:2024hos} or a kernel density estimate~\cite{Sadiq:2023zee}); or combining one-dimensional parametric models with flexible functions that describe correlations, for instance using a spline (rather than a linear function) to model evolution of a mass feature with redshift~\cite{2023arXiv231200993H}. 

Fitting physics-driven models to \ac{GW} catalogs requires a different strategy, based on population synthesis simulations.
The %physics-driven 
model may be described
%parameterized 
in terms of a mixture of formation channels, such as isolated binary evolution with common envelope, isolated binary evolution with stable mass transfer, chemically homogeneous evolution, triple star evolution, stellar evolution in young star clusters, dynamical assembly in globular clusters, dynamical assembly in the disks of active galactic nuclei, primordial black holes, and so on.
The relative mixture weights (also known as branching ratios) of different channels may be considered as free parameters, together with some of the uncertain astrophysical parameters within each channel~\cite{2015ApJ...810...58S,2017ApJ...846...82Z,2021ApJ...910..152Z,2021MNRAS.507.5224B}. 
Such parameters may include the mass accretion efficiency, common envelope description, black hole natal spins and natal kicks, the metallicity-specific star formation rate, and the mass and radius distribution of globular clusters~\cite{2008ApJ...672..479O,2021MNRAS.505.3873B,2022MNRAS.517.4034S,2023ApJ...950...80R,2023MNRAS.522.5546F,2023ApJ...955..127C}. 
For physics-driven models, it is computationally challenging to simulate enough populations that cover the resulting high-dimensional parameter space. %In order to alleviate this issue, 
Strategies to alleviate this issue such as interpolating between simulations and identifying where to run new simulations (e.g.\ using Gaussian processes) have been proposed~\cite{2018PhRvD..98h3017T,2019MNRAS.490.5228B}.
% One technique is to focus on individual GW events and perform ``backward population synthesis'' to map their source properties back to the population synthesis simulations that produce the highest rate of similar mergers~\cite{2021ApJ...914L..32A,2023ApJ...950..181W}.
% TD - these are nice papers, but I am not seeing how it helps us deal with the population ... any idea how to make the link better defined? 
Simulation-based inference may also be a promising technique for comparing physics-driven models to GW data~\cite{2020PNAS..11730055C,2023arXiv231112093L}. 
An even bigger challenge, however, is that many of the uncertainties in physics-driven models are systematic, stemming from the lack of an adequate model, and therefore defy parameterization~\cite{2021MNRAS.508.5028B,2022ApJ...925...69B}. 
This limitation might be tackled by combining physics-driven models with phenomenological or data-driven models: incorporating specific predictions from population synthesis simulations, such as the remnant mass function or the distribution of delay times between star formation and merger, into a phenomenological population model~\cite{2021ApJ...916L..16B,2023ApJ...957L..31F,2023arXiv231203973G}. This hybrid approach can be used to test one aspect of the population synthesis model at a time against the data; for example, checking whether the predicted delay time distribution is consistent with the inferred merger rate and the assumed star formation rate. 

As a technical aside, especially for data-driven or physics-driven \ac{CBC} population models, it can be tempting to fit the ``detected distribution'' (i.e., the distribution of GW sources conditioned on detection), rather than the underlying, astrophysical distribution. However, one has to be careful that doing so does not violate the conditional structure of the data-generating process~\cite{2023arXiv231002017E} to obtain a correct hierarchical Bayesian likelihood.
%. In other words, regardless of the method, the likelihood must be consistent with the correct hierarchical Bayesian likelihood of Eq.~\eqref{eq:inhom-pos-likelihood-sel} to avoid a biased inference. 

%Describe model comparison with Bayes factors, including Savage Dickey Density Ratios, and posterior predictive checks for goodness of fit as an alternative to model comparison. Outliers. 
\subsubsection{Population model comparisons and checking}
With so many population modeling choices, the question naturally arises of which model best describes the data. 
One answer %way of addressing this question 
is to compare models using their Bayes factors (the ratio of their evidences) or their posterior odds (see Section~\ref{ss:introBayes}).
Often, the population models we consider are nested, which means that we can easily compute a Bayes factor between two models with a Savage-Dickey density ratio~\cite{SDDR} rather than directly computing their evidences. 
A simple application is to mixture models: either a mixture between two formation channels in physics-driven models, or a mixture between (e.g.) power laws and Gaussians as in commonly-used phenomenological parametric models. Given the posterior on the branching ratio $f$ between two components, we can compute the posterior probability density at $f = 0$ compared to the prior probability density: this 
%Savage Dickey Density Ratio 
density ratio gives us the Bayes factor between the simpler one-component model and the two-component mixture model. % between the two components. 

However, in practice the Bayes factor or evidence ratio has various drawbacks.  In the mixture model example, if the data poorly constrain $f$, the posterior will closely resemble the prior, and the Bayes factor between the two models will be unity: the naive expectation that Bayesian model selection will penalize the more complex model 
fails when the additional parameters are not well-constrained by the data. 
More generally, the choice of hyper-prior can significantly affect the Bayes factor: this is easy to see in the mixture model example.  %Additionally, 
Such dependence on the hyper-priors can be problematic for interpreting population results: particularly for phenomenological models, there may not be a single well-motivated choice of hyper-prior. 
%for a population model's hyper-prior, particularly in phenomenological models.
For instance, two possible prior choices for the power-law slope of the BBH primary mass distribution would be a uniform distribution between limits $-10$ and $10$, or uniform between $-5$ and $5$; no more fundamental principle enables us to prefer one of these priors to the other. 
Nevertheless, over a higher-dimensional space of hyper-parameters, different prior choices can yield evidences differing by orders of magnitude for otherwise identical models.

An alternative to comparing two models is to merely check whether a model adequately fits the data, with a goodness-of-fit test. 
In the Bayesian context, such tests are often provided by posterior predictive checks~\cite{2020ApJ...891L..31F,2021ApJ...913L...7A,2022PASA...39...25R,Miller:2024sui}. 
After fitting a given population model to the data, one can generate mock observations from that population, which represent draws from the posterior predictive distribution. 
Ideally, the mock observations include both measurement uncertainty and selection effects, such that the posterior predictive distribution is a distribution over data-space $d$; a less powerful posterior predictive check can be carried out at the level of true source parameters $\pvec$ if one includes selection effects but not measurement uncertainty in generating mock observations. 
The predicted mock observations can then be checked for consistency with the actual observations: % \td{how? eg with K-S or other cdf tests?} 
in one dimension, this can be done by comparing the \ac{CDF} of the predicted versus observed source parameters, e.g.\ via a Kolmogorov-Smirnov (K-S) or Anderson-Darling test. 
In higher dimensions, one can %define a region of interest in source parameter space 
%(e.g.\ $m_1 > 45\,M_\odot$ and $z < 0.37$), and 
compare the predicted versus observed number of detections inside one or more regions of interest within source parameter space, e.g.~\cite{2021ApJ...912...98F}. 
For a choice of test statistic -- in the K-S test case the maximum distance between two \acp{CDF}, or the number of detections in a region of parameter space -- the tension between the population model and the data can be quantified by first constructing a null distribution of the test statistic by drawing many mock catalog realizations from the posterior predictive distribution. 
Then the ``posterior predictive p-value'' is calculated as the fraction of mock catalogs that are more extreme than the observed catalog with respect to the test statistic, quantifying where the observed value falls in the null distribution~\cite{2020ApJ...891L..31F}.
In addition to goodness-of-fit tests, posterior predictive checks have been used to classify population outliers in the GW literature~\cite{2020ApJ...891L..31F,2020PhRvD.102d3015A,2021ApJ...913L...7A,2022ApJ...926...34E}.

In general, posterior predictive p-values are also sensitive to the choice of model prior, insofar as this choice affects the posterior \ac{PDF}, and consequently also the posterior predictive distribution.
However, they are less sensitive than Bayes factors, which are affected by changes only in the prior (e.g.\ changes in prior boundaries) even in the case where the posterior is unaffected.

%%%%%%%%%%%%%%%%%%%%%%%%%%%%%%%%%
\subsection{Matter effects and equation of state of dense matter}
\label{ss:matter_eos}

During the late stages of a binary coalescence that involves at least one neutron star, finite-size effects modify the expected binary evolution~\cite{Chatziioannou:2020pqz,Dietrich:2020eud}. The two lowest-order effects in the PN phase expansion are that of the spin-induced quadrupole moment at 2PN~\cite{Poisson:1997ha} and tidal interactions at 5PN~\cite{Flanagan:2007ix}. Despite the higher PN order, the tidal interactions are more prominent in the signal due to the larger numerical  %value of the corresponding term
prefactor~\cite{Hinderer:2009ca}, as well as the fact that the former effect depends on the neutron star spin quadratically~\cite{Harry:2018hke}. Tidal interactions originate from the fact that \acp{NS} have a finite size, and are therefore %tidally 
distorted by the gravitational field of their companion, with two effects on the binary orbit: (i) the tidal deformation drains orbital energy from the system, and (ii) the tidally-induced quadrupole moment acts as another source of gravitational radiation. Black holes do not exhibit such tidal interactions~\cite{Binnington:2009bb,Chia:2020yla}. 

The degree of tidal deformation depends on the internal composition of the \ac{NS} and the (unknown) equation of state of dense nuclear matter in beta equilibrium. For stable, cold neutron stars 
%that have cooled down to energies below the corresponding Fermi energy, 
the equation of state is a relation between the pressure $p$ and the energy density $\epsilon$ which is expected to be common among all \acp{NS}~\cite{Lattimer:2015nhk,Oertel:2016bki,Baym:2017whm}. In general, softer equations of state predict lower internal pressures and smaller \acp{NS} that are less deformable, while stiffer equations of state predict higher pressures, larger \acp{NS}, and stronger tidal interactions. Further nuclear phenomena such as the emergence of strange degrees of freedom or deconfined quarks and phase transitions~\cite{Han:2019bub} might complicate this picture or even predict the existence of ``twin stars'', i.e., \acp{NS} with the same mass but different radii.  

%However, this does not preclude the universality of the $p(\epsilon)$ relation \td{can this be rephrased?}, 
Inference for the common $p(\epsilon)$ equation of state is a sub-case of the hierarchical inference described above. Specifically, the generic hierarchical posterior from Eq.~\eqref{eq:hierlike} now becomes~\cite{Landry:2020vaw,Chatziioannou:2020pqz}
\begin{equation} %\label{eq:logLhierEoS}
p\left(\{ {\cal{E}}, \vec{\Lambda}\} | \left\{d_i \right\}_{i = 1}^N \right) = \frac{p(\{ {\cal{E}}, \vec{\Lambda}\} ) \prod_{i=1}^N \int p \left( d_i | \pvec_i\right) p\left( \pvec_i | \{ {\cal{E}}, \vec{\Lambda}\} \right)  d\pvec_i }{ p( \left\{ d_i \right\}_{i = 1}^N) }\,,
\end{equation}
where we have separated the equation of state ${\cal{E}}\equiv p(\epsilon)$ from the remaining population hyperparameters $\vec{\Lambda}$. The latter could now represent the mass or spin distribution of \acp{NS}~\cite{Ozel:2016oaf}; this distribution must be inferred simultaneously with ${\cal{E}}$, as their uncertainties can be correlated~\cite{Wysocki:2020myz,Golomb:2021tll}. Restricting $\vec{\Lambda}$ to %inference of the equation of state and 
the \ac{NS} mass distribution, the relevant event-level parameters are the binary masses and tidal deformabilities $\pvec_i =\{m_i,\Lambda_i\}$ and the population model (or source prior) is now
\begin{align} \label{eq:logLhierEoS}
p\left( m_i,\Lambda_i | \{ {\cal{E}}, \vec{\Lambda}\} \right) &= p\left( \Lambda_i | m_i, \{ {\cal{E}}, \vec{\Lambda}\} \right) p\left( m_i| \{ {\cal{E}}, \vec{\Lambda}\} \right) \nonumber\\
&= \delta\left( \Lambda_i - {\cal{E}}(m_i) \right) p\left( m_i| \{ {\cal{E}}, \vec{\Lambda}\} \right) \,,
\end{align}
where $\delta(x)$ is the Dirac delta function.
The equation of state term %part of the population model, i.e., the 
$\delta\!\left( \Lambda_i - {\cal{E}}(m_i) \right)$ %term 
represents the fact that the tidal deformability is unambiguously defined given an equation of state and a \ac{NS} mass. The $p\left( m_i| \{ {\cal{E}}, \vec{\Lambda}\} \right)$ term couples the mass distribution parametrized by $\vec{\Lambda}$ and the equation of state encoded in ${\cal{E}}$, and in general reflects assumptions about \ac{NS} astrophysics. For example, a simple uniform %neutron star 
mass distribution between limits $M_l$ and $M_h$,
\begin{equation}
p\left( m_i| \{ {\cal{E}}, \vec{\Lambda}\} \right) = \frac{\Theta(M_l \leq m_i) \Theta( m_i \leq M_h )}{M_h-M_l}\,,
\end{equation}
could contain an implicit dependence on the equation of state, $M_h=M_{\mathrm{max}}({\cal{E}})$, if we assume that the maximum \ac{NS} mass realized in the Universe is determined by nuclear physics, $M_{\mathrm{max}}({\cal{E}})$ being the maximum mass of nonrotating \acp{NS} predicted by ${\cal{E}}$. If we instead assume that the astrophysical conditions dictating the formation of neutron stars limit neutron star masses to be below $M_{\mathrm{max}}({\cal{E}})$, $M_h$ would be independent of ${\cal{E}}$; the effects of such assumptions are explored further in~\cite{Legred:2021hdx}.  
%for a demonstration of the effects of these assumptions on current data.

Equation~\eqref{eq:logLhierEoS} requires some functional form for $\cal{E}$.  This may be achieved via parametric or nonparametric priors that prescribe candidate $p(\epsilon)$ curves (or equivalent functions). For example, the \emph{piecewise-polytropic} parametrization expresses the pressure as a piecewise function of the baryon density $\rho$~\cite{Read:2008iy,OBoyle:2020qvf}, expressed in its simplest form, as
\begin{equation}
    p(\rho;p_1=p(\rho_1),\Gamma_1,\Gamma_2,\Gamma_3) = \begin{cases}
        K_1 \rho^{\Gamma_1} : \rho < \rho_1 \\
        K_2 \rho^{\Gamma_2} : \rho_1 < \rho < \rho_2 \\
        K_3 \rho^{\Gamma_3} : \rho_2 < \rho\,,
    \end{cases}
\end{equation}
where $\rho_1$ and $\rho_2$ are fixed by optimization against nuclear models~\cite{Read:2008iy}, while $K_1$ and $K_2$ are fixed by continuity. 
%Extensions of this form that add more parameters are also possible~\cite{Steiner:2010fz,Steiner:2015aea,Raithel:2016bux,OBoyle:2020qvf}. 
%\td{I doubt we need to add a lot more citations if the inference method is basically the same: choose the most influential one or two to cite with Read 2008?}  
A closely related form is the \emph{spectral} parametrization that instead expands the polytropic index in a polynomial of the pressure~\cite{Lindblom:2010bb,Lindblom:2013kra}
\begin{equation}
  p(\rho;\gamma_i ) = \rho^{\Gamma(x)}\,, \qquad  \Gamma(x) = \sum_{i=0}^3 \gamma_i \log(x)^i \,,
\end{equation}
where $x\equiv p/p_0$ and $p_0$ is the pressure where the equation of state transitions to a fixed crust model.  An alternative approach instead parameterizes the dependence of the speed of sound~\cite{Tews:2018kmu,Greif:2018njt}. 
Broad and uninformative priors are typically chosen for the parameters in each case, cf.~\cite{Carney:2018sdv,Wysocki:2020myz}.

A parallel method expresses the equation of state in a nonparametric way, a term used to refer to methods such as Gaussian processes~\cite{Landry:2018prl,Essick:2019ldf,Miller:2021qha,Gorda:2022jvk} that parametrize the correlations between the values of a function at different points 
rather than the function itself. This approach offers the flexibility of directly selecting whether the equation of state prior includes strong or weak correlations~\cite{Legred:2022pyp} across different density scales while simultaneously imposing physical constraints such as thermodynamic stability and causality through the formulation of the Gaussian process.
%\td{These 2 sentences seem repetitive - cut down?} Maximum flexibility (modulo physical constraints such as thermodynamic stability and causality) such that constraints about some density range have minimal effect about other densities can be achieved~\cite{Legred:2022pyp}, however it is not guaranteed, as Gaussian processes with stronger correlations can be trivially constructed as well~\cite{Miller:2021qha}.

Both of the above phenomenological approaches can be augmented to include information from nuclear theory, and the associated % though such information can be subject to additional 
uncertainties.  %related to nuclear frameworks and calculations. 
Specific examples include the use of chiral effective field theory~\cite{Lonardoni:2019ypg,Tews:2019cap,Tews:2018kmu,Drischler:2020hwi,Drischler:2020yad} results at lower densities and perturbative quantum chromodynamics~\cite{Gorda:2022jvk,Somasundaram:2022ztm,Komoltsev:2021jzg} results at high densities. 
%, while nonparametric models can be conditioned on the same results in the relevant density regime.

The improving understanding about \acp{NS} properties in the recent years is the outcome of combining GW data~\cite{LIGOScientific:2017vwq,LIGOScientific:2018cki,LIGOScientific:2018hze,LIGOScientific:2020aai} with other probes: radio~\cite{Antoniadis:2013pzd,NANOGrav:2019jur,Fonseca:2021wxt} and X-ray~\cite{Miller:2019cac,Riley:2019yda,Miller:2021qha,Riley:2021pdl} observations of galactic neutron stars, observations of the electromagnetic counterpart of GW170817~\cite{Coughlin:2018miv,Coughlin:2018fis,Radice:2018ozg,Margalit:2019dpi}, nuclear experiments~\cite{Essick:2020flb,RocaMaza:2011pm,Adhikari:2021phr,Reed:2021nqk,Essick:2021ezp}, and nuclear theory~\cite{Machleidt:2011zz,Tews:2018kmu,Drischler:2020yad,Drischler:2020hwi}.  Discussion of %the different probes and 
the intricacies involved in using such diverse data sets is beyond the scope of this review, however the emerging picture suggests that typical $1.4\,M_{\odot}$ \acp{NS} have radii around $12-12.5$\,km (with uncertainties of a few kilometers) 
and maximum masses above $2\,M_{\odot}$~\cite{Miller:2019cac,Miller:2021qha,Raaijmakers:2019qny,Raaijmakers:2021uju,Raaijmakers:2019dks,Landry:2020vaw,Legred:2021hdx,Dietrich:2020efo,Pang:2021jta,Biswas:2021yge,Essick:2023fso}. Many questions remain about the maximum mass, the possibility of phase transitions, the effect of \acp{NS} rotation, as well as the nature of the $2-3\,M_{\odot}$ objects observed with gravitational waves~\cite{Abbott:2020khf,Essick:2020ghc,Fattoyev:2020cws,Tews:2020ylw,Dexheimer:2020rlp,Biswas:2020xna} that upcoming GW observations have the potential to offer information about.

%%%%%%%%%%%%%%%%%%%%%%%%%%%%%%%%%%
\subsection{Tests of General Relativity with multiple sources}

The binary dynamics described in Sec.~\ref{sec:Newtonian_binary} are derived assuming \ac{GR}: modified gravity theories can introduce both qualitative and quantitative changes in observable binary coalescence signals.  \ac{GR} is commonly expected to be the low-energy limit of an as yet unknown full quantum theory of gravity, implying modifications at some, also unknown, energy or length scale~\cite{Yunes:2013dva}. An additional motivation for testing the theory is the fact that exploration of beyond-\ac{GR} phenomena and dynamics can illuminate the properties of the theory itself. \ac{GW} data offer the only probe of the theory in the strongly nonlinear and dynamical regime of compact binary mergers~\cite{LIGOScientific:2016lio,Yunes:2016jcc,LIGOScientific:2019fpa,LIGOScientific:2020tif,LIGOScientific:2021sio}.

Efforts to test \ac{GR} are hampered by the fact that no self-consistent alternative theory has been studied in comparable depth or thoroughness.
% rigor of Sec.~\ref{sec:Newtonian_binary}. 
Therefore, most tests %of \ac{General Relativity 
are characterized as ``phenomenological'' or ``model-dependent'', meaning that though they might be motivated by specific beyond-\ac{GR} phenomena, they are not tuned to any specific high-energy completion of gravity. Tests of \ac{GR} may be classified according to various criteria. The first criterion distinguishes between ``generation'' \emph{vs.} ``propagation'' phenomena: those relating to how waves are generated at the source, or how they propagate to the detectors respectively. An alternative classification splits tests into those that target the fundamental or universal properties of the theory and of wave solutions (e.g., speed of propagation or existence of two transverse polarization modes), \emph{vs}. ``quantitative'' results (e.g., measuring the phase evolution of binary emission at some \ac{PN} order). Beyond these model-independent tests, theory-specific tests have also been pursued in certain cases, e.g.~\cite{Perkins:2021mhb,Nair:2019iur}.

Combining information about tests of \ac{GR} with multiple sources proceeds again under the same hierarchical framework, with the likelihood given in Eq.~\eqref{eq:hierlike} where $\vec{\Lambda}$ are now the hyperparameters that describe the population distribution of the beyond-\ac{GR} effect(s). The population model, $p( \pvec_i | \vec{\Lambda} )$ depends on the test under consideration, with two limiting cases commonly employed. The first corresponds to tests that introduce event-level parameters that are common to all events: it is equivalent to the traditional practice of ``multiplying the likelihoods of individual events''~\cite{Zimmerman:2019wzo}. Examples here would be a massive graviton and other propagation effects, or a theory-specific test that is formulated directly in terms of the theory coupling constant(s). The population model now simplifies to a Dirac delta function
\begin{equation}
 p( \pvec_i | \vec{\Lambda_0} ) = \delta (\vec{\Lambda_0}- \pvec_i)\,,
\end{equation}
where $\vec{\Lambda_0}$ is a common parameter that is shared by all events. The second approach instead corresponds to tests that introduce a \emph{new} event-level parameter \emph{for each new event}, and it is equivalent to the other traditional practice of ``multiplying the Bayes factors of individual events''~\cite{Zimmerman:2019wzo}. The Bayes Factors here compare the \ac{GR} and beyond-\ac{GR} hypothesis and are obtained by marginalizing over each event's astrophysical parameters (such as masses and spins) and (in the case of parametrized \ac{GR} tests) the beyond-\ac{GR} parameters themselves. In this case the population model has no hyperparameters and corresponds to the fixed prior used when computing the Bayes Factor
\begin{equation}
 p( \pvec_i | \vec{\Lambda_0} ) = p( \pvec_i )\,.
\end{equation}
The assumption of completely independent event-level parameters corresponds to gravity theories that introduce a new coupling constant for each event, akin to a charge. This is a fairly strong assumption, as it prevents the analysis from ``inferring'' the population prior and assumes it is fixed to what was selected during \ac{PE} inference, as described in Sec.~\ref{sec:pe}. If this assumption is violated, the analysis could result in incorrect physical conclusions~\cite{Isi:2022cii}.

Between those two limiting cases lies the situation where the beyond-\ac{GR} parameter values are not common among events, but come from a common, unknown underlying population distribution: a non-trivial population model $p( \pvec_i | \vec{\Lambda} )$ is thus required. This is the most common situation, encompassing cases such as the parametrized post-Einsteinian inspiral test~\cite{Yunes:2009ke,Chatziioannou:2012rf,Li:2011cg,Agathos:2013upa,Meidam:2017dgf,Cornish:2011ys,Sampson:2013lpa,Sampson:2013jpa}, the inspiral-merger-ringdown consistency test~\cite{Ghosh:2016qgn,Ghosh:2017gfp,Cabero:2017avf,Isi:2020tac}, parametrized ringdowns~\cite{Carullo:2019flw,Isi:2019aib,Isi:2021iql,Finch:2021qph}, and even the generic residual test~\cite{Ghonge:2020suv}. Since these tests introduce phenomenological deviations to the GW signals predicted by \ac{GR}~\cite{Johnson-McDaniel:2021yge}, defining a first-principles population model is impossible. To first order, then, analyses attempt to extract the mean and the standard deviation of the population model, with \ac{GR} predicting a zero value for both.

Then without loss of generality, the population model can be taken as a Gaussian distribution for the beyond-\ac{GR} parameters~\cite{Isi:2019asy}, motivated from the fact that the Gaussian distribution maximizes the entropy (i.e., minimizes the information) for a fixed mean and standard deviation. Moreover, it is desirable to consider the population distribution of beyond-\ac{GR} and `within-\ac{GR}' parameters (for example, component masses and spins) simultaneously as their measurement uncertainties are typically correlated~\cite{Payne:2023kwj}, similar to the discussion of Sec.~\ref{ss:populations_cosmo}.
Results from both beyond-\ac{GR}-only and joint analyses find no evidence for a nonzero mean or standard deviation~\cite{Isi:2019asy,LIGOScientific:2020tif,LIGOScientific:2021sio}, though the latter are in general more constraining as they incorporate more information about the astrophysical properties of the detected binaries~\cite{Payne:2023kwj}.
A generalization of the Gaussian population model also considers the possibility that the population consists of a mixture of two types of binary: those described within, and those beyond \ac{GR}~\cite{Saleem:2021vph}. In this case the population model is a mixture of a delta function at zero, for binaries that obey \ac{GR}, and a Gaussian, for binaries that violate it.

Finally, population constraints on beyond-\ac{GR} effects are subject to selection effects, that can be quantified and included in the analysis following the steps discussed in Sec.~\ref{subsec:sel} by simulating beyond-\ac{GR} waveforms and analyzing them with existing detection pipelines. 
In the context of parametrized tests of binary GW phase evolution, Ref.~\cite{Magee:2023muf} showed that current constraints after $\sim\!100$ detections are sufficiently constraining that the impact of selection effects is subdominant. However, this does not rule out the possibility of a ``hidden'' subpopulation of binaries with large deviations from \ac{GR}, as population constraints derived from the detected events would not be applicable if binaries originate from multiple populations.
Unmodeled burst searches~\cite{KAGRA:2021tnv} that do not make use of \ac{GR} compact binary models could alleviate the concern that binaries exhibiting strong deviations from \ac{GR} would not be detectable in the first place. Efforts to quantify these effects for specific beyond-\ac{GR} models, and to extend existing search methods, are underway, e.g.,~\cite{Chia:2020psj,Coogan:2022qxs,Narola:2022aob}.

\newpage
\section{Outlook}
\label{sec:outlook}

\subsection{Current and upcoming generations of compact binary observations}

The $\sim$\,$100$ transient \ac{GW} signals detected during the first three LIGO-Virgo observing runs are all consistent with a compact binary origin: %they correspond to 
the coalescence of two black holes, two neutron stars, or a neutron star and a black hole.
\begin{figure}[b]
 \includegraphics[width=0.9\textwidth]{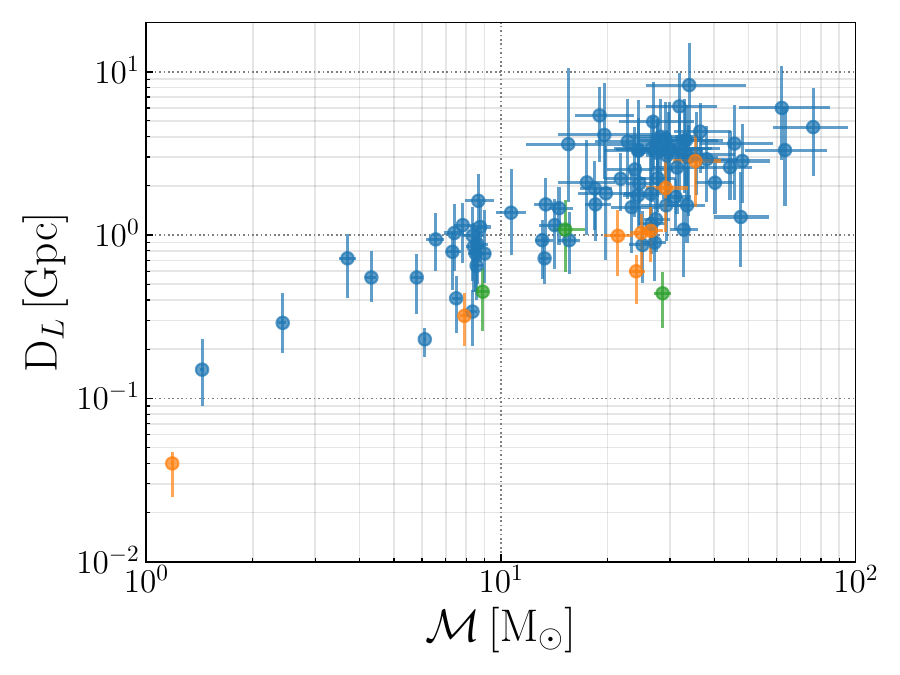}
 \caption{Source-frame chirp mass and luminosity distances of detected compact binaries during O3 (blue), O2 (orange), O1 (green).  The selection effect due to the amplitude scaling of \ac{CBC} signals for a given detector network is apparent in the ``exclusion region'' at low masses and high distances. Increasing detector sensitivity allows for detection of more distant (and potentially rare) events.}
 \label{fig:MassDistanceDetection}
\end{figure}
Figure~\ref{fig:MassDistanceDetection} summarizes some key properties of these events: their source-frame chirp mass, motivated and defined in Sec.~\ref{sec:cbc}, and their distance from Earth. 
This catalog arose from an ever-expanding network of kilometre-scale advanced
ground-based detectors operating in the $\sim$\,$10-1000$\,Hz regime which now includes, besides the twin LIGO~\cite{LIGOScientific:2014pky} detectors, the European Virgo~\cite{VIRGO:2014yos} and the Japanese cryogenic KAGRA~\cite{10.1093/ptep/ptab018} detector.

As interferometric detectors sense GW amplitude, rather than energy, a two-fold increase in sensitivity
leads approximately to an eight-fold (volumetric) increase in the detection rate. During the first and second observing runs (O1 and O2) approximately 1 compact binary coalescence was detected per month, while the third observing run (O3) brought approximately 1.5 events per week~\cite{LIGOScientific:2020ibl}, 
see %events colored by observing campaign in 
Fig.~\ref{fig:MassDistanceDetection}. Future estimates depend on the exact sensitivity achieved,
but the fourth observing run (O4) that started on May 24, 2023, scheduled to last for 20 months, is yielding 1 event per few days, which will eventually yield a catalog of hundreds.  
%[SAY SOMETHING ABOUT A-SHARP AND O5 AND LIGO-INDIA? ...]
The next steps for the current detector network include a fifth observing run starting around 2027~\cite{KAGRA:2013rdx,VIRGO:2023elp}, the addition of the LIGO-Aundha detector in India~\cite{Saleem:2021iwi}, and proposals for further LIGO improvements toward the A\# and/or Voyager~\cite{LIGO:2020xsf} configurations.
The ultimate sensitivity of the detector network is subject to upgrade plans, but it is reasonably expected to raise the total number of detections to the thousands by the end of its lifetime.
%Looking even further ahead, proposed next-generation detectors target a ten-fold sensitivity increase
%which will enable them to detect \emph{all} stellar mass black hole mergers and about half the neutron star
%mergers occurring in the Universe, see Fig.~\ref{fig:PSD}.

\subsection{Next generation challenges}
The next generation of ground-based gravitational wave observatories, planned for operation in the 2030s and beyond, are expected
to have a ten-fold improvement in sensitivity compared to the design sensitivity of the current generation~\cite{Punturo:2010zz,LIGOScientific:2016wof}: this translates naively to
a thousand-fold increase in the detection rate -- with, though, significant corrections for cosmological evolution~\cite{Chen:2017wpg}.   
Such a network would be capable of detecting nearly \emph{all} binary black hole mergers, and roughly half of binary neutron star mergers in the observable universe (see e.g.~\cite{Sachdev:2020bkk,Borhanian:2022czq}). 
In addition to the sheer number of detections, the reduction in noise level, while undoubtedly a boon for \ac{GW} astronomy, will
present novel data analysis challenges for future compact binary work. The greatly increased \ac{SNR} of the loudest signals means
that improvements will be needed both in waveform modelling, and in noise modelling, to fully exploit future observatories.

The Einstein Telescope~\cite{Punturo:2010zz} and Cosmic Explorer~\cite{CE} detector designs anticipate a greatly reduced low frequency noise
`wall', allowing the early inspiral at frequencies of a few Hz to be observable in the data; thus, the lightest \ac{BNS} signals will be visible for many hours and millions of signal cycles. 
With a possible detection rate of $10^3$ events per day, signals will inevitably overlap in the sensitive band of the detectors,
violating the usual model of a single signal buried in detector noise. Fortunately, unlike in the case of overlapping white dwarf binaries
in LISA, the rapidly evolving CBC signal will not result in true confusion noise~\cite{Johnson:2024foj}, but parameter inference can be affected when the coalescence times of two signals are within $\sim 0.1$\,s of each other~\cite{Samajdar:2021egv,Pizzati:2021apa,Relton:2021cax,Johnson:2024foj}. 
%except in the rare cases when the coalescence times of two signals differ by less than $\sim 0.1$\,s~\cite{Samajdar:2021egv,Pizzati:2021apa,Relton:2021cax,Johnson:2024foj}
%\kc{These studies do not say that you have ``confusion noise" when the signals are within $\sim 0.1$\,s of each other (you cannot have confusion noise from 2 signals). They say that you can have parameter biases when the signals merge closely. Suggested rephrase: the rapidly evolving CBC signals will not result in true Gaussian confusion noise~ . 
The residual power in overlapping signals may produce weak systematic biases in parameter estimation that could affect constraints on environmental and non-GR effects~\cite{Antonelli:2021vwg,Hu:2022bji}, and, unless overlapping signals are subtracted, can also impact detection efficiency~\cite{Wu:2022pyg}.
% by appearing as excess/unmodeled power. 
%kc{can we replace noise with ``unmodeled power" or something like that? We do not have enough signals to get ``noise" in the CLT sense.}.
% TD - does 'residual power' - whatever that means - already cover it? 

The greatly increased population of detections, and therefore the amount of available data, will inflate the computational cost
of inference for hierarchical models which marginalise over individual event uncertainties, as explained in \ref{sec:hier}. 
%\kc{cite Essick+Farr and Talbot+Golomb?}. 
More complicated population models will probably also be required to explain, or even just accurately represent, as-yet-unknown structures in the astrophysical distribution of sources. Non-parametric methods, discussed in Sec.~\ref{ss:populations_cosmo}, then offer 
a potentially useful means of modelling structure without %resorting to 
specific assumptions for the functional form of distributions,  
as do classical statistical methods valid in the limit of high event count, e.g.\ density estimation~\cite{Sadiq:2023zee}.

For high-\ac{SNR} signals, systematic errors in the waveform models themselves will become even more important~\cite{Purrer:2019jcp}: this motivates the further improvement of such models, informed both by \ac{NR} simulations and advances in analytic methods, as well as methods to mitigate remaining waveform
errors and uncertainties within the analysis stage~\cite{Moore:2015sza,Williams:2019vub}.  The future development of modelling and analysis tools to address these challenges will uncover a range of possible new observables, including deviations from general relativity~\cite{Pang:2018hjb,Moore:2021eok}, and effects of the source's environment
on the signal -- for instance, the presence of a third body in a coalescing system may produce an observable net acceleration along the line of sight (e.g.~\cite{Inayoshi:2017hgw}). 
%\jv{[CITES - more environmental and non-GR effects?] }
Louder signals will also require yet more accurate modeling of subdominant waveform effects, such as higher mode content, orbital precession, and orbital eccentricity, e.g.~\cite{Ramos-Buades:2021adz,Nagar:2021gss}.
The greater distance reach of future networks will reveal sources at higher redshifts (up to $z\sim 10$), which, along with improved low-frequency sensitivity, will
result in detection of many more signals with chirp masses $\gtrsim$\,$\mathcal{O}(100)\,\Msun$, whether from intermediate mass %black hole binaries
or redshifted stellar-mass black hole binaries.  Such massive mergers will enable detailed investigation of the merger and ringdown portions of the signal~\cite{Berti:2005ys,Isi:2021iql}.

The greatly increased length of low-mass signals will expand the quantity of data covered by 
%that can be usefully included in 
both search and parameter estimation analyses.
Thanks to the chirping nature of the \ac{CBC} signal, information does not arrive uniformly throughout the signal: to limit computing cost, one may thus use a reduced
data rate in the early inspiral for a multi-band analysis~\cite{Morisaki:2021ngj}, or use a reference template to perform heterodyning %/ relative binning 
of the data, reducing its effective bandwidth~\cite{Cornish:2021lje,Zackay:2018qdy}. 
Long signal durations also offer the ability to provide early warning detection and pre-merger localisation of the source, to enable electromagnetic observations from the moment of coalescence onward (see \cite{Magee:2021xdx,Nitz:2020vym} and discussion in Sec.~\ref{ss:low_latency}); realization of this goal requires sufficiently rapid localization,
and an associated low latency infrastructure to distribute pre-merger alerts~\cite{Hu:2023hos}.

Machine learning methods offer an avenue for performing extremely rapid inference which has been successfully demonstrated
for signal durations of a few seconds, typical of \ac{BBH} observations in the current detector network~(see Sec.~\ref{sss:pe-ml}). 
These methods typically use the data stream as conditioning information
for a neural network; however, as the data volume increases, the amount of information per datapoint will be reduced, motivating the development of compressed representations of the data that can efficiently encode the signal information.  Large-scale use of advanced machine learning inference in production will also require more standardized and dedicated platforms to make efficient use of available processors, while retaining sufficient reliability and flexibility/scalability (e.g.\ Inference-as-a-Service~\cite{Gunny_Nature}).

As a fundamental consideration, any method of identifying a modelled source, whether based on ML or a classical template bank to cover the signal space,
must contend with the fact that the number of independent templates will explode with longer signal durations, as seen from
the fact that the fractional precision of chirp mass measurement scales inversely with the number of observed signal cycles. The presence of higher modes,
precession and eccentricity also increase the number of independent templates that must be considered by an analysis
This overall increase in our ability to resolve, and extract astrophysical information from, 
the complexity of \ac{CBC} signals, implies that analysis methods are sure to continue 
development over the coming years. 

It is not only \ac{CBC} signals that will pose increasing challenges in a next-generation network: the characterization of detector noise will likely require more complex and sophisticated methods. 
We cannot predict what types of non-ideal behaviour will occur in detectors whose locations are not yet decided, but with their broad-band sensitivity pushing the limits from many disparate types of noise in different frequency ranges, there are more possible sources of glitches or non-stationary disturbance.  Although such effects, for instance Newtonian noise (reviewed in~\cite{Harms:2019dqi,Trozzo:2022tar}), are not limiting on our current observations, with increased sensitivity weaker glitches may become visible above the stationary noise floor.  Also in general, the longer signal durations implied by larger bandwidths imply a higher probability that a glitch will intersect with any given signal, placing an increasing emphasis on modelling and subtraction of artefacts.
Finally, the persistent presence of signals in the data stream suggests that the detector noise properties will need to be estimated concurrently with the astrophysical signals~\cite{Plunkett:2022zmx}, as no ``offsource" data might exist.  Such considerations will become yet more pertinent with the advent of \ac{GW} observatories in space, and the prospect of joint observations between Earth- and space-based interferometers~\cite{LISA:2017pwj,Sesana:2016ljz,Gerosa:2019dbe,Klein:2022rbf} -- pointing to a yet broader potential science scope, which will go hand in hand with these increased technical challenges.

\subsection{Conclusion}
The field of gravitational wave astronomy as an observational science has exploded in recent years, with compact binaries now being observed on a daily basis. Detecting and analysing \ac{CBC} signals is the foundation of performing astrophysics with this wealth of new observations, and \acp{CBC} will continue to dominate the sources observed in future detectors.

Adjacent areas such as detector development\cite{Pitkin:2011yk,Freise:2009sf,KAGRA:2013rdx}, binary astrophysics~\cite{Faber:2012rw,Mandel:2021smh}, theoretical understanding of \ac{CBC} sources~\cite{Futamase:2007zz,Blanchet:2013haa,Cardoso:2014uka,Bishop:2016lgv,Sasaki:2003xr}, and their potential electromagnetic counterparts~\cite{Metzger:2019zeh,Kyutoku:2021icp}, strongly interact with the observation and analysis of \acp{CBC}, and deserve their own reviews. However, in this article we have aimed to describe the foundations of practical \ac{CBC} analysis, from the modelling of the signals, through detection and parameter estimation, to performing population inference with a large number of signals. We hope the reader finds this comprehensive review useful as an entry point into the field, or as a reference for advanced research addressing the challenges that the future will hold.

\begin{acknowledgements}
The authors acknowledge many helpful and enlightening interactions with members of the LSC, Virgo and KAGRA Compact Binary Coalescence working group on the topics of this work, as well as with the wider community; we thank Jolien Creighton, Fr{\'e}d{\'e}rique Marion, and Ben Farr for comments on a previous version of the manuscript. 
% NSF boilerplate is included within the GWOSC block at the end. 
%This material is based upon work supported by NSF's LIGO Laboratory which is a major facility fully funded by the National Science Foundation. 
%
K.C.~was supported by NSF Grant PHY-2110111 and the Department of Energy under award number DE-SC0023101. 
This work has received financial support from the Xunta de Galicia (CIGUS Network of research centers), the European Union, and the ``Mar{\'i}a de Maeztu'' Units of Excellence program CEX2020-001035-M.
TD is supported by research grant PID2020-118635GB-I00 from the Spanish Ministerio de Ciencia e Innovaci{\'o}n.
MF acknowledges support from the Natural Sciences and Engineering Research Council of Canada (NSERC) RGPIN-2023-05511.
VR was supported by the UK Science and Technology Facilities Council grant ST/V005618/1.
JV was supported by the UK Science and Technology Facilities Council grant ST/V005634/1.

This research has made use of data or software obtained from the Gravitational Wave Open Science Center (gwosc.org), a service of the LIGO Scientific Collaboration, the Virgo Collaboration, and KAGRA. This material is based upon work supported by NSF's LIGO Laboratory which is a major facility fully funded by the National Science Foundation, as well as the Science and Technology Facilities Council (STFC) of the United Kingdom, the Max-Planck-Society (MPS), and the State of Niedersachsen/Germany for support of the construction of Advanced LIGO and construction and operation of the GEO600 detector. Additional support for Advanced LIGO was provided by the Australian Research Council. Virgo is funded, through the European Gravitational Observatory (EGO), by the French Centre National de Recherche Scientifique (CNRS), the Italian Istituto Nazionale di Fisica Nucleare (INFN) and the Dutch Nikhef, with contributions by institutions from Belgium, Germany, Greece, Hungary, Ireland, Japan, Monaco, Poland, Portugal, Spain. KAGRA is supported by Ministry of Education, Culture, Sports, Science and Technology (MEXT), Japan Society for the Promotion of Science (JSPS) in Japan; National Research Foundation (NRF) and Ministry of Science and ICT (MSIT) in Korea; Academia Sinica (AS) and National Science and Technology Council (NSTC) in Taiwan.
\end{acknowledgements}

\newpage 
\bibliographystyle{spmpsci}      % mathematics and physical sciences
\bibliography{references.bib}   % name your BibTeX data base

\end{document}